\documentclass[ejs]{imsart}

\doi{10.1214/154957804100000000}
\pubyear{0000}
\volume{0}
\firstpage{0}
\lastpage{0}

\usepackage{amssymb, amsmath, mathrsfs, amsfonts, graphicx, subfigure, psfrag, natbib}
\usepackage{adjustbox}
\RequirePackage[colorlinks, citecolor=blue, urlcolor=blue, breaklinks=true]{hyperref}

\numberwithin{equation}{section}

\newtheorem{them}{Theorem}[section]

\newtheorem{definition}{Definition}[section]

\usepackage{epsfig}
\usepackage{epstopdf}
\usepackage{url}
\usepackage[usenames, dvipsnames]{color}

\newcommand{\calM}{\mathcal{M}}

\newcommand{\ISE}{\mathrm{ISE}}

\newcommand{\bh}{\mathbf{h}}

\newcommand{\EISE}{\mathrm{EISE}}
\newcommand{\ds}{\displaystyle}

\begin{document}

\begin{frontmatter}

\title{Conditional density estimation with covariate measurement error}
\runtitle{Density estimation with covariate measurement error}

\author{\fnms{Xianzheng} \snm{Huang} \corref{} \ead[label=e1]{huang@stat.sc.edu}}
\address{Department of Statistics, University of South Carolina, Columbia, SC 29208, USA \\ \printead{e1}}
\and
\author{\fnms{Haiming} \snm{Zhou} \ead[label=e2]{zhouh@niu.edu}}
\address{Division of Statistics, Northern Illinois University, DeKalb, IL 60115, USA \\ \printead{e2}}

\runauthor{X. Huang and H. Zhou}

\begin{abstract}
We consider estimating the density of a response conditioning on an error-prone covariate. Motivated by two existing kernel density estimators in the absence of covariate measurement error, we propose a method to correct the existing estimators for measurement error. Asymptotic properties of the resultant estimators under different types of measurement error distributions are derived. Moreover, we adjust bandwidths readily available from existing bandwidth selection methods developed for error-free data to obtain bandwidths for the new estimators. Extensive simulation studies are carried out to compare the proposed estimators with naive estimators that ignore measurement error, which also provide empirical evidence for the effectiveness of the proposed bandwidth selection methods. A real-life data example is used to illustrate implementation of these methods under practical scenarios. An R package, \texttt{lpme}, is developed for implementing all considered methods, which we demonstrate via an R code example in Appendix H. 
\end{abstract}

\begin{keyword}[class=MSC]
\kwd[Primary ]{62G08}
\kwd[; secondary ]{62G20}
\end{keyword}

\begin{keyword}
\kwd{Bandwidth}
\kwd{bias}
\kwd{cross validation}
\kwd{deconvoluting kernel}
\end{keyword}



\end{frontmatter}

\section{Introduction}
\label{s:intro}
The conditional density of a continuous response $Y$ given a covariate $X$, denoted by $p(y|x)$, provides a complete picture of the association between $Y$ and $X$ that is valuable for data visualization and exploration. \citet{rosenblatt1969conditional} is one of the pioneers who considered kernel density estimators for $p(y|x)$. \citet{Hyndman.etal1996} further studied properties of the kernel density estimator based on a random sample, $\{(X_j, Y_j)\}_{j=1}^n$, given by
\begin{equation}\label{eq:condden1}
\hat p_1(y|x) = \frac{\ds\frac{1}{nh_1 h_2} \sum_{j=1}^n K_1\left(\frac{X_j-x}{h_1}\right)K_2\left(\frac{Y_j-y}{h_2}\right)}{\ds\frac{1}{nh_1} \sum_{j=1}^n K_1\left(\frac{X_j-x}{h_1}\right)}, 
\end{equation}
where $K_1(t)$ and $K_2(t)$ are kernels, $h_1$ and $h_2$ are bandwidths. This estimator originates from two other well studied density estimators. The denominator of (\ref{eq:condden1}) is the kernel density estimator for the probability density function (pdf) of $X$, denoted by $f_{\hbox {\tiny $X$}}(x)$; and the numerator is the kernel density estimator for the joint pdf of $(X, Y)$, denoted by $p(x, y)$. \citet{Fan.etal1996} followed the idea of local polynomial estimation of a mean function \citep[][Chapter 3]{Fan&Gijbels1996} to construct a class of local polynomial estimators for $p(y|x)$. The estimator $\hat p_1(y|x)$ in (\ref{eq:condden1}) belongs to this class, referred to as the local constant estimator. \citet{Hyndman.Yao2002} revised the local polynomial estimators to guarantee non-negativity.  Besides $\hat p_1(y|x)$, \citet{Hyndman.etal1996} proposed another estimator for $p(y|x)$ with a different estimator for $p(x,y)$ in the numerator, leading to 
\begin{equation}\label{eq:condden2}
\hat{p}_2(y|x) = \frac{\ds\frac{1}{nh_1 h_2} \sum_{j=1}^n K_1\left(\frac{X_j-x}{h_1}\right)K_2\left\{\frac{Y_j-\hat{m}(X_j)-y+\hat{m}(x)}{h_2}\right\}}{\ds\frac{1}{nh_1} \sum_{j=1}^n K_1\left(\frac{X_j-x}{h_1}\right)},
\end{equation}
where $\hat m(x)$ is an estimator for $m(x)=E(Y|X=x)$, such as a local polynomial estimator. If one replaces $\hat m(\cdot)$ with $m(\cdot)$ in (\ref{eq:condden2}), one obtains the regular kernel density estimator for the density of $e=Y-m(X)$ given $X$, denoted by $f_{e|\hbox {\tiny $X$}}(e|x)$, which relates to $p(y|x)$ via $p(y|x)=f_{e|\hbox {\tiny $X$}}\{y-m(x)|x\}$. This relationship motivates the construction of $\hat p_2(y|x)$ in (\ref{eq:condden2}). \citet{Hyndman.etal1996} showed that $\hat p_2(y|x)$ has a smaller asymptotic mean integrated squared error (MISE) when compared with $\hat p_1(y|x)$ under some situations commonly encountered in practice. \citet{hansen2004nonparametric} studied $\hat p_2(y|x)$ more closely, who referred to $\hat p_2(y|x)$ as a two-step estimator to stress the estimation of $m(x)$ that is not needed for $\hat p_1(y|x)$, a one-step estimator in contrast. 

It is common in practice that a covariate of interest cannot be measured directly or precisely. This motivates our work presented in this article, where we aim to estimate $p(y|x)$ when $X$ is prone to measurement error. Due to error contamination, the observed data are $\{(W_j, Y_j)\}_{j=1}^n$ as opposed to $\{(X_j, Y_j)\}_{j=1}^n$, where $W_j$ is an unbiased surrogate of $X_j$, for $j=1, \ldots, n$. We assume in this study 
a classical additive measurement error model \citep[][Section 1.2]{Carroll&etal2006} that relates the observed covariate $W$ and the true covariate $X$ via
\begin{equation}
W_j=X_j+U_j, \label{eq:memodel}
\end{equation}
where $U_j$ represents measurement error with mean zero and variance $\sigma^2_u$, following a distribution specified by the pdf $f_{\hbox {\tiny $U$}}(u)$, and is independent of $(X_j, Y_j)$, for $j=1, \ldots, n$. For reasons related to identifiability issues, we assume $f_{\hbox {\tiny $U$}}(u)$ known in the majority of the study, and discuss treatments for unknown error distribution in Section~\ref{s:real}. \citet{robins1995semiparametric} considered estimating unknown parameters in $p(y|x)$ that belongs to a pre-specified parametric family when covariates are missing or measured with error. We are not aware of existing works on estimating $p(y|x)$ nonparametrically in the presence of measurement error. This article presents solutions to this fundamentally important problem, supplemented with an R package \texttt{lpme} \citep{zhouR2017} for easy implementation of the proposed methods.

Using the error contaminated data in $\hat p_1(y|x)$ and $\hat p_2(y|x)$ leads to two naive estimators for $p(y|x)$ that ignore covariate measurement error, 
\begin{align}
\tilde{p}_1(y|x) & =  \frac{\ds\frac{1}{nh_1 h_2} \sum_{j=1}^n K_1\left(\frac{W_j-x}{h_1}\right)K_2\left(\frac{Y_j-y}{h_2}\right)}{\ds\frac{1}{nh_1} \sum_{j=1}^n K_1\left(\frac{W_j-x}{h_1}\right)}, \label{eq:naive1} \\
\tilde{p}_2(y|x) & =  \frac{\ds\frac{1}{nh_1 h_2} \sum_{j=1}^n K_1\left(\frac{W_j-x}{h_1}\right)K_2\left\{\frac{Y_j-\hat{m}^*(W_j)-y+\hat{m}^*(x)}{h_2}\right\}}{\ds\frac{1}{nh_1} \sum_{j=1}^n K_1\left(\frac{W_j-x}{h_1}\right)}, \label{eq:naive2}
\end{align}
where $\hat {m}^*(x)$ is an estimator for $m^*(x)=E(Y|W=x)$. These naive estimators are sensible estimators for the conditional density of $Y$ given $W=x$, denoted by $p^*(y|x)$, but are usually inadequate estimators for $p(y|x)$. 

In Section~\ref{s:methods}, we correct the above naive estimators for measurement error, producing two non-naive estimators for $p(y|x)$. Asymptotic properties of the proposed estimators are presented in Section~\ref{s:theory}. In Section~\ref{s:band} we develop methods for selecting bandwidths involved in these estimators. Finite sample performance of these estimators are demonstrated in comparison with the two naive estimators in simulation studies in Section~\ref{s:simulation}. Practical considerations for implementing the proposed methods are discussed in Section~\ref{s:real}, where we entertain a real-life data example. Lastly, in Section~\ref{s:discussion}, we summarize the contribution of our work and discuss future research directions. 

\section{Proposed estimators}
\label{s:methods}
\subsection{The rationale}
Denote by $p^*(x, y)$ the joint density of $(W, Y)$ evaluated at $(x, y)$. Given the measurement error model in (\ref{eq:memodel}), one can show that $p^*(x, y)$ is equal to the convolution of $f_{\hbox {\tiny $U$}}(u)$ and $p(x, y)$ with respect to the first argument, that is, 
\begin{equation}
p^*(x, y)=\int p(v, y) f_{\hbox {\tiny $U$}}(x-v)dv =\{p(\cdot, y)*f_{\hbox {\tiny $U$}}\}(x).
\label{eq:twojoint}
\end{equation}
where ``$*$" in the last expression is the convolution operator. The range of integration in all integrals in this article is the entire real line, unless specified otherwise. Denote by $\phi_g(t)$ the Fourier transform of a function $g(\cdot)$ or the characteristic function of a random variable $g$. Applying Fourier transform on both sides of (\ref{eq:twojoint}) yields $\phi_{p^*(\cdot, y)} (t) = \phi_{p(\cdot, y)}(t) \phi_{\hbox {\tiny $U$}}(t),$ which is equivalent to $\phi_{p(\cdot, y)}(t) =\phi_{p^*(\cdot, y)} (t)/\phi_{\hbox {\tiny $U$}}(t)$, assuming $\phi_{\hbox {\tiny $U$}}(t)\ne 0$ for all $t$. Applying inverse Fourier transform on both sides of the preceding identity gives
\begin{equation}
p(y|x) f_{\hbox {\tiny $X$}}(x) = \frac{1}{2\pi} \int e^{-itx} \frac{\phi_{p^*(\cdot, y)}(t)}{\phi_{\hbox {\tiny $U$}}(t)} dt,
\label{eq:connection}
\end{equation}
where $i$ is the imaginary unit. 

Putting a ``hat" on top of each unknown quantity in (\ref{eq:connection}) to represent an estimator for this quantity, we obtain a general form of estimators for $p(y|x)$ that account for covariate measurement error, 
\begin{equation}
\hat p(y|x) = \hat f^{-1}_{\hbox {\tiny $X$}}(x)\cdot\frac{1}{2\pi} \int e^{-itx} \frac{\phi_{\hat p^*(\cdot, y)}(t)}{\phi_{\hbox {\tiny $U$}}(t)} dt. 
\label{eq:linkest}
\end{equation}
With $\hat p(x, y)=\hat p(y|x)\hat f_{\hbox {\tiny $X$}}(x)$ being an estimator for $p(x, y)$, (\ref{eq:linkest}) relates $\hat p(x, y)$ to $\hat p^*(x, y)$, which is a naive estimator for $p(x, y)$ that is suitable for estimating $p^*(x,y)$. The numerators in $\tilde p_1(y|x)$ and $\tilde p_2(y|x)$ are examples of $\hat p^*(x, y)$. Even though, by construction, the integral in (\ref{eq:connection}) is real as long as all integrals leading to (\ref{eq:connection}) are well defined, the integral in (\ref{eq:linkest}) can be complex with $p^*(\cdot, y)$ now replaced by $\hat p^*(\cdot, y)$. A sensible treatment when the right-hand side of (\ref{eq:linkest}) returns a complex quantity is to use the real part as an estimator of $p(y|x)$, and argue that the imaginary part is merely a consistent estimator of zero by showing that (\ref{eq:linkest}) is a consistent estimator of the real-valued $p(y|x)$.  

As for the estimator for $f_{\hbox {\tiny $X$}}(x)$ in (\ref{eq:linkest}), we adopt the deconvoluting density estimator \citep{carroll1988optimal, Stefanski&Raymond1990},
\begin{equation}
\hat f_{\hbox {\tiny $X$}}(x)=\ds\frac{1}{nh_1} \sum_{j=1}^n K^*_1\left(\frac{W_j-x}{h_1}\right),
\label{eq:fxhat}
\end{equation}
where 
\begin{equation}
K^*_1(t)=\frac{1}{2\pi}\int e^{-its}\frac{\phi_{\hbox {\tiny $K_1$}}(s)}{\phi_{\hbox {\tiny $U$}}(-s/h_1)}ds 
\label{eq:K1*}
\end{equation}
is referred to as the deconvoluting kernel. Under conditions (K1) given in Section~\ref{s:conditions}, \citet{Stefanski&Raymond1990} showed that 
\begin{equation}
E\left\{K_1^*\left(\left.\frac{W-x}{h_1}\right)\right| X\right\}=K_1\left(\frac{X-x}{h_1}\right),
\label{eq:K1*K1}
\end{equation}
suggesting that $\hat f_{\hbox {\tiny $X$}}(x)$ has the same bias as the ordinary kernel density estimator for $f_{\hbox {\tiny $X$}}(x)$ that appears as the common denominator of $\hat p_1(y|x)$ and $\hat p_2(y|x)$ in (\ref{eq:condden1}) and (\ref{eq:condden2}).

\subsection{Two estimators accounting for measurement error}
\label{s:p3p4}
Using the numerator of $\tilde p_1(y|x)$ as $\hat p^*(x, y)$ in (\ref{eq:linkest}), one can show via straightforward algebra that (\ref{eq:linkest}) reduces to 
\begin{equation}\label{eq:est3}
\hat{p}_3(y|x) = \frac{\ds\frac{1}{nh_1 h_2} \sum_{j=1}^n K^*_1\left(\frac{W_j-x}{h_1}\right)K_2\left(\frac{Y_j-y}{h_2}\right)}{\ds\frac{1}{nh_1} \sum_{j=1}^n K^*_1\left(\frac{W_j-x}{h_1}\right)}.
\end{equation}
Looking back at its naive counterpart, $\tilde p_1(y|x)$, makes the construction of $\hat{p}_3(y|x)$ in (\ref{eq:est3}) transparent. Since the naive estimator $\tilde p_1(y|x)$ depends on $W_j$ only via $K_1\{(W_j-x)/h_1\}$, (\ref{eq:K1*K1}) suggests that replacing $K_1\{(W_j-x)/h_1\}$ with $K^*_1\{(W_j-x)/h_1\}$, for $j=1, \ldots, n$, suffices to correct $\tilde p_1(y|x)$ for measurement error. This substitution yields $\hat p_3(y|x)$.

To correct $\tilde p_2(y|x)$ for measurement error is more involved because it depends on $\{W_j\}_{j=1}^n$ in a more complicated way than $\tilde p_1(y|x)$ does, and the trick of replacing the regular kernel with a deconvoluting kernel that leads to (\ref{eq:est3}) (and also (\ref{eq:fxhat})) does not work here. Indeed, if one sets $\hat p^*(x, y)$ in (\ref{eq:linkest}) as the numerator of $\tilde p_2(y|x)$, an estimator for $p^*(x,y)$ denoted by $\tilde p_2(x, y)$, one obtains an estimator for $p(y|x)$ given by 
\begin{equation}
\hat p_4(y|x) = \hat f^{-1}_{\hbox {\tiny $X$}}(x)\cdot\frac{1}{2\pi} \int e^{-itx} \frac{\phi_{\tilde p_2(\cdot, y)}(t)}{\phi_{\hbox {\tiny $U$}}(t)} dt,
\label{eq:p4}
\end{equation}
which cannot be further simplified. For concreteness, in the majority of our study, we use the local linear estimator for $m^*(x)$ as $\hat m^*(x)$ in $\tilde p_2(x, y)$, with kernel $K_3(t)$ and bandwidth $h_3$. Considerations of other estimators for $m^*(x)$ are discussed in Sections~\ref{s:simulation} and \ref{s:real}.

In the absence of measurement error, \citet{Hyndman.etal1996} showed that, under certain conditions (to be presented in Section~\ref{s:theorems4}), the two-step estimator $\hat p_2(y|x)$ often has a lower MISE than the one-step estimator $\hat p_1(y|x)$. In the presence of measurement error, we show next that, after correcting $\tilde p_2(y|x)$ and $\tilde p_1(y|x)$ for measurement error, the comparison between $\hat p_4(y|x)$ and $\hat p_3(y|x)$ becomes more involved, but $\hat p_4(y|x)$ still improves over $\hat p_3(y|x)$ under similar conditions. 

\section{Asymptotic properties}
\label{s:theory}
\subsection{Preamble}
\label{s:conditions}
We study properties of $\hat p_3(y|x)$ and $\hat p_4(y|x)$ under two types of measurement error distributions, namely ordinary smooth distributions and super smooth distributions \citep{fan1991asymptotic, fan1991global, fan1991optimal}. Their definitions are given next. 
\begin{definition}
\label{def:ordsm}
The distribution of $U$ is ordinary smooth of order $b$ if
$$\lim_{t\to +\infty} \left|t^b \phi_{\hbox {\tiny $U$}}(t)\right|=c \textrm{ and } \lim_{t\to +\infty} \left|t^{b+1} \phi'_{\hbox {\tiny $U$}}(t)\right|=cb$$ 
for some positive constants $b$ and $c$. 
\end{definition}
\begin{definition}
\label{def:supsm}
The distribution of $U$ is super smooth of order $b$ if
$$d_0|t|^{b_0}\exp(-|t|^b/d_2)\le |\phi_{\hbox {\tiny $U$}}(t)| \le d_1|t|^{b_1}\exp(-|t|^b/d_2),  \textrm { as $|t|\to \infty$,}$$ 
for some positive constants $b$, $b_0$, $b_1$, $d_0$, $d_1$, and $d_2$.
\end{definition}

Laplace and gamma distributions are examples of ordinary smooth distributions. Normal and Cauchy distributions are super smooth, for instance. Technical conditions imposed on different functions for the study of asymptotics are listed below. The first set of conditions are solely regarding measurement error $U$. \\

\noindent
{\bf Conditions U:} 
\begin{itemize}
\item[](U1) For all $t$, $\phi_{\hbox {\tiny $U$}}(t) \ne 0$.
\item[](U2) $|\phi'_{\hbox {\tiny $U$}}(t)|_{\infty} < \infty$.
\end{itemize}
Condition (U1) is needed to reach (\ref{eq:connection}) and for the validity of $K_1^*(t)$ in (\ref{eq:K1*}). Condition (U2) is imposed to guarantee finite variance for $\hat f_{\hbox {\tiny $X$}}(x)$ when $U$ is ordinary smooth, which is a condition that can be relaxed when $U$ is super smooth due to a stronger condition on $K_1(t)$ given in (K5) included in the following set of conditions.\\

\noindent 
{\bf Conditions K:}
\begin{itemize}
\item[](K1) $|\phi_{\hbox {\tiny $K_1$}}(t)/\phi_{\hbox {\tiny $U$}}(-t/h_1)|_\infty<\infty$,   $\int |\phi_{\hbox {\tiny $K_1$}}(t)/\phi_{\hbox {\tiny $U$}}(-t/h_1)|dt<\infty$. 
\item[](K2) $|\phi_{\hbox {\tiny $K_1$}}(t)|_\infty <\infty$ and $|\phi'_{\hbox {\tiny $K_1$}}(t)|_\infty <\infty$.
\item[](K3) $\int |t|^b |\phi_{\hbox {\tiny $K_1$}}(t)|dt< \infty$, $\int |t|^{2b} |\phi_{\hbox {\tiny $K_1$}}(t)|^2dt< \infty$.
\item[](K4) $\int (|t|^b+|t|^{b-1})( |\phi'_{\hbox {\tiny $K_1$}}(t)|+|\phi_{\hbox {\tiny $K_1$}}(t)|)dt< \infty$. 
\item[](K5) The support of $\phi_{\hbox {\tiny $K_1$}}(t)$ is $[-1, 1]$. 
\end{itemize}
Condition (K1) is needed to establish (\ref{eq:K1*K1}). Conditions (K2)--(K4) are regularity conditions for the variance of $\hat p_3(y|x)$ to exist when $U$ is ordinary smooth, whereas only (K5) is needed for this purpose when $U$ is super smooth. Besides conditions on $K_1(t)$ stated above, we choose all three kernels, $K_1(t)$, $K_2(t)$, and $K_3(t)$, to be real and even functions with finite second moments. When $\hat p_4(y|x)$ is concerned, once Conditions K are imposed on $K_1(t)$, intuition suggests that having a bounded $K_2(t)$ should suffice to guarantee finite first two moments for $\hat p_4(y|x)$ although one should exercise care in formulating more concrete conditions relating to $K_2(t)$. We will come back to this point with more discussions in Section~\ref{s:theorems4}. For numerical stability and simplicity, we choose both $K_1(t)$ and $K_2(t)$ to be the same kernel in $\hat p_4(y|x)$ as done in \citet{masry1993strong} for instance.  Lastly, it is assumed that $f_{\hbox {\tiny $X$}}(x)$ does not vanish over the support of $X$, and it is twice differentiable.  

\subsection{Properties of the one-step estimator $\hat p_3(y|x)$}
\label{s:theorems3}
We derive in Appendix A the asymptotic bias and variance of $\hat p_3(y|x)$, summarized in the following theorem.
\begin{them}
\label{thm:bias3}
When $U$ is ordinary smooth of order $b$, if $nh_1^{1+2b}h_2(h_1^2+h_2^2)^2 \to \infty$ as $n\to \infty$, $h_1, h_2 \to 0$, then 
\begin{align}
& \hat p_3(y|x)-p(y|x) \nonumber \\
= & \textrm{DB}_3(x, y, h_1, h_2)+O(h_1^4)+O(h_2^4)+O(h_1^2h_2^2)+O_p\left(\frac{1}{\sqrt{nh_1^{1+2b}h_2}}\right), \label{eq:bias3ord}
\end{align}
where 
\begin{align}
\textrm{DB}_3(x, y, h_1, h_2)=\frac{1}{2f_{\hbox {\tiny $X$}}(x)} \left[\left\{p_{xx}(x,y)-p(y|x)f''_{\hbox {\tiny $X$}}(x) \right\}\mu_{2,1}h_1^2 + p_{yy}(x,y)\mu_{2,2}h_2^2\right] \label{eq:db3}
\end{align}
is the dominating bias, in which $f''_{\hbox {\tiny $X$}}(x)$ is the second derivative of $f_{\hbox {\tiny $X$}}(x)$, $p_{xx}(x,y)=(\partial^2/\partial x^2) p(x, y)$, $p_{yy}(x,y)=(\partial^2/\partial y^2) p(x, y)$, and $\mu_{2,\ell}=\int t^2 K_\ell(t) dt$, for $\ell=1, 2$. 
When $U$ is super smooth of order $b$, if $nh_1^{1-2b_2}h_2 \exp(-2h_1^{-b}/d_2)(h_1^2+h_2^2)^2\to \infty$ as $n\to \infty$, $h_1, h_2\to 0$, then   
\begin{align}
 & \hat p_3(y|x)-p(y|x) \nonumber\\
= &  \textrm{DB}_3(x, y, h_1, h_2)+O(h_1^4)+O(h_2^4)+O(h_1^2h_2^2)+O_p\left\{\frac{\exp(h_1^{-b}/d_2)}{\sqrt{n h_1^{1-2b_2}h_2}}\right\},
\label{eq:bias3sup}
\end{align}
where $b_2=b_0I(b_0<0.5)$.
\end{them}

In contrast to $\hat p_3(y|x)$,  \citet[][Section 3.1]{Hyndman.etal1996} obtained the following result for the error-free counterpart estimator $\hat p_1(y|x)$,
\begin{align}
 & \hat p_1(y|x)-p(y|x) \nonumber\\
= & \frac{1}{2} \left[ \left\{\frac{\partial^2 p(y|x)}{\partial x^2}+2 \frac{\partial p(y|x)}{\partial x}\frac{f'_{\hbox {\tiny $X$}}(x)}{f_{\hbox {\tiny $X$}}(x)}  \right\}\mu_{2,1}h_1^2 +\frac{\partial^2 p(y|x)}{\partial y^2}\mu_{2,2}h_2^2 \right] \nonumber \\
& +O(h_1^4)+O(h_2^4)+O(h_1^2h_2^2)+O_p\left(\frac{1}{\sqrt{nh_1h_2}}\right), \label{eq:p1bias}
\end{align}
where $f'_{\hbox {\tiny $X$}}(x)$ is the first derivative of $f_{\hbox {\tiny $X$}}(x)$. Straightforward algebra reveal that the dominating bias in (\ref{eq:p1bias}) is equal to the dominating bias of $\hat p_3(y|x)$ given in (\ref{eq:db3}). Although exhibiting same asymptotic bias, the asymptotic variance of $\hat p_3(y|x)$ is inflated due to measurement error when compared to $\hat p_1(y|x)$, with more substantial inflation when $U$ is super smooth than when it is ordinary smooth.

\subsection{Properties of the two-step estimator $\hat p_4(y|x)$}
\label{s:theorems4}
For a generic bivariate function, $g(w, y)$, we define the following double integral transform of $g(w, y)$ via the operator $\mathscr{T}_x(\cdot)$, assuming $|\phi_{g(\cdot, y)}(t)/\phi_{\hbox {\tiny $U$}}(t)|_\infty<\infty$ and $\int|\phi_{g(\cdot, y)}(t)/\phi_{\hbox {\tiny $U$}}(t)|dt<\infty$ for each $y$, 
\begin{equation}
\mathscr{T}_x\left\{g(\cdot, y)\right\}= \frac{1}{2\pi}\int e^{-itx}\frac{ \phi_{g(\cdot, y)}(t)}{\phi_{\hbox {\tiny $U$}}(t)}dt. \label{eq:Tx}
\end{equation}

In Appendix B, we establish the following results regarding $\hat p_4(y|x)$. 
\begin{them}
\label{thm:bias4}
When $U$ is ordinary smooth of order $b$, if $nh_1^{1+2b}h_2(h_1^2+h_2^2)^2 \to \infty$ and $h_3=O(h_2)$ as $n\to \infty$, $h_1, h_2, h_3\to 0$, then 
\begin{align}
 & \hat p_4(y|x)-p(y|x) \nonumber \\
= & \textrm{DB}_4(x, y, h_1, h_2)+O(h_1^4)+O(h_2^4)+O(h_1^2h_2^2)+O_p\left(\frac{1}{\sqrt{nh_1^{1+2b}h_2}}\right),
\label{eq:bias4ord}
\end{align}
where
\begin{align}
  & \textrm{DB}_4(x, y, h_1, h_2)\nonumber \\
= & \frac{1}{2f_{\hbox {\tiny $X$}}(x)} \left(\left[p_{xx}(x, y)+\sum_{k=2}^4\mathscr{T}_x\{I_k(\cdot, y)\}-p(y|x)f''_{\hbox {\tiny $X$}}(x) \right]\mu_{2,1}h_1^2+ p_{yy}(x,y)\mu_{2,2}h_2^2\right)
\label{eq:db4}
\end{align}
is the dominating bias, in which
\begin{equation}
\label{eq:3Is}
\left\{
\begin{aligned}
I_2(w, y)& = \left\{\frac{d^2}{dw^2} m^*(w)\right\} \int p_y(v, y) f_{\hbox {\tiny $U$}}(w-v) dv, \\
I_3(w, y)& = \left\{\frac{d}{dw} m^*(w)\right\}^2 \int  p_{yy}(v, y) f_{\hbox {\tiny $U$}}(w-v) dv, \\
I_4(w, y)& = 2 \left\{\frac{d}{dw} m^*(w)\right\}\int  p_{xy}(v, y) f_{\hbox {\tiny $U$}}(w-v) dv. 
\end{aligned}
\right.
\end{equation}
When $U$ is super smooth of order $b$, if $nh_1^{1-2b_2}h_2 \exp(-2h_1^{-b}/d_2)(h_1^2+h_2^2)^2\to \infty$ and $h_3=O(h_2)$ as $n\to \infty$, $h_1, h_2, h_3\to 0$, then   
\begin{align}
& \hat p_4(y|x)-p(y|x) \nonumber\\
= &  \textrm{DB}_4(x, y, h_1, h_2) +O(h_1^4)+O(h_2^4)+O(h_1^2h_2^2)+O_p\left\{\frac{\exp(h_1^{-b}/d_2)}{\sqrt{n h_1^{1-2b_2}h_2}}\right\}.
\label{eq:bias4sup}
\end{align}
\end{them}

Similar to the one-step estimator $\hat p_3(y|x)$, correcting for measurement error results in a higher variance for the two-step estimator $\hat p_4((y|x)$ than its error-free counterpart $\hat p_2(y|x)$. In addition, Theorem~\ref{thm:bias4} indicates that, as long as $h_3=O(h_2)$, the effects of estimating $m^*(x)$ on $\hat p_4(y|x)$ are negligible in regard to both bias and variance. 

In what follows, we compare the dominating bias of $\hat p_4(y|x)$ with those of $\hat p_2(y|x)$ and $\hat p_3(y|x)$. In the absence of measurement error, \citet{hansen2004nonparametric} established the following result regarding the two-step estimator $\hat p_2(y|x)$, 
\begin{align*}
 & \hat p_2(y|x)-p(y|x)  \\
= & \frac{1}{2}\left[ \left\{\frac{\partial^2 f_{e|\hbox {\tiny $X$}}(e|x)}{\partial x^2} +2 \frac{\partial f_{e|\hbox {\tiny $X$}}(e|x)}{\partial x}\frac{f'_{\hbox {\tiny $X$}}(x)}{f_{\hbox {\tiny $X$}}(x)} \right\} \mu_{2,1}h_1^2+ \frac{\partial^2 f_{e|\hbox {\tiny $X$}}(e|x)}{\partial e^2}\mu_{2,2}h_2^2\right] \\
& +O(h_1^4)+O(h_2^4)+O(h_1^2h_2^2)+O_p\left(\frac{1}{\sqrt{nh_1h_2}}\right). 
\end{align*}
Elaborations of derivatives of $f_{e|\hbox {\tiny $X$}}(e|x)$ reveal that Hansen's result suggests the following dominating bias of $\hat p_2(y|x)$, 
\begin{align}
& \textrm{DB}_2(x, y, h_1, h_2)\nonumber \\
= & \frac{1}{2f_{\hbox{\tiny $X$}}(x)}\left[\left\{f_{\hbox{\tiny $X$}, e, 11}^{(2)}(x, e)-p(y|x)f_{\hbox{\tiny $X$}}''(x)  \right\} \mu_{2,1}h_1^2+p_{yy}(x,y)\mu_{2,2}h_2^2\right],\label{eq:p2bias}
\end{align}
where $f_{\hbox{\tiny $X$}, e}(x, e)$ is the joint density of $X$ and $e=Y-m(X)$, and $f_{\hbox{\tiny $X$}, e, 11}^{(2)}(x, e)=(\partial^2/\partial x^2)f_{\hbox {\tiny $X$}, e}(x, e)$. An interesting finding here is that Hansen's dominating bias of the two-step estimator for $p(y|x)$ in the absence of measurement error is generally not equal to the dominating bias of our proposed two-step estimator accounting for measurement error given in (\ref{eq:db4}). Starting from $p(x, y)=f_{\hbox {\tiny $X$}, e}\{x, y-m(x)\}$, one can derive $p_{xx}(x, y)$ and show that 
\begin{align}
& p_{xx}(x, y) \nonumber \\
= &  f^{(2)}_{\hbox {\tiny $X$}, e, 11}(x, e)-m''(x)f^{(1)}_{\hbox {\tiny $X$}, e, 2}(x, e)+\left\{  m'(x)\right\}^2f^{(2)}_{\hbox {\tiny $X$}, e, 22}(x, e)-2m'(x)f^{(2)}_{\hbox {\tiny $X$}, e, 21}(x, e),
\label{eq:longpxx}
\end{align}
where $m'(x)$ and $m''(x)$ are the first and second derivatives of $m(x)$, respectively, $f^{(1)}_{\hbox {\tiny $X$}, e, 2}(x, e)=(\partial/\partial e)f_{\hbox {\tiny $X$}, e}(x, e)$, $f^{(2)}_{\hbox {\tiny $X$}, e, 22}(x, e)=(\partial^2/\partial e^2)f_{\hbox {\tiny $X$}, e}(x, e)$, and $f^{(2)}_{\hbox {\tiny $X$}, e, 21}(x, e)=(\partial^2/\partial x \partial e)f_{\hbox {\tiny $X$}, e}(x, e)$. Substituting $p_{xx}(x, y)$ in (\ref{eq:db4}) with (\ref{eq:longpxx}), one can see that   
\begin{align}
& \textrm{DB}_4(x, y, h_1, h_2)\nonumber \\
= & \textrm{DB}_2(x, y, h_1, h_2)+
\frac{\mu_{2,1}h_1^2}{2f_{\hbox {\tiny $X$}}(x)} \left[\sum_{k=2}^4\mathscr{T}_x\left\{I_k(\cdot, y)\right\}-\right. \nonumber\\
& m''(x)f^{(1)}_{\hbox {\tiny $X$}, e, 2}(x, e)+\left\{  m'(x)\right\}^2f^{(2)}_{\hbox {\tiny $X$}, e, 22}(x, e)-2m'(x)f^{(2)}_{\hbox {\tiny $X$}, e, 21}(x, e)
\Bigg].
\label{eq:db4v2}
\end{align}
Even though there exists an interesting connection between the three functions defined in (\ref{eq:3Is}) and the last three terms in (\ref{eq:db4v2}), (\ref{eq:db4v2}) does not provide much insight on how $\hat p_4(y|x)$ compares with $\hat p_2(y|x)$. We next consider three special cases under which (\ref{eq:db4v2}) can be further simplified in order to gain more insight on the dominating bias associated with different estimators. 

The first special case is when $m(x)$ is a constant function, in which case one can show that $m^*(w)$ is also a constant function. Now, by (\ref{eq:3Is}), all terms in (\ref{eq:db4v2}) following $\textrm{DB}_2(x, y, h_1, h_2)$ reduce to zero. If fact, by (\ref{eq:db3}) and (\ref{eq:longpxx}), $\textrm{DB}_2(x, y, h_1, h_2)= \textrm{DB}_3(x, y, h_1, h_2)=\textrm{DB}_4(x, y, h_1, h_2)$ when $m(x)$ is free of $x$. The second special case is when there is no measurement error, under which we show in Section B.3 of Appendix B that terms insides the square brackets in (\ref{eq:db4v2}) also reduce to zero, suggesting $\textrm{DB}_4(x, y, h_1, h_2)=\textrm{DB}_2(x, y, h_1, h_2)$, as it should be in the absence of measurement error. The third special case results from imposing the conditions stated in \citet{Hyndman.etal1996}, under which they concluded that $\hat p_2(y|x)$ is superior than $\hat p_1(y|x)$. These conditions include that (H1) the covariate is locally uniform near $x$ so that $f'_{\hbox {\tiny $X$}}(x)\approx 0$ and $f''_{\hbox {\tiny $X$}}(x)\approx 0$, (H2) $e\perp X$ so that $p(y|x)=f_e\{  y-m(x)\}$, and (H3) $m(x)$ is locally linear near $x$ so that $m''(x)\approx 0$. Under Conditions (H1)--(H3), we simplify (\ref{eq:p2bias}), (\ref{eq:db3}), and (\ref{eq:db4}) in Section B.3 of Appendix B and find that 
\begin{align*}
\textrm{DB}_3(x, y, h_1, h_2) \approx &\ \textrm{DB}_2(x, y, h_1, h_2)+0.5 f_e''(e)\left\{m'(x)\right\}^2\mu_{2,1}h_1^2, \\
\textrm{DB}_4(x, y, h_1, h_2) \approx &\ \textrm{DB}_3(x, y, h_1, h_2)\\
& + 0.5 f_e''(e)\left[\left\{\frac{d}{dx}m^*(x)\right\}^2-2m'(x)\frac{d}{dx}m^*(x)\right]\mu_{2,1}h_1^2. 
\end{align*}
It has been observed in many measurement error model settings that $(d/dx)m^*(x)$ attenuates towards zero compared to $m'(x)$, with the exact attenuation factor derived for the case when $m(x)$ is linear, and $X$ and $U$ are normally distributed \citep[][Section 1.1]{fuller2009measurement}. To be more specific, if $m(x)=\beta_0+\beta_1 x$, where $\beta_0$ and $\beta_1$ are the intercept and slope parameters, then it has been shown in this case that $m^*(x)=\alpha_0+\alpha_1 x$, where $\alpha_0$ and $\alpha_1$ are the intercept and slope parameters in the naive regression, in which $\alpha_1=\lambda \beta_1$, with $\lambda=\sigma^2_x/(\sigma^2_x+\sigma^2_u)$ known as the reliability ratio \citep[][Section 3.2.1]{Carroll&etal2006},  and $\sigma^2_x$ being the variance of $X$. This gives $\textrm{DB}_3(x, y, h_1, h_2) \approx 0.5 f_e''(e)(\beta_1^2\mu_{2,1}h_1^2+ \mu_{2,2} h_2^2)$, in contrast to $\textrm{DB}_4(x, y, h_1, h_2) \approx 0.5 f_e''(e)\{(1-\lambda)^2\beta_1^2\mu_{2,1}h_1^2+ \mu_{2,2} h_2^2\}$. In summary, under the third special case, one would usual expect the following trend of comparisons, $|\textrm{DB}_2(x, y, h_1, h_2)|\le   |\textrm{DB}_4(x, y, h_1, h_2)| \le|\textrm{DB}_3(x, y, h_1, h_2)|$. Therefore, under the same set of conditions considered in \citet{Hyndman.etal1996}, the proposed two-step estimator $\hat p_4(y|x)$ is still asymptotically superior than the one-step estimator $\hat p_3(y|x)$. 

We are now in the position to reflect on the findings that the duo of $\hat p_2(y|x)$ and $\hat p_4(y|x)$ do not share the same dominating bias, whereas the other duo, $\hat p_1(y|x)$ and $\hat p_3(y|x)$, do. Looking back at the construction of the two proposed estimators accounting for measurement error in Section~\ref{s:methods}, one can see that they only differ in the estimator of $p^*(x, y)$ used in (\ref{eq:linkest}) to obtain an estimator for the joint density $p(x, y)$ via the integral transform defined in (\ref{eq:Tx}). Denote by $\hat p_1(x, y)$ and $\hat p_2(x, y)$ the numerators of (\ref{eq:condden1}) and (\ref{eq:condden2}), respectively, which are two estimators for $p(x, y)$ in the absence of measurement error. Denote by $\tilde p_1(x, y)$ and $\tilde p_2(x, y)$ the numerators of (\ref{eq:naive1}) and (\ref{eq:naive2}), respectively, which are two estimators for $p^*(x, y)$, viewed as naive estimators for $p(x, y)$ in the presence of measurement error. In the one-step estimator $\hat p_3(y|x)$, the estimator for $p(x, y)$ can be expressed as 
\begin{align}
 \mathscr{T}_x\left\{\tilde p_1(\cdot, y)\right\} 
= & \frac{1}{nh_1h_2}\sum_{j=1}^n \mathscr{T}_x\left\{K_1\left(\frac{W_j-\cdot}{h_1}\right)\right\}K_2\left(\frac{Y_j-y}{h_2}\right) \label{eq:Txp1}\\
= & \frac{1}{nh_1h_2}\sum_{j=1}^n K^*_1\left(\frac{W_j-x}{h_1}\right)K_2\left(\frac{Y_j-y}{h_2}\right), \textrm{ by (\ref{eq:K1*})}\nonumber 
\end{align}
which has the same expectation as that of $\hat p_1(x, y)$ according to (\ref{eq:K1*K1}). This explains why $\hat p_3(y|x)$ and $\hat p_1(y|x)$ have the same dominating bias. In contrast, in the two-step estimator $\hat p_4(y|x)$, the estimator for $p(x, y)$ is
\begin{align}
 \mathscr{T}_x\left\{\tilde p_2(\cdot, y)\right\} 
= \frac{1}{nh_1h_2}\sum_{j=1}^n \mathscr{T}_x\left[K_1\left(\frac{W_j-\cdot}{h_1}\right)K_2\left\{ \frac{Y_j-\hat m^*(W_j)-y+\hat m^*(\cdot)}{h_2}  \right\}\right], \label{eq:Txp2}
\end{align}
of which the expectation is typically not equal to $E\{\hat p_2(x, y)\}$. Hence, it is not surprising that, after correcting the naive two-step estimator $\tilde p_2(y|x)$ for measurement error, $\hat p_4(y|x)$ does not have the same dominating bias as that of $\hat p_2(y|x)$. 

Contrasting (\ref{eq:Txp2}) with (\ref{eq:Txp1}) also brings awareness that more involved conditions are needed for $\mathscr{T}_x\{\tilde p_2(\cdot, y)\}$ to be well-defined. According to (\ref{eq:Txp1}), $\mathscr{T}_x\{\tilde p_1(\cdot, y)\}$ is well-defined because $h^{-1}_1\mathscr{T}_x[K_1\{(W-\cdot)/h_1\}]$ is, thanks to Condition (K1). By (\ref{eq:Txp2}), $\mathscr{T}_x\{\tilde p_2(\cdot, y)\}$ is well-defined if $(h_1h_2)^{-1}\mathscr{T}_x(K_1\{(W-\cdot)/h_1\}K_2[\{ Y-\hat m^*(W)-y+\hat m^*(\cdot)\}/h_2])$ is, for which sufficient conditions formulated in the same spirit as those in Condition (K1) are that $|\textrm{CR}(t, Y, W, y)|_\infty<\infty$ and $\int |\textrm{CR}(t, Y, W, y)| dt<\infty$ with probability one for each $y$, where 
\begin{align}
& \textrm{CR}(t, Y, W, y) \nonumber \\
=& \frac{\displaystyle{(h_1h_2)^{-1}\int e^{itw} K_1\left(\frac{W-w}{h_1}\right)K_2\left\{ \frac{Y-\hat m^*(W)-y+\hat m^*(w)}{h_2}  \right\}dw}}{\phi_{\hbox {\tiny $U$}}(t)}.
\label{eq:CR}
\end{align}
These sufficient conditions formulated in terms of $\textrm{CR}(t, Y, W, y)$ essentially imply that the Fourier transform of the product kernel, $K_1(t)K_2\{s(t)\}$, tails off to zero much faster than $\phi_{\hbox {\tiny $U$}}(t)$ does as $|t|\to \infty$ so that the norm of the complicated ratio in (\ref{eq:CR}) is integrable, where $s(\cdot)$ denotes some function of $t$, introduced here to signify that arguments in $K_1(\cdot)$ and $K_2(\cdot)$ in (\ref{eq:CR}) both involve $w$. Because imposing Condition (K1) already guarantees that the Fourier transform of $K_1(t)$ diminishes fast enough, compared with how fast $\phi_{\hbox {\tiny $U$}}(t)$ diminishes as $|t|$ diverges, we conjecture that the aforementioned conditions in terms of $\textrm{CR}(t, Y, W, y)$ are satisfied when $K_2(\cdot)$ is of the same order as $K_1(\cdot)$ so that the Fourier transform of the product kernel appearing in (\ref{eq:CR}) tends to zero no slower than $\phi_{\hbox {\tiny $K_1$}}(t)$ does as $|t|\to \infty$. Indeed, when implementing the proposed two-stage estimation method, we set $K_2$ the same as $K_1$ and encounter little numerical complication in obtaining $\hat p_4(y|x)$ in the simulation study.

More general analytic comparisons between $\hat p_3(y|x)$ and $\hat p_4(y|x)$ outside of the aforementioned special cases are unattainable. Empirical evidence from simulation study can shed more light on how they compare with each other and also with the naive estimators. In order to implement the proposed methods, strategies for choosing bandwidths are needed. This is the topic of the next section. 

\section{Bandwidths selection}
\label{s:band}
\subsection{Relevant strategies}
\label{s:review}
The choice of bandwidths in kernel density estimators has a great impact on the estimators. There are two main streams in the literature on bandwidth selection, one relating to the so-called plug-in methods, the other in line with cross validation (CV). Both veins of methodology development start from a criterion that assesses the quality of an estimator, such as the integrated squared error (ISE) of a density estimator, or the MISE. Oftentimes one invokes asymptotic approximations or imposes parametric assumptions, or does both, to simplify a criterion. If the resultant (approximated) criterion can be optimized with respect to a bandwidth explicitly, an asymptotically optimal choice of this bandwidth can be derived. Plug-in methods are based on so-obtained bandwidths, such as the normal reference rule \citep{Silverman1986, scott2015multivariate}. For more complex criteria, a cross validation strategy is often used to estimate the criterion and search for bandwidths that optimize the estimated criterion. Besides plug-in methods and CV methods, \citet{jones1996brief} reviewed other bandwidth selection methods for density estimation, including the ones that involve bootstrap estimation of a criterion. 

The main challenge bandwidth selection methods attempt to overcome is estimation of the aforementioned criteria. Criteria like ISE, MISE, or asymptotic MISE (AMISE) depend on complicated functionals of unknown densities, and estimating these functionals is often a harder problem than the original problem of density estimation. This challenge is even more formidable in the presence of measurement error. To select bandwidths for marginal density estimation in the presence of measurement error, \citet{delaigle2004bootstrap, delaigle2004practical} developed plug-in methods and bootstrap methods based on MISE or AMISE, which require estimation of functionals such as the integrated squared density derivatives using error-prone data. \citet{delaigle2002estimation} constructed estimators for these functionals, which again involve bandwidths selection. 

Later, \citet{Delaigle.Hall2008} combined cross validation with the strategy of simulation extrapolation \citep[SIMEX,][]{Cook&Stefanski1994, stefanski1995simulation} to choose bandwidths in the presence of measurement error. Their CV-SIMEX method entails estimating a CV criterion and finding a bandwidth twice using error-contaminated data (at two levels of contamination) in the same way one would do when data are error-free. The resulting two bandwidths together lead to a bandwidth accounting for measurement error via an extrapolation step. Compared to methods considered in \citet{delaigle2004bootstrap, delaigle2004practical}, one novelty of the CV-SIMEX method is that it avoids direct estimation of a CV criterion accounting for measurement error. This is achieved at the price of increased computational burden caused by the combination of CV and SIMEX, each of which is computationally expensive on its own. Moreover, what extrapolant function should be used at the extrapolation step is rarely known \citep[][Section 5.3.2]{Carroll&etal2006}. Indeed, the extrapolant used in \citet{Delaigle.Hall2008} is only asymptotically justified, i.e., for large sample, under the assumption that error contamination is close to none. For a given application, it is difficult to gauge if the sample size is large enough, relative to the amount of error contamination, for the extrapolation step to yield a bandwidth improving over a naive bandwidth one chooses while ignoring measurement error. A more realistic goal one can achieve by applying the CV-SIMEX method with caution is to somewhat adjust a naive bandwidth in the right direction. This direction is usually upward when measurement errors compromise naive estimation, because, intuitively, a wider bandwidth is needed when measurement errors blur the underlying pattern of association between two variables. Indeed, we observe that a bandwidth used in the proposed estimators that is larger than the naive bandwidth typically yields more satisfactory results in our extensive simulation study.

\subsection{Bandwidth selection for $\hat p_3(y|x)$}
\label{s:band3}
The one-step estimator $\hat p_3(y|x)$ depends on two bandwidths in $\bh=(h_1, h_2)$. We propose to choose $\bh$ by adjusting the naive bandwidths, denoted by $\bh^{(1)}_{\textrm{nv}}=(h^{(1)}_{\textrm{nv},1}, \, h^{(1)}_{\textrm{nv},2})$, obtained via a CV method for estimating $p^*(y|x)$ using $\tilde p_1(y|x)$. In particular, we employ the CV method proposed by \citet{Fan.Yim2004} and \citet{Hall.etal2004} to obtain  $\bh^{(1)}_{\textrm{nv}}$.  

As an estimator for $p^*(y|x)$, the authors considered the ISE of $\tilde p_1(y|x)$ given by  
\begin{equation}\label{eq:ISE}
	\begin{aligned}
		\ISE (\tilde p_1)= & \iint \{\tilde p_1(y|x)-p^*(y|x)\}^2 f_{\hbox {\tiny $W$}}(x)\omega(x)dxdy\\
		=&\iint \left\{\tilde p_1(y|x)\right\}^2 f_{\hbox {\tiny $W$}}(x)\omega(x)dxdy - 2 \iint \tilde p_1(y|x) p^*(x,y)\omega(x) dxdy\\
		& + \int \left\{p^*(y|x)\right\}^2 f_{\hbox {\tiny $W$}}(x)\omega(x)dxdy,
	\end{aligned}
\end{equation}
where $f_{\hbox {\tiny $W$}}(x)$ is the pdf of $W$, and $\omega(x)$ is a nonnegative weight function used to avoid estimating $p^*(y|x)$ at an $x$ around which data are scarce. Observing that the third integral above does not depend on bandwidths, the authors defined a CV criterion based on the following estimator of the first two integrals in (\ref{eq:ISE}), 
\begin{equation}\label{eq:CV}
	\textrm{CV}(\tilde p_1) = \frac{1}{n}\sum_{j=1}^{n}\omega(W_j)\int\left\{\tilde p_{1,-j}(y|W_j)\right\}^2dy - \frac{2}{n}\sum_{j=1}^{n}\omega(W_j)\tilde p_{1, -j}(Y_j|W_j),
\end{equation} 
where $\tilde p_{1,-j}(y|W_j)$ results from computing the estimator $\tilde p_1(y|W_j)$ using all observed data except the $j$th data point, $(W_j, Y_j)$. We set $K_2(t)$ as the Gaussian kernel in $\tilde p_1(y|x)$, and thus in $\hat p_3(y|x)$ as well. Thanks to this choice of $K_2(t)$, the integral in (\ref{eq:CV}) can be derived explicitly, as shown in Appendix C, resulting in an elaborated expression of $\textrm{CV}(\tilde p_1)$ provided there. As for the other kernel, $K_1(t)$, in $\tilde p_1(y|x)$, and thus also in $\hat p_3(y|x)$, we set 
\begin{equation}\label{eq:sec-order}
K_1(t) = \frac{48\cos t}{\pi t^4} \left(1-\frac{15}{t^2}\right) - \frac{144\sin t}{\pi t^5} \left(2-\frac{5}{t^2}\right),
\end{equation}
of which the characteristic function is $\phi_{\hbox {\tiny $K_1$}}(s)=(1-s^2)^3I(-1\le s \le 1)$, which satisfies Conditions K listed in Section~\ref{s:conditions}. Other choices of $K_1(t)$ one may consider that also satisfy Conditions K include the sinc kernel, and the kernel used in \citet{delaigle2009design}, of which the characteristic function is $\phi_{\hbox {\tiny $K_1$}}(s)=(1-s^2)^8I(-1\le s \le 1)$. As commented in Section~\ref{s:conditions}, (K5) in Conditions K can be relaxed when $U$ is ordinary smooth. We keep our choice of $K_1(t)$ to fulfill condition (K5) even when $U$ is ordinary smooth mainly for the numerical stability it renders when computing the deconvoluting kernel $K_1^*(t)$.

Following the CV method, we search bandwidths that minimize $\textrm{CV}(\tilde p_1)$, resulting in $\bh^{(1)}_{\textrm{nv}}=(h^{(1)}_{\textrm{nv},1}, \, h^{(1)}_{\textrm{nv},2})$. Denote by $\bh^{(1)}=(h^{(1)}_1, \, h^{(1)}_2)$ the bandwidths we choose for $\hat p_3(y|x)$ to estimate $p(y|x)$. Since $Y$ is observed without error, we set $h^{(1)}_2=h^{(1)}_{\textrm{nv},2}$; and to account for covariate measurement error, we set
\begin{equation}
h^{(1)}_1= \left(1+|\rho_{wy}|\sqrt{1-\hat{\lambda}}\right)h^{(1)}_{\textrm{nv},1},
\label{eq:h1}
\end{equation}
where $\rho_{wy}$ is the sample correlation between $W$ and $Y$, and $\hat \lambda=1-\sigma^2_u/s^2_w$ is an estimate of the reliability ratio $\lambda$, in which $s^2_w$ is the sample variance of $W$. The adjustment of $h^{(1)}_{\textrm{nv},1}$ given in (\ref{eq:h1}) is motivated by the following considerations. When there is no measurement error, certainly no adjustment is needed, which is exactly what (\ref{eq:h1}) indicates when $\sigma^2_u=0$ (yielding $\hat \lambda=1$). When there exists measurement error but $X$ and $Y$ are independent, $W$ and $Y$ are also independent because $U$ is independent of $(X, Y)$. In this case, since both $p(y|x)$ and $p^*(y|x)$ reduce to the marginal density of $Y$, accounting for measurement error when estimating $p(y|x)$ is not necessary, and thus neither is adjusting bandwidths for measurement error, which is also what (\ref{eq:h1}) suggests with $\rho_{wy}$ consistently estimating the zero correlation. In the presence of measurement error, if $X$ and $Y$ are dependent, it is sensible to inflate the naive bandwidth associated with $X$ to adjust for measurement error, with the adjustment depending on the severity of error contamination and the strength of dependence between $X$ and $Y$, which can be partially assessed by the correlation between them. In summary, (\ref{eq:h1}) suggests use of the naive bandwidth when no adjustment for measurement error is necessary, and it leads to a different bandwidth by adjusting the naive bandwidth in the right direction otherwise.

\subsection{Bandwidth selection for $\hat p_4(y|x)$}
\label{s:band4}
To select bandwidths in $\bh=(h_1, h_2)$ for the two-step estimator $\hat p_4(y|x)$, we also begin with some naive bandwidths, denoted by $\bh^{(2)}_{\textrm{nv}}=(h^{(2)}_{\textrm{nv},1}, \, h^{(2)}_{\textrm{nv},2})$, obtained from the CV method for estimating $p^*(y|x)$ using $\tilde p_2(y|x)$. Here, the CV criterion is 
\begin{equation}\label{eq:CV2}
	\textrm{CV}(\tilde p_2) = \frac{1}{n}\sum_{j=1}^{n}\omega(W_j)\int\tilde p_{2,-j}(y|W_j)^2dy - \frac{2}{n}\sum_{j=1}^{n}\omega(W_j)\tilde p_{2, -j}(Y_j|W_j).
\end{equation} 
Even though this criterion is similar to (\ref{eq:CV}), there are two complications. 

First, when $m^*(y|x)$ is estimated by a local polynomial estimator, as done in the majority of our study, $\tilde p_2(y|x)$ involves an additional bandwidth $h_3$ in $\hat m^*(y|x)$. In this case, we use the plug-in method for local polynomial regression \citep[][Chapter 3]{Fan&Gijbels1996} implemented by the R function \texttt{locpol} to obtain $h_3$, with $K_3(t)$ being the Gaussian kernel. For the other two kernels $K_1(t)$ and $K_2(t)$, we set them both as the kernel in (\ref{eq:sec-order}) for ease of numerical implementation as commented in Section~\ref{s:conditions}. This choice of $K_2(t)$ causes the second complication, which is that the integral in (\ref{eq:CV2}) for $\textrm{CV}(\tilde p_2)$ now cannot be derived explicitly. To avoid direct evaluation of this integral, we put the Gaussian kernel back for $K_2(t)$ in (\ref{eq:CV2}), and proceed with the CV method to choose $\bh=(h_1, h_2)$. This produces an elaborated expression of $\textrm{CV}(\tilde p_2)$ provided in equation (\ref{eq:CV:formula2}) in Appendix C that involves residuals defined by $e_j^*=Y_j-\hat m^*(W_j)$. Denote by $\bh^{(2)*}_{\textrm{nv}}=(h^{(2)*}_{\textrm{nv},1}, \, h^{(2)*}_{\textrm{nv},2})$ the bandwidths that minimize (\ref{eq:CV:formula2}). We then set $\bh^{(2)}_{\textrm{nv}}=(h^{(2)*}_{\textrm{nv},1}, \, 0.403 h^{(2)*}_{\textrm{nv},2})$ to acknowledge that the kernel used as $K_2(t)$ in the actual $\tilde p_2(y|x)$ is not the Gaussian kernel. The factor $c=0.403$ used in this adjustment for the bandwidth associated with $K_2(t)$ is deduced as follows. Consider generically estimating the density of a random variable $V$, $f_{\hbox {\tiny $V$}}(v)$, via a kernel density estimator with $K(t)$ as the kernel. \citet[][page 45]{Silverman1986} suggested the following reference rule for choosing bandwidth, 
\begin{equation}
h= \left[ \frac{8\sqrt{\pi}\int K^2(t)dt}{3\left\{\int t^2 K(t)dt\right\}^2} \right]^{1/5}s_v n^{-1/5}, \label{eq:h2*}
\end{equation}
where $s_v$ is the sample standard deviation of $V$. For a given sample of size $n$, (\ref{eq:h2*}) provides a relationship between $h$ and $K(t)$. If $K(t)$ is the Gaussian kernel, (\ref{eq:h2*}) suggests the reference rule of $h=1.06s_v n^{-1/5}$; and if $K(t)$ is given by (\ref{eq:sec-order}), one has $h=0.427s_v n^{-1/5}$. The ratio of the latter reference rule over the former gives $c=0.403$, a sensible scale factor to use when one changes from a Gaussian kernel to the kernel in (\ref{eq:sec-order}). 

Lastly, once we have $\bh^{(2)}_{\textrm{nv}}$, we use $\bh^{(2)}=(h^{(2)}_1, \, h^{(2)}_2)$ in $\hat p_4(y|x)$, where $h^{(2)}_2=h^{(2)}_{\textrm{nv},2}$ and
\begin{equation}
h^{(2)}_1= \left(1+|\rho_{we^*}|\sqrt{1-\hat{\lambda}}\right)h^{(2)}_{\textrm{nv},1}, \label{eq:h1p4}
\end{equation}
in which $\rho_{we^*}$ is the sample correlation between $W$ and $e^*$. The adjustment in (\ref{eq:h1p4}) is in the same spirit as (\ref{eq:h1}), although we use $\rho_{we^*}$ in place of $\rho_{wy}$. This replacement is motivated by the fact that $\tilde p_2(y|x)$ and $\hat p_4(y|x)$ are essentially estimating the conditional density of a mean residual given the corresponding covariate. 
 
\section{Simulation study}
\label{s:simulation}
\subsection{Simulation design}
We are now in the position to compare finite sample performance of the naive estimators, $\tilde p_1(y|x)$ and $\tilde p_2(y|x)$, and their non-naive counterparts, $\hat p_3(y|x)$ and $\hat p_4(y|x)$. In the simulation experiments, we consider the following three models of $Y$ given $X$:
\begin{enumerate}
	\item[(C1)] $[Y|X=x]\sim N\left(m(x), \, \sigma^2(x)\right)$, where $m(x)=\sin(\pi x/2)$ and $\sigma(x)=\exp(1-x/3)/8$;
	\item[(C2)] $[Y|X=x]\sim 0.5N\left(m(x)-1, \, \sigma^2(x)\right)+0.5N\left(m(x)+1,\,  \sigma^2(x)\right)$,  where $m(x)=\sin(\pi x/2)$ and $\sigma(x)=\exp(1-x/3)/12$;
	\item[(C3)] $[Y|X=x]\sim N\left(m(x), \, \sigma^2(x)\right)$, where $m(x)=x$ and $\sigma(x)=\exp(1-x/3)/8$.
\end{enumerate}
The three primary (conditional) models are formulated to create two contrasting scenarios under which we compare the four density estimators. One scenario is having a unimodal conditional density (as in (C1) and (C3)) versus a multimodal density (as in (C2)); the other scenario is having a nonlinear conditional mean (as in (C1) and (C2)) versus a linear mean (as in (C3)). The designs of these primary models partly follow the illustrative examples in \citet{sugiyama2010conditional} with heteroscedastic noise.

In conjunction with each of the three primary models, we vary the true covariate distribution, the measurement error distribution, and the reliability ratio to create four configurations of secondary models: (a) $X\sim N(0,1)$, $U\sim \textrm{Laplace}(0, \, \sigma_u/\sqrt{2})$, $\lambda=0.8$; (b) $X\sim N(0,1)$, $U\sim \textrm{Laplace}(0, \, \sigma_u/\sqrt{2})$, $\lambda=0.9$; (c) $X\sim N(0,1)$, $U\sim N(0, \, \sigma_u^2)$, $\lambda=0.8$; (d) $X\sim \textrm{Uniform}(-2, \, 2)$, $U\sim \textrm{Laplace}(0, \, \sigma_u/\sqrt{2})$, $\lambda=0.8$. Contrasting (a) and (b) allows comparison under different severity of error contamination in the covariate. Comparing estimates under (a) and (c) can shed light on effects of different types of measurement error on considered estimators. In particular, the Laplace distribution for $U$ under (a) is an example of ordinary smooth error distributions, whereas the normal distribution for $U$ under (c) provides an example of super smooth distributions. Finally, the contrast of (a) and (d) provides a testbed for inspecting the performance of estimators when the true covariate has an unbounded support compared to when it has a bounded support.  

Putting the three primary models with the four secondary model configurations lead to twelve true model settings, according to each of which we generate 200 Monte Carlo (MC) replicates of size $n=500$. Given each simulated data set, we carry out two rounds of density estimation. In the first round, to mitigate the confounding effect of data-driven bandwidth selection on the estimation quality, we use the approximated theoretical optimal bandwidths associated with each of the four estimators. Generically denote by $\hat p(y|x)$ one of the estimators, the approximated theoretical optimal $\bh=(h_1, h_2)$ associated with $\hat{p}(y|x)$ is obtained (through a grid search) by minimizing the empirical integrated squared error (EISE),
\begin{equation}\label{eq:EISE:dens}
	\EISE = \sum_{j=1}^{\calM'}\sum_{k=0}^{\calM} \left\{ \hat{p}(y_j|x_k)-p(y_j|x_k) \right\}^2 f_{\hbox {\tiny $X$}}(x_k)\Delta\Delta',
\end{equation}
where $\{x_k=x_{\hbox {\tiny $L$}}+k\Delta\}_{k=0}^{\calM}$, $\Delta$ is the partition resolution, $\calM$ is the largest integer no greater than $(x_{\hbox {\tiny $U$}}-x_{\hbox {\tiny $L$}})/\Delta$, in which $x_{\hbox {\tiny $U$}}=-2$ and $x_{\hbox {\tiny $L$}}=2$; and $\{y_j\}_{j=1}^{\calM'}$ is a sequence of  grid points equally spaced over the observed sample range of $Y$,  with $y_{j+1} - y_j=\Delta'$. The additional bandwidth, $h_3$, in $\tilde p_2(y|x)$ and $\hat p_4(y|x)$ is obtained by minimizing 
\begin{equation}\label{eq:EISE:mean}
	\EISE_m = \sum_{k=0}^{\calM} \left\{ \hat{m}^*(x_k)-m^*(x_k) \right\}^2 f_{\hbox {\tiny $X$}}(x_k)\Delta.
\end{equation}
In the second round, we use the proposed methods in Sections~\ref{s:band3} and \ref{s:band4} to obtain bandwidths for $\hat p_3(y|x)$ and $\hat p_4(y|x)$, and apply the CV method in the absence of measurement error to choose bandwidths for $\tilde p_1 (y|x)$ and $\tilde p_2(y|x)$. In the CV criteria used for these methods, we set the weight function $\omega(x)=I(x_{\hbox {\tiny $L$}}\le x \le x_{\hbox {\tiny $U$}})$, where $x_{\hbox {\tiny $U$}}$ and $x_{\hbox {\tiny $L$}}$ are the 2.5th and 97.5th percentiles of the observed covariate data, respectively. A similar weight function, as an indicator function over the interval of interest regarding the covariate, is used in \citet[][Section 3.3]{Fan.Yim2004}. Besides the practical consideration in regard to covariate values of interest, one may also choose a weight function to avoid numerical difficulties caused by dividing by numbers close or equal to zero when computing the conditional density estimate as discussed in \citet[][Section 2]{Hall.etal2004}.

As stated in Section~\ref{s:review}, there exists many different bandwidth selection strategies in the context of density estimation. To have a more focused simulation experiment presented in this article, we avoid going beyond comparing our proposed data-driven bandwidths selection methods with the approximated theoretical optimal approaches, although we did compare the former with their naive counterparts (with results omitted here to save space for other findings) and observe noticeable gain in accuracy of density estimation from adopting the proposed methods. More comprehensive comparisons between various bandwidth selection methods in conjunction with different density estimators besides the four considered here deserve a manuscript dedicated to reporting simulation study of a larger scale. 

\subsection{Simulation results}
\begin{figure}
	\centering
	\setlength{\linewidth}{0.45\linewidth}
	\subfigure[]{ \includegraphics[width=\linewidth]{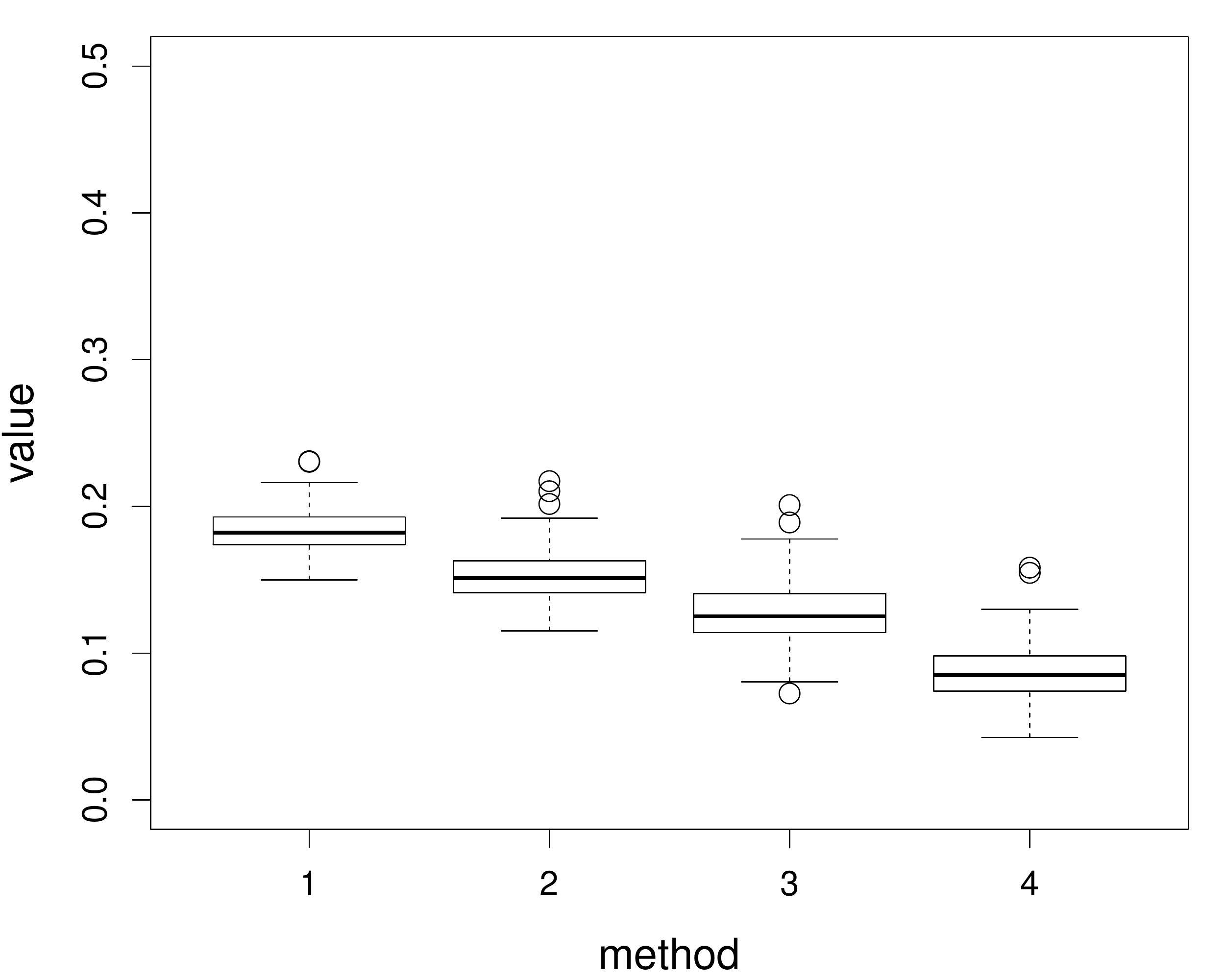} }
	\subfigure[]{ \includegraphics[width=\linewidth]{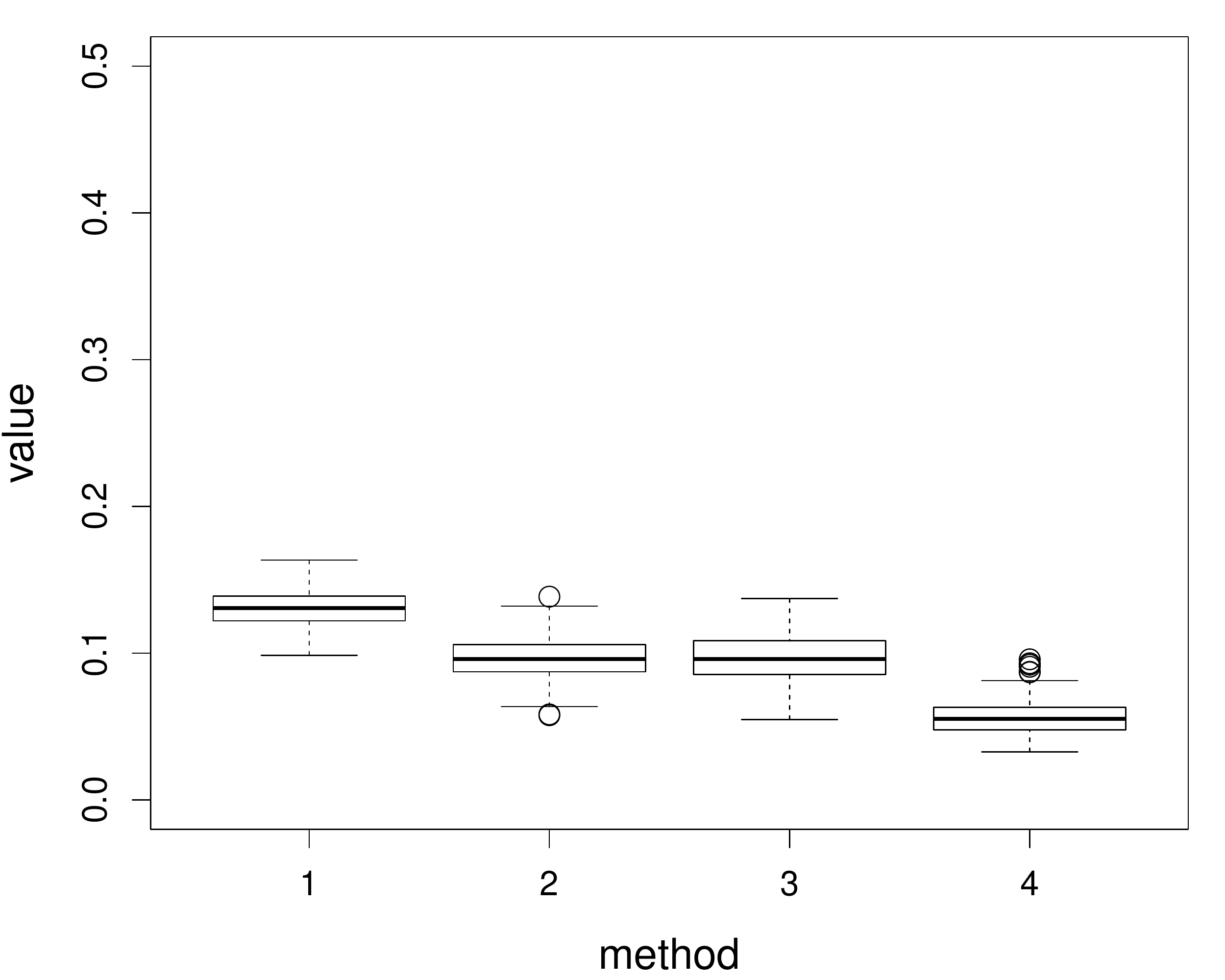} }\\
	\subfigure[]{ \includegraphics[width=\linewidth]{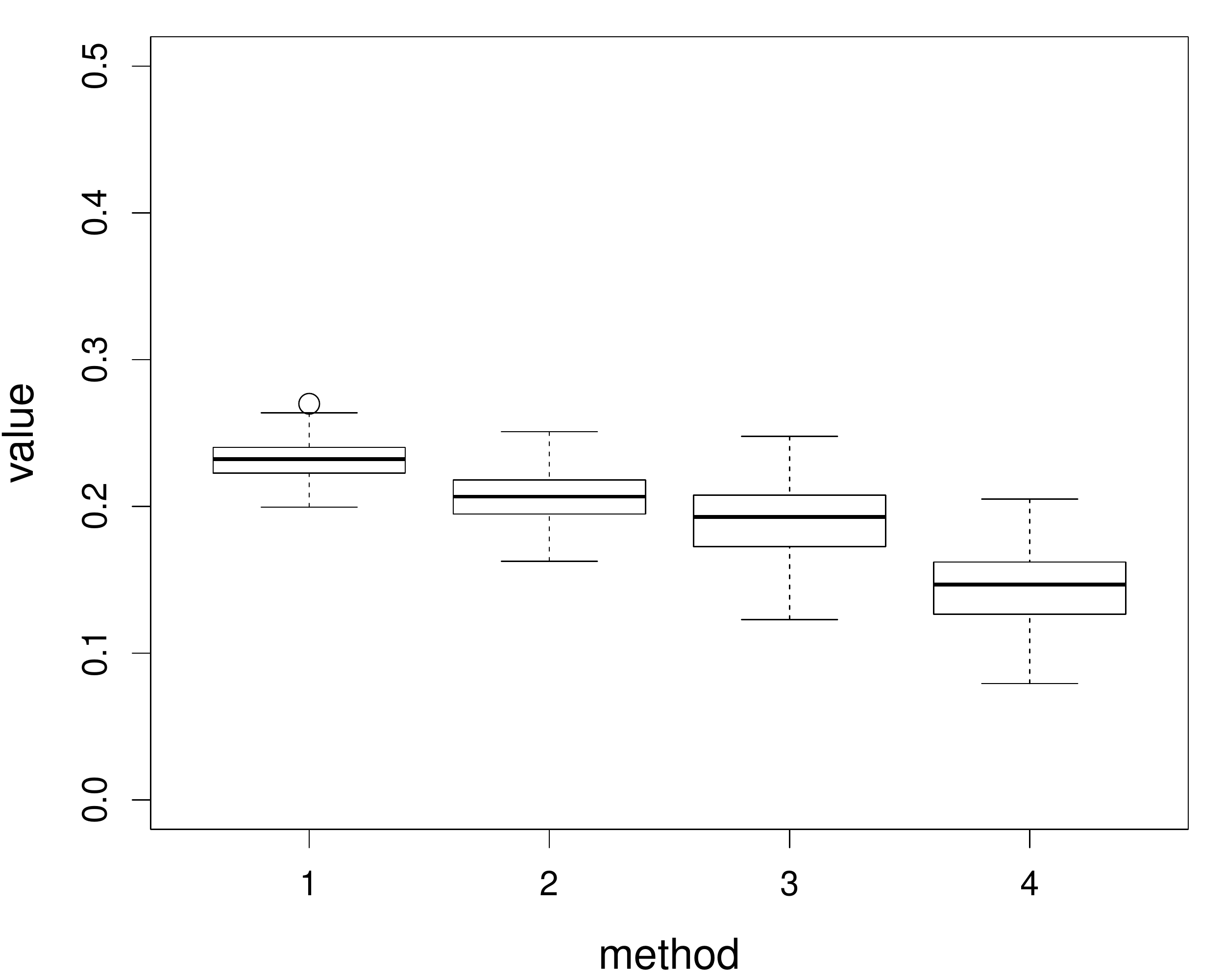} }
	\subfigure[]{ \includegraphics[width=\linewidth]{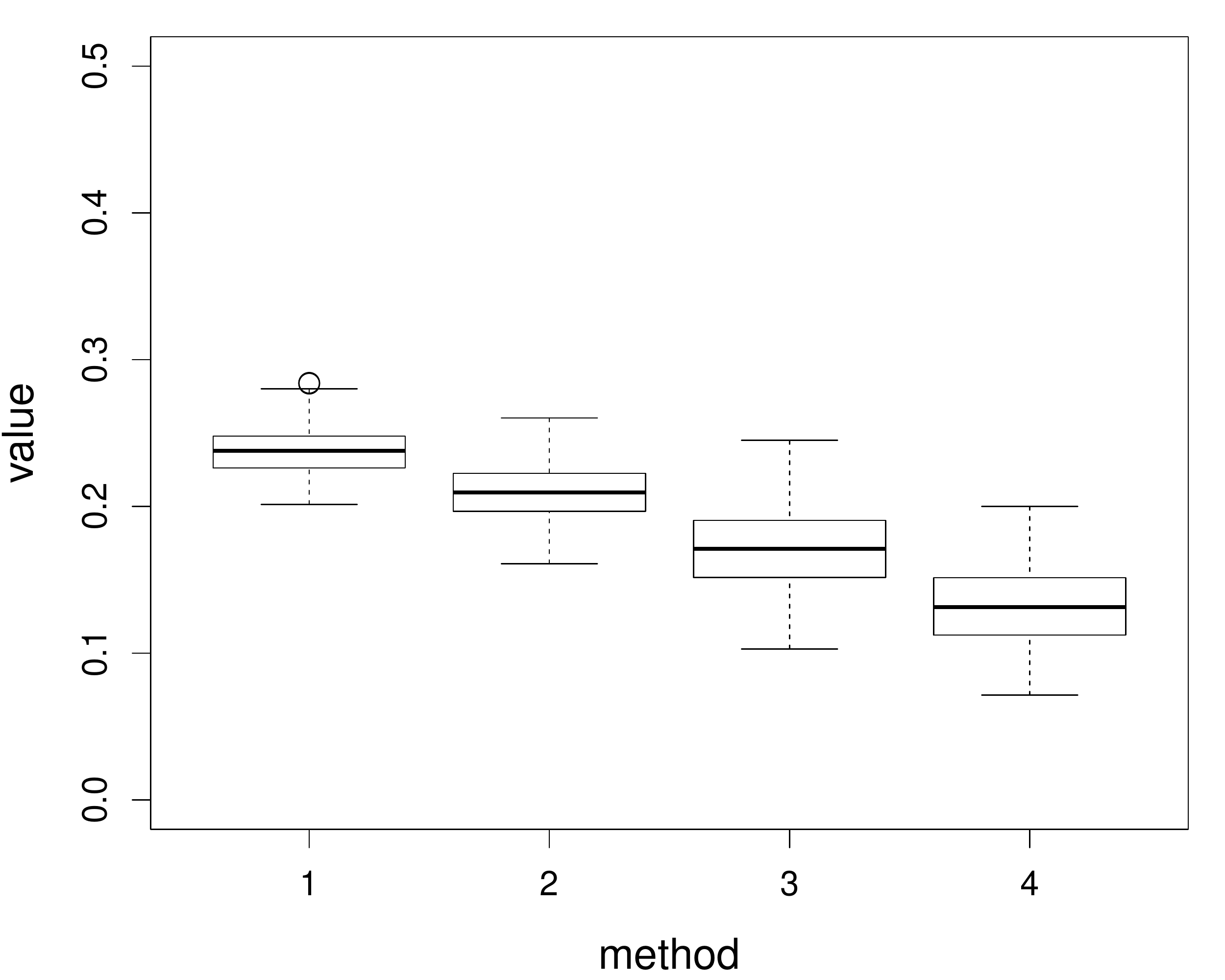} }
	\caption{Boxplots of EISE using the approximated theoretical optimal bandwidths when the primary model is (C1) and the secondary models are (a) $X\sim N(0,1)$, $U\sim \textrm{Laplace}(0, \,\sigma_u/\sqrt{2})$, $\lambda=0.8$; (b) $X\sim N(0,1)$, $U\sim \textrm{Laplace}(0, \,\sigma_u/\sqrt{2})$, $\lambda=0.9$; (c) $X\sim N(0,1)$, $U\sim N(0, \, \sigma_u^2)$, $\lambda=0.8$; (d) $X\sim \textrm{Uniform}(-2, 2)$, $U\sim \textrm{Laplace}(0, \,\sigma_u/\sqrt{2})$, $\lambda=0.8$. Method 1, 2, 3, 4 correspond to $\tilde p_1(y|x)$, $\tilde p_2(y|x)$, $\hat p_3(y|x)$, and $\hat p_4(y|x)$, respectively.} 
	\label{Sim1:box}
\end{figure}

\begin{figure}
	\centering
	\setlength{\linewidth}{0.45\linewidth}
	\subfigure[]{ \includegraphics[width=\linewidth]{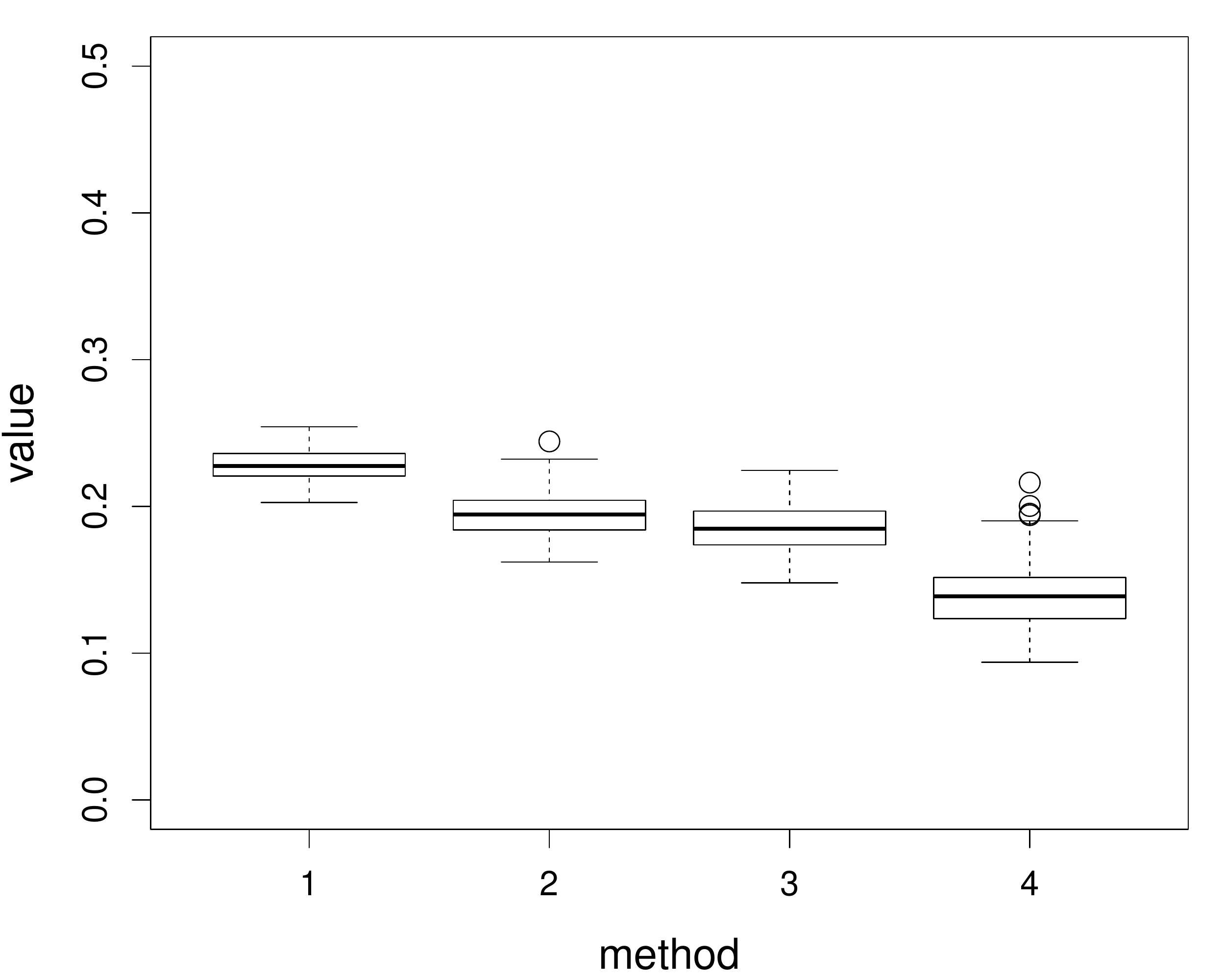} }
	\subfigure[]{ \includegraphics[width=\linewidth]{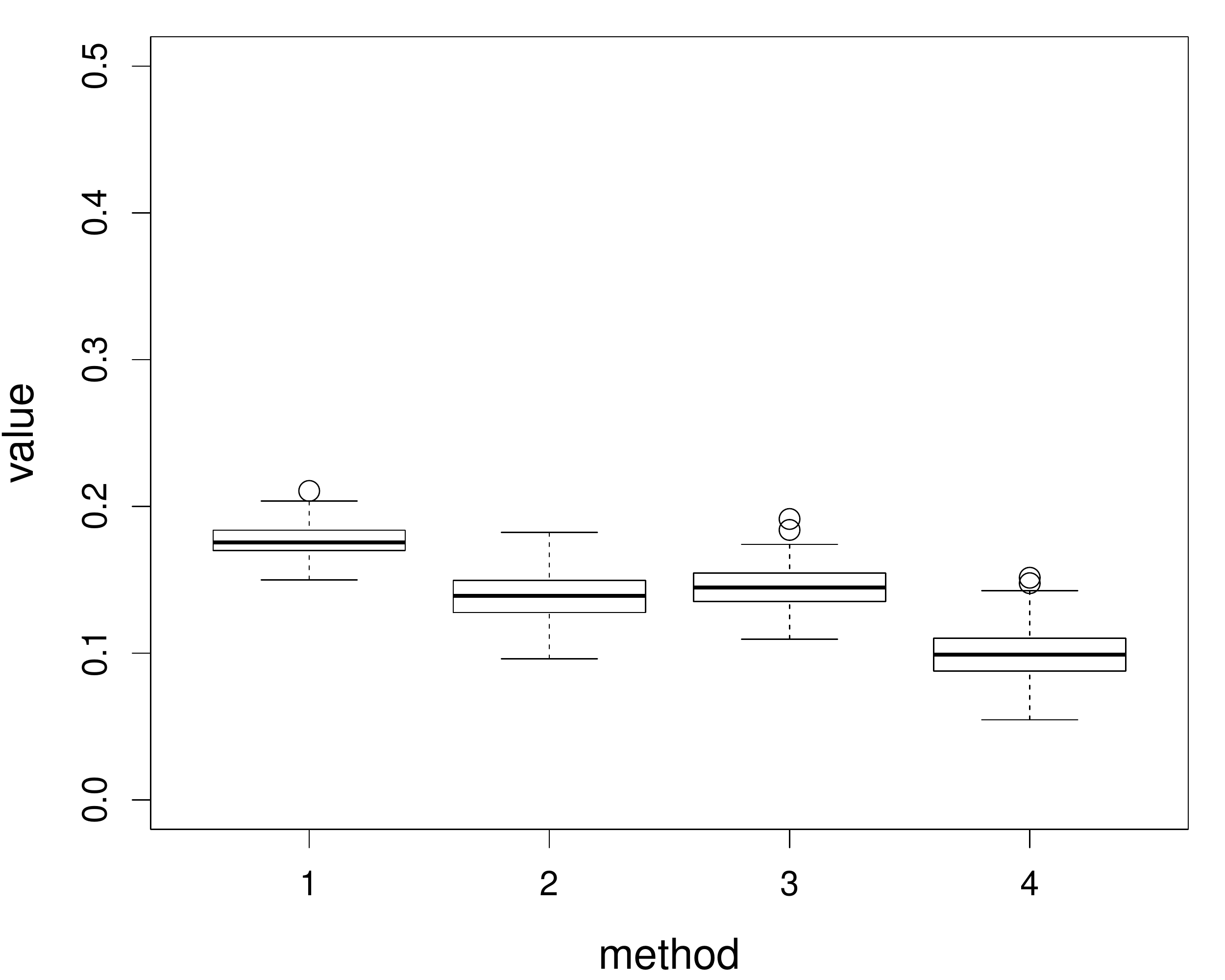} }\\
	\subfigure[]{ \includegraphics[width=\linewidth]{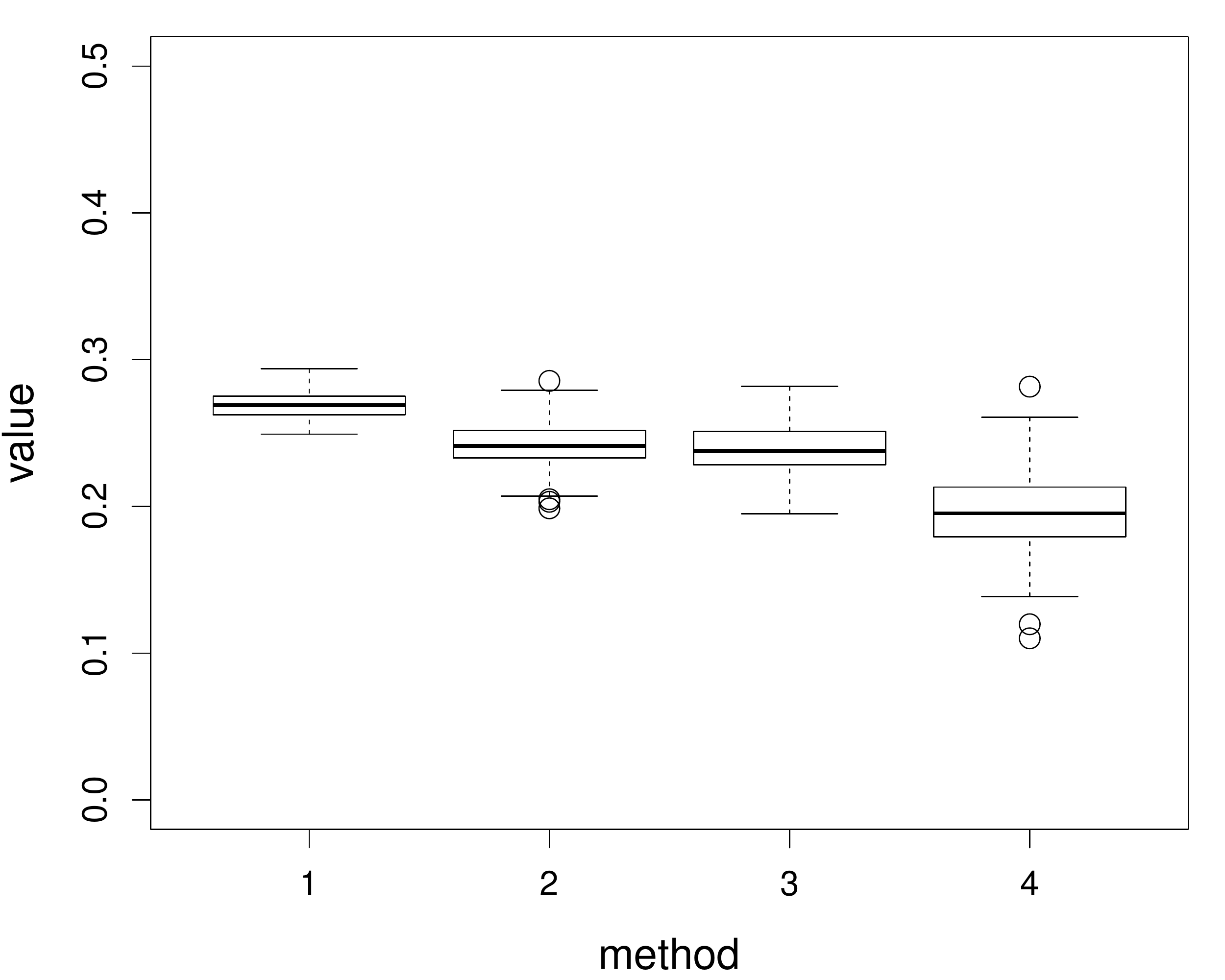} }
	\subfigure[]{ \includegraphics[width=\linewidth]{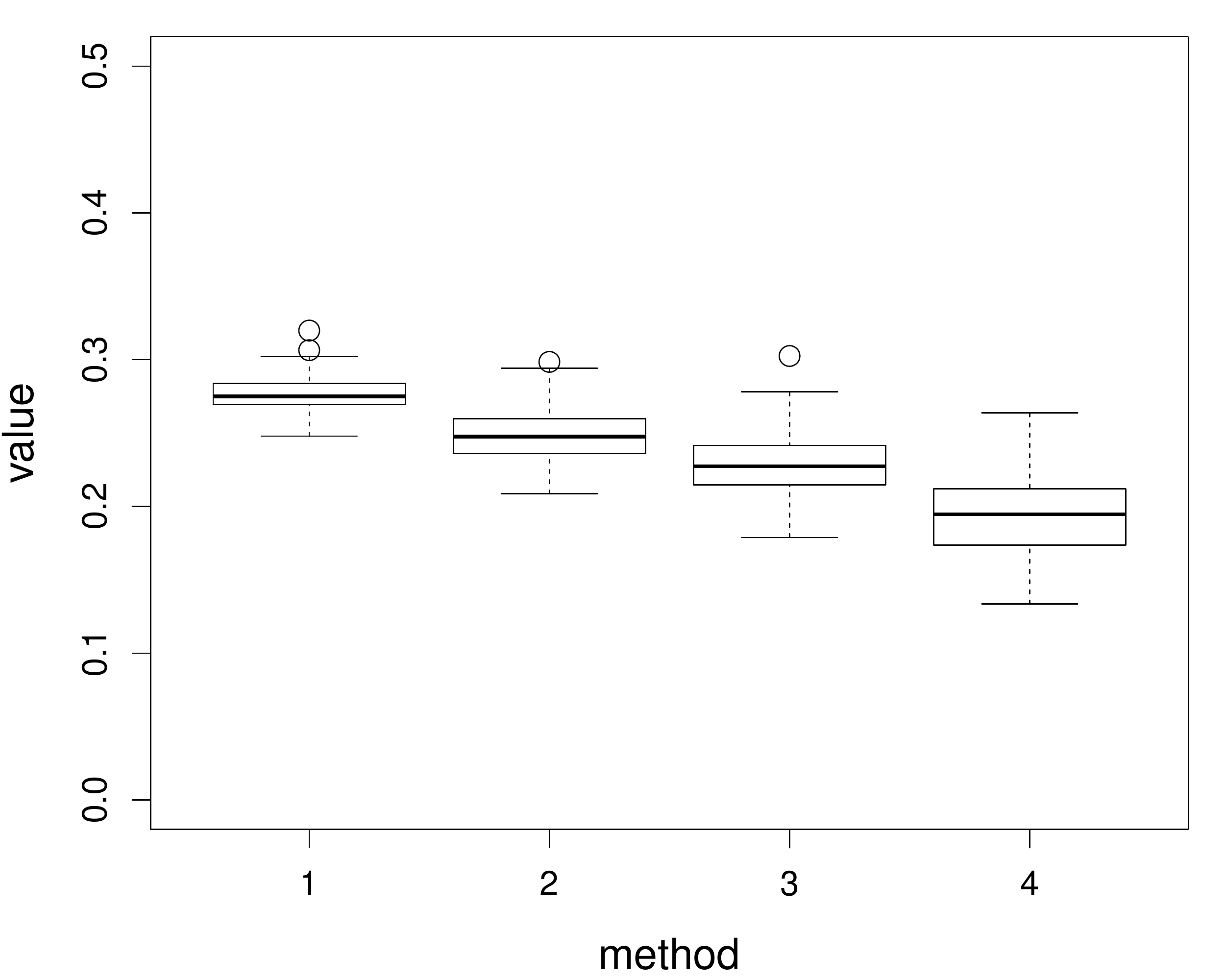} }
	\caption{Boxplots of EISE using the approximated theoretical optimal bandwidths when the primary model is (C2) and the secondary models are (a) $X\sim N(0,1)$, $U\sim \textrm{Laplace}(0, \,\sigma_u/\sqrt{2})$, $\lambda=0.8$; (b) $X\sim N(0,1)$, $U\sim \textrm{Laplace}(0, \,\sigma_u/\sqrt{2})$, $\lambda=0.9$; (c) $X\sim N(0,1)$, $U\sim N(0, \, \sigma_u^2)$, $\lambda=0.8$; (d) $X\sim \textrm{Uniform}(-2, 2)$, $U\sim \textrm{Laplace}(0, \,\sigma_u/\sqrt{2})$, $\lambda=0.8$. Method 1, 2, 3, 4 correspond to $\tilde p_1(y|x)$, $\tilde p_2(y|x)$, $\hat p_3(y|x)$, and $\hat p_4(y|x)$, respectively.} 
	\label{Sim2:box}
\end{figure}

\begin{figure}
	\centering
	\setlength{\linewidth}{0.45\linewidth}
	\subfigure[]{ \includegraphics[width=\linewidth]{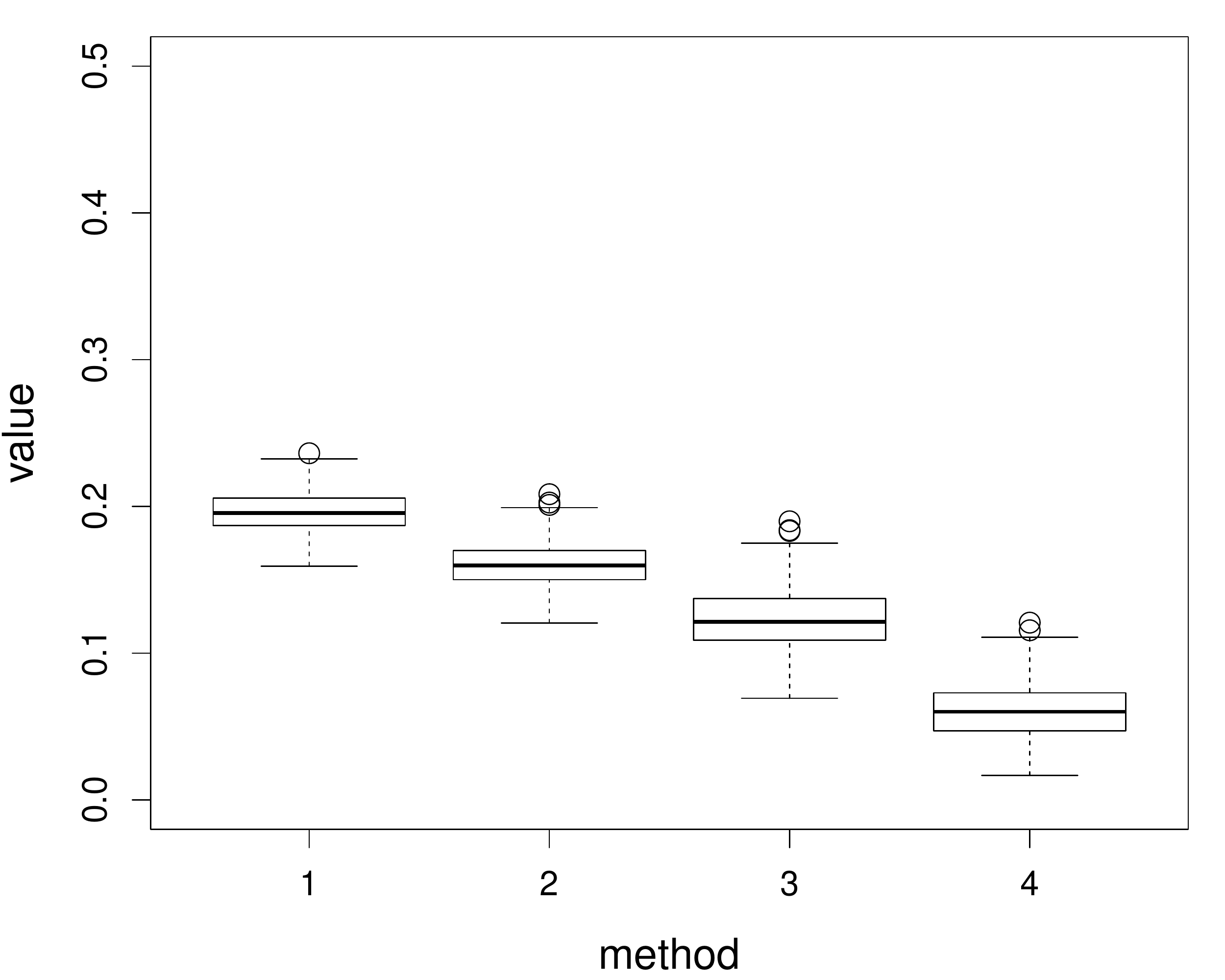} }
	\subfigure[]{ \includegraphics[width=\linewidth]{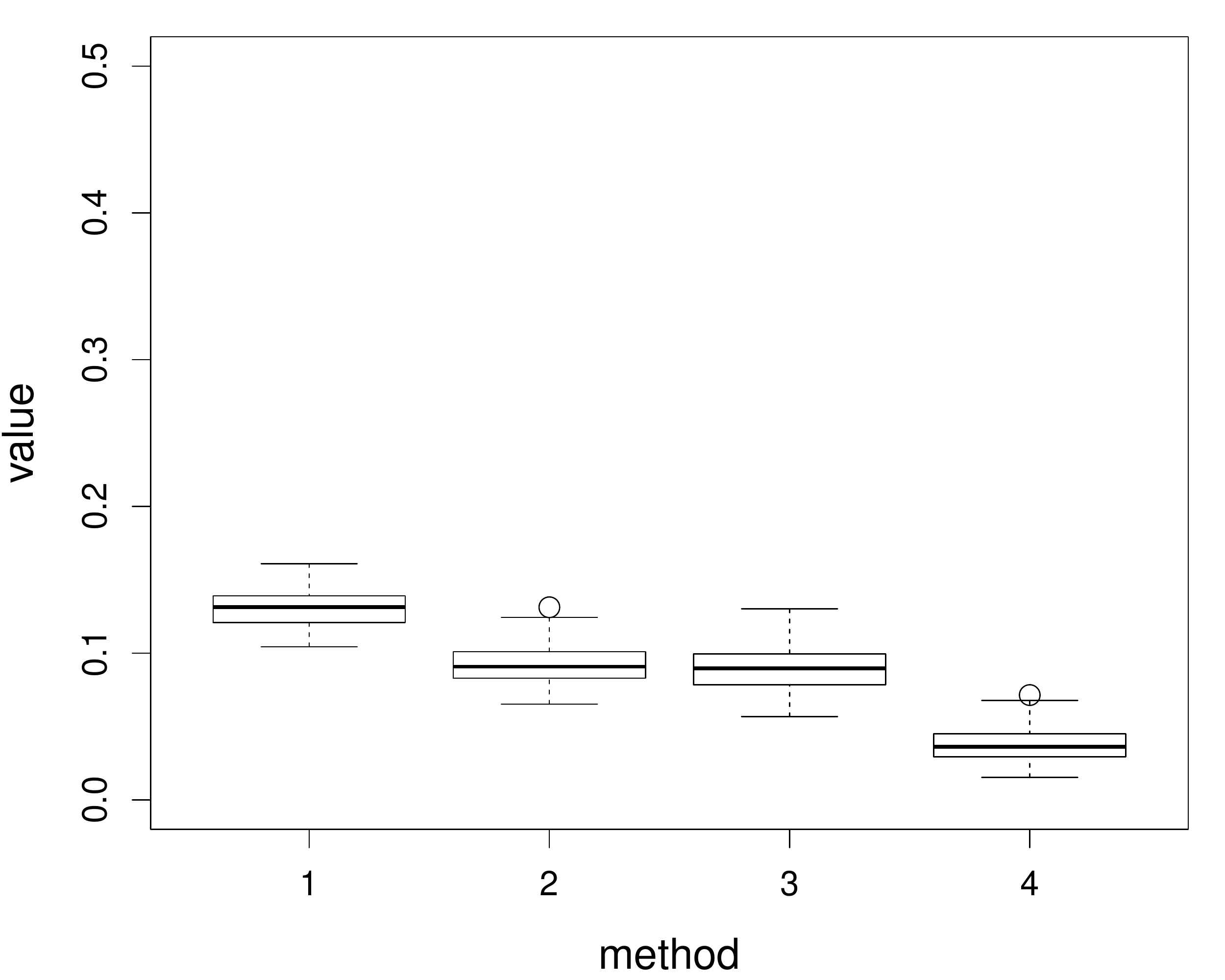} }\\
	\subfigure[]{ \includegraphics[width=\linewidth]{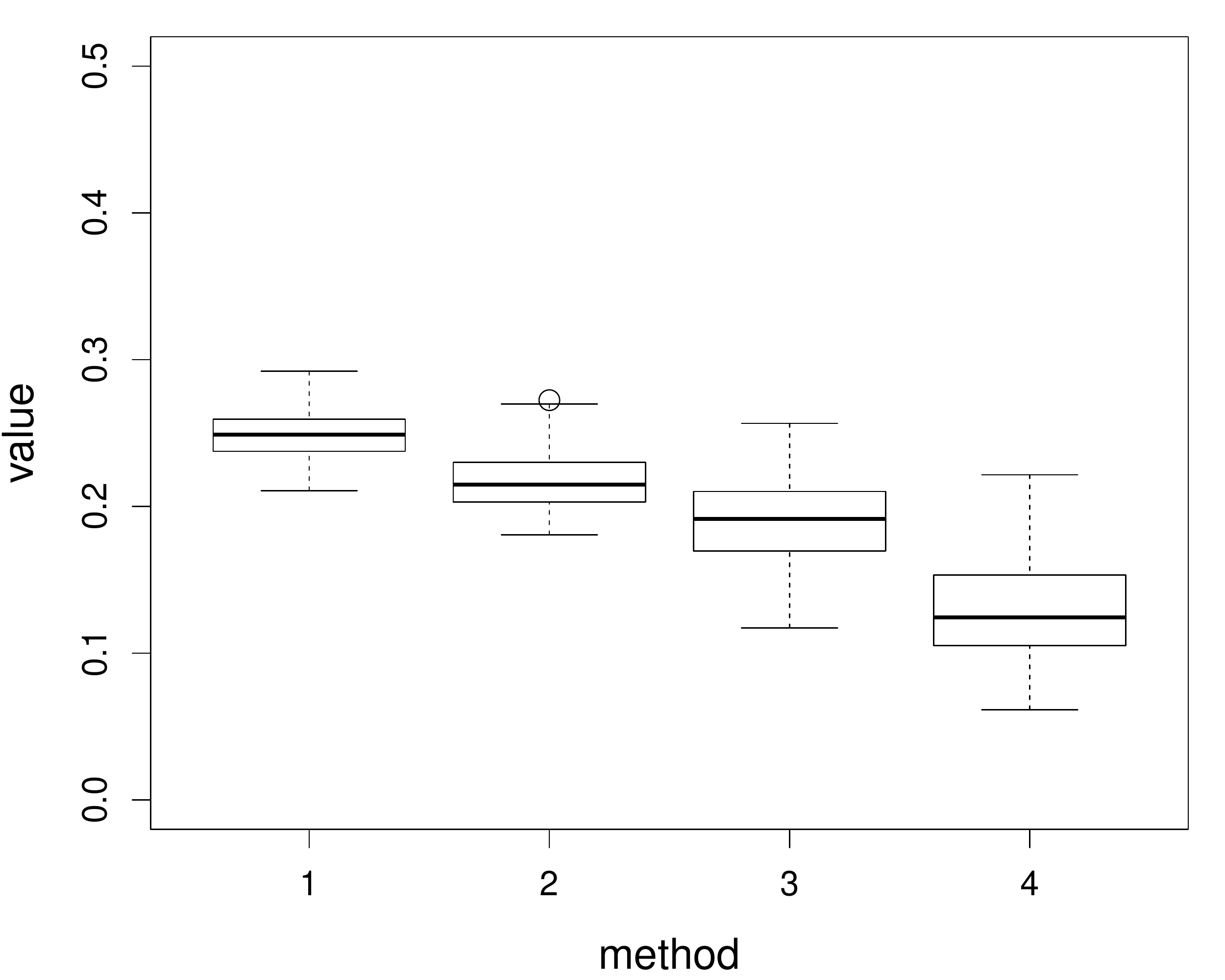} }
	\subfigure[]{ \includegraphics[width=\linewidth]{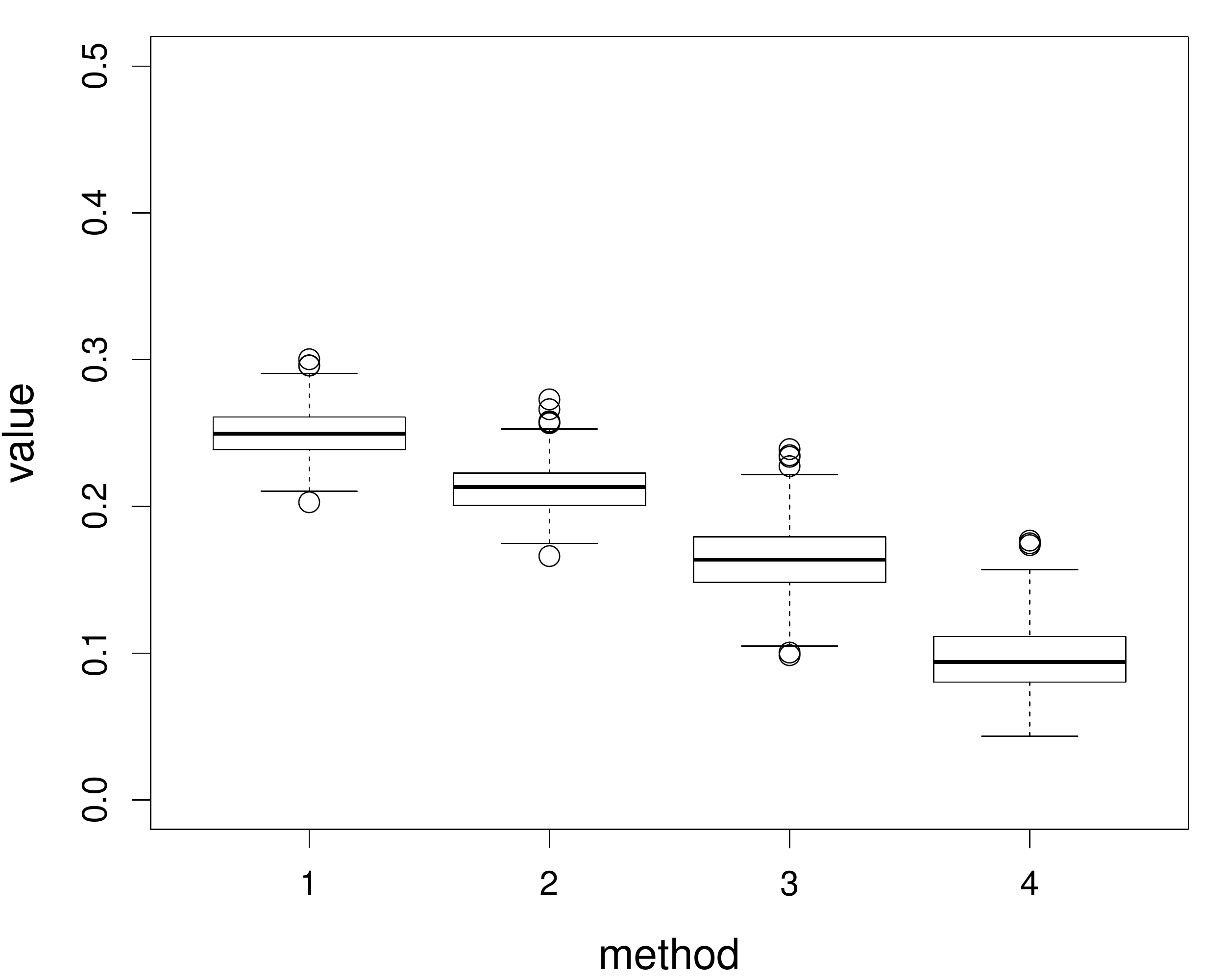} }
	\caption{Boxplots of EISE using the approximated theoretical optimal bandwidths when the primary model is (C3) and the secondary models are (a) $X\sim N(0,1)$, $U\sim \textrm{Laplace}(0, \,\sigma_u/\sqrt{2})$, $\lambda=0.8$; (b) $X\sim N(0,1)$, $U\sim \textrm{Laplace}(0, \,\sigma_u/\sqrt{2})$, $\lambda=0.9$; (c) $X\sim N(0,1)$, $U\sim N(0, \, \sigma_u^2)$, $\lambda=0.8$; (d) $X\sim \textrm{Uniform}(-2, 2)$, $U\sim \textrm{Laplace}(0, \,\sigma_u/\sqrt{2})$, $\lambda=0.8$. Method 1, 2, 3, 4 correspond to $\tilde p_1(y|x)$, $\tilde p_2(y|x)$, $\hat p_3(y|x)$, and $\hat p_4(y|x)$, respectively.} 
	\label{Sim3:box}
\end{figure}

To quantitatively compare different density estimators, we use the EISE defined in (\ref{eq:EISE:dens}) as the metric to assess the quality of estimates. Figures~\ref{Sim1:box}--\ref{Sim3:box} present boxplots of EISE under three primary model configurations when  the approximated theoretical optimal bandwidths are used. When comparing a naive estimator with a non-naive one, one can see that adjusting for measurement error clearly leads to estimates of better quality in terms of EISE. On the other hand, $\tilde p_2(y|x)$ is less compromised by measurement error than $\tilde p_1(y|x)$ is. This can be mostly explained by the findings in \citet{Hyndman.etal1996} and \citet{hansen2004nonparametric}, which suggest that $\tilde p_2(y|x)$ often outperforms $\tilde p_1(y|x)$ as estimators for $p^*(y|x)$. Even though estimating $p^*(y|x)$ well typically does not imply reliable estimation of $p(y|x)$, a less satisfactory estimator for the former usually leads to less reliable estimation for the latter. Intuition suggests that correcting a better estimator of $p^*(y|x)$ for measurement error can yield a better non-naive estimator of $p(y|x)$. This intuition is supported by the observations from Figures~\ref{Sim1:box}--\ref{Sim3:box} that the most reliable estimator for $p(y|x)$ among the four is $\hat p_4(y|x)$ in all considered simulation settings. The benefit of the two-step estimator $\hat p_4(y|x)$ compared to $\hat p_3(y|x)$ is more evident when the mean function is linear (see panel (d) in Figure~\ref{Sim1:box} in contrast to panel (d) in Figure~\ref{Sim3:box}). This can serve as evidence for that adjusting for the mean in the first step then estimating the residual conditional density in the second step leads to better estimates for $p(y|x)$ than a one-step estimator; and this improvement is more noticeable when the dependence of $Y$ on the covariate is mostly explained by the conditional mean that can be well estimated in the first step. As pointed out in Section~\ref{s:theorems4}, $\hat p_4(y|x)$ does not offer any gain asymptotically when compared with $\hat p_3(y|x)$ if $m(x)$ is a constant function of $x$. This is clearly also the case in terms of their finite sample performance. To demonstrate this point, we include in Appendix D boxplots of EISE associated with these two estimators and their naive counterparts when data are generated according to a primary model with a constant $m(x)$. From there, one can see that $\hat p_3(y|x)$ behaves very similarly as $\hat p_4(x|y)$, and the former is less variable than the latter when the fully data-driven bandwidths are used.

\begin{figure}
	\centering
	\setlength{\linewidth}{0.45\linewidth}
	\psfrag{Sample}{Monte Carlo Replicate}
	\subfigure[]{ \includegraphics[width=\linewidth]{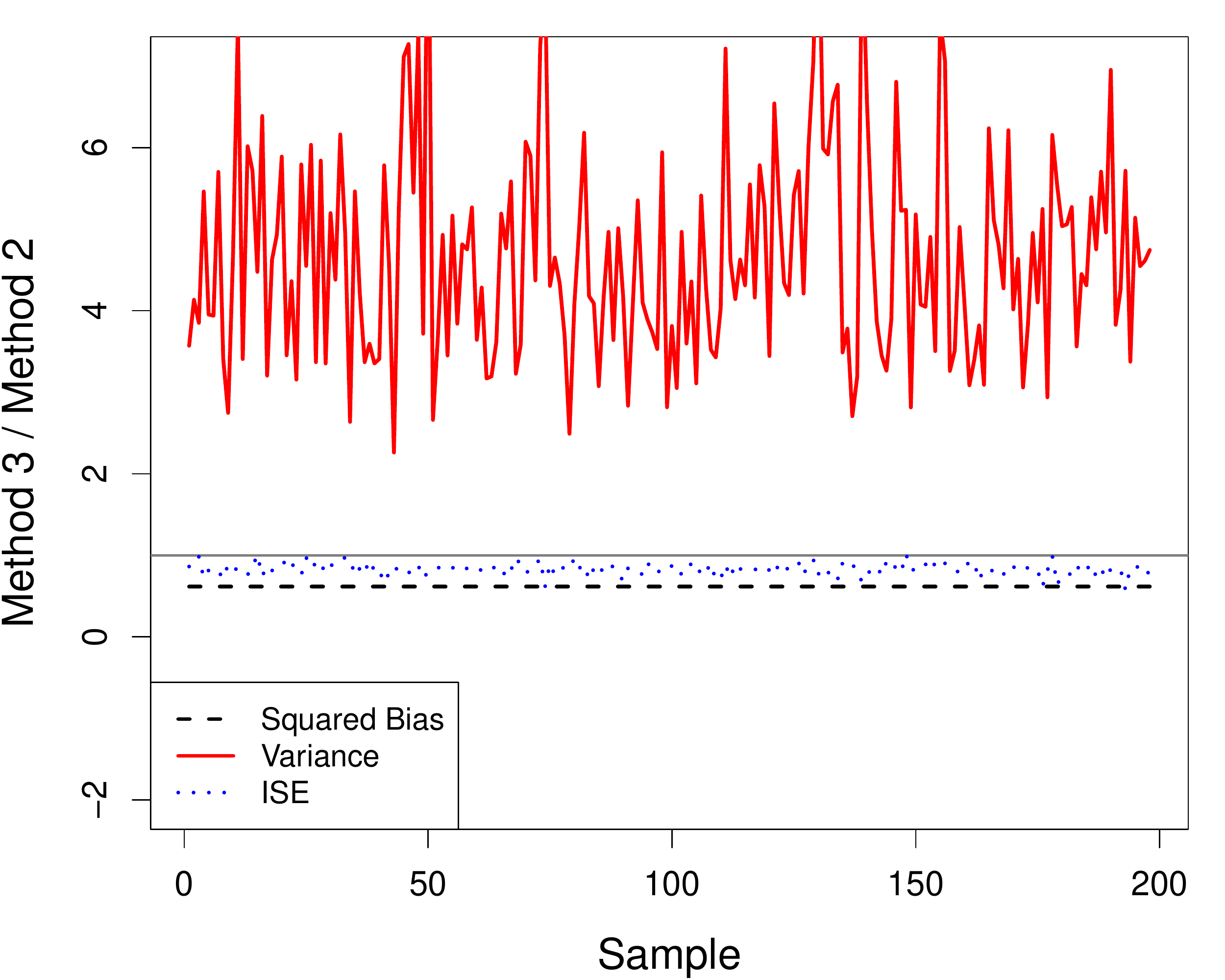} }
	\subfigure[]{ \includegraphics[width=\linewidth]{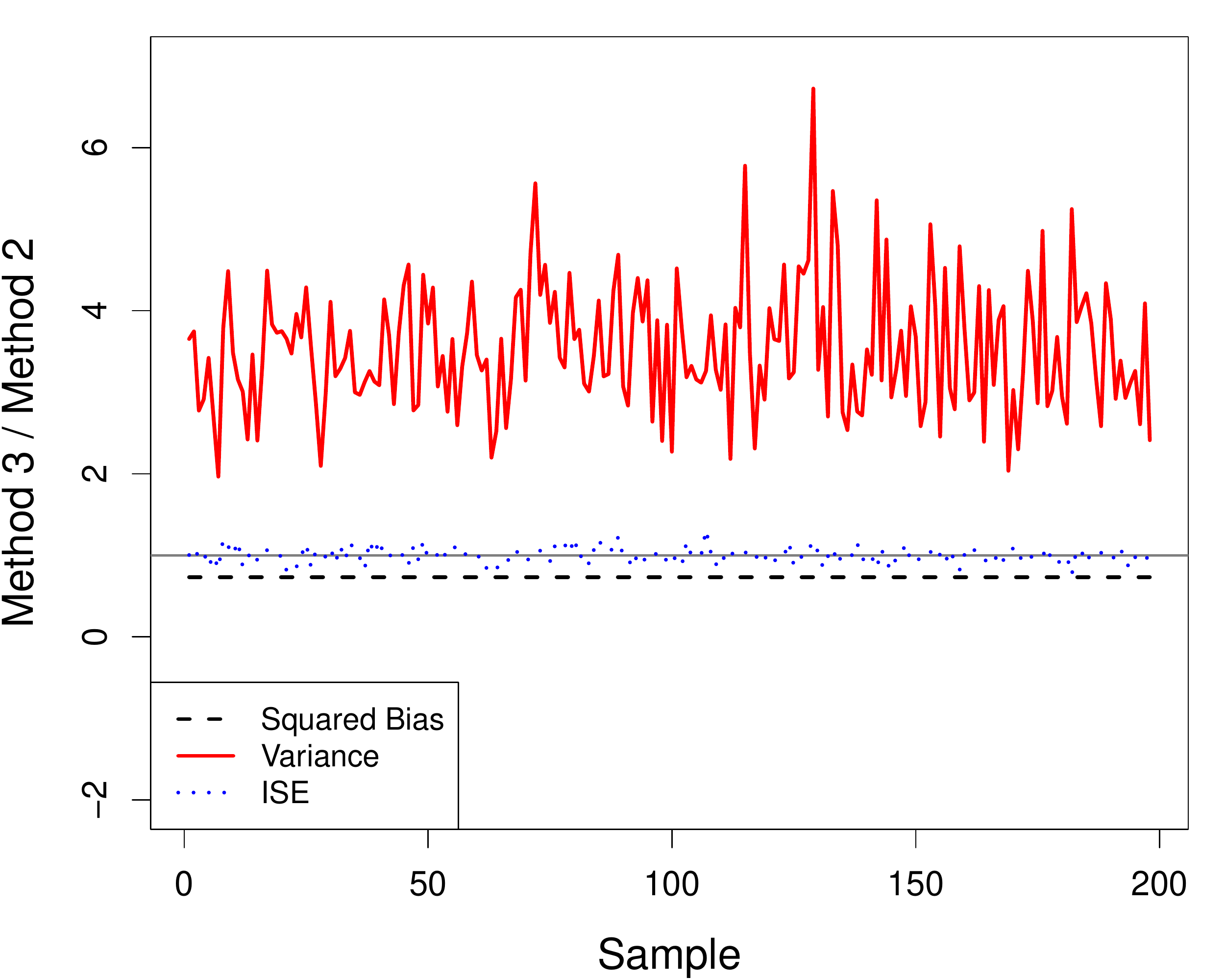} }
	\caption{The ratio of the empirical integrated squared bias (black dashed lines) associated with $\hat p_3(y|x)$ across 200 Monte Carlo replicates over that associated with $\tilde p_2(y|x)$, the ratio of the empirical integrated variance (red solid lines) between them, and the ratio of EISE (blue dotted lines) between them, using the approximated theoretical optimal bandwidths when the primary model is (C1) and the secondary models are (a) $X\sim N(0,1)$, $U\sim \textrm{Laplace}(0, \,\sigma_u/\sqrt{2})$, $\lambda=0.8$; (b) $X\sim N(0,1)$, $U\sim \textrm{Laplace}(0, \,\sigma_u/\sqrt{2})$, $\lambda=0.9$. Method 2 and 3 correspond to $\tilde p_2(y|x)$ and $\hat p_3(y|x)$,  respectively. The black horizontal solid lines are the reference lines at value one.} 
	\label{f:decomp}
\end{figure}

Although $\hat p_3(y|x)$ substantially improves over $\tilde p_1(y|x)$, $\tilde p_2(y|x)$ can perform similarly as $\hat p_3(y|x)$ in terms of EISE, especially when error contamination is mild (see, for instance, panel (b) in Figures~\ref{Sim1:box}--\ref{Sim3:box} where $\lambda=0.9$). To compare $\tilde p_2(y|x)$ and $\hat p_3(y|x)$ more closely in regard to bias and variance, we decompose EISE in (\ref{eq:EISE:dens}) as follows, where the additional subscript, $\textrm{MC}(\in \{1, \ldots, 200\})$, is added to signify that, under each simulation setting, there are 200 EISE's recorded for a density estimator, and, for each point $(x_k, y_j)$ at which the density estimate and the true densities are evaluated, there are 200 realizations of a density estimator, 
\begin{align}
\EISE_{\hbox {\tiny MC}} & = \sum_{j=1}^{\calM'}\sum_{k=0}^{\calM} \left\{ \hat{p}_{\hbox {\tiny MC}}(y_j|x_k)-p(y_j|x_k) \right\}^2 f_{\hbox {\tiny $X$}}(x_k)\Delta\Delta' \nonumber\\
& = \sum_{j=1}^{\calM'}\sum_{k=0}^{\calM} \left\{ \hat{p}_{\hbox {\tiny MC}}(y_j|x_k)-\bar{p}(y_j|x_k) \right\}^2 f_{\hbox {\tiny $X$}}(x_k)\Delta\Delta' \label{eq:intvar}\\
& + \sum_{j=1}^{\calM'}\sum_{k=0}^{\calM} \left\{ \bar{p}(y_j|x_k)-p(y_j|x_k) \right\}^2 f_{\hbox {\tiny $X$}}(x_k)\Delta\Delta'\label{eq:intbias2} \\
& + 2\sum_{j=1}^{\calM'}\sum_{k=0}^{\calM} \left\{ \hat{p}_{\hbox {\tiny MC}}(y_j|x_k)-\bar p(y_j|x_k) \right\}\left\{ \bar{p}(y_j|x_k)-p(y_j|x_k) \right\} f_{\hbox {\tiny $X$}}(x_k)\Delta\Delta' \nonumber,
\end{align}
where $\bar p(y_j|x_k)=\sum_{\hbox {\tiny MC}=1}^{200}\hat{p}_{\hbox {\tiny MC}}(y_j|x_k)/200$ for each point $(x_k, y_j)$. With $\bar p(y_j|x_k)$ being the empirical mean of an estimator evaluated at $(x_k, y_j)$, (\ref{eq:intvar}) can be interpreted as an empirical integrated variance (EIV) associated with a considered estimator, and (\ref{eq:intbias2}) can be viewed as an empirical integrated squared bias (EISB) of the estimator. By construction, the EIV in (\ref{eq:intvar}) varies across different MC replicates, whereas the EISB in (\ref{eq:intbias2}) does not. Figure~\ref{f:decomp} shows the ratio of the EISB of $\hat p_3(y|x)$ over that of $\tilde p_2(y|x)$ under the model setting for panels (a) and (b) in Figure~\ref{Sim1:box}. The ratio of EIV of $\hat p_3(y|x)$ over that of $\tilde p_2(y|x)$, and the ratio of the two EISE's are also depicted in Figure~\ref{f:decomp}. Recall that the true model settings under panels (a) and (b) in each aforementioned figure are the same except for the reliability ratio $\lambda$, with $\lambda=0.8$ in (a) and $\lambda=0.9$ in (b). Under both levels of error contamination, one can see in Figure~\ref{f:decomp} that $\textrm{EISB}(\hat p_3)/\textrm{EISB}(\tilde p_2)<1$ and $\textrm{EIV}(\hat p_3)/\textrm{EIV}(\tilde p_2)>1$, suggesting that $\hat p_3(y|x)$ does eliminate some bias in the naive estimator $\tilde p_2(y|x)$ at the price of an inflated variance. This price is lower when the error contamination is milder, yielding lower ratios of EIV in (b) compared to those in (a); although milder error contamination also diminishes the amount of bias reduction in $\hat p_3(y|x)$ compared to $\tilde p_2(y|x)$ since the latter is less compromised in the presence of less measurement error. These comparisons between the two estimators in EISB and EIV explain the resemblance of the estimators in terms of EISE, resulting in $\textrm{EISE}(\hat p_3)/\textrm{EISE}(\tilde p_2) \approx 1$ when $\lambda=0.9$.   

\begin{figure}
	\centering
	\setlength{\linewidth}{0.45\linewidth}
	\psfrag{Sample}{Monte Carlo Replicate}
	\subfigure[]{ \includegraphics[width=\linewidth]{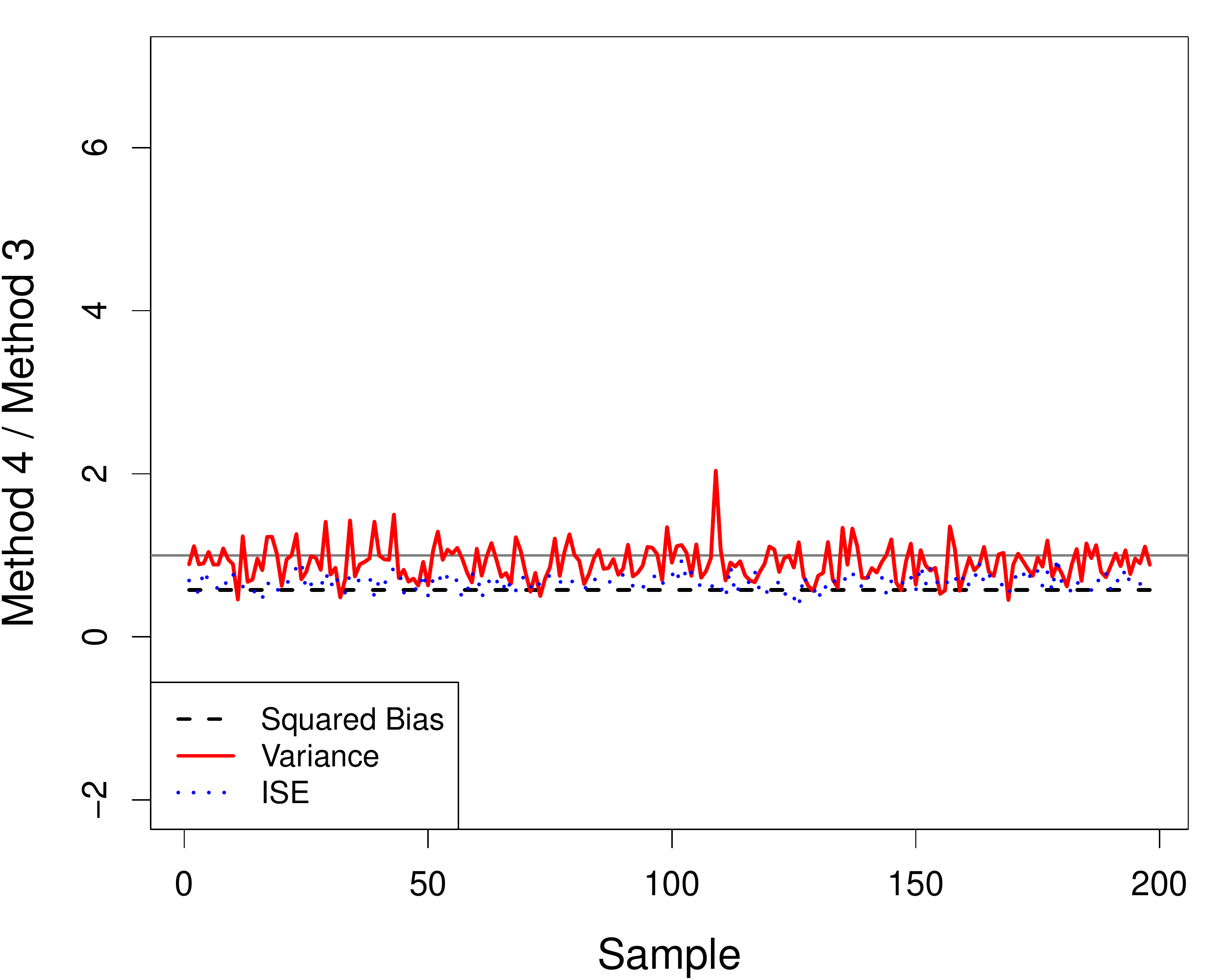} }
	\subfigure[]{ \includegraphics[width=\linewidth]{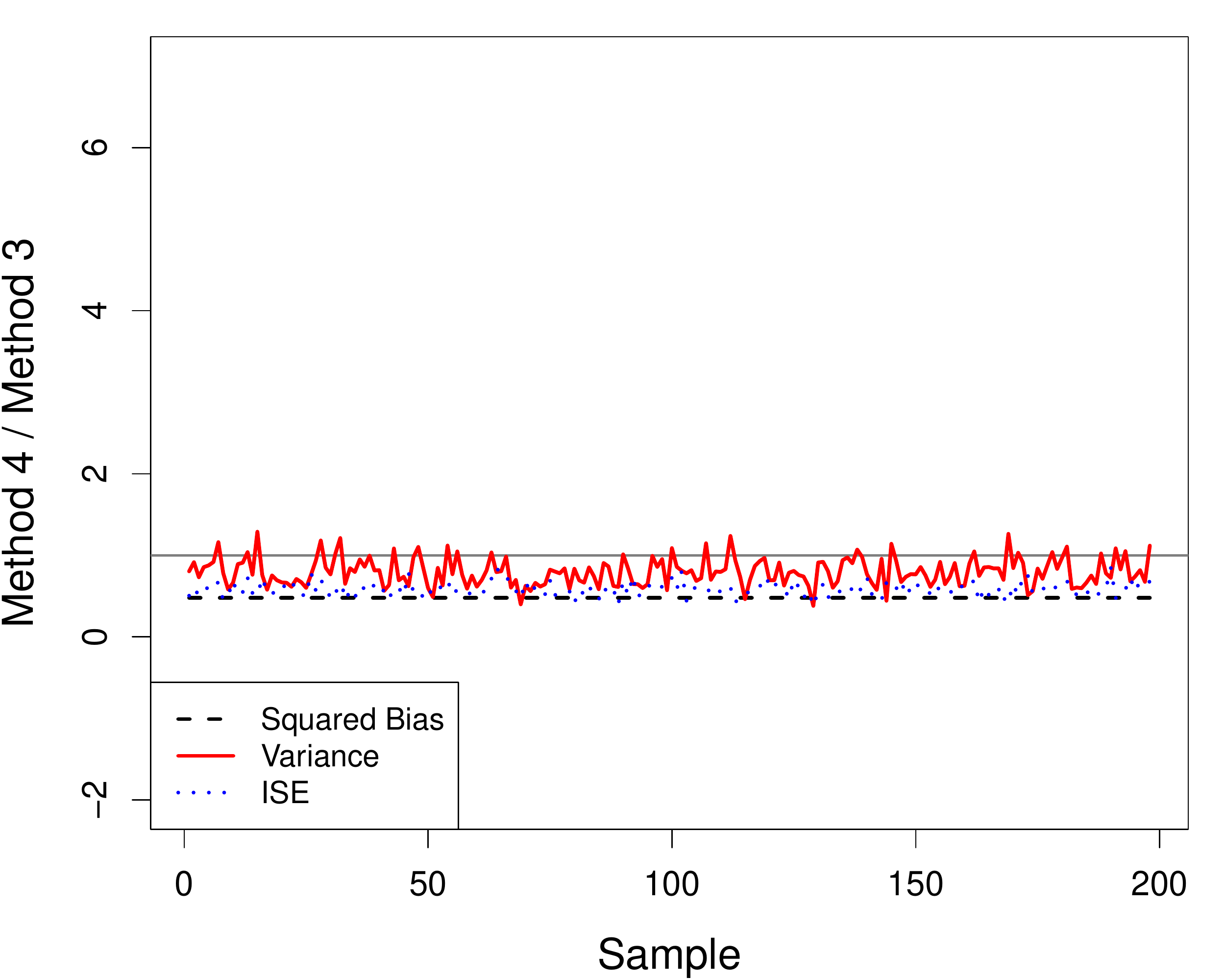} }
	\caption{The ratio of the empirical integrated squared bias (black dashed lines) associated with $\hat p_4(y|x)$ across 200 Monte Carlo replicates over that associated with $\hat p_3(y|x)$, the ratio of the empirical integrated variance (red solid lines) between them, and the ratio of EISE (blue dotted lines) between them, using the approximated theoretical optimal bandwidths when the primary model is (C1) and the secondary models are (a) $X\sim N(0,1)$, $U\sim \textrm{Laplace}(0, \,\sigma_u/\sqrt{2})$, $\lambda=0.8$; (b) $X\sim N(0,1)$, $U\sim \textrm{Laplace}(0, \,\sigma_u/\sqrt{2})$, $\lambda=0.9$. Method 3 and 4 correspond to $\hat p_3(y|x)$ and $\hat p_4(y|x)$,  respectively. The black horizontal solid lines are the reference lines at value one.} 
	\label{f:decomp34}
\end{figure}

To compare $\hat p_3(y|x)$ and $\hat p_4(y|x)$ in regard to bias and variance separately as in Figure~\ref{f:decomp}, we create Figure~\ref{f:decomp34} to present the ratios of the EISB and EIV of $\hat p_4(y|x)$ over those of $\hat p_3(y|x)$ under the model setting for panels (a) and (b) in Figure~\ref{Sim1:box}. Figure~\ref{f:decomp34} clearly suggests that bias reduction is achieved by $\hat p_4(y|x)$ compared to $\hat p_3(y|x)$ even outside of the special cases considered in Section~\ref{s:theorems4}, under which we analytically show the superiority of $\hat p_4(y|x)$ over $\hat p_3(y|x)$. 

\begin{figure}
	\centering
	\setlength{\linewidth}{0.45\linewidth}
	\subfigure[]{ \includegraphics[width=\linewidth]{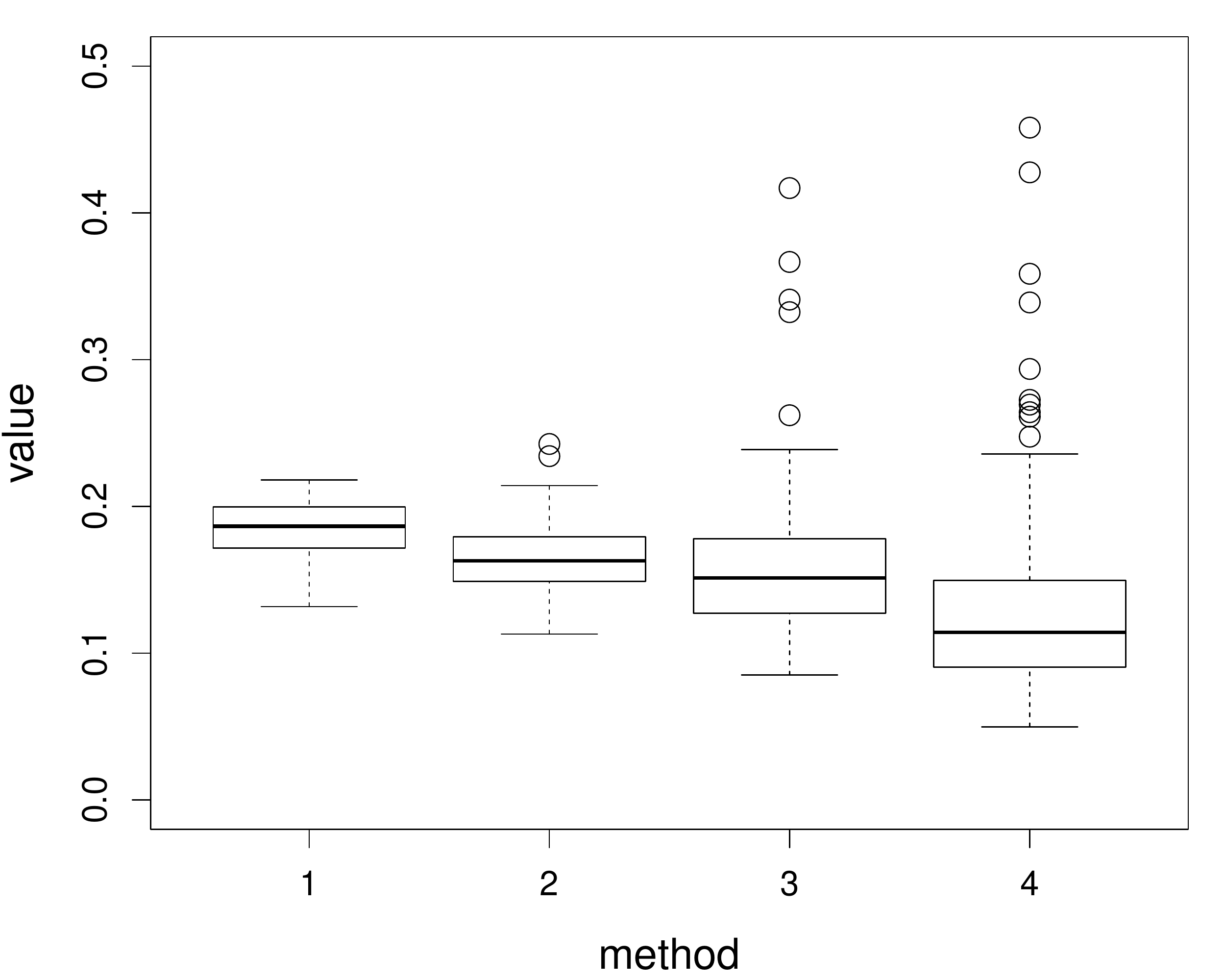} }
	\subfigure[]{ \includegraphics[width=\linewidth]{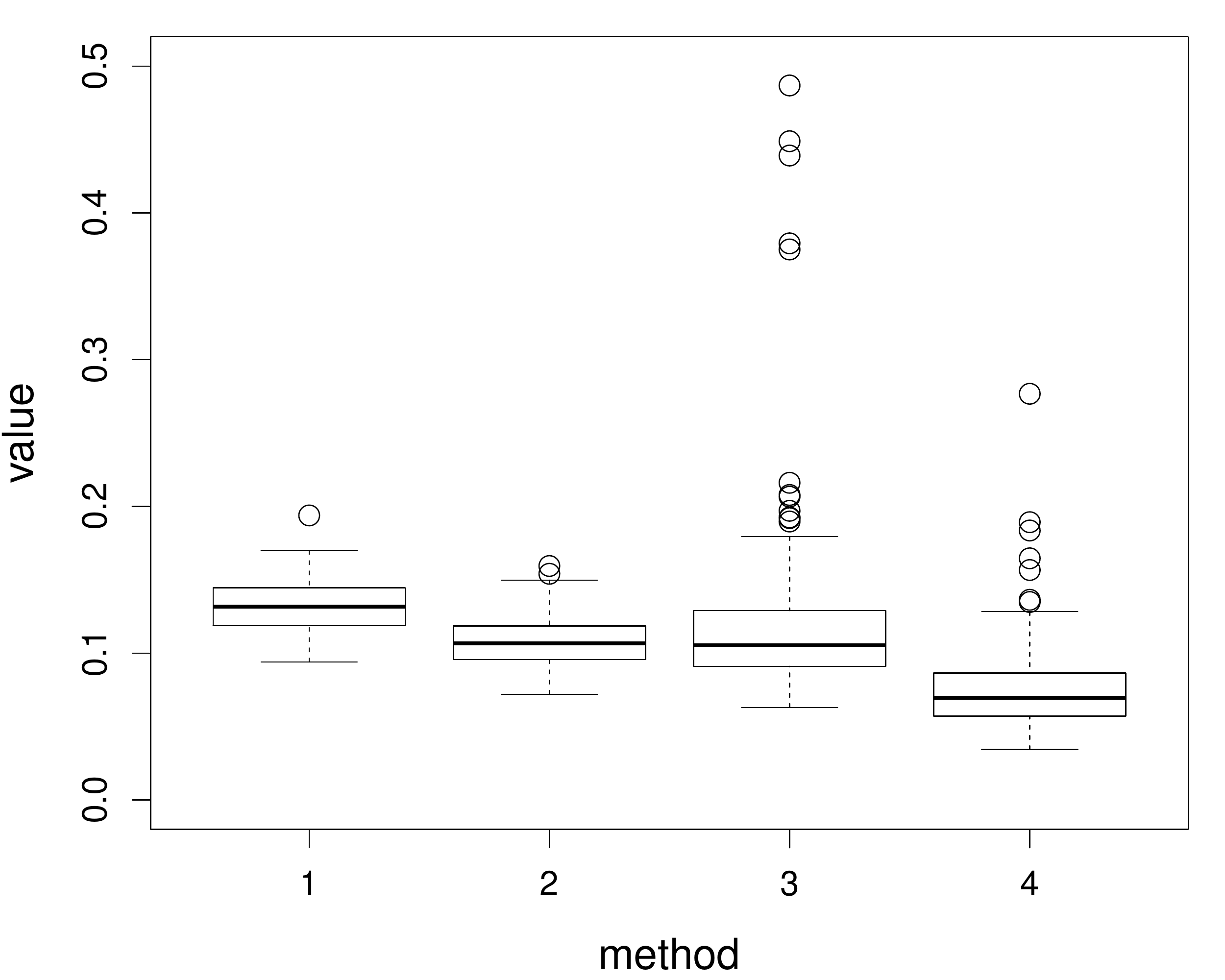} }\\
	\subfigure[]{ \includegraphics[width=\linewidth]{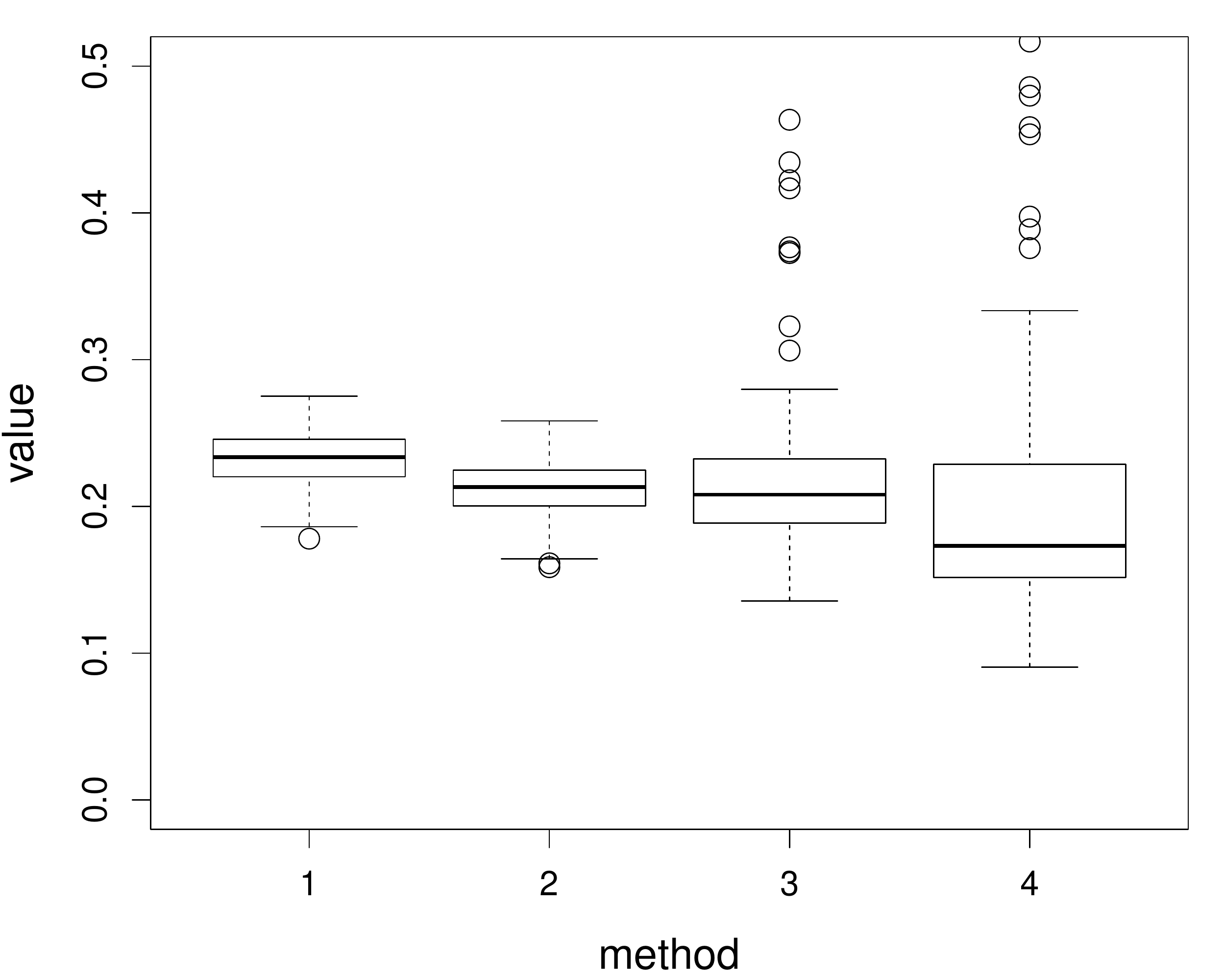} }
	\subfigure[]{ \includegraphics[width=\linewidth]{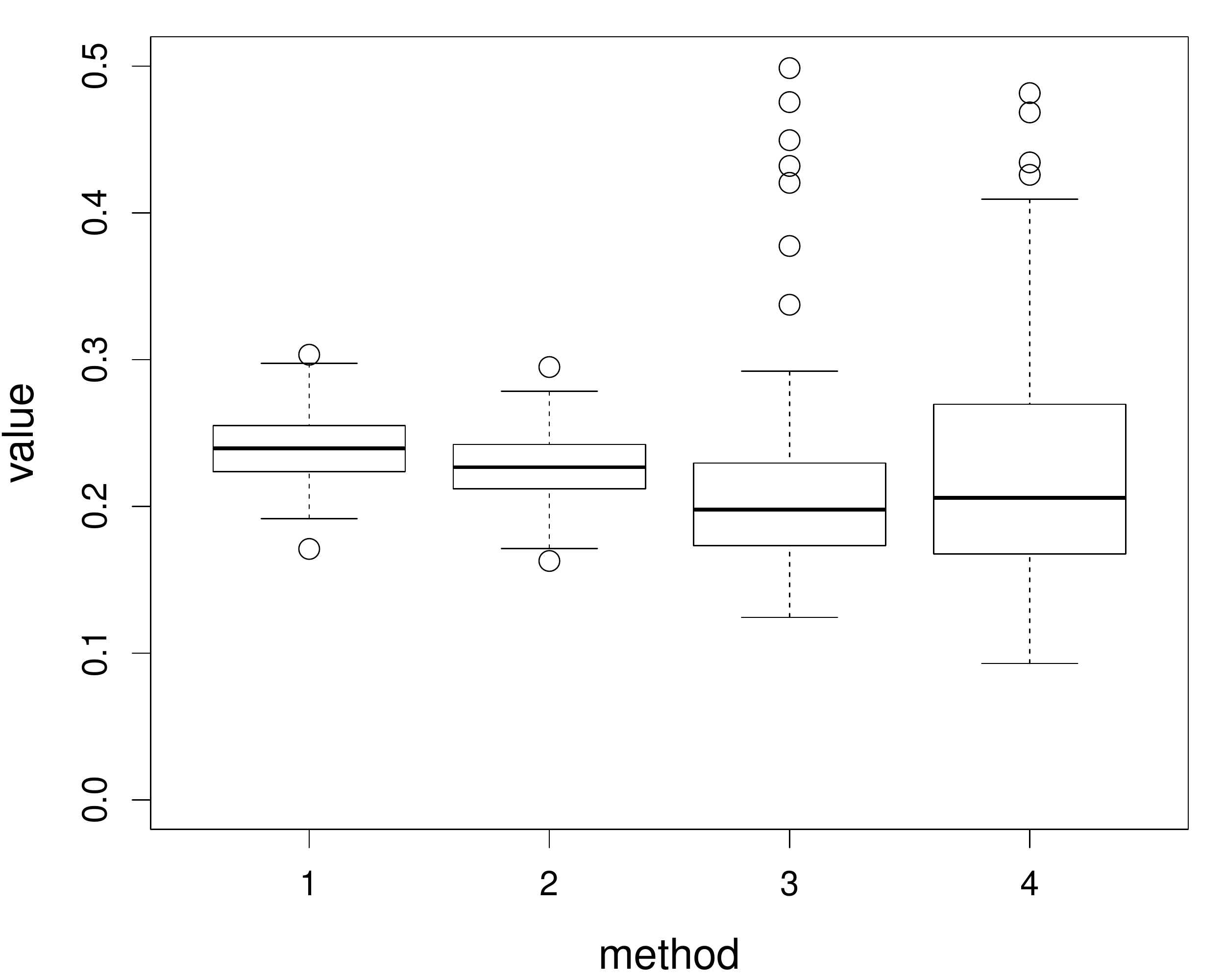} }
	\caption{Boxplots of EISE using the fully data-driven bandwidths when the primary model is (C1) and the secondary models are (a) $X\sim N(0,1)$, $U\sim \textrm{Laplace}(0, \,\sigma_u/\sqrt{2})$, $\lambda=0.8$; (b) $X\sim N(0,1)$, $U\sim \textrm{Laplace}(0, \,\sigma_u/\sqrt{2})$, $\lambda=0.9$; (c) $X\sim N(0,1)$, $U\sim N(0, \, \sigma_u^2)$, $\lambda=0.8$; (d) $X\sim \textrm{Uniform}(-2, 2)$, $U\sim \textrm{Laplace}(0, \,\sigma_u/\sqrt{2})$, $\lambda=0.8$. Method 1, 2, 3, 4 correspond to $\tilde p_1(y|x)$, $\tilde p_2(y|x)$, $\hat p_3(y|x)$, and $\hat p_4(y|x)$, respectively.} 
	\label{Sim1:box:data}
\end{figure}

\begin{figure}
	\centering
	\setlength{\linewidth}{0.45\linewidth}
	\subfigure[]{ \includegraphics[width=\linewidth]{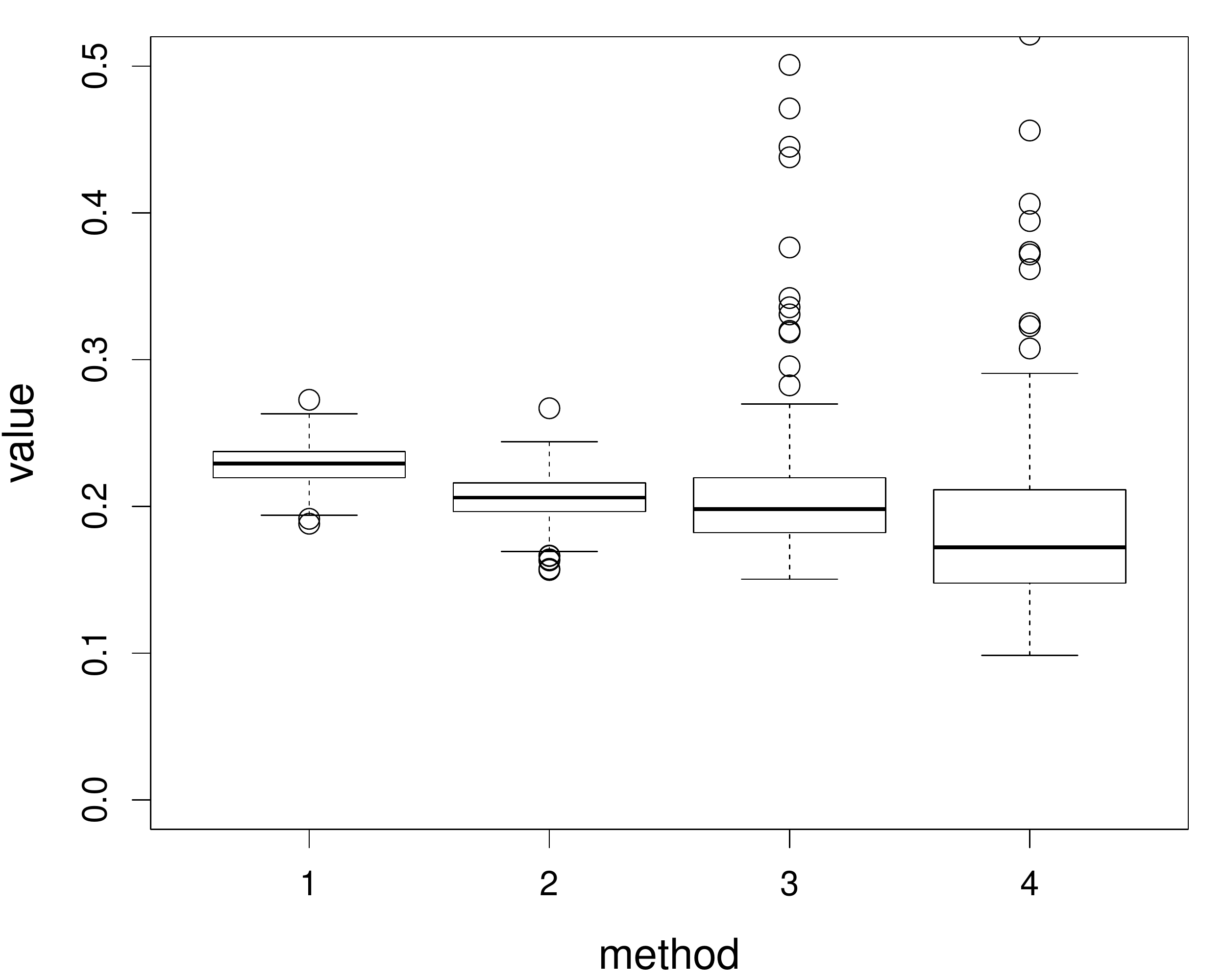} }
	\subfigure[]{ \includegraphics[width=\linewidth]{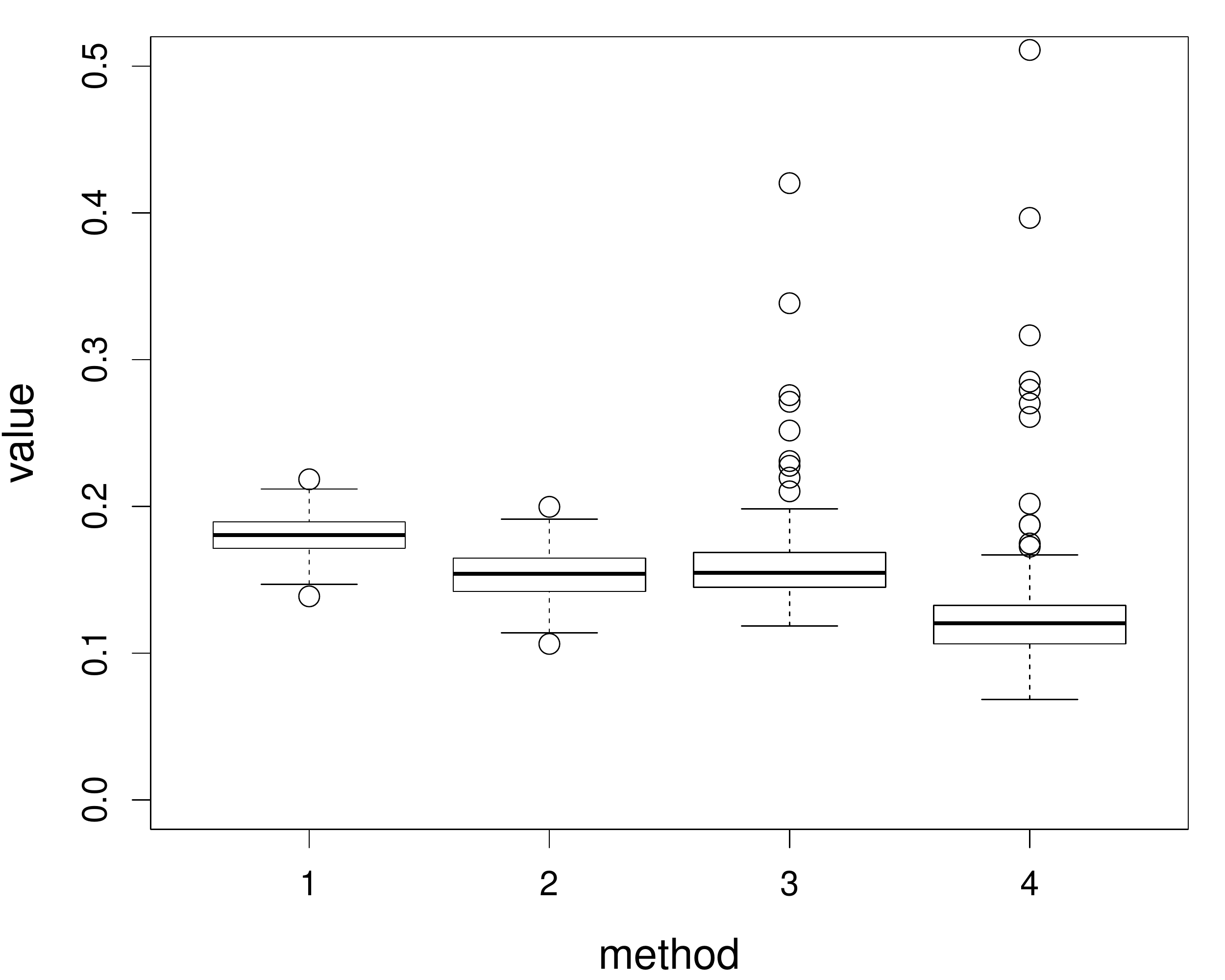} }\\
	\subfigure[]{ \includegraphics[width=\linewidth]{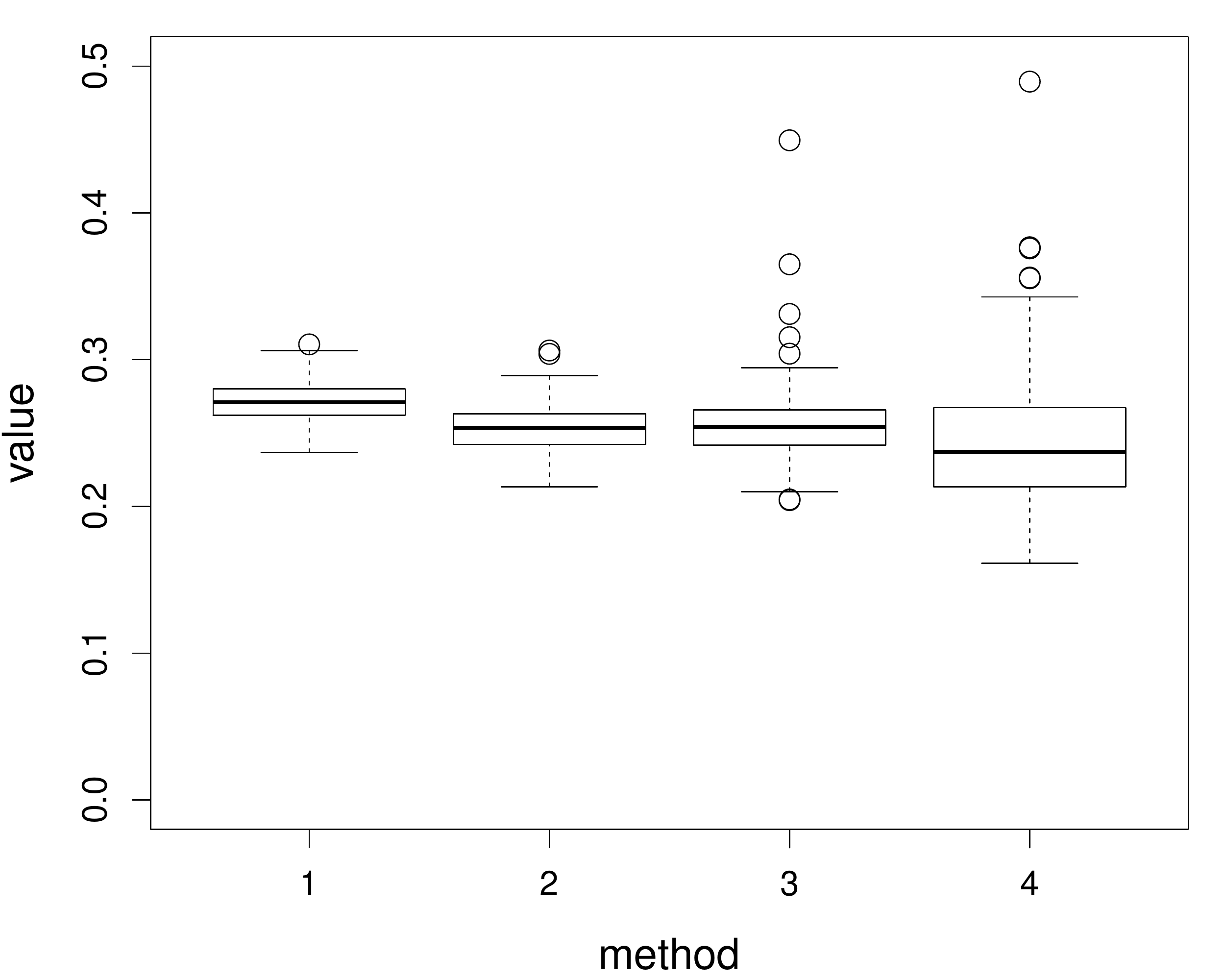} }
	\subfigure[]{ \includegraphics[width=\linewidth]{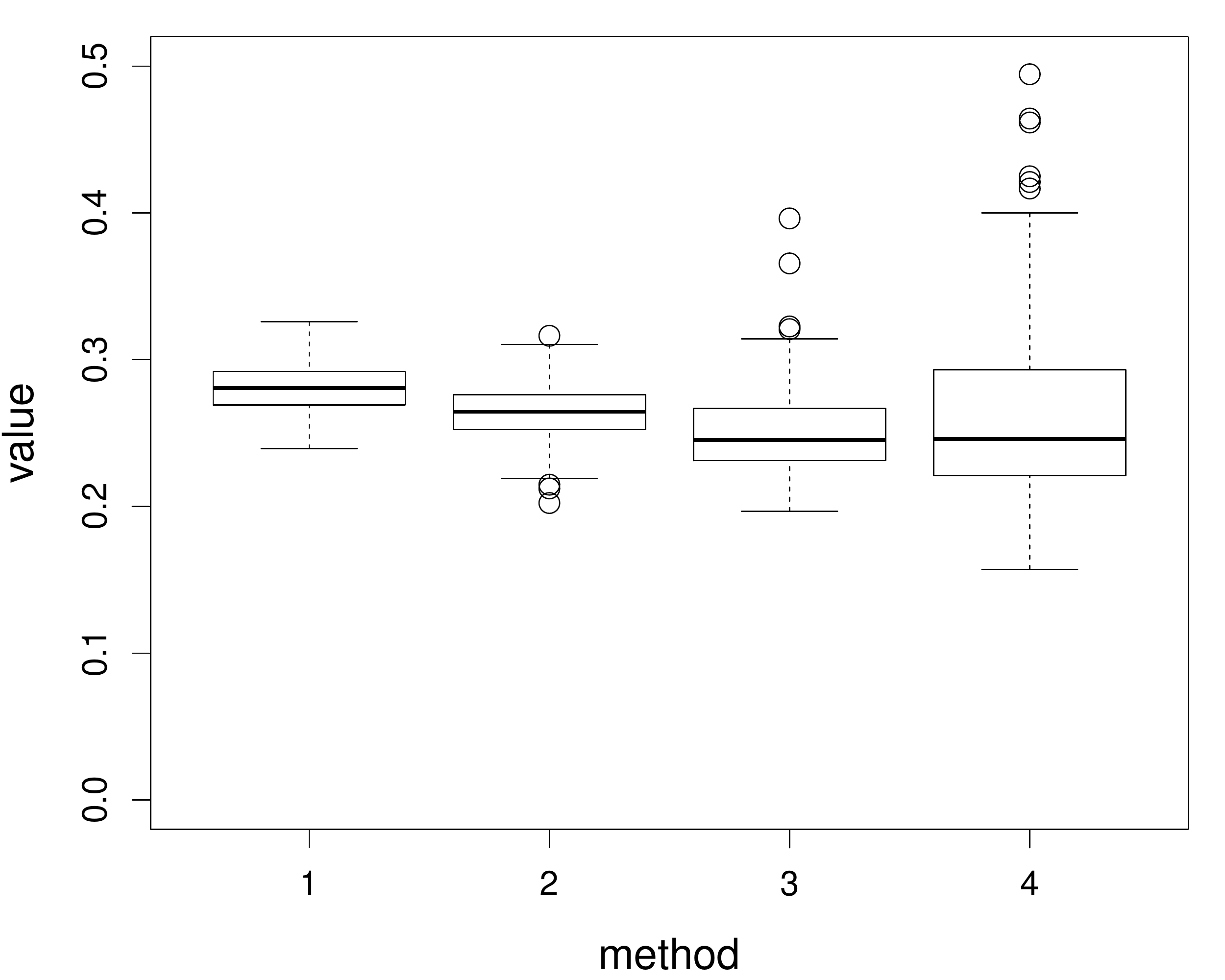} }
	\caption{Boxplots of EISE using the fully data-driven bandwidths when the primary model is (C2) and the secondary models are (a) $X\sim N(0,1)$, $U\sim \textrm{Laplace}(0, \,\sigma_u/\sqrt{2})$, $\lambda=0.8$; (b) $X\sim N(0,1)$, $U\sim \textrm{Laplace}(0, \,\sigma_u/\sqrt{2})$, $\lambda=0.9$; (c) $X\sim N(0,1)$, $U\sim N(0, \, \sigma_u^2)$, $\lambda=0.8$; (d) $X\sim \textrm{Uniform}(-2, 2)$, $U\sim \textrm{Laplace}(0, \,\sigma_u/\sqrt{2})$, $\lambda=0.8$. Method 1, 2, 3, 4 correspond to $\tilde p_1(y|x)$, $\tilde p_2(y|x)$, $\hat p_3(y|x)$, and $\hat p_4(y|x)$, respectively.} 
	\label{Sim2:box:data}
\end{figure}

\begin{figure}
	\centering
	\setlength{\linewidth}{0.45\linewidth}
	\subfigure[]{ \includegraphics[width=\linewidth]{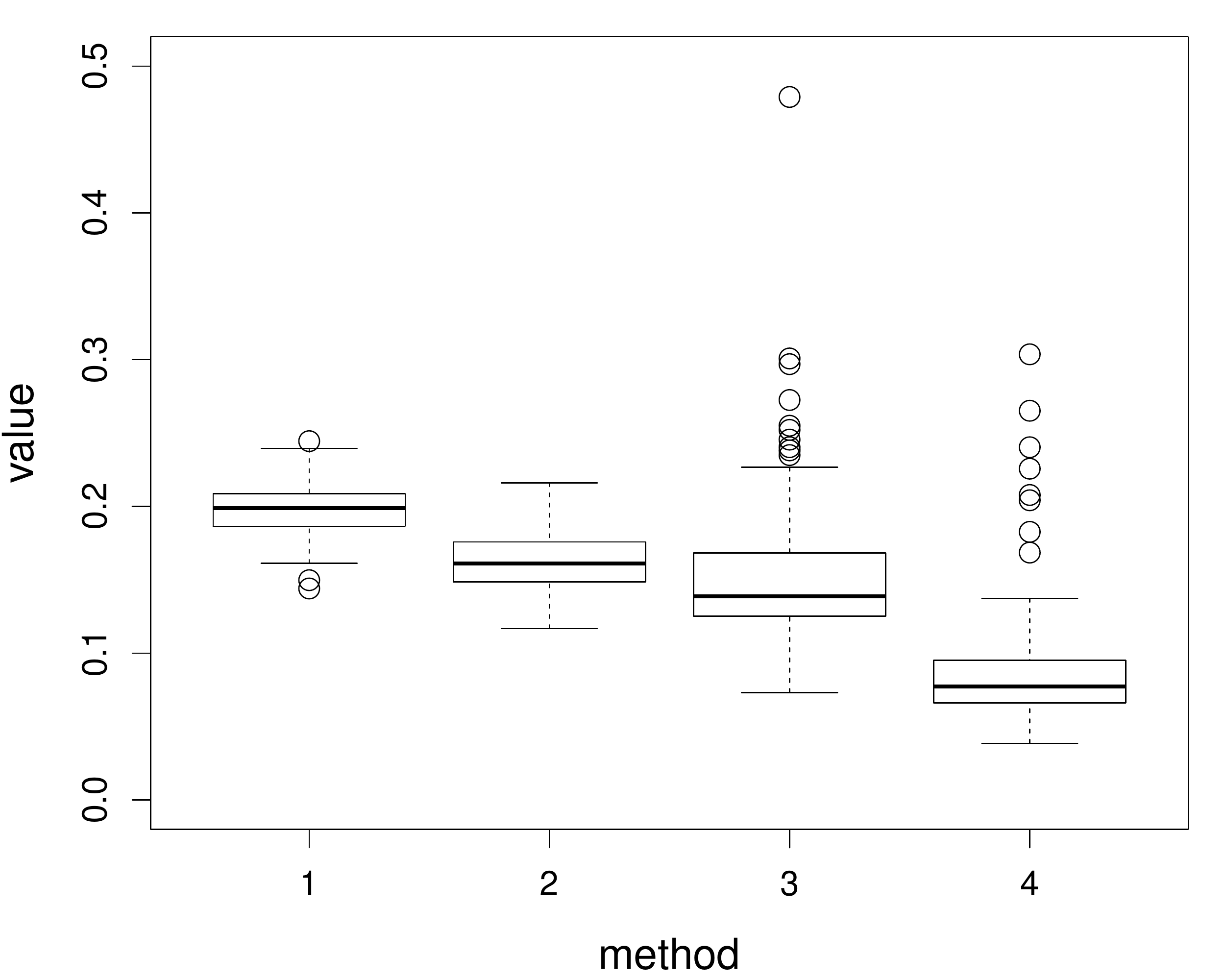} }
	\subfigure[]{ \includegraphics[width=\linewidth]{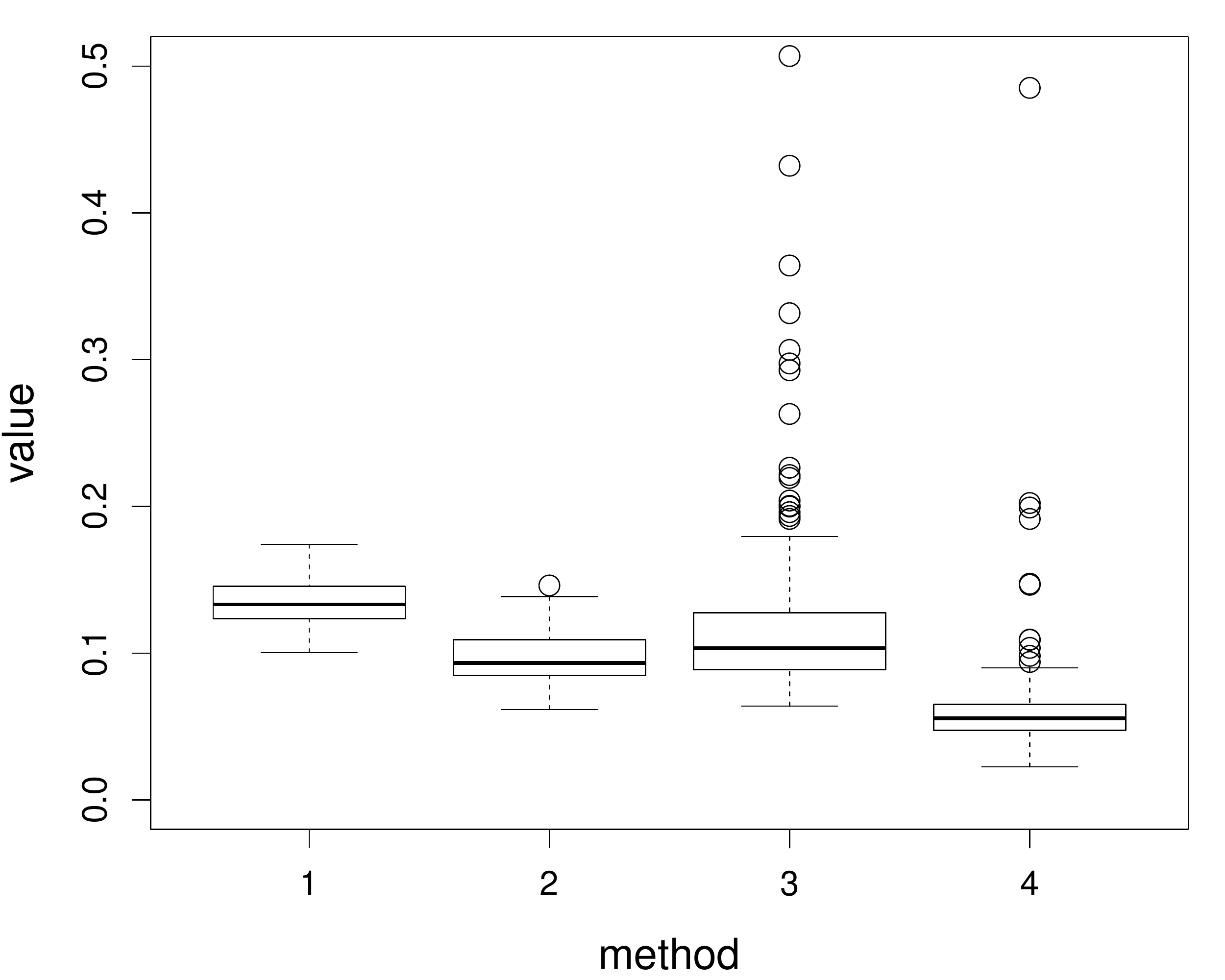} }\\
	\subfigure[]{ \includegraphics[width=\linewidth]{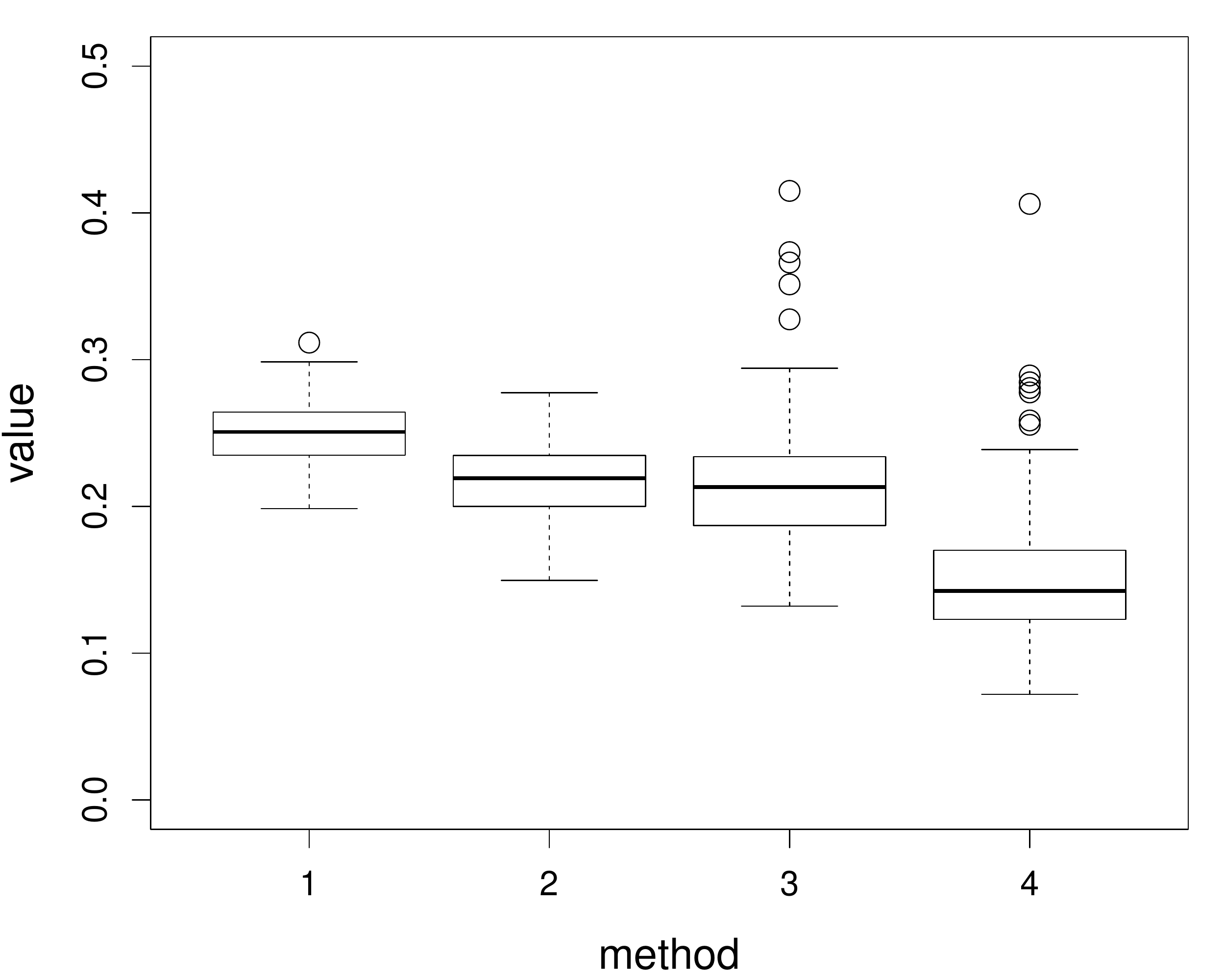} }
	\subfigure[]{ \includegraphics[width=\linewidth]{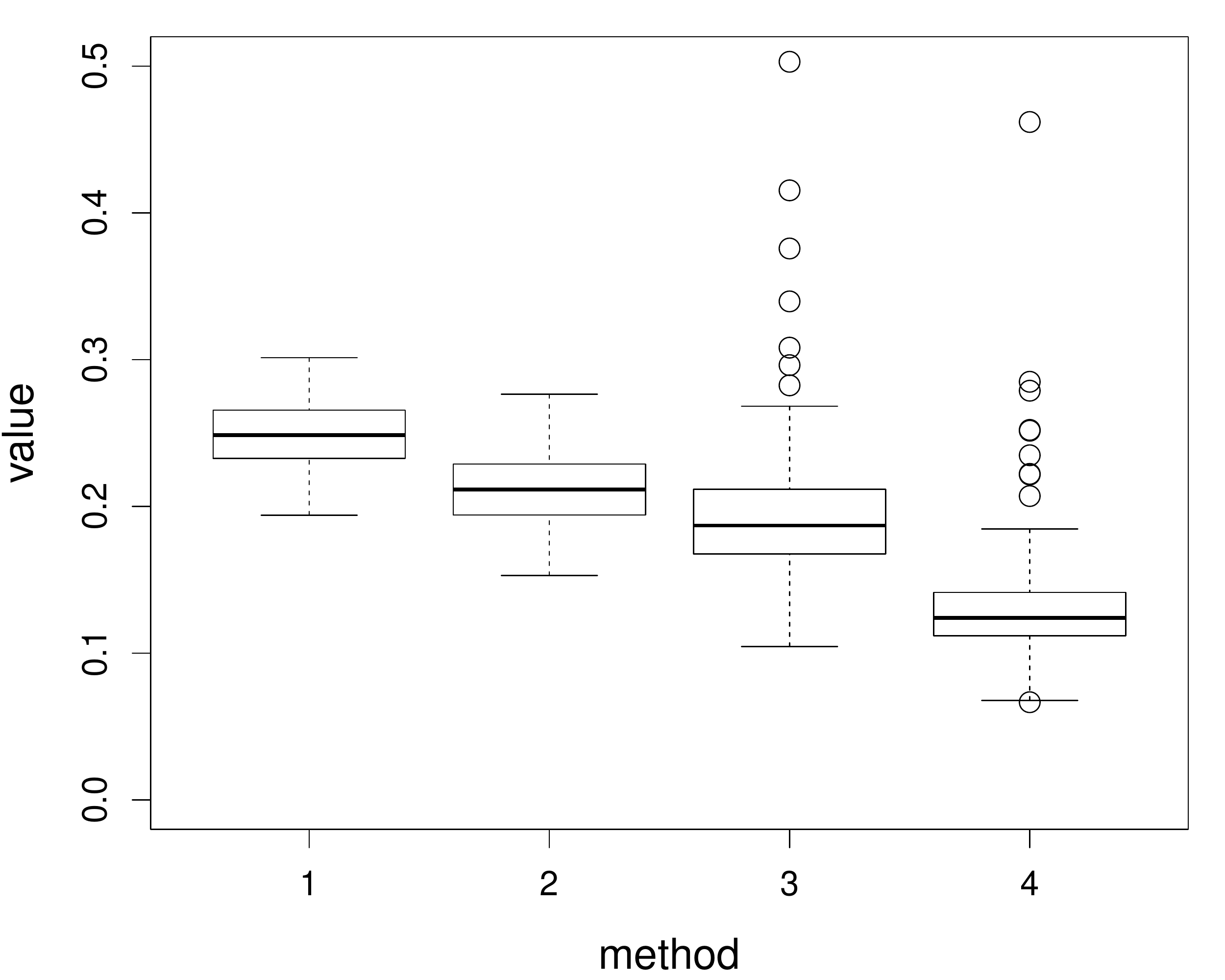} }
	\caption{Boxplots of EISE using the fully data-driven bandwidths when the primary model is (C3) and the secondary models are (a) $X\sim N(0,1)$, $U\sim \textrm{Laplace}(0, \,\sigma_u/\sqrt{2})$, $\lambda=0.8$; (b) $X\sim N(0,1)$, $U\sim \textrm{Laplace}(0, \,\sigma_u/\sqrt{2})$, $\lambda=0.9$; (c) $X\sim N(0,1)$, $U\sim N(0, \, \sigma_u^2)$, $\lambda=0.8$; (d) $X\sim \textrm{Uniform}(-2, 2)$, $U\sim \textrm{Laplace}(0, \,\sigma_u/\sqrt{2})$, $\lambda=0.8$. Method 1, 2, 3, 4 correspond to $\tilde p_1(y|x)$, $\tilde p_2(y|x)$, $\hat p_3(y|x)$, and $\hat p_4(y|x)$, respectively.} 
	\label{Sim3:box:data}
\end{figure}

Figures~\ref{Sim1:box:data}--\ref{Sim3:box:data} provide boxplots of EISE associated with density estimates when the fully data-driven bandwidths are used. Table~\ref{Sim1:table:data} presents medians and interquartile ranges of the EISE depicted in these figures. All patterns described earlier are also observed here, implying great potential of the proposed bandwidth selection methods to approximate theoretical optimal bandwidths. It is not surprising to see increased variability across all estimates now, with more uncertainty involved in bandwidth selection, and even more fluctuation when attempts are made to adjust bandwidths for measurement error. 

\begin{figure}
	\centering
	\setlength{\linewidth}{0.45\linewidth}
	\subfigure[]{ \includegraphics[width=\linewidth]{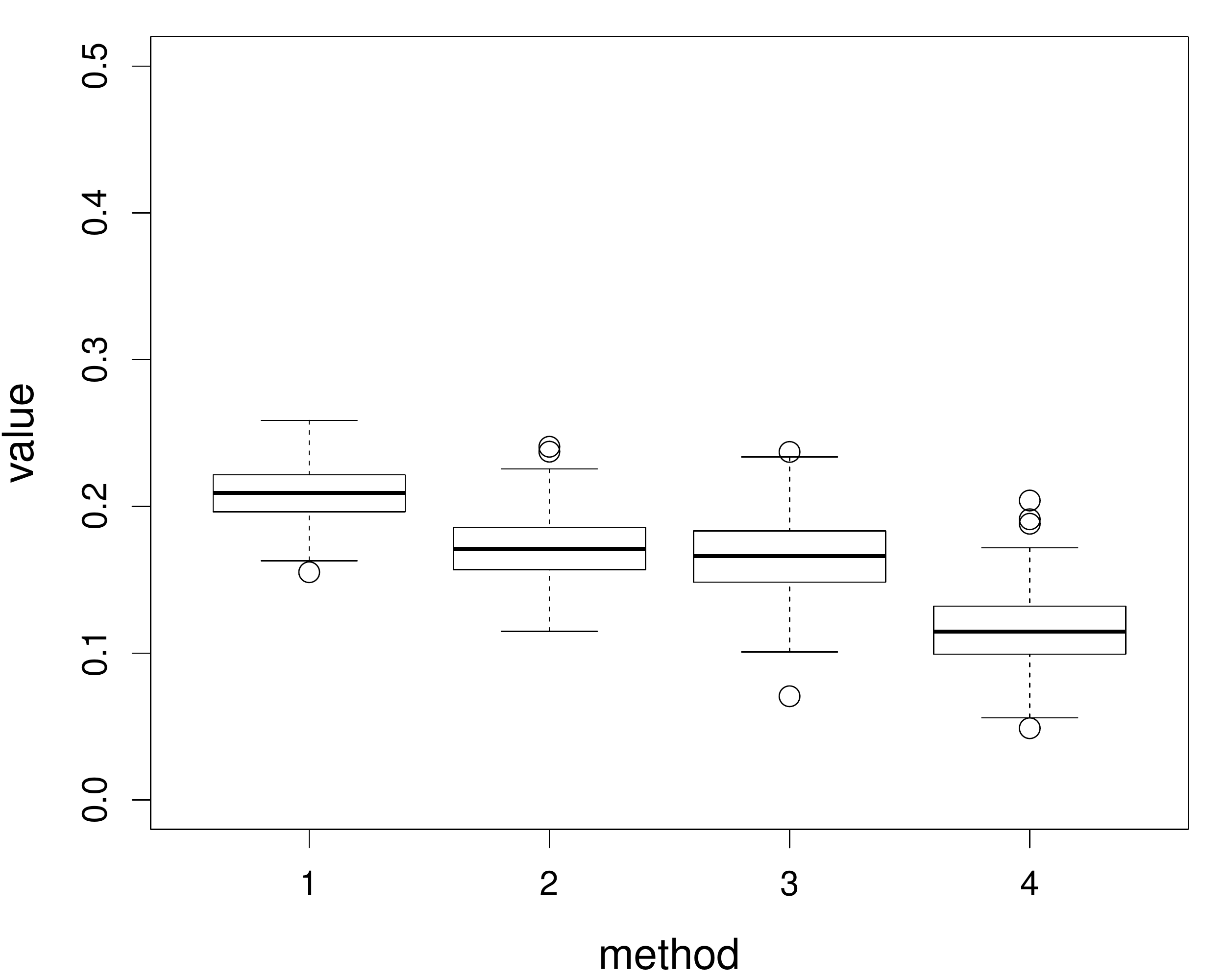} }
	\subfigure[]{ \includegraphics[width=\linewidth]{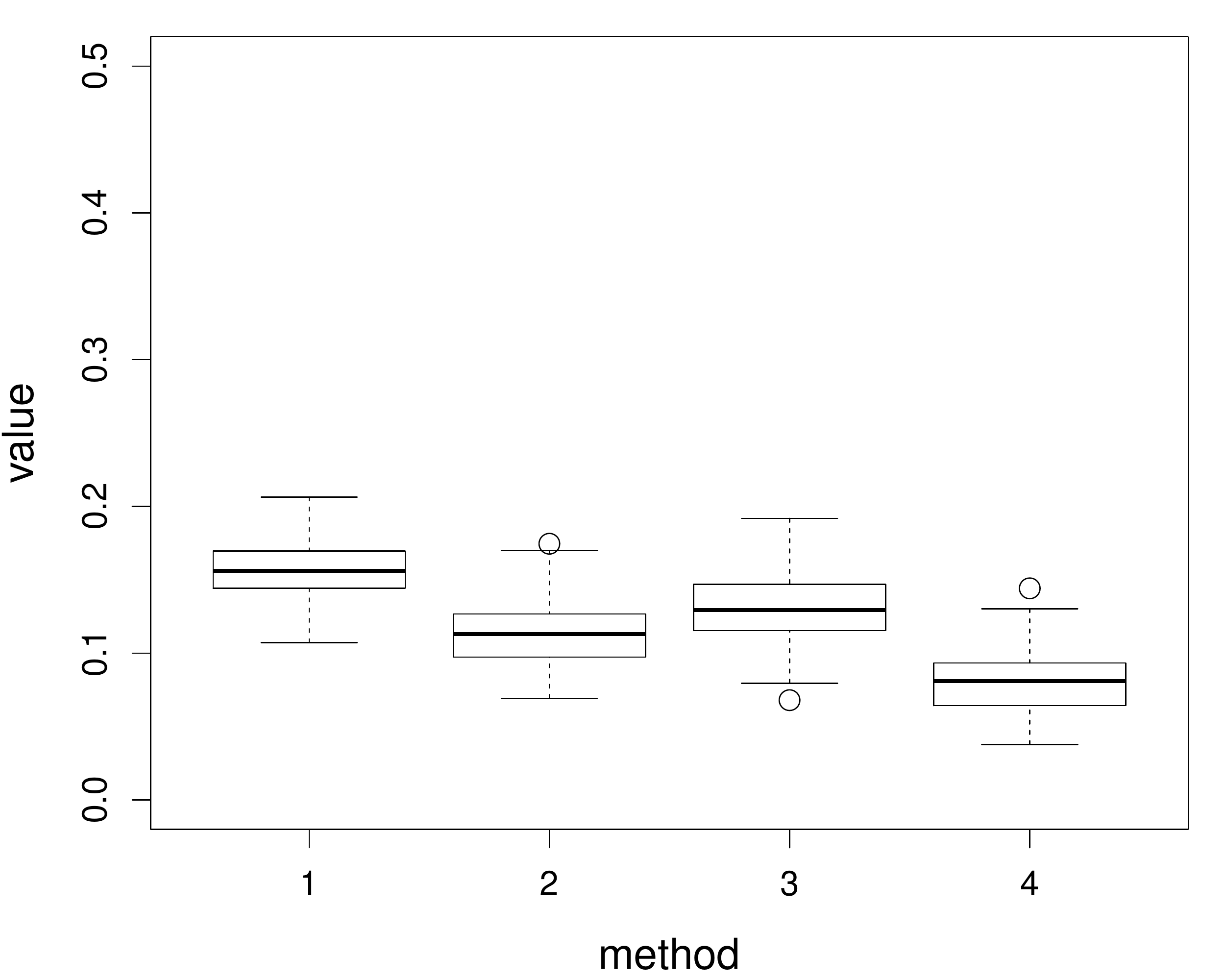} }\\
	\subfigure[]{ \includegraphics[width=\linewidth]{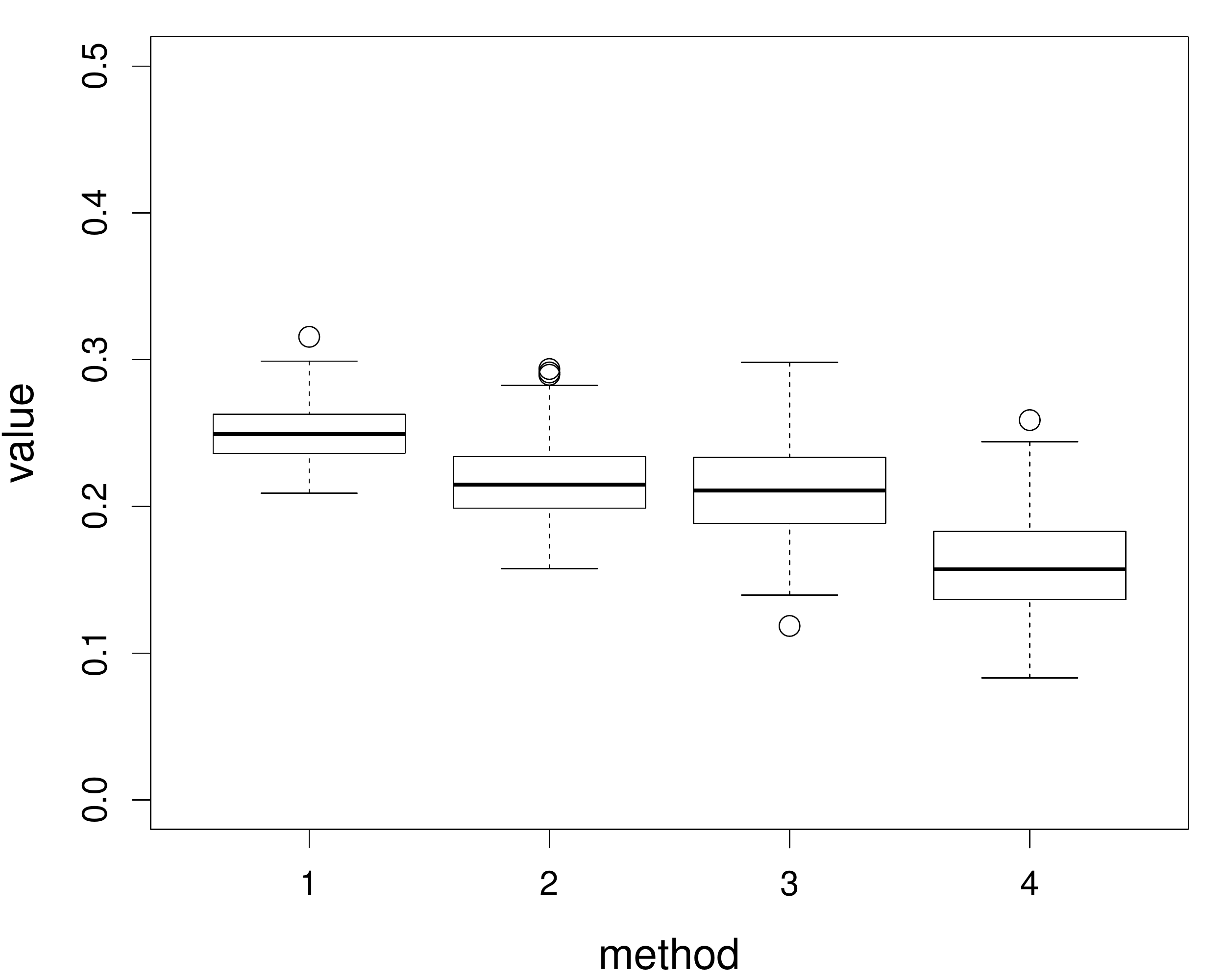} }
	\subfigure[]{ \includegraphics[width=\linewidth]{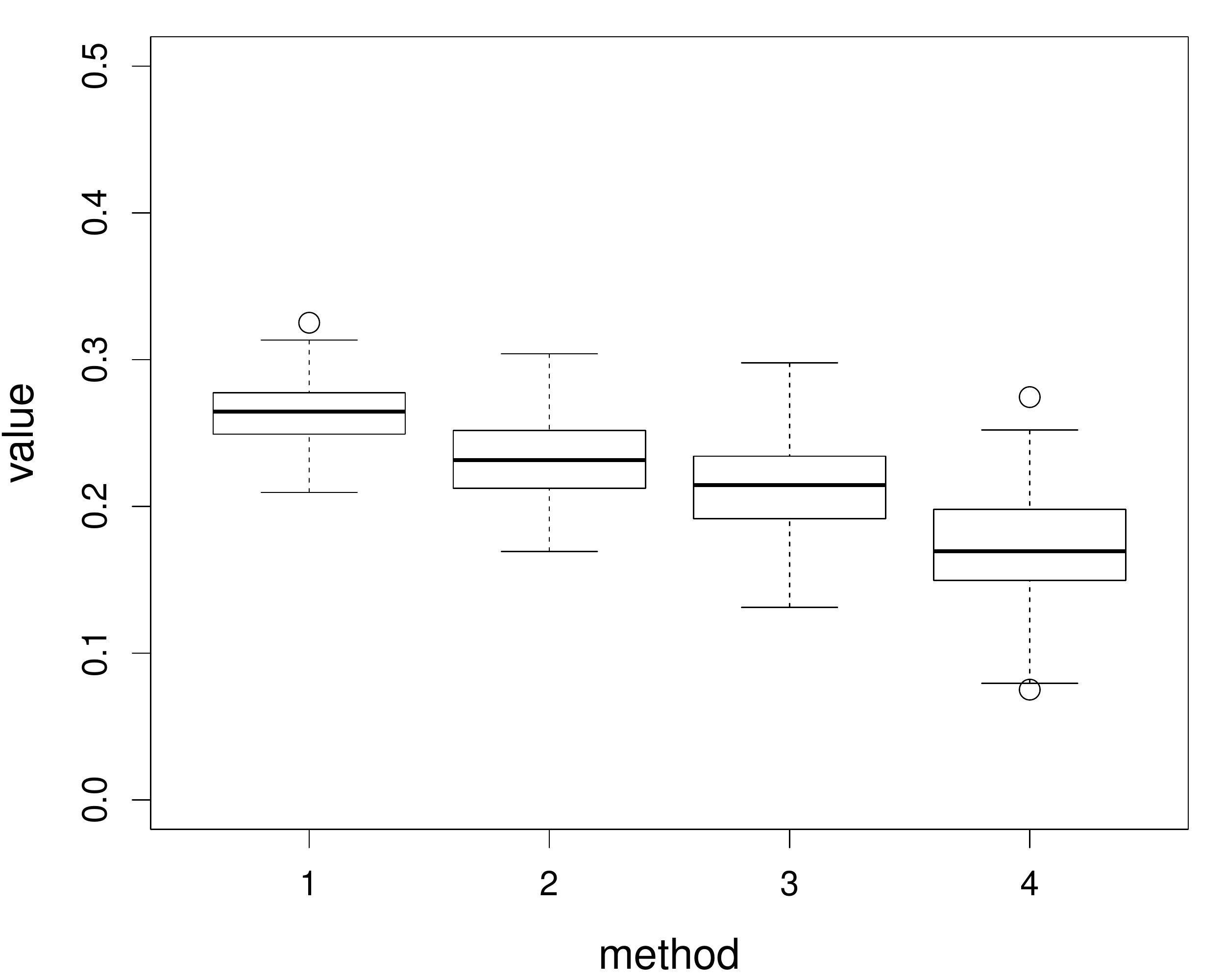} }
	\caption{Boxplots of EISE using the approximated theoretical optimal bandwidths when the primary model is (C1) and the secondary models are (a) $X\sim N(0,1)$, $U\sim \textrm{Laplace}(0, \,\sigma_u/\sqrt{2})$, $\lambda=0.8$; (b) $X\sim N(0,1)$, $U\sim \textrm{Laplace}(0, \,\sigma_u/\sqrt{2})$, $\lambda=0.9$; (c) $X\sim N(0,1)$, $U\sim N(0, \, \sigma_u^2)$, $\lambda=0.8$; (d) $X\sim \textrm{Uniform}(-2, 2)$, $U\sim \textrm{Laplace}(0, \,\sigma_u/\sqrt{2})$, $\lambda=0.8$. Method 1, 2, 3, 4 correspond to $\tilde p_1(y|x)$, $\tilde p_2(y|x)$, $\hat p_3(y|x)$, and $\hat p_4(y|x)$, respectively. The sample size is $n=200$.} 
	\label{Sim1:box:n200}
\end{figure}

\begin{figure}
	\centering
	\setlength{\linewidth}{0.45\linewidth}
	\subfigure[]{ \includegraphics[width=\linewidth]{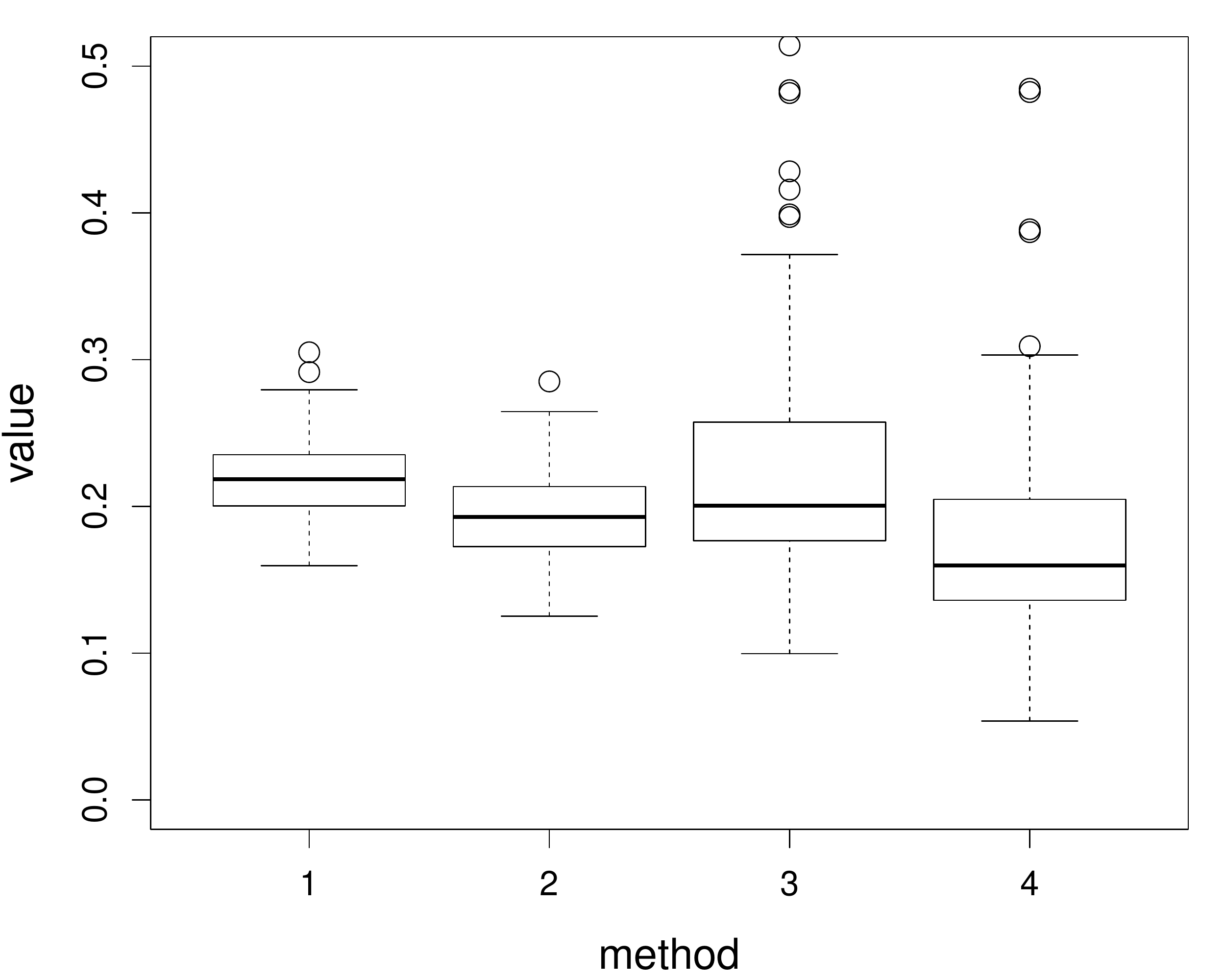} }
	\subfigure[]{ \includegraphics[width=\linewidth]{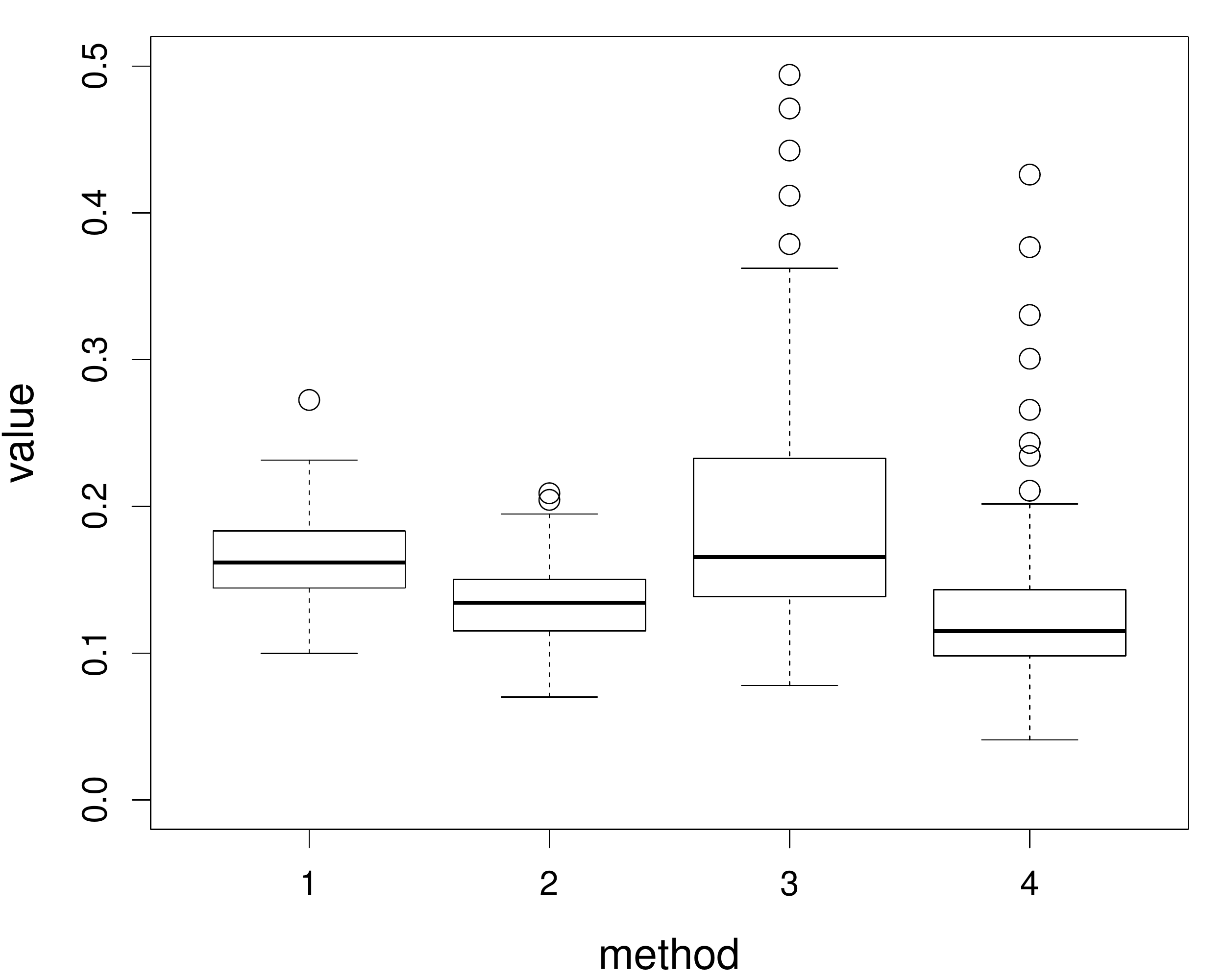} }\\
	\subfigure[]{ \includegraphics[width=\linewidth]{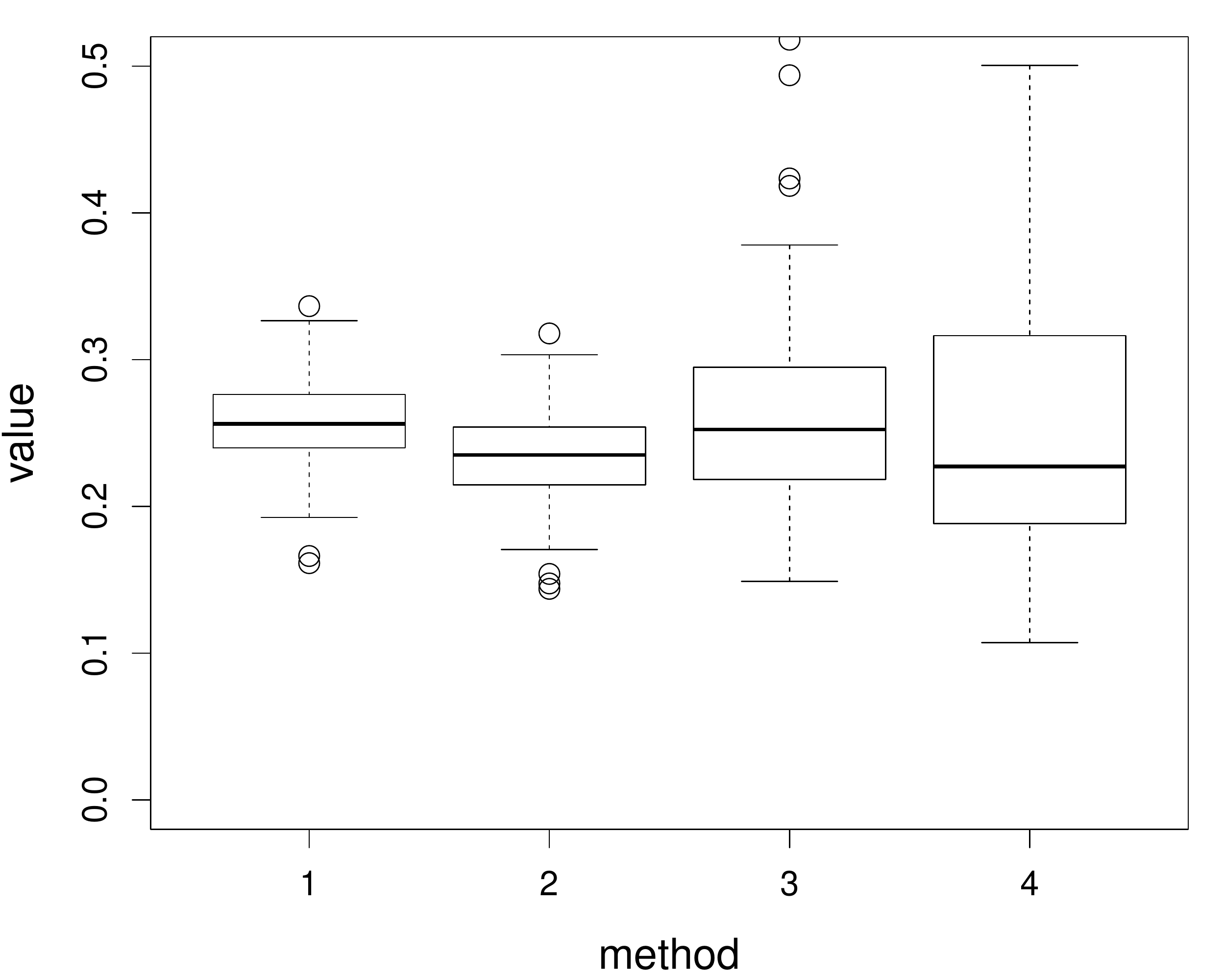} }
	\subfigure[]{ \includegraphics[width=\linewidth]{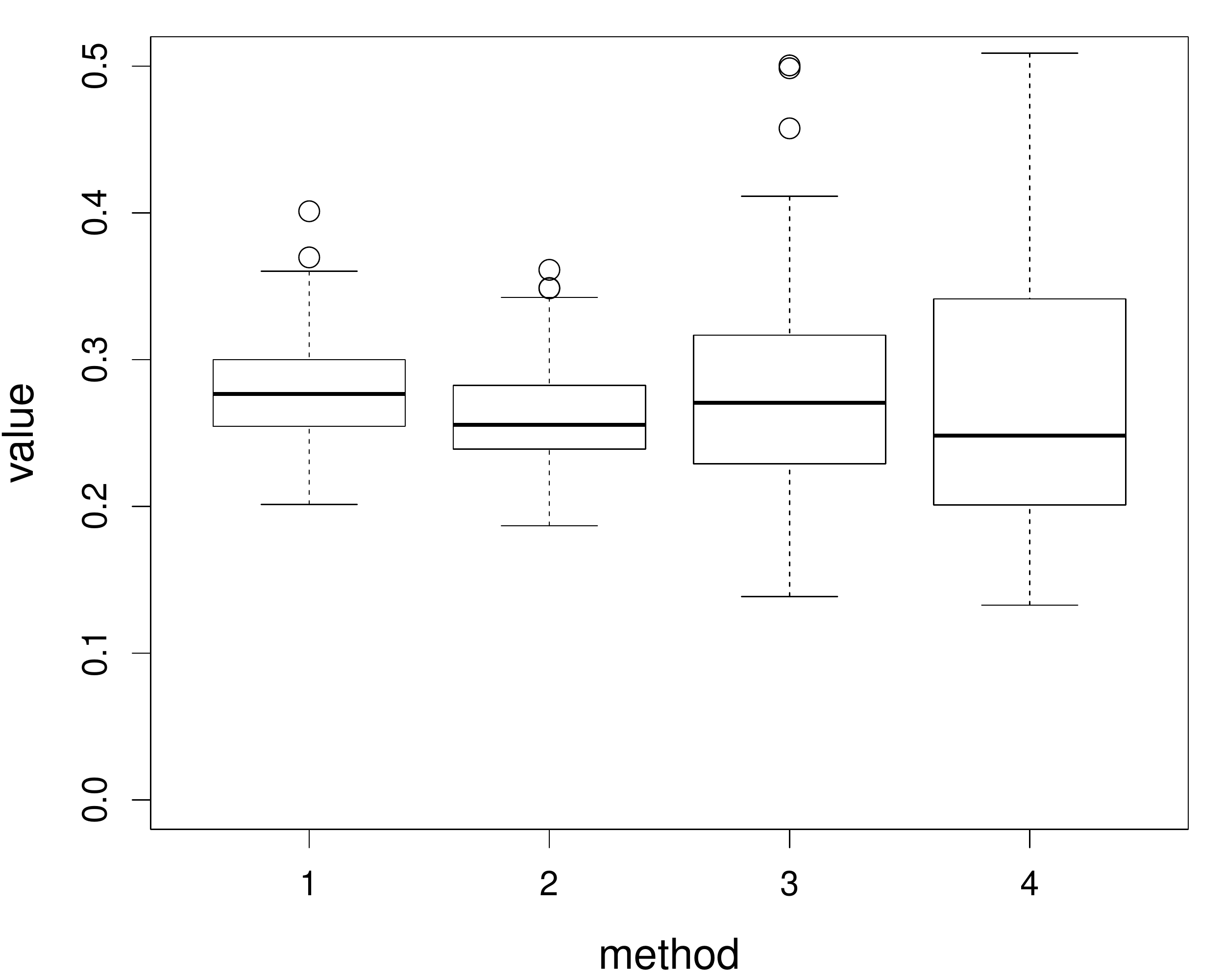} }
	\caption{Boxplots of EISE using the fully data-driven bandwidths when the primary model is (C1) and the secondary models are (a) $X\sim N(0,1)$, $U\sim \textrm{Laplace}(0, \,\sigma_u/\sqrt{2})$, $\lambda=0.8$; (b) $X\sim N(0,1)$, $U\sim \textrm{Laplace}(0, \,\sigma_u/\sqrt{2})$, $\lambda=0.9$; (c) $X\sim N(0,1)$, $U\sim N(0, \, \sigma_u^2)$, $\lambda=0.8$; (d) $X\sim \textrm{Uniform}(-2, 2)$, $U\sim \textrm{Laplace}(0, \,\sigma_u/\sqrt{2})$, $\lambda=0.8$. Method 1, 2, 3, 4 correspond to $\tilde p_1(y|x)$, $\tilde p_2(y|x)$, $\hat p_3(y|x)$, and $\hat p_4(y|x)$, respectively. The sample size is $n=200$.} 
	\label{Sim1:box:data:n200}
\end{figure}

In both rounds of simulation experiments, EISE associated with each of the four considered estimates is higher when the underlying conditional density is bimodal compared to when it is unimodal, or when error contamination is more severe. These are all expected since multimodal densities or noisier data create more unwieldy situations for statistical inference in general. Asymptotic results in Sections~\ref{s:theorems3} and~\ref{s:theorems4} suggest higher variability for the proposed estimators in the presence of super smooth $U$ than when $U$ is ordinary smooth. The observation that EISE's under panel (c) (with normal $U$) are higher than those in panel (a) (with Laplace $U$) in each of Figures~\ref{Sim1:box}--\ref{Sim3:box:data} indicates that the comparison of finite sample variance concurs with the large sample variance comparison. Finally, to demonstrate the effect of sample size, we repeat the simulation study using a much smaller sample size. Figures~\ref{Sim1:box:n200} and \ref{Sim1:box:data:n200} show simulation results obtained under the setting with the primary model in (C1) with $n=200$. Comparing with Figures~\ref{Sim1:box} and \ref{Sim1:box:data} where $n=500$, one can still see similar patterns the estimates exhibit, although more variable EISE are observed for all estimators, especially when the fully data-driven bandwidths are used.

As discussed in \citet{Hyndman.etal1996}, besides local polynomial estimators, other nonparametric estimators deemed suitable for estimating $m^*(x)$ can be employed in the two-step estimators, such as spline-based estimators. Properties of $\hat p_4(y|x)$ in Theorem~\ref{thm:bias4}, as well as properties of $\tilde p_2(y|x)$ established in \citet{Hyndman.etal1996} and \citet{hansen2004nonparametric}, remain valid provided that the adopted $\hat m^*(x)$ converges to $m^*(x)$ faster than the kernel density estimator for the joint density of $(W, \, e^*)$ converges to the truth. As an example, we use cubic spline estimates for $m^*(x)$ in $\tilde p_2(y|x)$ and $\hat p_4(y|x)$, and repeat the second round of the simulation experiments. As counterpart plots of Figures~\ref{Sim1:box:data}--\ref{Sim3:box:data}, Appendix E provides these additional boxplots of EISE, which are mostly comparable with Figures~\ref{Sim1:box:data}--\ref{Sim3:box:data}.
 
\begin{table}[h]
	\caption{Medians and interquartile ranges (in parenthesis) of EISE associated with four estimators across 200 Monte Carlo replicates when the fully data-driven bandwidths are used. Data are generated according to primary models (C1)--(C3) along with secondary models (a)--(d) formulated in Section 5.1. Method 1, 2, 3, 4 correspond to $\tilde p_1(y|x)$, $\tilde p_2(y|x)$, $\hat p_3(y|x)$, and $\hat p_4(y|x)$, respectively}
	\label{Sim1:table:data}
	\centering
	{
		\begin{tabular}{cccccc}
		
			\hline\noalign{\smallskip}
			Model &  Method & (a) & (b)		& (c)       & (d)  \\
			\noalign{\smallskip}\hline\noalign{\smallskip}
			(C1)&  $1$& 0.186 (0.028)	&   0.134 (0.021)    & 0.234 (0.023)       & 0.239 (0.028) \\
			~	& 2      & 0.163 (0.030)	& 0.108 (0.022)      & 0.215 (0.027)       & 0.225 (0.029)	\\
			~ 	& 3 	 & 0.151 (0.049)	& 0.112 (0.044) 	 & 0.205 (0.033)	   & 0.194 (0.057)	\\
			~ 	& 4 	 & 0.114 (0.059)	& 0.075 (0.029) 	 & 0.181 (0.051)	   & 0.195 (0.154)	\\
			\noalign{\smallskip}
			(C2)&  $1$& 0.230 (0.020)	&   0.179 (0.015)    & 0.270 (0.017) 		& 0.276 (0.021)\\
			~	& 2      & 0.206 (0.023)	& 0.150 (0.020)      & 0.254 (0.020) 		& 0.263 (0.022)	\\
			~ 	& 3 	 & 0.196 (0.036)	& 0.153 (0.021) 	 & 0.254 (0.025) 		& 0.245 (0.038)	\\
			~ 	& 4 	 & 0.167 (0.059)	& 0.115 (0.030) 	 & 0.240 (0.062) 		& 0.251 (0.086)	\\
			\noalign{\smallskip}
			(C3)&  $1$& 0.200 (0.026)	&   0.135 (0.022)    & 0.250 (0.025) 		& 0.251 (0.031)\\
			~	& 2      & 0.164 (0.029)	& 0.096 (0.023)      & 0.220 (0.027) 		& 0.213 (0.037)	\\
			~ 	& 3 	 & 0.145 (0.042)	& 0.105 (0.036) 	 & 0.209 (0.045) 		& 0.182 (0.049)	\\
			~ 	& 4 	 & 0.079 (0.025)	& 0.056 (0.018) 	 & 0.143 (0.044) 		& 0.126 (0.042)	\\
			\noalign{\smallskip}\hline
			
		\end{tabular}
	}
\end{table}


\section{Application to dietary data}
\label{s:real}
\begin{figure}
	\centering
	\setlength{\linewidth}{0.45\linewidth}
	\subfigure[]{ \includegraphics[width=\linewidth]{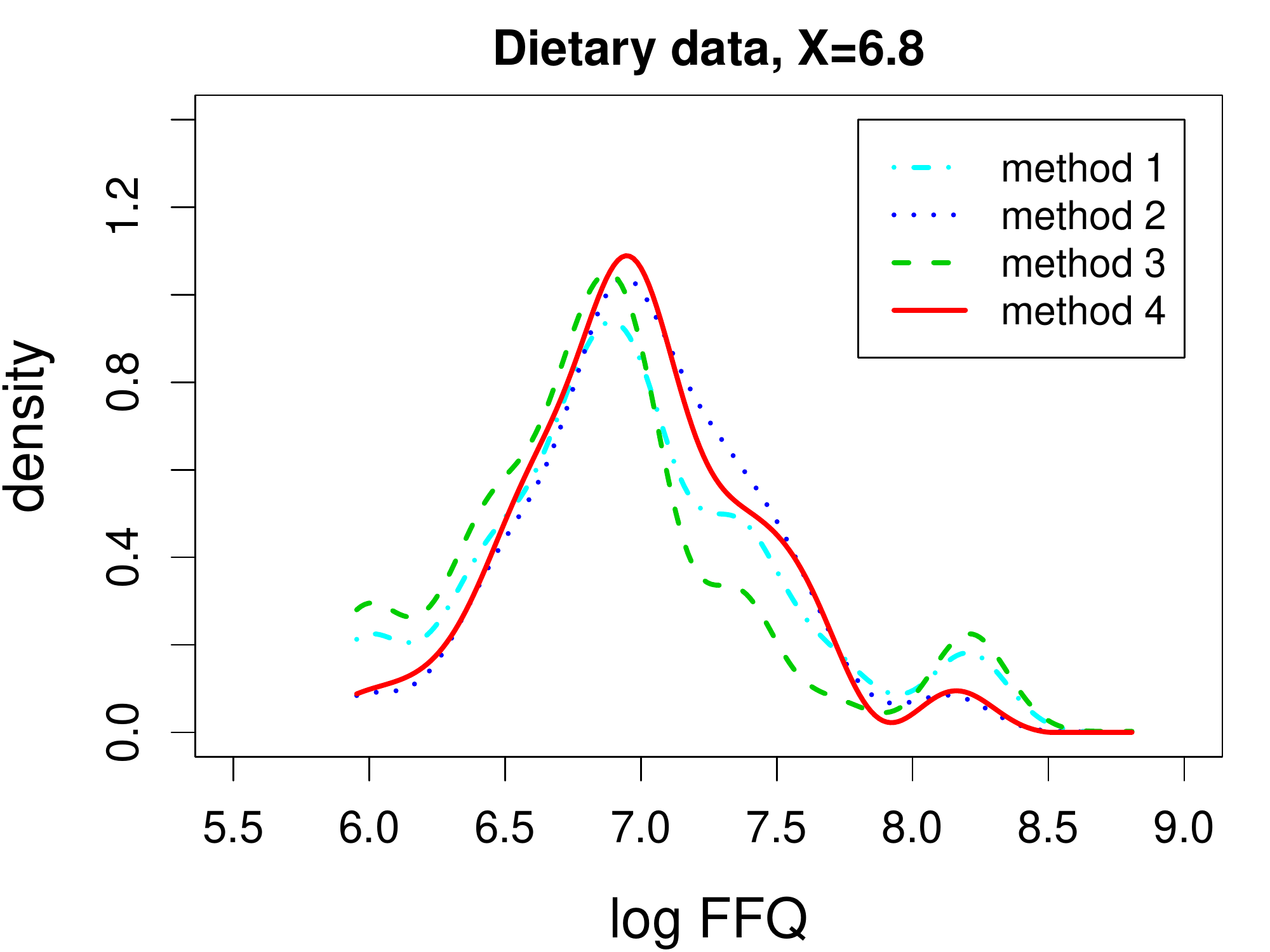} }
	\subfigure[]{ \includegraphics[width=\linewidth]{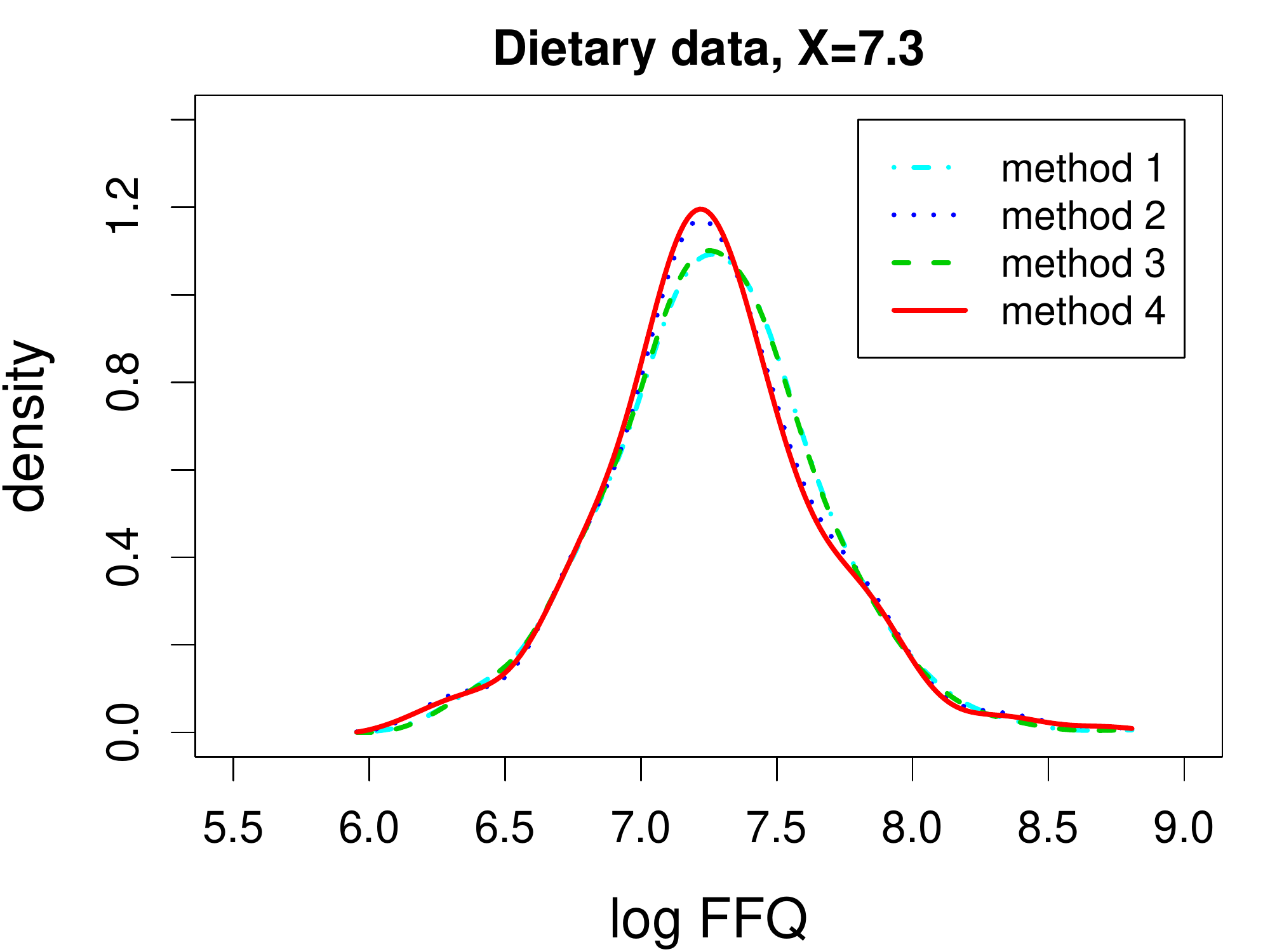} }\\
	\subfigure[]{ \includegraphics[width=\linewidth]{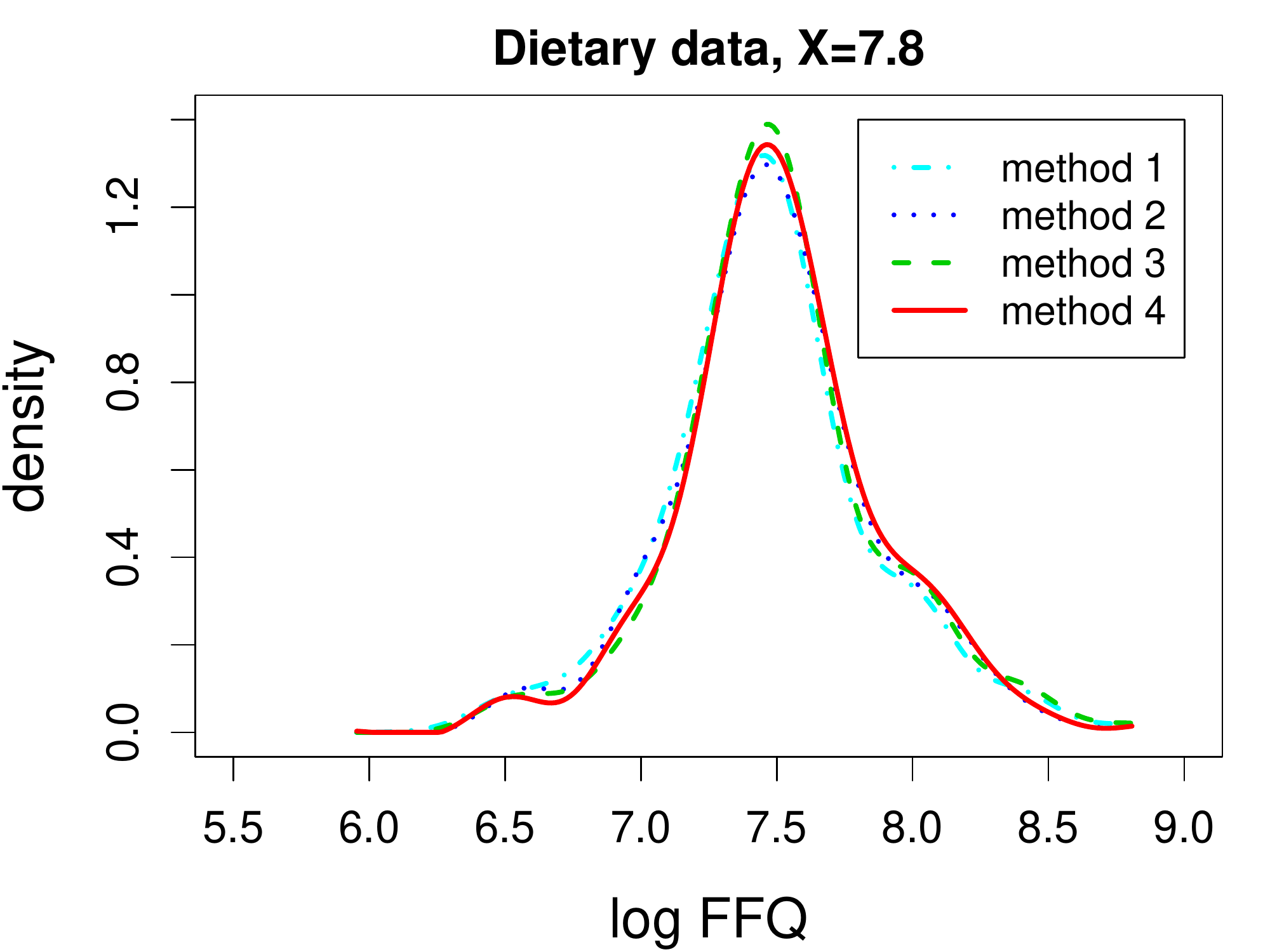} }
	\subfigure[]{ \includegraphics[width=\linewidth]{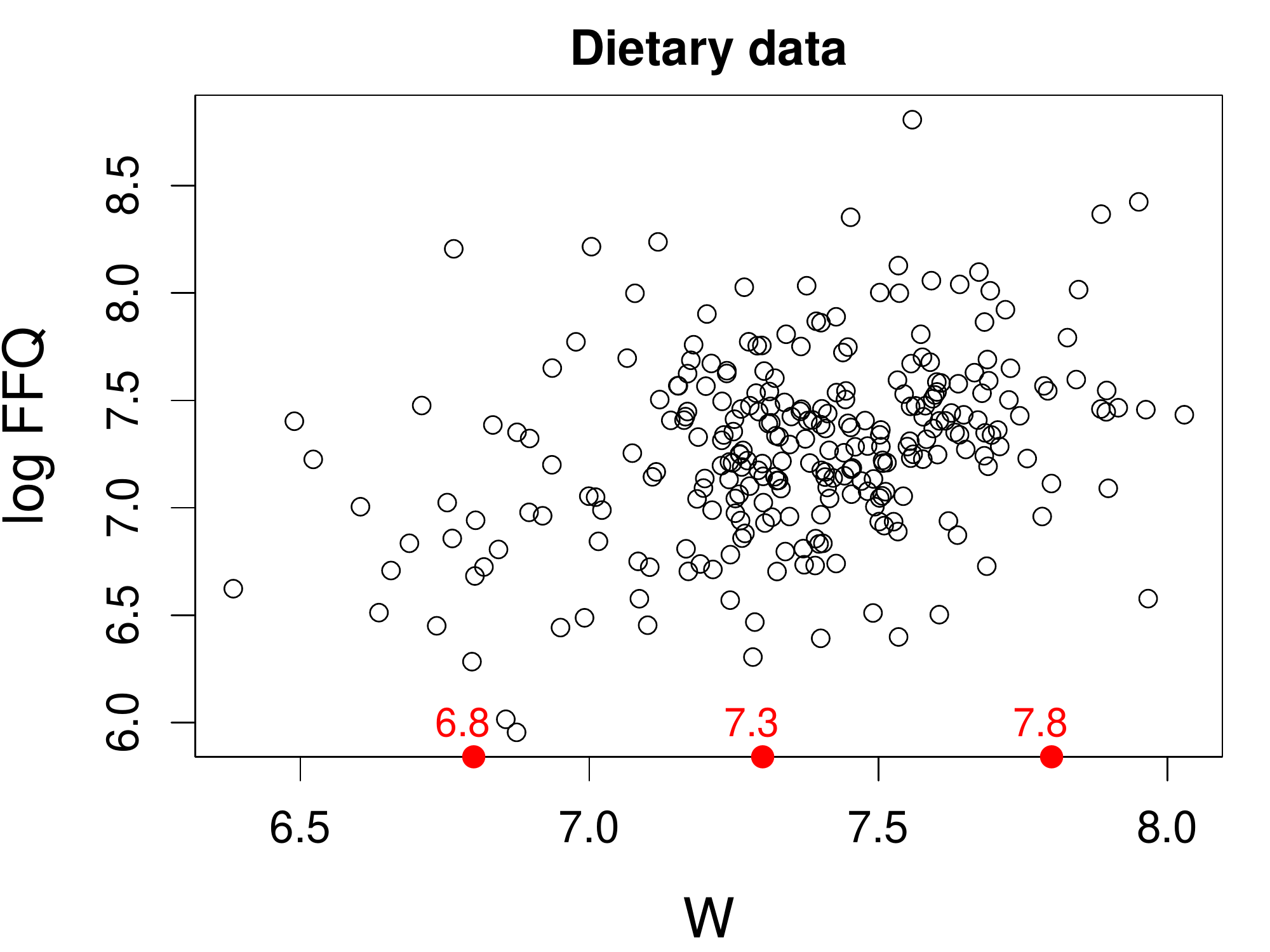} }
	\caption{Naive estimates of the conditional density of the logarithm of FFQ intake corresponding to $\tilde p_1(y|x)$ (cyan dash-dotted lines) and $\tilde p_2(y|x)$ (blue dotted lines), and two non-naive density estimates, $\hat p_3(y|x)$ (green dashed lines) and $\hat p_4(y|x)$ (red solid lines) when $x=6.8$ (in panel (a)), 7.3 (in panel (b)), and 7.8 (in panel (c)), respectively. In each panel of (a)--(c), method 1, 2, 3, 4 correspond to $\tilde p_1(y|x)$, $\tilde p_2(y|x)$, $\hat p_3(y|x)$, and $\hat p_4(y|x)$, respectively. The scatter plot of the observed response versus the observed covariate values is shown in panel (d), where the three values of $x$ at which $p(y|x)$ is estimated are highlighted in red dots on the horizontal axis.} 
	\label{f:dietary}
\end{figure}

The data set to be analyzed in this section is from the Women's Interview Survey of Health, which contains the food frequency questionnaire (FFQ) intake, measured as percent calories from fat, and six 24-hour food recalls from $271$ subjects. It is of interest to estimate the density of the logarithm of FFQ intake ($Y$) conditioning on one's long-term usual intake ($X$). The covariate of interest, the long-term usual intake, cannot be observed directly. A common practice in epidemiology studies is to use data from 24-hour food recalls to construct a surrogate ($W$) of the true covariate. For instance, \citet{liang2005partially} used the average of two 24-hour food recalls from a subject as $W$ and studied the mean of the log-FFQ intake conditioning on $X$ and other error-free covariates; \citet{wang2012corrected} used the average of six 24-hour food recalls as $W$ and estimated conditional quantiles of the log-FFQ intake. We follow the construction of $W$ in \citet{wang2012corrected}, associated with which the estimated reliability ratio is 0.737. Panel (d) in Figure~\ref{f:dietary} shows the scatter plot of the log-FFQ versus the so-constructed $W$ from this data set. 

For illustration purposes, we estimate the conditional density of the log-FFQ when the long-term usual intake is equal to 6.8, 7.3, and 7.8, respectively. Panels (a)--(c) in Figure~\ref{f:dietary} depict four estimated density curves, $\tilde p_1(y|x)$, $\tilde p_2(y|x)$, $\hat p_3(y|x)$, and $\hat p_4(y|x)$, at each of the three covariate values. At $x=6.8$, the two two-step estimates, $\tilde p_2(y|x)$ and $\hat p_4(y|x)$, are similar but the latter exhibits more distinct peak features, which can be a sign that $\hat p_4(y|x)$ corrects $\tilde p_2(y|x)$ for measurement error to recover the height around modes of the underlying density. The other non-naive estimate, $\hat p_3(y|x)$, resembles $\hat p_4(y|x)$ around the highest peak more than the two naive estimates do, and it also differs noticeably from its naive counterpart $\tilde p_1(y|x)$ at other regions of the support of $Y$. At $x=7.3$, around which data are denser, the difference among the four estimated density curves appears to be mostly due to whether one uses two-step estimates or one-step estimates. This can be viewed as an example where the effect of measurement error is mild and the two-step estimates lead to improved estimates compared to the one-step estimates. Finally, at $x=7.8$, around which data become scarce and the association between the response and the covariate may be weaker, the four estimated density curves are less distinguishable. The similarity among the four estimates can be due to low correlation between the response and the true covariate, or that the conditional mean of the response is nearly constant, or lack of  sufficient data information for the non-naive estimates effectively correct the naive ones. 

We repeat the estimation based on $\tilde p_2(y|x)$ and $\hat p_4(y|x)$ using the cubic spline estimate for $m^*(\cdot)$ and obtain comparable results in terms of how four estimated densities compare. A figure showing these estimated density curves is given in Appendix F. Unlike in simulation studies, here, we actually do not know the measurement error variance $\sigma_u^2$ or the distribution family for the measurement error. We resolved this complication by estimating $\sigma_u^2$ via equation (4.3) in \citet{Carroll&etal2006} using repeated measurements (i.e., six 24-hour food recalls from each subject) while assuming Laplace measurement error. Other approaches for estimating $\sigma^2_u$ are discussed in \citet{carroll2014measurement}, including that based on correlated repeated measurements \citep{wang1996quasilikelihood} and those based on validation data or instrumental variables \citep{buzas2014measurement}. This treatment gives rise to two practical concerns we address next. The first concern relates to misspecification of $\sigma^2_u$ in the proposed estimators since an estimated error variance in place of its truth is now used in these estimators. Appendix G presents additional numerical experiments where we repeat part of the simulation studies described in Section~\ref{s:simulation} but with $\sigma_u^2$ set at values different from its truth when obtaining $\hat p_3(y|x)$ and $\hat p_4(y|x)$.  Besides via $\phi_{\hbox {\tiny $U$}}(t)$, these two estimators also depend on $\sigma^2_u$ via bandwidths chosen by the data-driven methods proposed in Section~\ref{s:band}. Despite the two sources of dependence on $\sigma_u^2$, realizations of $\hat p_3(y|x)$ and $\hat p_4(y|x)$ from the experiments tend to exhibit smaller EISE than those associated with their naive counterparts even when a wrong error variance is used. Hence, although the proposed estimators for $p(y|x)$ are compromised by a misspecified error variance, they remain more superior than the naive estimators provided that the misspecification is not close to ignoring measurement error, e.g., as a result of substantially underestimating $\sigma_u^2$. The downside of overestimating $\sigma^2_u$ is inflated variability of the non-naive estimators as evidenced in the simulation presented in Appendix G. The second concern is in regard to the assumed error distribution. It has been reported in abundant existing studies that nonparametric inference are often fairly robust to distributional assumptions on measurement error \citep[e.g.,][]{delaigle2009design, zhou2016nonparametric, huang2017alternative}. \citet{meister2004effect} and \citet{delaigle2008alternative} provided more theoretical insight on this robustness. If one feels uneasy at assuming a measurement error distribution, one may estimate the characteristic function of $U$ using repeated measurements as proposed by \citet{delaigle2008deconvolution}, and use this estimate in place of $\phi_{\hbox {\tiny $U$}}(u)$ in the density estimators. We conjecture that theoretical properties of the resulting density estimators that involve such estimated $\phi_{\hbox {\tiny $U$}}(u)$ can be derived following similar lines of arguments in \citet{delaigle2008deconvolution}, which are beyond the scope of the current study.

\section{Discussions}
\label{s:discussion}
\begin{figure}[h]
	\centering
	\setlength{\linewidth}{0.45\linewidth}
	\subfigure[]{ \includegraphics[width=\linewidth]{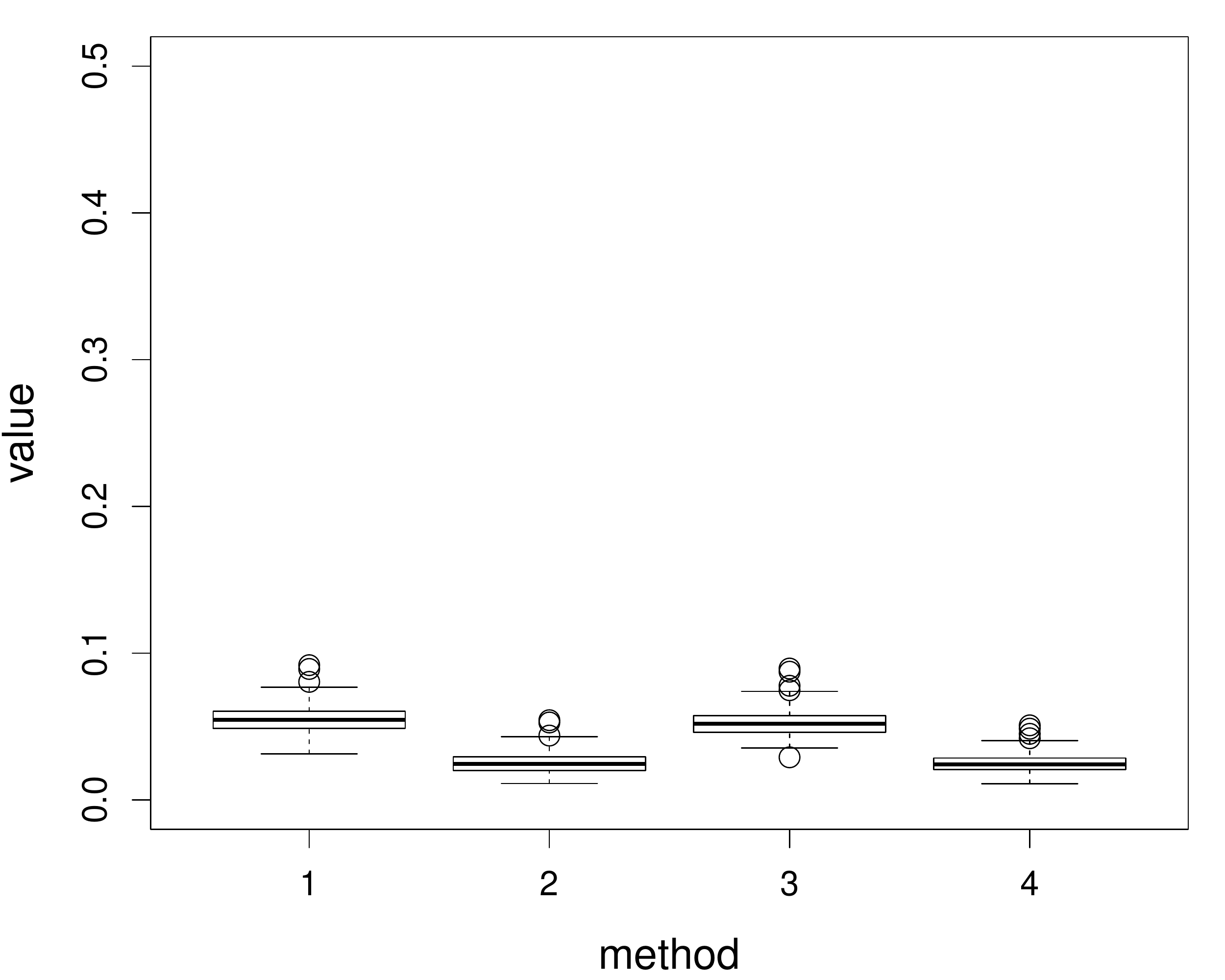} }
	\subfigure[]{ \includegraphics[width=\linewidth]{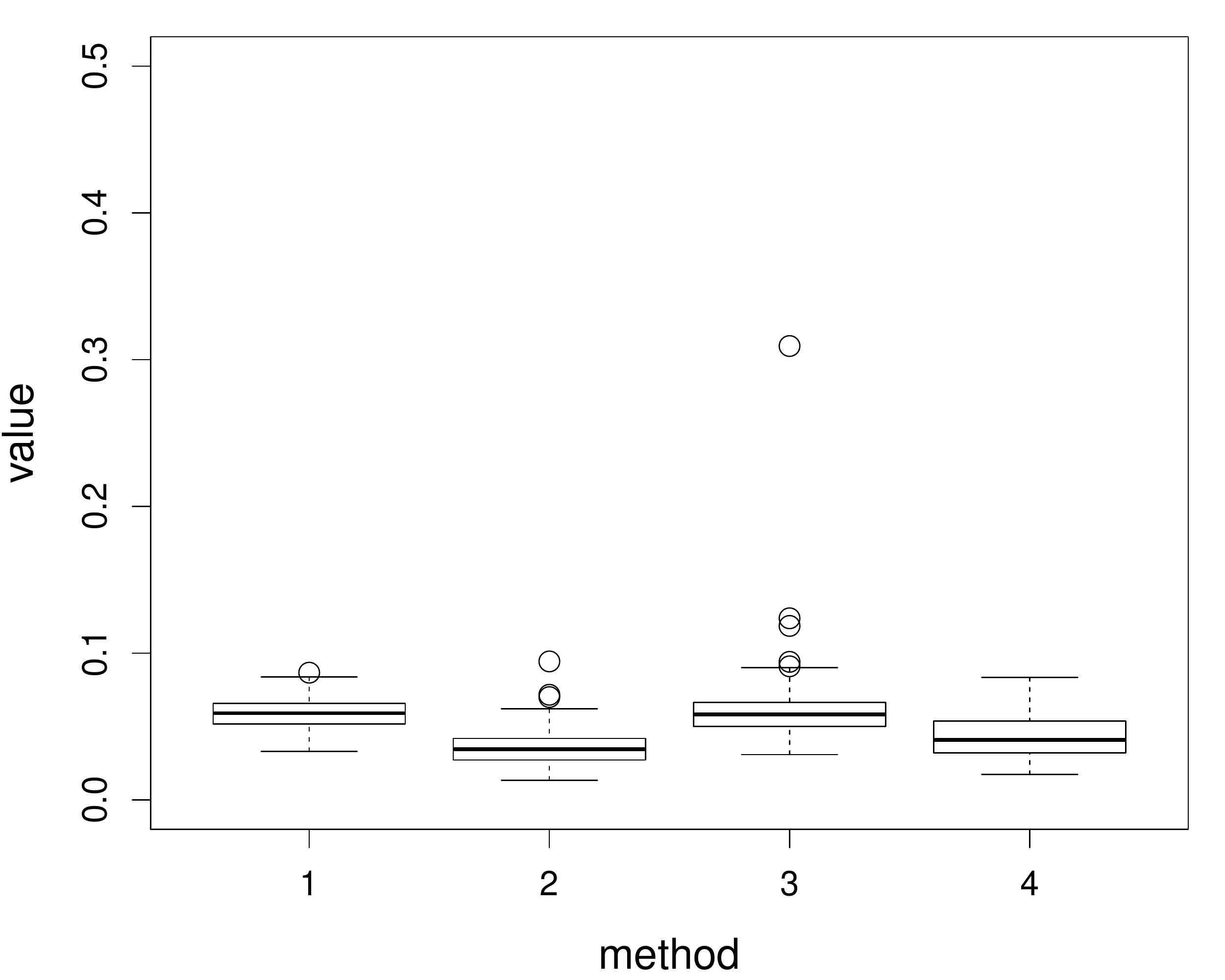} }
	\caption{Boxplots of EISE using the approximated theoretical optimal bandwidths (in panel (a)) and boxplots of EISE using the fully data-driven bandwidths (in panel (b)) when the primary model is (C1) and the secondary model is $X\sim N(0,1)$, $U\sim \textrm{Laplace}(0, \,\sigma_u/\sqrt{2})$, $\lambda=0.99$. Method 1, 2, 3, 4 correspond to $\tilde p_1(y|x)$, $\tilde p_2(y|x)$, $\hat p_3(y|x)$, and $\hat p_4(y|x)$, respectively. The sample size is $n=500$.}
	\label{Sim1:box:lambda99}
\end{figure}

In this study we propose two conditional density estimators that account for covariate measurement error by correcting two existing kernel density estimators developed for error-free data. An R code example is provided in Appendix H to demonstrate use of the R package \texttt{lpme} to obtain all four density estimates. When the conditional mean of the response contributes a lot to explaining the dependence of the response on the covariate, the two-step estimators $\tilde p_2(y|x)$ and $\hat p_4(y|x)$ can substantially benefit from first estimating the mean function. This strategy can even alleviate to some extent the adverse effect of measurement error on naive estimation, even though it can bring in more variability given a finite sample. As one may expect, there will be little return in the effort to account for measurement error when the error contamination is very small. Figure~\ref{Sim1:box:lambda99} provides comparisons between the four estimators considered in Section~\ref{s:simulation} in such a scenario, where data for responses are generated according to the primary model in (C1), and $X\sim N(0,1)$, $U\sim \textrm{Laplace}(0, \,\sigma_u/\sqrt{2})$, with $\lambda=0.99$. When the approximated optimal theoretical bandwidths are used, one can see in this figure high resemblance between $\tilde p_1(y|x)$ and $\hat p_3(y|x)$, as well as between $\tilde p_2(y|x)$ and $\hat p_4(y|x)$. In addition, Figure~\ref{Sim1:box:lambda99} indicates that, when one makes the extra effort to select bandwidths using the fully data-driven methods proposed in Section~\ref{s:band}, the proposed non-naive estimators exhibit higher variability than their naive counterparts, making the proposed estimators less appealing without gaining noticeable bias reduction.

On the theoretical side, in addition to deriving the asymptotic bias and variance of each proposed estimator, we provide an in-depth comparison between the proposed estimators and their error-free counterparts in regard to the dominating bias. We believe that asymptotic normality of $\hat p_3(y|x)$ can be established by proving the Lyapunov's conditions \citep{billingsley2008probability} under additional regularity conditions following arguments similar to those in \cite{huang2017alternative}, although showing the same conditions for $\hat p_4(y|x)$ can be much more formidable due to the higher (than two) order moments of (\ref{eq:Txp2}) arising in the proof. Given the substantial content of this article, we set aside the study of asymptotic normality of proposed estimators and impacts of the smoothness of the covariate and measurement error distributions on their asymptotic distributions for a separate technical note. 

The construction of the proposed estimators can be generalized to estimate a multivariate conditional density given a multivariate covariate, some or all elements of which are prone to measurement error. But kernel-based estimators are less well received when there are many variables involved due to the curse of dimensionality \citep[][Chapter 7]{scott2015multivariate} among several other reasons. The use of the integral transform in (\ref{eq:Tx}) to account for measurement error in some variables, as done in (\ref{eq:Txp1}) and (\ref{eq:Txp2}), only magnifies the challenges in implementing kernel-based density estimation in high dimensional settings. New strategies for nonparametric conditional density estimation are needed in these settings. 

Bandwidth selection has been a hurdle for which no unified solution seems to exist that is numerically convenient and effective for most kernel-based estimation problems. We develop strategies for our proposed estimators aiming to, first, take advantage of existing well accepted bandwidth selection methods in the absence of measurement error, and second, adjust the bandwidths for measurement error in the right direction. Achieving the first goal frees one from estimating a CV criterion using error-prone data. We reach the second goal by a simple adjustment of naive bandwidths that depends on the severity of measurement error and the correlation between the covariate and the response or a mean residual. A more refined adjustment demands systematic investigation on relationships between naive bandwidths and theoretically optimal bandwidths accounting for measurement error. 

\setcounter{equation}{0}
\setcounter{figure}{0}
\renewcommand{\theequation}{A.\arabic{equation}}
\renewcommand{\thefigure}{A.\arabic{figure}}
\renewcommand{\thesection}{A.\arabic{section}}
\renewcommand{\thesubsection}{A.\arabic{subsection}}

\section*{Appendix A: Proof of Theorem 3.1}
The construction of $\hat p_3(y|x)$ can be viewed as $\hat p_3(y|x)=\hat f^{-1}_{\hbox {\tiny $X$}}(x)\hat p_3(x, y)$, where $\hat f_{\hbox {\tiny $X$}}(x)$ is the deconvoluting kernel estimator of $f_{\hbox {\tiny $X$}}(x)$ in (\ref{eq:fxhat}), and 
\begin{equation}
\hat p_3(x, y)  =\frac{1}{n h_1h_2}\sum_{j=1}^n K^*_1\left(\frac{W_j-x}{h_1}\right)K_2\left(\frac{Y_j-y}{h_2}\right)\label{eq:Tme}
\end{equation}
is an estimator of $p(x, y)$. 

We next approximate $\hat f_{\hbox {\tiny $X$}}(x)$ and $\hat p_3(x, y)$ via the decomposition, $A=E(A)+O_p\{\sqrt{\textrm{Var}(A)}\}$, for a random variable $A$ under regularity conditions. 

\subsection{Approximation of $\hat f_{\hbox {\tiny $X$}}(x)$}
\label{s:appfxinv}
Because the mean of $\hat f_{\hbox {\tiny $X$}}(x)$ is the same as the mean of the regular kernel density estimator of $f_{\hbox {\tiny $X$}}(x)$ in the absence of measurement error, which is well established \citep[][equation (6.16)]{scott2015multivariate}, one has 
\begin{equation}
E\left\{\hat f_{\hbox {\tiny $X$}}(x)\right\} =
f_{\hbox {\tiny $X$}}(x) +0.5 f''_{\hbox {\tiny $X$}}(x) \mu_{2,1} h_1^2 +O(h_1^4), \label{eq:meanSn}
\end{equation}
where $f''_{\hbox {\tiny $X$}}(x)$ is the second derivative of $f_{\hbox {\tiny $X$}}(x)$, and $\mu_{2,\ell}=\int t^2 K_\ell(t)dt$, for $\ell=1, 2$.

Also similar to the variance result for the ordinary kernel density estimator of $f_{\hbox {\tiny $X$}}(x)$ \citep[][equation (6.17)]{scott2015multivariate}, one can show that
\begin{equation}
\textrm{Var}\left\{\hat f_{\hbox {\tiny $X$}}(x)\right\} = \frac{f_{\hbox {\tiny $W$}}(x) R(K_1^*)}{nh_1}+O(n^{-1}), 
\label{eq:varS}
\end{equation}
where $f_{\hbox {\tiny $W$}}(\cdot)$ is the density of $W$, and $R(K_1^*)=\int  \{K^*_1(t)\}^2dt$. In the sequel, we use $R(g)$ to denote $\int g^2(t) dt$ for a square integrable function $g(t)$. Note that $R(K_1^*)$ depends on $h_1$ since $K^*_1(t)$ depends on $h_1$, and by Lemmas B.4 and B.9 in \citet{delaigle2009design}, under Conditions U and Conditions K in the main article,  
\begin{equation}
R(K_1^*)=
\left\{
\begin{array}{ll}
O(h_1^{-2b}), & \textrm{if $U$ is ordinary smooth}, \\
O\left\{h_1^{2b_2}\exp(2h_1^{-b}/d_2)\right\}, & \textrm{if $U$ is super smooth},
\end{array}
\right.
\label{eq:roughK1*}
\end{equation}
where $b_2=b_0I(b_0<0.5)$. 

By (\ref{eq:meanSn})--(\ref{eq:roughK1*}), one has, when $U$ is ordinary smooth, 
\begin{equation}
\hat f_{\hbox {\tiny $X$}}(x)=f_{\hbox {\tiny $X$}}(x)+0.5 f''_{\hbox {\tiny $X$}}(x)\mu_{2,1} h_1^2 +O(h_1^4)+O_p\left(\frac{1}{\sqrt{nh_1^{1+2b}}} \right),\label{eq:appSnl1}
\end{equation}
and, when $U$ is super smooth, 
\begin{equation}
\hat f_{\hbox {\tiny $X$}}(x)=f_{\hbox {\tiny $X$}}(x)+0.5 f''_{\hbox {\tiny $X$}}(x)\mu_{2,1} h_1^2 +O(h_1^4)+O_p\left\{\frac{\exp(h_1^{-b}/d_2)}{\sqrt{nh_1^{1-2b_2}}}\right\}.\label{eq:appSnl2}
\end{equation}

Following (\ref{eq:appSnl1}) and (\ref{eq:appSnl2}), one has, for ordinary smooth $U$,  
\begin{equation}
\hat f^{-1}_{\hbox {\tiny $X$}}(x)  = f^{-1}_{\hbox {\tiny $X$}}(x)-0.5f^{-2}_{\hbox {\tiny $X$}}(x)f''_{\hbox {\tiny $X$}}(x)\mu_{2,1}h_1^2 +O(h_1^4)+O_p\left(\frac{1}{\sqrt{nh_1^{1+2b}}}\right), \label{eq:Sn00ord}
\end{equation}
and, for super smooth $U$, 
\begin{equation}
\hat f^{-1}_{\hbox {\tiny $X$}}(x)  = f^{-1}_{\hbox {\tiny $X$}}(x)-0.5f^{-2}_{\hbox {\tiny $X$}}(x)f''_{\hbox {\tiny $X$}}(x)\mu_{2,1}h_1^2 +O(h_1^4)+O_p\left\{\frac{\exp(h_1^{-b}/d_2)}{\sqrt{nh_1^{1-2b_2}}}\right\}. \label{eq:Sn00sup}
\end{equation}

\subsection{Approximation of $\hat p_3(x, y)$}
\label{s:aboutp3}
Because the mean of $\hat p_3(x, y)$ is the same as the mean of the regular bivariate kernel density estimator of $p(x, y)$ in the absence of measurement error, which has been established \citep[][equation (6.40)]{scott2015multivariate}, one has 
\begin{align}
& E\left\{\hat p_3(x,y)\right\} \nonumber \\
= & p(x,y)+0.5\{p_{xx}(x,y)\mu_{2,1}h_1^2+p_{yy}(x,y) \mu_{2,2}h_2^2\}+O(h_1^4)+O(h_2^4)+O(h_1^2h_2^2), \label{eq:Tmean0}
\end{align}
where $p_{xx}(x,y)=(\partial^2/\partial x^2)p(x,y)$ and $p_{yy}(x,y)=(\partial^2/\partial y^2)p(x,y)$.

Following similar derivations leading to the asymptotic variance of a regular bivariate kernel density estimator in the absence of measurement error \citep[][equation (6.41)]{scott2015multivariate}, one can show that 
\begin{equation}
\textrm{Var}\left\{\hat p_3(x, y)\right\} = \frac{p^*(x, y)R(K_1^*)R(K_2)}{nh_1h_2} +O(n^{-1}). 
\label{eq:varT}
\end{equation}

By (\ref{eq:Tmean0}), (\ref{eq:varT}), and (\ref{eq:roughK1*}), one has, for ordinary smooth $U$, 
\begin{align}
\hat p_3(x,y) & =  p(x,y)+0.5\{p_{xx}(x,y)\mu_{2,1}h_1^2+p_{yy}(x,y) \mu_{2,2}h_2^2\}\nonumber \\
& +O(h_1^4)+O(h_2^4)+O(h_1^2h_2^2)+O_p\left(\frac{1}{\sqrt{nh_1^{1+2b}h_2}}\right),  \label{eq:Tmean0final1}
\end{align}
and, for super smooth $U$, 
\begin{align}
\hat p_3(x,y) & = p(x,y)+0.5\{p_{xx}(x,y)\mu_{2,1}h_1^2+p_{yy}(x,y) \mu_{2,2}h_2^2\} \nonumber \\
 & +O(h_1^4)+O(h_2^4)+O(h_1^2h_2^2)+O_p\left\{\frac{\exp(h_1^{-b}/d_2)}{\sqrt{nh_1^{1-2b_2}h_2}}\right\}.  \label{eq:Tmean0final2}
\end{align}

The result in Theorem 3.1 regarding $\hat p_3(y|x)-p(y|x)$ is obtained from (\ref{eq:Sn00ord}) multiplying (\ref{eq:Tmean0final1}) for ordinary smooth $U$, and (\ref{eq:Sn00sup}) multiplying (\ref{eq:Tmean0final2}) for super smooth $U$. 

\setcounter{equation}{0}
\setcounter{figure}{0}
\setcounter{subsection}{0}
\renewcommand{\theequation}{B.\arabic{equation}}
\renewcommand{\thefigure}{B.\arabic{figure}}
\renewcommand{\thesection}{B.\arabic{section}}
\renewcommand{\thesubsection}{B.\arabic{subsection}}

\section*{Appendix B: Proof of Theorem 3.2}
Because the integral transform $\mathscr{T}_x(\cdot)$ defined in (\ref{eq:Tx}) is a linear operator by construction, it can commute with another linear operator, such as expectation. In addition, $\mathscr{T}_x\left\{g(\cdot, y)\right\}=g(x, y)$ if $\phi_{\hbox {\tiny $U$}}(t)=1$ for all $t$.

The construction of $\hat p_4(y|x)$ originates from $\hat p_4(y|x)=\hat f^{-1}_{\hbox {\tiny $X$}}(x) \hat p_4(x, y)$, where $\hat p_4(x, y)$ is an estimator of $p(x, y)$ obtained via the aforementioned integral transform of the following estimator for $p^*(x, y)$, 
\begin{equation}
\tilde p_2(x, y) =\frac{1}{nh_1h_2}\sum_{j=1}^n K_1\left(\frac{W_j-x}{h_1}\right)K_2\left\{\frac{Y_j-\hat m^*(W_j)-y+\hat m^*(x)}{h_2}\right\}.
\label{eq:p2org}
\end{equation}
More specifically, 
\begin{align}
\hat p_4(x,y) & = \frac{1}{2\pi}\int e^{-itx}\frac{\phi_{\hbox {\tiny $\tilde p_2(\cdot, y)$}}(t)}{\phi_{\hbox {\tiny $U$}}(t)}dt \nonumber \\
& = \frac{1}{2\pi}\int e^{-itx}\frac{ \int e^{itw}\tilde p_2(w, y) dw}{\phi_{\hbox {\tiny $U$}}(t)}dt \nonumber \\
& = \mathscr{T}_x\left\{\tilde p_2(\cdot, y)  \right\}.
\label{eq:p4p2}
\end{align}
We next use the mean and variance results for $\tilde p_2(x, y)$ to obtain those for $\hat p_4(x, y)$.

\citet{hansen2004nonparametric} showed that, despite the extra estimation of $m^*(\cdot)$, the asymptotic variance of the two-step estimator for $p^*(x,y)$, namely $\tilde p_2(x, y)$, is of the same order as that of the one-step estimator given by
\begin{equation}
\tilde p_1(x, y) =\frac{1}{nh_1h_2}\sum_{j=1}^n K_1\left(\frac{W_j-x}{h_1}\right)K_2\left(\frac{Y_j-y}{h_2}\right).
\label{eq:p1org}
\end{equation}
Since $\hat p_3(x,y) = \mathscr{T}_x\{\tilde p_1(\cdot, y)\}$, which is the same as how $\hat p_4(x, y)$ relates to $\tilde p_2(x, y)$ in (\ref{eq:p4p2}), the asymptotic variance of $\hat p_4(x, y)$ is also of the same order as that of $\hat p_3(x, y)$, which is provided in Section~\ref{s:aboutp3}. 

In our study, we set $\hat m^*(\cdot)$ as the local linear estimator of $m^*(\cdot)$ with kernel $K_3(t)$ and bandwidth $h_3$. Following the proof in \citet[][Section 10]{hansen2004nonparametric}, one can show that, if $h_3=O(h_2)$ as $h_2$ and $h_3$ tend to zero, 
\begin{equation}
\label{eq:naivejnt}
\begin{aligned}
E\left\{\tilde p_2(x, y)\right\}=& p^*(x, y)+ 0.5 \left\{g_1(x,y)\mu_{2, 1} h_1^2+ g_2(x,y) \mu_{2, 2} h_2^2\right\} + \\
& O(h_1^4)+O(h_2^4)+O(h_1^2h_2^2), 
\end{aligned}
\end{equation}
where 
\begin{align}
g_1(x, y)& =f^{(2)}_{\hbox {\tiny $W$}, e^*, 11}(x, e^*)=\left.\left[\frac{\partial^2}{\partial w^2}f_{\hbox {\tiny $W$}, e^*}(w, e^*)\right]\right\vert_{w=x, e^*=y-m^*(x)},\label{eq:g1}\\
g_2(x,y) & = f^{(2)}_{\hbox {\tiny $W$}, e^*, 22}(x, e^*) =\left.\left[\frac{\partial^2}{\partial e^{*2}}f_{\hbox {\tiny $W$}, e^*}(w, e^*)\right]\right\vert_{w=x, e^*=y-m^*(x)}, \label{eq:g2}
\end{align}
in which $f_{\hbox {\tiny $W$}, e^*}(w, e^*)$ is the joint density of $W$ and $e^*=Y-m^*(W)$. 
It follows that, by commuting the operations of expectation and $\mathscr{T}_x$,
\begin{align*}
& E\{\hat p_4(x,y)\}\\
= & \mathscr{T}_x [E\{\tilde p_2(\cdot, y)\}]  \\
= & \mathscr{T}_x\{p^*(\cdot, y)\}+0.5 \left[\mathscr{T}_x\{g_1(\cdot,y)\}\mu_{2,1}h_1^2+
\mathscr{T}_x\{g_2(\cdot, y)\}\mu_{2,2}h_2^2 \right]\\ 
& +O(h_1^4)+O(h_2^4)+O(h_1^2h_2^2), 
\end{align*}
where 
\begin{align}
\mathscr{T}_x\{p^*(\cdot, y)\} & =  p(x, y), \label{eq:p*xypxy} \\
\mathscr{T}_x\{g_1(\cdot,y)\} & =  p_{xx}(x, y)+\sum_{k=2}^4\mathscr{T}_x\{I_k(\cdot, y)\}, \label{eq:fwefxe} \\
\mathscr{T}_x\{g_2(\cdot, y)\} & =  p_{yy}(x,y), \label{eq:fwepyy}
\end{align}
in which $I_k(w, y)$, for $k=2, 3, 4$, are defined in (\ref{eq:3Is}).
Among (\ref{eq:p*xypxy})--(\ref{eq:fwepyy}), (\ref{eq:p*xypxy}) can be proved using (\ref{eq:twojoint}) in the main article, (\ref{eq:fwefxe}) and (\ref{eq:fwepyy}) are proved in Section~\ref{s:twolemmas}. In conclusion, we have 
\begin{equation}
\label{eq:p4mean}
\begin{aligned}
& E\{\hat p_4(x,y)\}\\
= & p(x, y)+0.5 \left(\left[p_{xx}(x, y)+\sum_{k=2}^4\mathscr{T}_x\{I_k(\cdot, y)\}\right]\mu_{2,1}h_1^2+p_{yy}(x,y)\mu_{2,2}h_2^2 \right) + \\
& O(h_1^4)+O(h_2^4)+O(h_1^2h_2^2). 
\end{aligned}
\end{equation}

Combining (\ref{eq:p4mean}) with the variance rate of $\hat p_4(x, y)$, we have, for ordinary smooth $U$, 
\begin{equation}
\label{eq:p4final1}
\begin{aligned}
& \hat p_4(x,y) \\
=& p(x,y)+0.5 \left(\left[p_{xx}(x, y)+\sum_{k=2}^4\mathscr{T}_x\{I_k(\cdot, y)\}\right]\mu_{2,1}h_1^2+p_{yy}(x,y)\mu_{2,2}h_2^2 \right) \\
 & +O(h_1^4)+O(h_2^4)+O(h_1^2h_2^2)+O_p\left(\frac{1}{\sqrt{nh_1^{1+2b}h_2}}\right),  
\end{aligned}
\end{equation}
and, for super smooth $U$, 
\begin{equation}
\label{eq:p4final2}
\begin{aligned}
& \hat p_4(x,y)\\
 = & p(x,y)+0.5 \left(\left[p_{xx}(x, y)+\sum_{k=2}^4\mathscr{T}_x\{I_k(\cdot, y)\}\right]\mu_{2,1}h_1^2+p_{yy}(x,y)\mu_{2,2}h_2^2 \right)\\
& +O(h_1^4)+O(h_2^4)+O(h_1^2h_2^2)+O_p\left\{\frac{\exp(h_1^{-b}/d_2)}{\sqrt{nh_1^{1-2b_2}h_2}}\right\}.  
\end{aligned}
\end{equation}

The result in Theorem 3.2 regarding $\hat p_4(y|x)-p(y|x)$ is obtained from (\ref{eq:Sn00ord}) multiplying (\ref{eq:p4final1}) for ordinary smooth $U$, and (\ref{eq:Sn00sup}) multiplying (\ref{eq:p4final2}) for super smooth $U$. 

\subsection{Proof of (\ref{eq:fwefxe}) and (\ref{eq:fwepyy})}
\label{s:twolemmas}
In order to derive the two transforms on the left-hand side of (\ref{eq:fwefxe}) and (\ref{eq:fwepyy}), we need to elaborate the two functions, $g_1(x, y)$ and $g_2(x,y)$, defined in (\ref{eq:g1}) and (\ref{eq:g2}). For notational clarity, we first derive partial derivatives of $f_{\hbox {\tiny $W$}, e^*}(w, e^*)$, viewing it as a function of $w$ and $e^*$, before evaluating the partial derivatives at $w=x$ and $e^*=y-m^*(x)$ to obtain $g_1(x, y)$ and $g_2(x,y)$.

Because 
\begin{align}
f_{\hbox {\tiny $W$}, e^*}(w, e^*)  = p^*(w,y) = \int p(v, y) f_{\hbox {\tiny $U$}}(w-v) dv, \label{eq:fweint}
\end{align}
where $y=m^*(w)+e^*$, one has 
\begin{align*}
& \frac{\partial }{\partial w}f_{\hbox {\tiny $W$}, e^*}(w, e^*) \\
= &\left\{\frac{d}{dw}m^*(w)\right\} 
\int p_y(v,y) f_{\hbox {\tiny $U$}}(w-v) dv+\int p(v,y) f'_{\hbox {\tiny $U$}}(w-v) dv \\
= &\left\{\frac{d}{dw}m^*(w)\right\} 
\int p_y(v,y) f_{\hbox {\tiny $U$}}(w-v) dv+\int p_x(v,y) f_{\hbox {\tiny $U$}}(w-v) dv, 
\end{align*}
where integration-by-part is used to obtain the last integral, $p_x(v,y)$ is equal to $(\partial/\partial x)p(x, y)$ evaluated at $(x,y)=(v, y)$, and $p_y(v,y)$ is equal to $(\partial/\partial y)p(x, y)$ evaluated at $(x,y)=(v, y)$. It follows that 
\begin{align*}
& \frac{\partial^2 }{\partial w^2}f_{\hbox {\tiny $W$}, e^*}(w, e^*) \\
= & \left\{\frac{d^2}{dw^2} m^*(w)\right\} \int p_y(v, y) f_{\hbox {\tiny $U$}}(w-v) dv+
\left\{\frac{d}{dw}m^*(w)\right\}^2\int p_{yy}(v,y)f_{\hbox {\tiny $U$}}(w-v) dv+\\
& \left\{\frac{d}{dw}m^*(w)\right\} \int p_y(v,y) f'_{\hbox {\tiny $U$}}(w-v) dv+
 \left\{\frac{d}{dw}m^*(w)\right\} \int p_{xy}(v,y) f_{\hbox {\tiny $U$}}(w-v) dv+\\
& \int p_x(v,y) f'_{\hbox {\tiny $U$}}(w-v) dv\\
= & \left\{\frac{d^2}{dw^2} m^*(w)\right\} \int p_y(v, y) f_{\hbox {\tiny $U$}}(w-v) dv+
\left\{\frac{d}{dw}m^*(w)\right\}^2\int p_{yy}(v,y)f_{\hbox {\tiny $U$}}(w-v) dv+\\
& \left\{\frac{d}{dw}m^*(w)\right\} \int p_{xy}(v,y) f_{\hbox {\tiny $U$}}(w-v) dv+
 \left\{\frac{d}{dw}m^*(w)\right\} \int p_{xy}(v,y) f_{\hbox {\tiny $U$}}(w-v) dv+\\
& \int p_{xx}(v,y) f_{\hbox {\tiny $U$}}(w-v) dv\\
= &  \int p_{xx}(v,y) f_{\hbox {\tiny $U$}}(w-v) dv+\left\{\frac{d^2}{dw^2} m^*(w)\right\} \int p_y(v, y) f_{\hbox {\tiny $U$}}(w-v) dv+\\
& \left\{\frac{d}{dw}m^*(w)\right\}^2\int p_{yy}(v,y)f_{\hbox {\tiny $U$}}(w-v) dv+
2\left\{\frac{d}{dw}m^*(w)\right\} \int p_{xy}(v,y) f_{\hbox {\tiny $U$}}(w-v) dv.\\
\end{align*}
Evaluating the last expression at $w=x$ and $e^*=y-m^*(x)$ gives $g_1(x, y)=\sum_{k=1}^4 I_k(x,y)$, 
where $I_1(x, y)$ is equal to $\int p_{xx}(v,y) f_{\hbox {\tiny $U$}}(w-v) dv$ evaluated at $(w,y)=(x, y)$, and $I_2(x,y)$, $I_3(x,y)$, $I_4(x,y)$ are the three functions defined in (\ref{eq:3Is}) evaluated at $(w, y)=(x, y)$, respectively. It is worth pointing out that, in the absence of measurement error, (\ref{eq:fweint}) can be simply viewed as $f_{\hbox {\tiny $W$}, e^*}(w, e^*)=p^*(w, y)=p(x, y)$, which is symbolically equivalent to viewing $\int p(v, y)f_{\hbox {\tiny $U$}}(w-v)dv$ as $p(x, y)$. Following this viewpoint, one has the following definitions of $I_k(x, y)$ in the absence of measurement error, for $k=2, 3, 4$, 
\begin{equation}
\label{eq:3Isnome}
\left\{
\begin{aligned}
I_2(x, y)& = m''(x) p_y(x, y), \\
I_3(x, y)& = \left\{m'(x)\right\}^2 p_{yy}(x, y), \\
I_4(x, y)& = 2m'(x)p_{xy}(x, y). 
\end{aligned}
\right.
\end{equation}

To this end, one has $\mathscr{T}_x\{g_1(\cdot, y)\}=\sum_{k=1}^4 \mathscr{T}_x\{I_k(\cdot,y)\}$, where 
\begin{align*}
\mathscr{T}_x\{I_1(\cdot,y)\}& = \frac{1}{2\pi}\int e^{-itx} \frac{1}{\phi_{\hbox {\tiny $U$}}(t)}
\int e^{itw}\int p_{xx}(v,y) f_{\hbox {\tiny $U$}}(w-v) dv dw dt \\
  &= \frac{1}{2\pi}\int e^{-itx} \frac{1}{\phi_{\hbox {\tiny $U$}}(t)}
\int e^{itv} p_{xx}(v,y) \int e^{it(w-v)}f_{\hbox {\tiny $U$}}(w-v)dw dvdt \\
	&= \frac{1}{2\pi}\int e^{-itx} \frac{1}{\phi_{\hbox {\tiny $U$}}(t)}
\int e^{itv} p_{xx}(v,y) \phi_{\hbox {\tiny $U$}}(t) dvdt \\
& = p_{xx}(x, y). 
\end{align*}
This proves (\ref{eq:fwefxe}), where the latter three transforms, $\mathscr{T}_x\{I_k(\cdot,y)\}$, for $k=2, 3, 4$, cannot be further simplified in the presence of measurement error without additional assumptions, such as those on $m^*(w)$.

To show (\ref{eq:fwepyy}), we first derive $g_2(x, y)$ defined in (\ref{eq:g2}). By (\ref{eq:fweint}), it is easy to see that $f^{(2)}_{\hbox {\tiny $W$}, e^*, 22}(w, e^*)=\int p_{yy}(v, y)f_{\hbox {\tiny $U$}}(w-v)dv$. Thus $g_2(w, y)=\int p_{yy}(v, y)f_{\hbox {\tiny $U$}}(w-v)dv$. It follows that  
\begin{align*}
\mathscr{T}_x\left\{g_2(\cdot, y)\right\} & = \frac{1}{2\pi}\int e^{-itx} \frac{1}{\phi_{\hbox {\tiny $U$}}(t)}
\int e^{itw}\int p_{yy}(v,y) f_{\hbox {\tiny $U$}}(w-v) dv dw dt \\
  &= \frac{1}{2\pi}\int e^{-itx} \frac{1}{\phi_{\hbox {\tiny $U$}}(t)}
\int e^{itv} p_{yy}(v,y) \int e^{it(w-v)}f_{\hbox {\tiny $U$}}(w-v)dw dvdt \\
	&= \frac{1}{2\pi}\int e^{-itx} \frac{1}{\phi_{\hbox {\tiny $U$}}(t)}
\int e^{itv} p_{yy}(v,y) \phi_{\hbox {\tiny $U$}}(t) dvdt \\
& = p_{yy}(x, y).
\end{align*} 
This completes the proof of (\ref{eq:fwepyy}).

\subsection{Consideration of two special cases in Section 3.3}
We state in Section 3.3 in the main article that, in the absence of measurement error, (\ref{eq:db4v2}) reduces to $\textrm{DB}_4(x, y, h_1, h_2)=\textrm{DB}_2(x, y, h_1, h_2)$. We first prove this statement in this subsection. 

Because $p(x, y)=f_{\hbox {\tiny $X$}, e}\{x, y-m(x)\}$, one has 
\begin{align*}
p_y(x, y) & = (\partial/\partial y)f_{\hbox {\tiny $X$}, e}\{x, y-m(x)\}=f^{(1)}_{\hbox {\tiny $X$}, e,2}(x, e), \\
p_{yy}(x, y) & = (\partial^2/\partial y^2)f_{\hbox {\tiny $X$}, e}\{x, y-m(x)\}=f^{(2)}_{\hbox {\tiny $X$}, e,22}(x, e), \\
p_{xy}(x, y) & = (\partial/\partial x)f^{(1)}_{\hbox {\tiny $X$}, e,2}\{x, y-m(x)\}=f^{(2)}_{\hbox {\tiny $X$}, e,21}(x, e)-m'(x)f^{(2)}_{\hbox {\tiny $X$}, e,22}(x, e).
\end{align*}
Using these three results in (\ref{eq:3Isnome}), one has that, in the absence of measurement error, 
\begin{align*}
& \sum_{k=2}^4 \mathscr{T}_x\{I_k(\cdot, y)\}\\
= & \sum_{k=2}^4 I_k(x, y) \\
= & m''(x) f^{(1)}_{\hbox {\tiny $X$}, e,2}(x, e)+ \left\{m'(x)\right\}^2 f^{(2)}_{\hbox {\tiny $X$}, e,22}(x, e)+2m'(x) \left\{ f^{(2)}_{\hbox {\tiny $X$}, e,21}(x, e)-m'(x)f^{(2)}_{\hbox {\tiny $X$}, e,22}(x, e) \right\}\\
= & m''(x)f^{(1)}_{\hbox {\tiny $X$}, e,2}(x, e)- \left\{m'(x)\right\}^2 f^{(2)}_{\hbox {\tiny $X$}, e,22}(x, e) +2m'(x) f^{(2)}_{\hbox {\tiny $X$}, e,21}(x, e), 
\end{align*}
which cancel with the last three terms in (\ref{eq:db4v2}) in the main article. This proves that $\textrm{DB}_4(x, y, h_1, h_2)=\textrm{DB}_2(x, y, h_1, h_2)$ in the absence of measurement error. 

In another special case considered in Section 3.3, we impose the following the conditions stated in \citet{Hyndman.etal1996}: (H1) the covariate is locally uniform near $x$ so that $f'_{\hbox {\tiny $X$}}(x)\approx 0$ and $f''_{\hbox {\tiny $X$}}(x)\approx 0$, (H2) $e\perp X$ so that $p(y|x)=f_e\{  y-m(x)\}$, and (H3) $m(x)$ is locally linear near $x$ so that $m''(x)\approx 0$. Next, we simplify the following dominating bias associated with $\hat p_2 (y|x)$, $\hat p_3 (y|x)$, and $\hat p_4 (y|x)$, respectively,
\begin{align}
\textrm{DB}_2(x, y, h_1, h_2)& =\frac{1}{2f_{\hbox{\tiny $X$}}(x)}\left[\left\{f_{\hbox{\tiny $X$}, e, 11}^{(2)}(x, e)-p(y|x)f_{\hbox{\tiny $X$}}''(x)  \right\} \mu_{2,1}h_1^2+p_{yy}(x,y)\mu_{2,2}h_2^2\right],\label{eq:db2}\\
\textrm{DB}_3(x, y, h_1, h_2) & =\frac{1}{2f_{\hbox {\tiny $X$}}(x)} \left[\left\{p_{xx}(x,y)-p(y|x)f''_{\hbox {\tiny $X$}}(x) \right\}\mu_{2,1}h_1^2 + p_{yy}(x,y)\mu_{2,2}h_2^2\right] \label{eq:db3app}\\
\textrm{DB}_4(x, y, h_1, h_2) & =  \frac{1}{2f_{\hbox {\tiny $X$}}(x)} \left(\left[p_{xx}(x, y)+\sum_{k=2}^4\mathscr{T}_x\{I_k(\cdot, y)\}-p(y|x)f''_{\hbox {\tiny $X$}}(x) \right]\mu_{2,1}h_1^2\right.\nonumber \\
&  + \left.p_{yy}(x,y)\mu_{2,2}h_2^2\right).
\label{eq:db4app}
\end{align}

First, by (H2),
\begin{equation} 
p_{yy}(x, y)=\frac{\partial^2}{\partial y^2} \{f_{\hbox {\tiny $X$}}(x)p(y|x)\}=
\frac{\partial^2}{\partial y^2} \left[f_{\hbox {\tiny $X$}}(x)f_e\{  y-m(x)\}\right]=f_{\hbox {\tiny $X$}}(x)f''_e(e).\label{eq:pyyapp}
\end{equation}
Second, by (H1) and (H2), $f_{\hbox{\tiny $X$}, e, 11}^{(2)}(x, e)$ and $f_{\hbox{\tiny $X$}}''(x)$ are approximately zero. Hence, (\ref{eq:db2}) reduces to 
$$ \textrm{DB}_2(x, y, h_1, h_2) \approx 0.5 f_e''(e)\mu_{2,2}h_2^2.$$
Third, 
\begin{equation}
\label{eq:pxxapp}
\begin{aligned}
p_{xx}(x, y) & =\frac{\partial^2}{\partial x^2} \{f_{\hbox {\tiny $X$}}(x)p(y|x)\}\\
& = \frac{\partial}{\partial x}\left[ f'_{\hbox {\tiny $X$}}(x)f_e(e)-f_{\hbox {\tiny $X$}}(x)f_e'(e)m'(x) \right], \textrm{ by (H2),} \\
& \approx -\frac{\partial}{\partial x}\left[ f_{\hbox {\tiny $X$}}(x)f_e'(e)m'(x) \right], \textrm{ by (H1),} \\
& = -f'_{\hbox {\tiny $X$}}(x)f_e'(e)m'(x)+f_{\hbox {\tiny $X$}}(x)f_e''(e)\{m'(x)\}^2-
f_{\hbox {\tiny $X$}}(x)f_e'(e)m''(x)\\
& \approx f_{\hbox {\tiny $X$}}(x)f_e''(e)\{m'(x)\}^2, \textrm{ by (H1) and (H3).}
\end{aligned}
\end{equation}
Hence, (\ref{eq:db3app}) simplifies to 
$$ \textrm{DB}_3(x, y, h_1, h_2) \approx 0.5 f_e''(e)\left[\left\{ m'(x) \right\}^2\mu_{2,1}h_1^2+\mu_{2,2}h_2^2\right].$$

Lastly, to simplify $\textrm{DB}_4(x, y, h_1, h_2)$, we need to look into the transform $\mathscr{T}_x\{I_k(\cdot, y)\}$, for $k=2, 3, 4$. The only reason that it is more difficult to obtain closed-form expressions of these three transforms than the same transform of $I_1(w, y)=\int p_{xx}(v,y) f_{\hbox {\tiny $U$}}(w-v) dv$ is the additional function of $w$ (as derivatives of $m^*(w)$) outside of the integrals in (\ref{eq:3Is}). If $m^*(w)$ is approximately linear so that $(d/dw)m^*(w)$ is approximately a constant, then this only obstacle disappears. This additional assumption can be satisfied for some measurement error models given Condition (H3). With this assumption on $m^*(w)$, one immediately has $I_2(w, y)\approx 0$ according to (\ref{eq:3Is}) and thus $\mathscr{T}_x\{I_2(\cdot, y)\}\approx 0$. Following the same idea behind the derivation for $\mathscr{T}_x\{I_1(\cdot, y)\}$ and the proof for (\ref{eq:fwepyy}) in Section~\ref{s:twolemmas}, one can show that,  
\begin{align*}
\mathscr{T}_x\{I_3(\cdot, y)\} & \approx \left\{ \frac{d}{dx}m^*(x) \right\}^2 p_{yy}(x,y)\\
& =\left\{ \frac{d}{dx}m^*(x) \right\}^2 f_{\hbox {\tiny $X$}}(x)f''_e(e), \textrm{ by (\ref{eq:pyyapp})},
\end{align*}
and 
\begin{align*}
\mathscr{T}_x\{I_4(\cdot, y)\}& \approx 2\left\{ \frac{d}{dx}m^*(x) \right\} p_{xy}(x,y) 
\\
& = 2\left\{ \frac{d}{dx}m^*(x) \right\} \frac{\partial}{\partial y} \left\{ \frac{\partial}{\partial x} f_{\hbox {\tiny $X$}}(x) f_e(e)\right\}, \textrm{ by (H2),}\\
& = 2\left\{ \frac{d}{dx}m^*(x) \right\} \frac{\partial}{\partial y} \left\{ -f_{\hbox {\tiny $X$}}(x) f'_e(e)m'(x)\right\}, \textrm{ by (H1),}\\
& =-2\left\{ \frac{d}{dx}m^*(x) \right\}f_{\hbox {\tiny $X$}}(x) f''_e(e)m'(x).
\end{align*}
Putting these results of $\mathscr{T}_x\{I_k(\cdot, y)\}$, for $k=2, 3, 4$, along with (\ref{eq:pxxapp}) and (\ref{eq:pyyapp}), back in (\ref{eq:db4app}), one has 
$$\textrm{DB}_4(x, y, h_1, h_2) \approx 0.5 f_e''(e)\left[\left\{m'(x)-\frac{d}{dx}m^*(x)\right\}^2\mu_{2,1}h_1^2+ \mu_{2,2} h_2^2\right]. 
$$

In summary, under the aforementioned special case, we have 
\begin{align*}
\textrm{DB}_2(x, y, h_1, h_2) & \approx 0.5 f_e''(e) \mu_{2,2} h_2^2, \\
\textrm{DB}_3(x, y, h_1, h_2) & \approx 0.5 f_e''(e)\left[\left\{m'(x)\right\}^2\mu_{2,1}h_1^2+ \mu_{2,2} h_2^2\right], \\
\textrm{DB}_4(x, y, h_1, h_2) & \approx 0.5 f_e''(e)\left[\left\{m'(x)-\frac{d}{dx}m^*(x)\right\}^2\mu_{2,1}h_1^2+ \mu_{2,2} h_2^2\right]. 
\end{align*}
These three approximations imply the comparison between $\textrm{DB}_3(x, y, h_1, h_2)$ and $\textrm{DB}_2(x, y, h_1, h_2)$, and that between $\textrm{DB}_4(x, y, h_1, h_2)$ and $\textrm{DB}_3(x, y, h_1, h_2)$ summarized in Section~\ref{s:theorems4}.

\setcounter{equation}{0}
\setcounter{figure}{0}
\renewcommand{\theequation}{C.\arabic{equation}}
\renewcommand{\thefigure}{C.\arabic{figure}}
\renewcommand{\thesection}{C.\arabic{section}}

\section*{Appendix C: Derivations of the CV criteria associated with $\tilde p_1(y|x)$ and $\tilde p_2(y|x)$}

The cross validation (CV) criterion proposed by \citet{Fan.Yim2004} and \citet{Hall.etal2004} for choosing bandwidths in the estimator of $p^*(y|x)$, $\tilde p_1(y|x)$, is given by (\ref{eq:CV}), the integral in which can be derived explicitly as follows when $K_2(t)$ is the Gaussian kernel.

By the definition of $\tilde p_{1,-j}(y|W_j)$, one has 
\begin{align*}
 &  \int \left\{\tilde p_{1,-j}(y|W_j)\right\}^2dy \\
= & \int \left\{
  \frac{\ds\frac{1}{(n-1)h_1h_2}\sum_{j'\ne j} K_1 \left(\frac{W_{j'}-W_j}{h_1} \right) K_2 \left(\frac{Y_{j'}-y}{h_2} \right)}
 {\ds\frac{1}{(n-1)h_1}\sum_{j'\ne j} K_1 \left(\frac{W_{j'}-W_j}{h_1} \right)}
  \right\}^2 dy \\
= & \left\{\sum_{j'\ne j} K_1 \left(\frac{W_{j'}-W_j}{h_1} \right)\right\}^{-2} 
\int  \frac{1}{h^2_2}  \sum_{j_1\ne j}\sum_{j_2\ne j} \left\{K_1 \left(\frac{W_{j_1}-W_j}{h_1}\right)\times\right.\\
& \left. K_1 \left(\frac{W_{j_2}-W_j}{h_1}\right)  K_2 \left(\frac{Y_{j_1}-y}{h_2}\right) K_2 \left(\frac{Y_{j_2}-y}{h_2}\right)\right\}  dy \\
= & \left\{\sum_{j'\ne j} K_1 \left(\frac{W_{j'}-W_j}{h_1} \right)\right\}^{-2} 
\frac{1}{h^2_2}  \sum_{j_1\ne j}\sum_{j_2\ne j} \left\{K_1 \left(\frac{W_{j_1}-W_j}{h_1}\right) \times\right. \\
& \left. K_1 \left(\frac{W_{j_2}-W_j}{h_1}\right)  \int K_2 \left(\frac{Y_{j_1}-y}{h_2}\right) K_2 \left(\frac{Y_{j_2}-y}{h_2}\right)dy\right\},
\end{align*}
in which 
\begin{align*}
& \int K_2 \left(\frac{Y_{j_1}-y}{h_2}\right) K_2 \left(\frac{Y_{j_2}-y}{h_2}\right)  dy\\
= & h_2 \int K_2 (t) K_2 \left(t-\frac{Y_{j_1}-Y_{j_2}}{h_2}\right)  dt\\
= & h_2 \int \frac{1}{\sqrt{2\pi}}\exp(-t^2/2) \cdot \frac{1}{\sqrt{2\pi}}\exp\left\{-\frac{1}{2}\left(t-\frac{Y_{j_1}-Y_{j_2}}{h_2} \right)^2\right\} dt \\
= & \frac{h_2}{\sqrt{4\pi}}\exp\left\{-\left( \frac{Y_{j_1}-Y_{j_2}}{2h_2} \right)^2\right\}.
\end{align*}
Hence, 
\begin{align*}
 &  \int \left\{\tilde p_{1,-j}(y|W_j)\right\}^2dy \\
=& \frac {\ds\frac{1}{\sqrt{4\pi}h_2} \sum_{j_1\ne j}\sum_{j_2\ne j} K_1 \left(\frac{W_{j_1}-W_j}{h_1}\right) K_1 \left(\frac{W_{j_2}-W_j}{h_1}\right) \exp\left\{-\left( \frac{Y_{j_1}-Y_{j_2}}{2h_2} \right)^2\right\} } {\ds\left\{\sum_{j'\ne j} K_1 \left(\frac{W_{j'}-W_j}{h_1} \right)\right\}^2}.
\end{align*}
Using this result in the first half of (\ref{eq:CV}), and using the definition of $\tilde p_{1,-j}(y|W_j)$ in the second half of (\ref{eq:CV}) leads to the following elaboration of (\ref{eq:CV}), 
\begin{equation}\label{eq:CV:formula1}
\begin{aligned}
& \textrm{CV}(\tilde p_1) \\
= & \frac{1}{nh_2}\sum_{j=1}^n\omega(W_j)\times \\
& \left[\frac{\ds \frac{1}{\sqrt{4\pi}}\sum_{j_1\neq j}\sum_{j_2\neq j}K_1\left(\frac{W_{j_1}-W_j}{h_1}\right)K_1\left(\frac{W_{j_2}-W_j}{h_1}\right)\exp\left\{-\left(\frac{Y_{j_1}-Y_{j_2}}{2h_2}\right)^2\right\}}{\left\{\ds\sum_{j'\neq j}K_1\left(\frac{W_{j'}-W_j}{h_1}\right)\right\}^2}\right.\\
& \left.-2 \frac{\ds\sum_{j'\neq j}^nK_1\left(\frac{W_{j'}-W_j}{h_1}\right)K_2\left(\frac{Y_{j'}-Y_{j}}{h_2}\right)}{\ds\sum_{j'\neq j}K_1\left(\frac{W_{j'}-W_j}{h_1}\right)} \right]. 
\end{aligned}
\end{equation}

Following similar derivations leading to (\ref{eq:CV:formula1}), one can show that, with $K_2(t)$ being the Gaussian kernel, (\ref{eq:CV2}) becomes 
\begin{equation}
\label{eq:CV:formula2}
\begin{aligned}
& \textrm{CV}(\tilde p_2) \\
 = & \frac{1}{nh_2}\sum_{j=1}^n\omega(W_j)\times \\
& \left[\frac{\ds \frac{1}{\sqrt{4\pi}}\sum_{j_1\neq j}\sum_{j_2\neq j}K_1\left(\frac{W_{j_1}-W_j}{h_1}\right)K_1\left(\frac{W_{j_2}-W_j}{h_1}\right)\exp\left\{-\left(\frac{e^*_{j_1}-e^*_{j_2}}{2h_2}\right)^2\right\}}{\left\{\ds\sum_{j'\neq j}K_1\left(\frac{W_{j'}-W_j}{h_1}\right)\right\}^2}\right.\\
& \left.-2 \frac{\ds\sum_{j'\neq j}^nK_1\left(\frac{W_{j'}-W_j}{h_1}\right)K_2\left(\frac{e^*_{j'}-e^*_{j}}{h_2}\right)}{\ds\sum_{j'\neq j}K_1\left(\frac{W_{j'}-W_j}{h_1}\right)} \right], 
\end{aligned}
\end{equation}
where $e_j^*=Y_j-\hat m^*(W_j)$.

\setcounter{equation}{0}
\setcounter{figure}{0}
\renewcommand{\theequation}{D.\arabic{equation}}
\renewcommand{\thefigure}{D.\arabic{figure}}
\renewcommand{\thesection}{D.\arabic{section}}

\section*{Appendix D: Boxplots of EISE associated with four density estimators when $m(x)\equiv 1$}
\begin{figure}
	\centering
	\setlength{\linewidth}{0.45\linewidth}
	\subfigure[]{ \includegraphics[width=\linewidth]{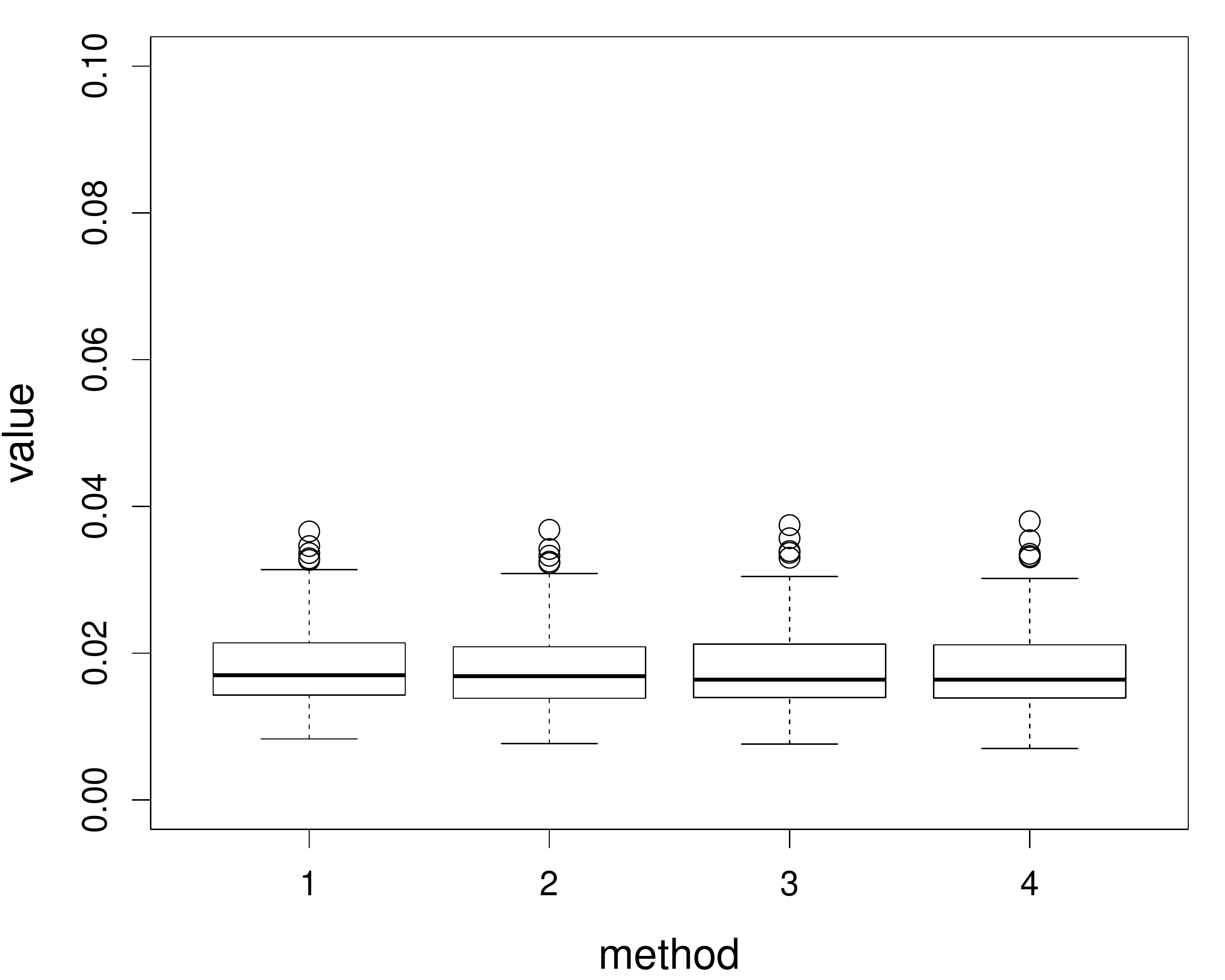} }
	\subfigure[]{ \includegraphics[width=\linewidth]{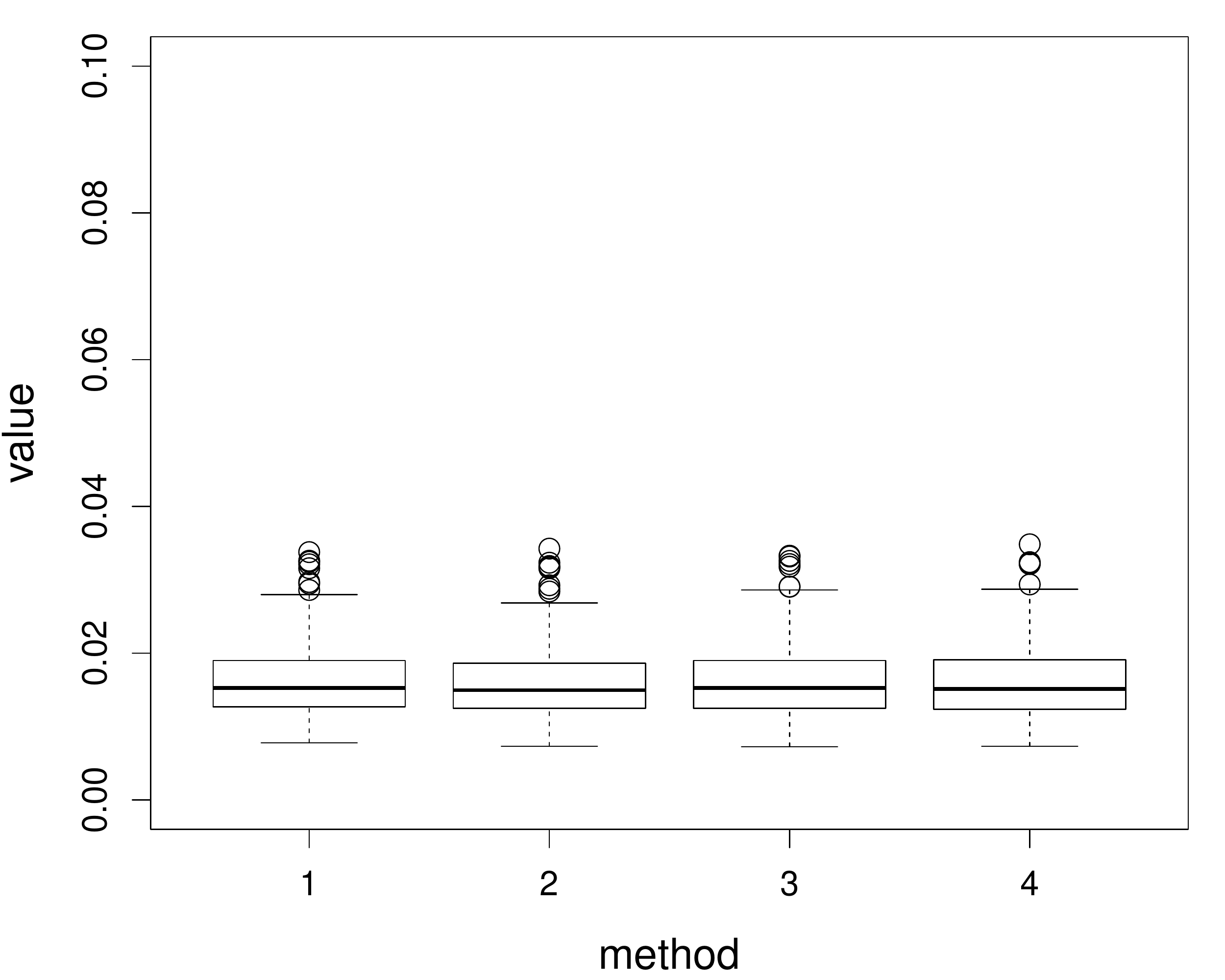} }\\
	\subfigure[]{ \includegraphics[width=\linewidth]{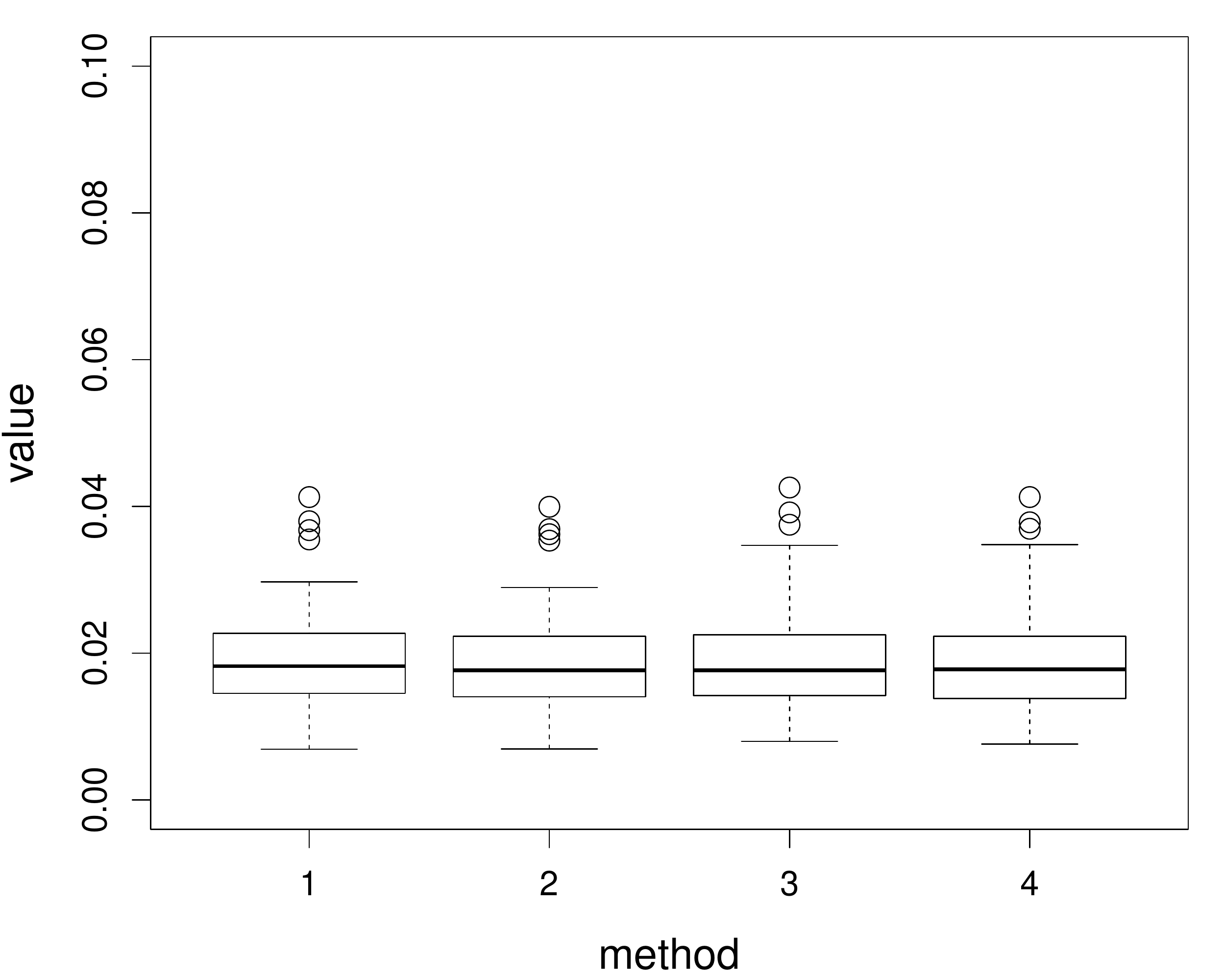} }
	\subfigure[]{ \includegraphics[width=\linewidth]{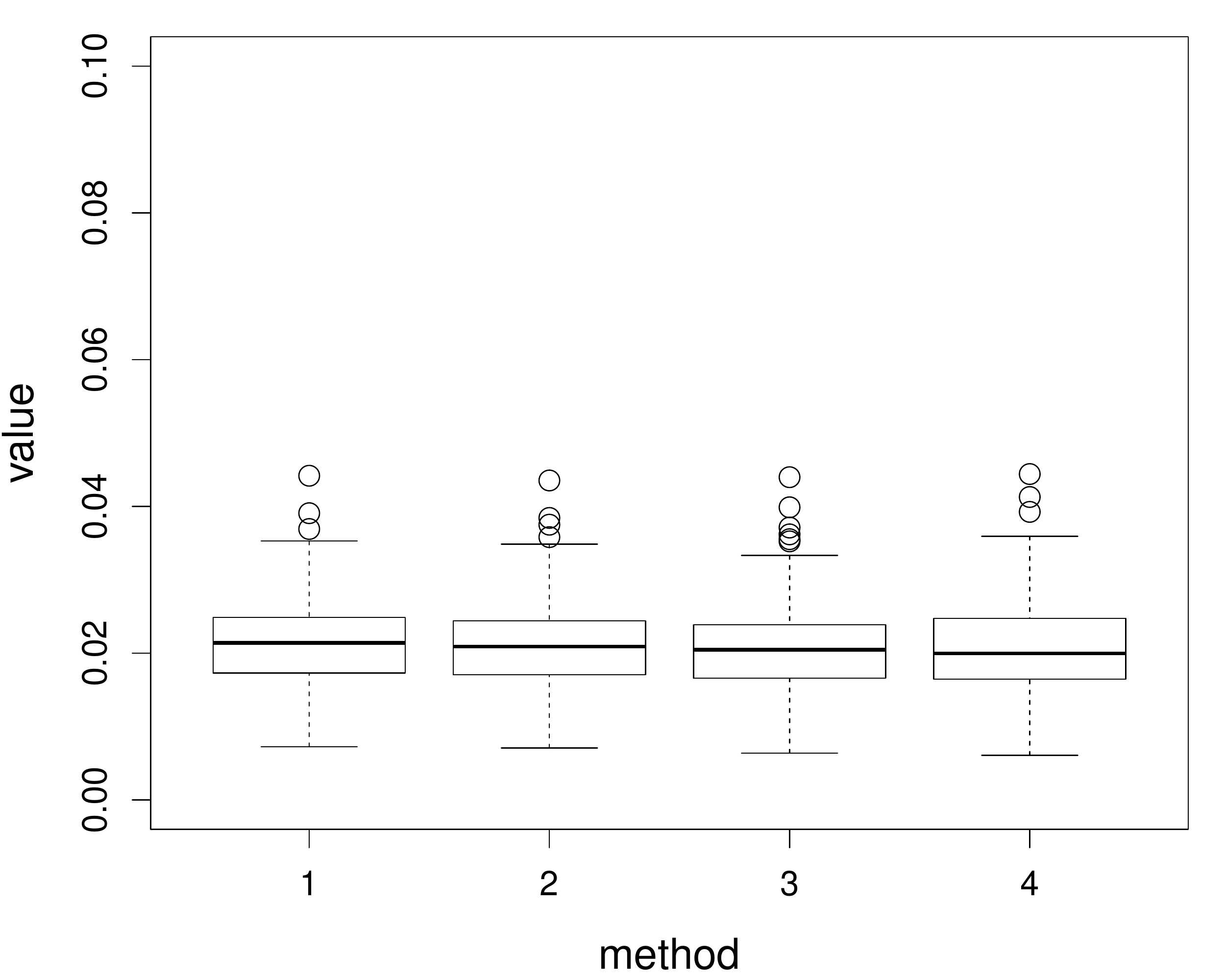} }
	\caption{Boxplots of EISE using the approximated theoretical optimal bandwidths when the primary model is (C4) and the secondary models are (a) $X\sim N(0,1)$, $U\sim \textrm{Laplace}(0, \,\sigma_u/\sqrt{2})$, $\lambda=0.8$; (b) $X\sim N(0,1)$, $U\sim \textrm{Laplace}(0, \,\sigma_u/\sqrt{2})$, $\lambda=0.9$; (c) $X\sim N(0,1)$, $U\sim N(0, \, \sigma_u^2)$, $\lambda=0.8$; (d) $X\sim \textrm{Uniform}(-2, 2)$, $U\sim \textrm{Laplace}(0, \,\sigma_u/\sqrt{2})$, $\lambda=0.8$. Method 1, 2, 3, 4 correspond to $\tilde p_1(y|x)$, $\tilde p_2(y|x)$, $\hat p_3(y|x)$, and $\hat p_4(y|x)$, respectively.} 
	\label{Sim4:box}
\end{figure}

\begin{figure}
	\centering
	\setlength{\linewidth}{0.45\linewidth}
	\subfigure[]{ \includegraphics[width=\linewidth]{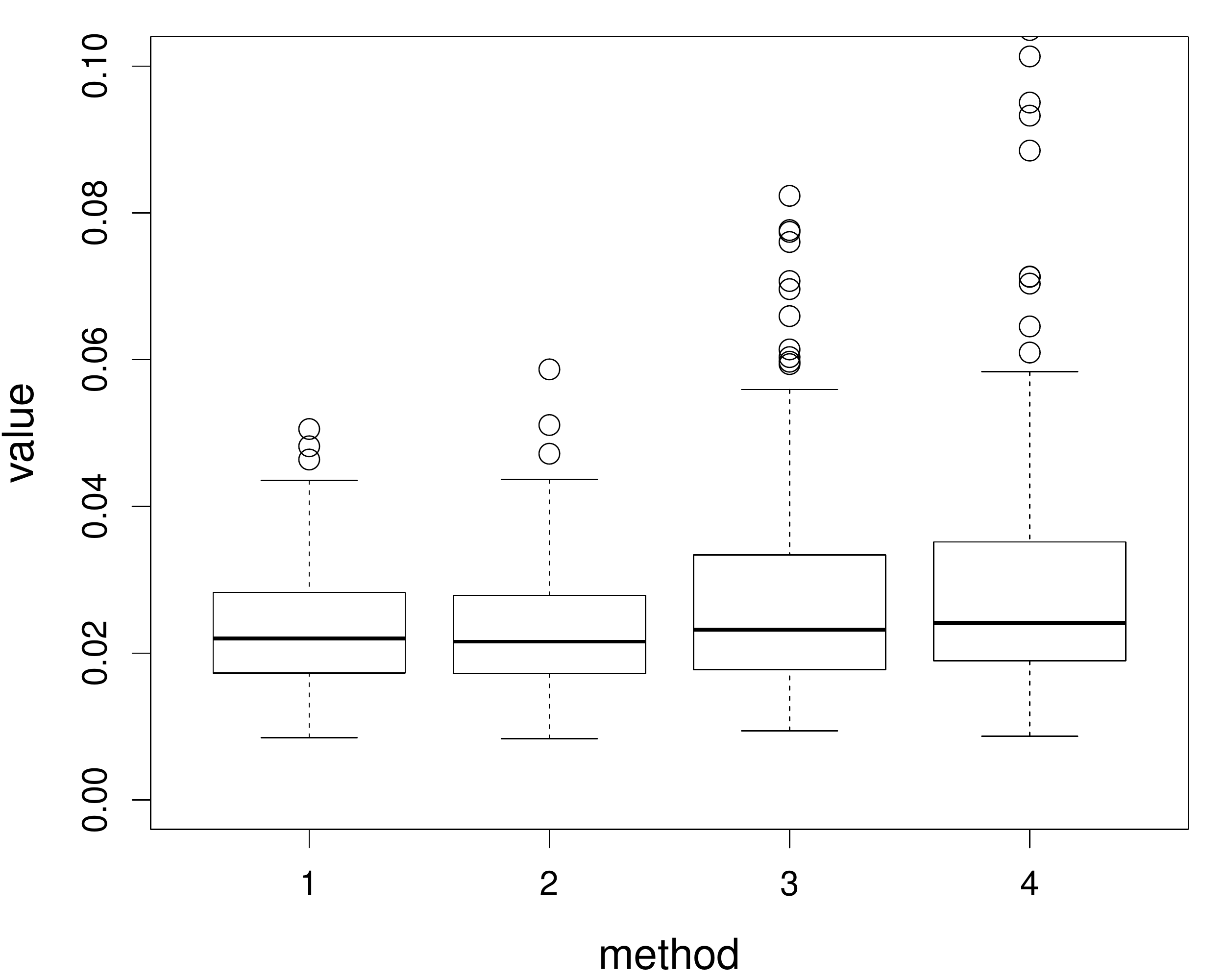} }
	\subfigure[]{ \includegraphics[width=\linewidth]{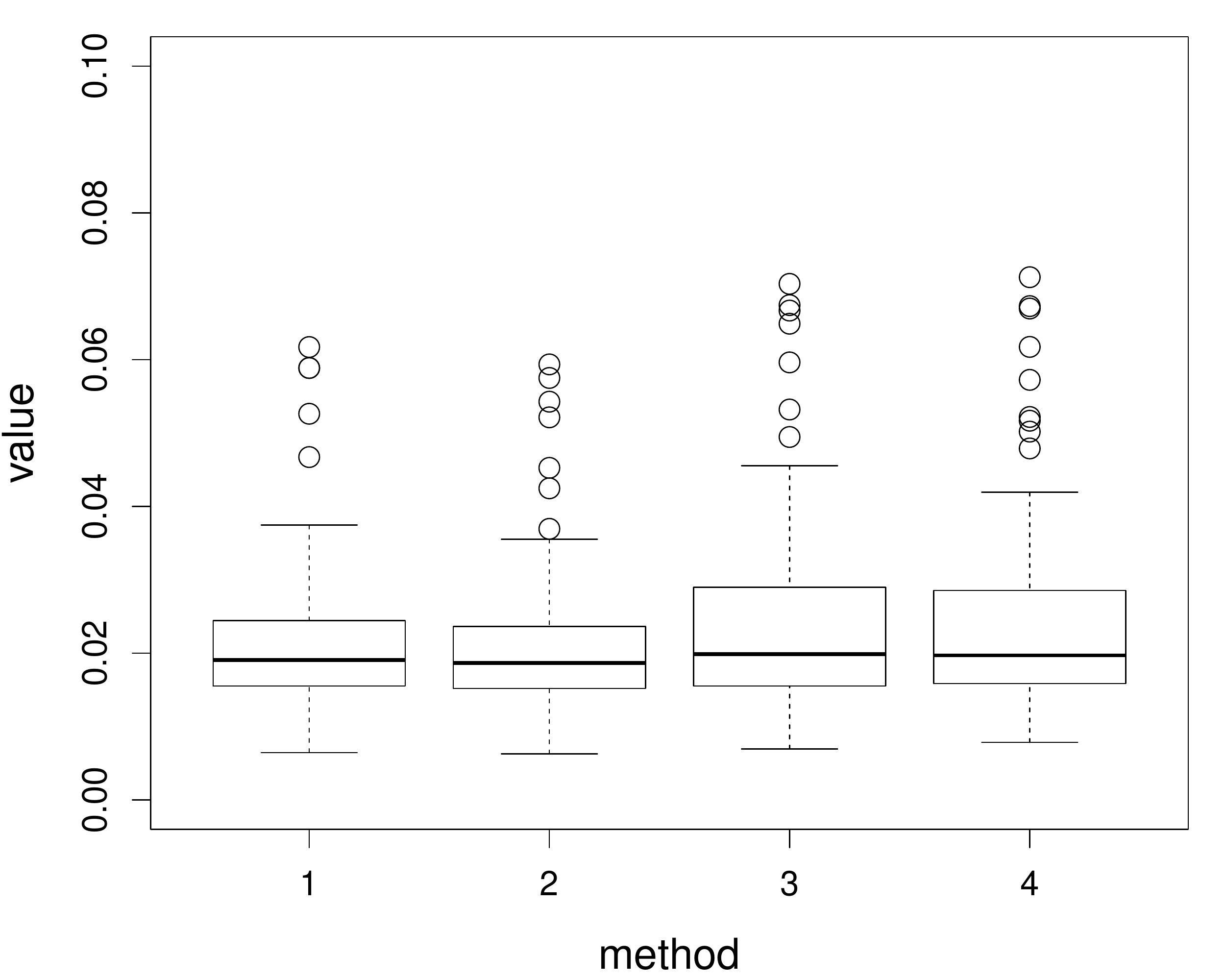} }\\
	\subfigure[]{ \includegraphics[width=\linewidth]{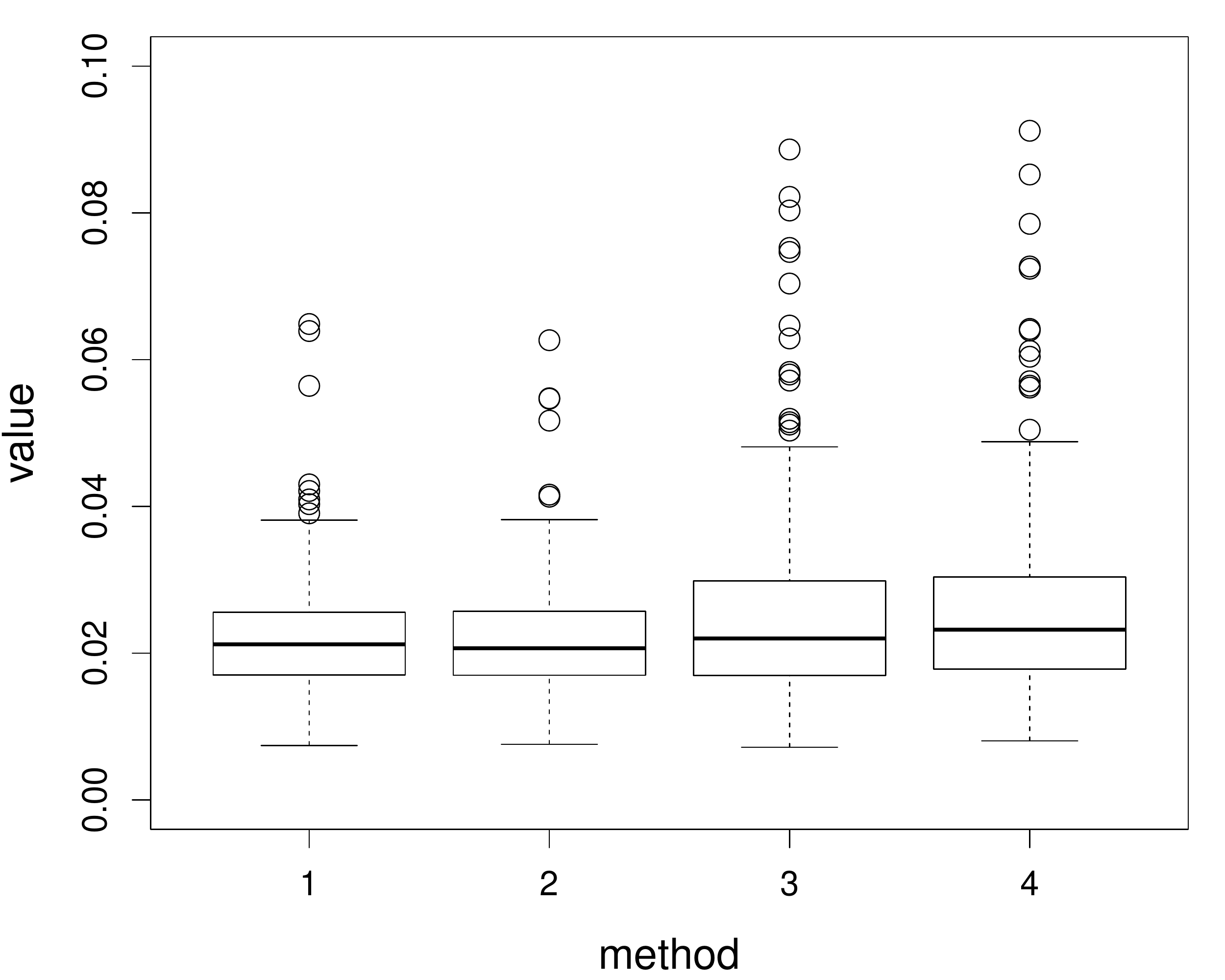} }
	\subfigure[]{ \includegraphics[width=\linewidth]{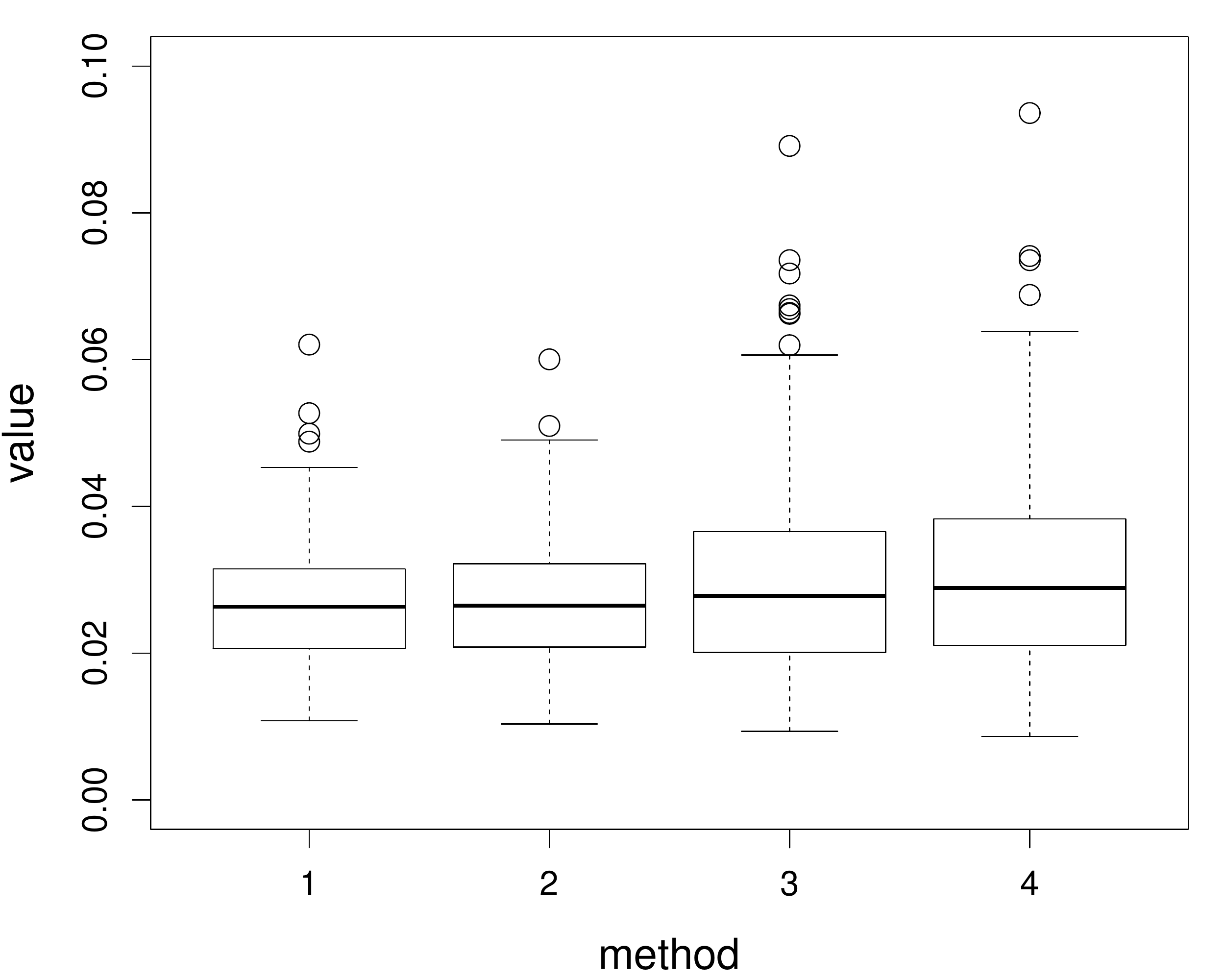} }
	\caption{Boxplots of EISE using the fully data-driven bandwidths when the primary model is (C4) and the secondary models are (a) $X\sim N(0,1)$, $U\sim \textrm{Laplace}(0, \,\sigma_u/\sqrt{2})$, $\lambda=0.8$; (b) $X\sim N(0,1)$, $U\sim \textrm{Laplace}(0, \,\sigma_u/\sqrt{2})$, $\lambda=0.9$; (c) $X\sim N(0,1)$, $U\sim N(0, \, \sigma_u^2)$, $\lambda=0.8$; (d) $X\sim \textrm{Uniform}(-2, 2)$, $U\sim \textrm{Laplace}(0, \,\sigma_u/\sqrt{2})$, $\lambda=0.8$. Method 1, 2, 3, 4 correspond to $\tilde p_1(y|x)$, $\tilde p_2(y|x)$, $\hat p_3(y|x)$, and $\hat p_4(y|x)$, respectively.} 
	\label{Sim4:box:data}
\end{figure}

We simplify the primary model setting (C1) in Section 5.1 in the main article to create the following primary model setting, 
\begin{itemize}
\item[(C4)] $[Y|X=x]\sim N\left(m(x), \, \sigma^2(x)\right)$, where $m(x)\equiv 1$ and $\sigma(x)=e^{1-x/3}/8$.
\end{itemize}
Along with the secondary model settings (a)--(d) stated in Section 5.1 in the main article, we now have four data generating processes,  according to each of which data of the form $\{(W_j, Y_j)\}_{j=1}^{500}$ are generated independently 200 times. Figure~\ref{Sim4:box} provides boxplots of EISE associated with $\tilde p_1(y|x)$, $\tilde p_2(y|x)$, $\hat p_3(y|x)$, and $\hat p_4(y|x)$ when the approximated theoretical optimal bandwidths are used, which suggest that all four estimators perform similarly. Figure~\ref{Sim4:box:data}
shows the same boxplots when the fully data-driven bandwidths are used. From there one can see that the two non-naive estimators are more variable than their naive counterparts, but are otherwise comparable. Between the two non-naive estimators, $\hat p_3(y|x)$ is slightly less variable than $\hat p_4(y|x)$. 

\setcounter{equation}{0}
\setcounter{figure}{0}
\renewcommand{\theequation}{E.\arabic{equation}}
\renewcommand{\thefigure}{E.\arabic{figure}}
\renewcommand{\thesection}{E.\arabic{section}}

\section*{Appendix E: Boxplots of EISE under the simulation settings in Section 5 when cubic spline estimates of the mean function are used}

\begin{figure}
	\centering
	\setlength{\linewidth}{0.45\linewidth}
	\subfigure[]{ \includegraphics[width=\linewidth]{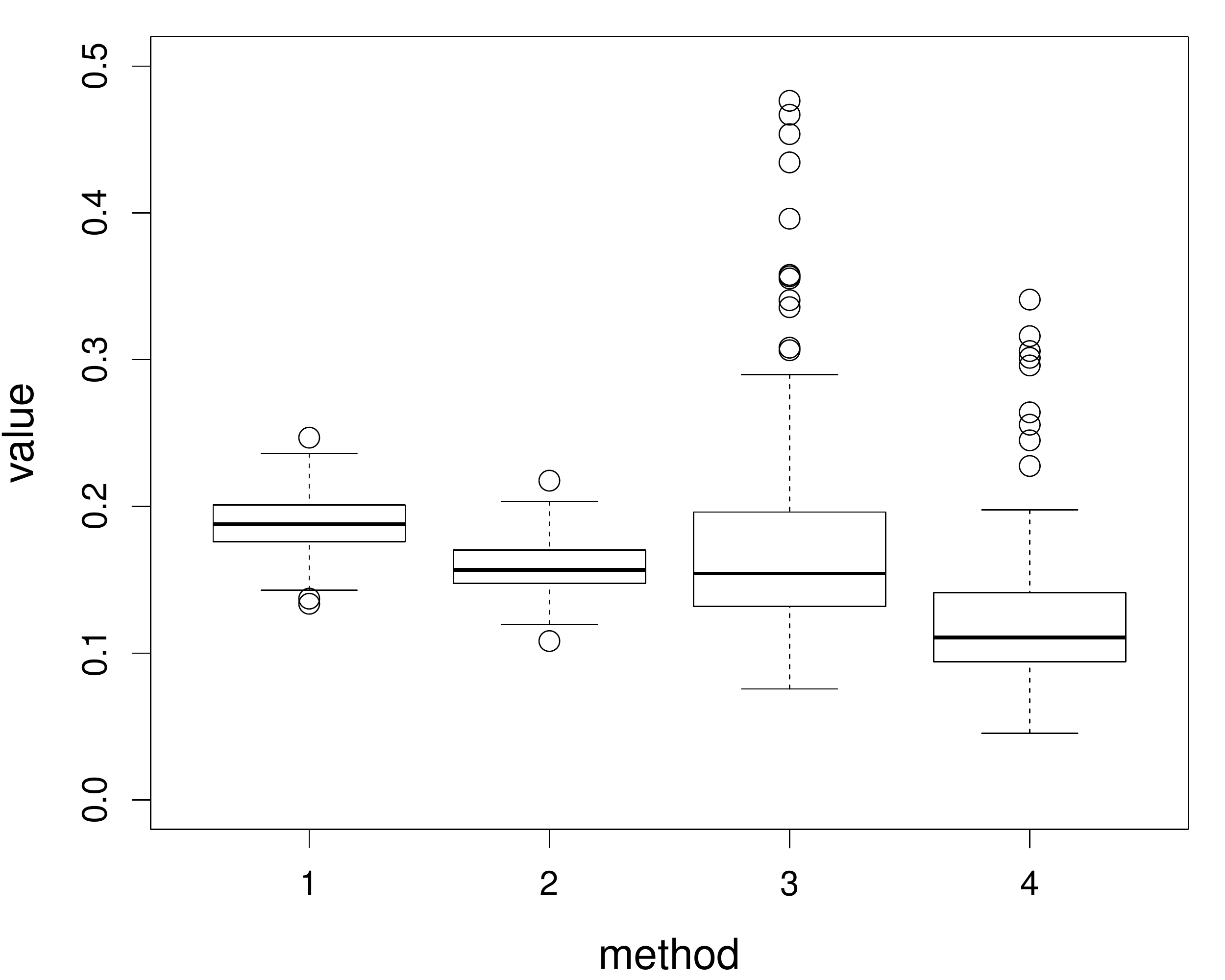} }
	\subfigure[]{ \includegraphics[width=\linewidth]{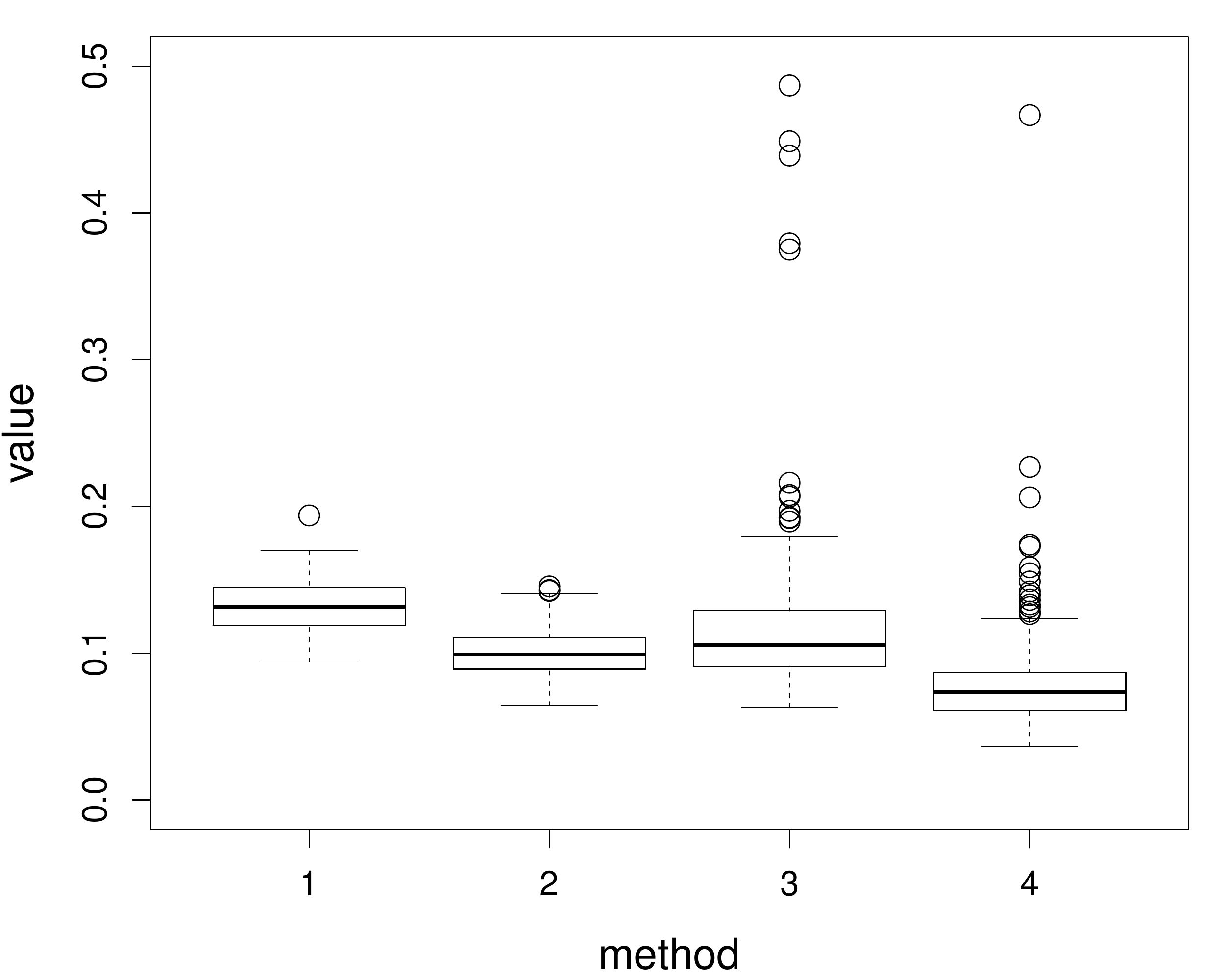} }\\
	\subfigure[]{ \includegraphics[width=\linewidth]{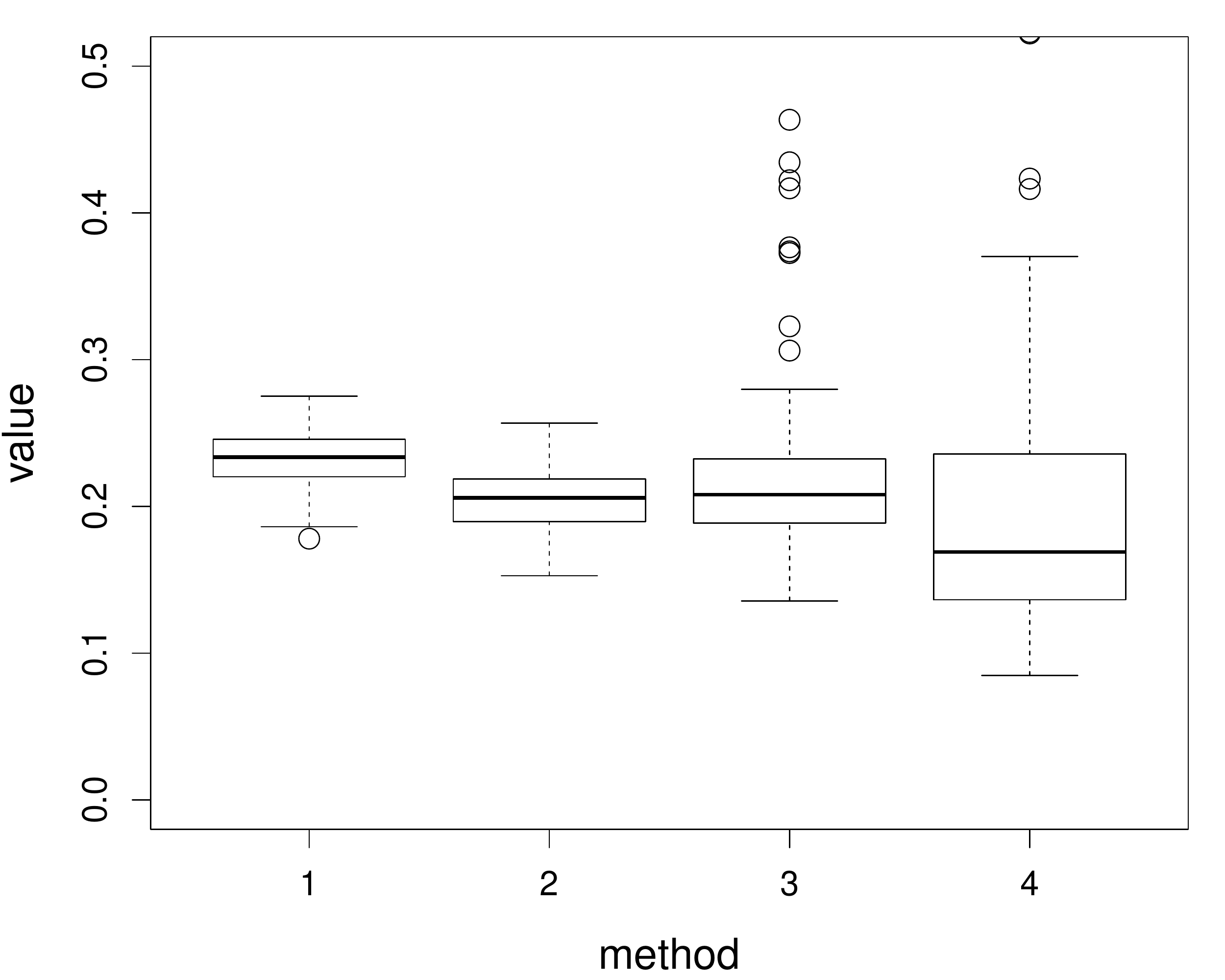} }
	\subfigure[]{ \includegraphics[width=\linewidth]{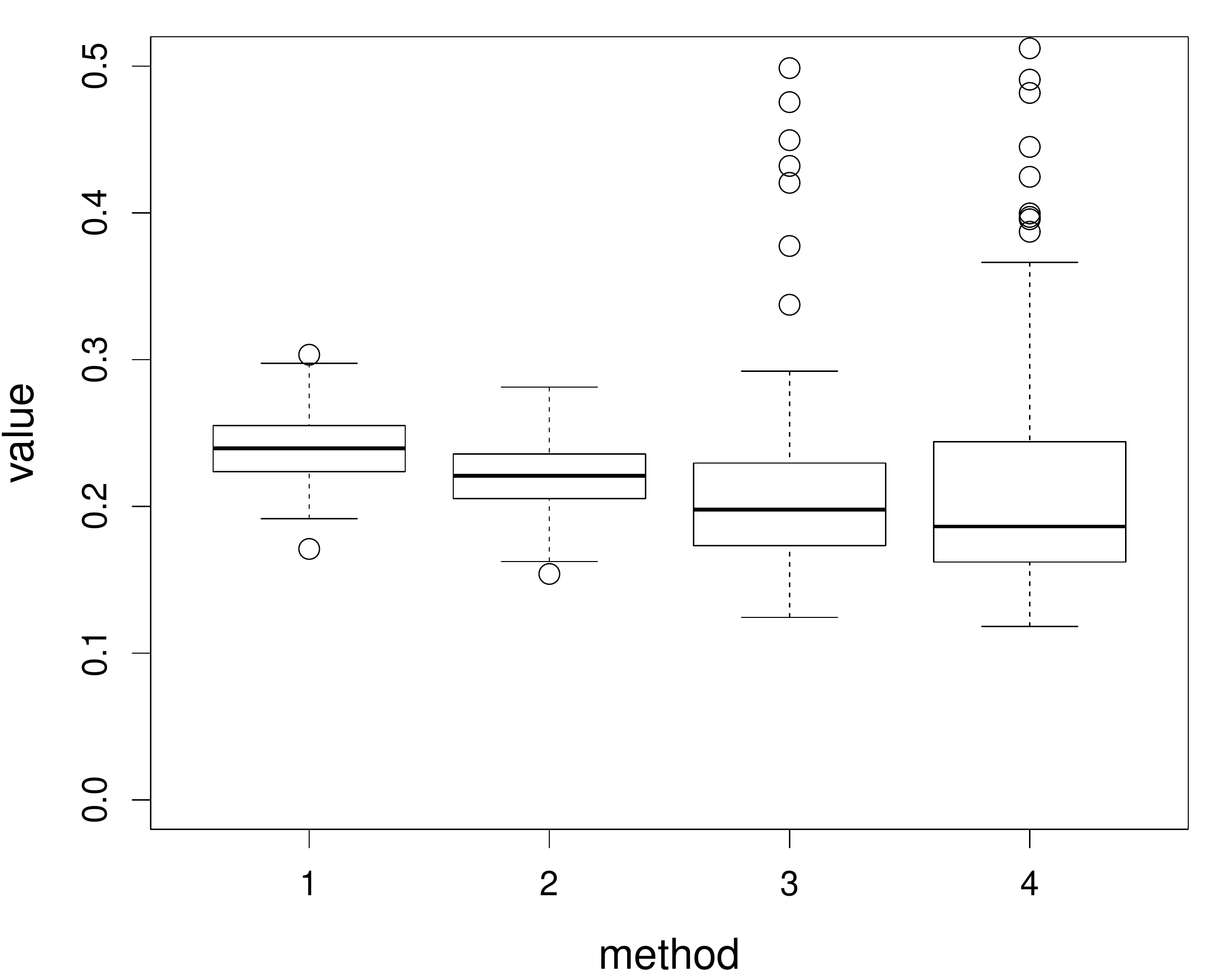} }
	\caption{Boxplots of EISE using the fully data-driven bandwidths with the cubic spline mean estimate when the primary model is (C1) and the secondary models are (a) $X\sim N(0,1)$, $U\sim \textrm{Laplace}(0, \,\sigma_u/\sqrt{2})$, $\lambda=0.8$; (b) $X\sim N(0,1)$, $U\sim \textrm{Laplace}(0, \, \sigma_u/\sqrt{2})$, $\lambda=0.9$; (c) $X\sim N(0,1)$, $U\sim N(0, \, \sigma_u^2)$, $\lambda=0.8$; (d) $X\sim \textrm{Uniform}(-2, 2)$, $U\sim \textrm{Laplace}(0, \,\sigma_u/\sqrt{2})$, $\lambda=0.8$. Method 1, 2, 3, 4 correspond to $\tilde p_1(y|x)$, $\tilde p_2(y|x)$, $\hat p_3(y|x)$, and $\hat p_4(y|x)$, respectively.} 
	\label{Sim1:box:data:spline}
\end{figure}

\begin{figure}
	\centering
	\setlength{\linewidth}{0.45\linewidth}
	\subfigure[]{ \includegraphics[width=\linewidth]{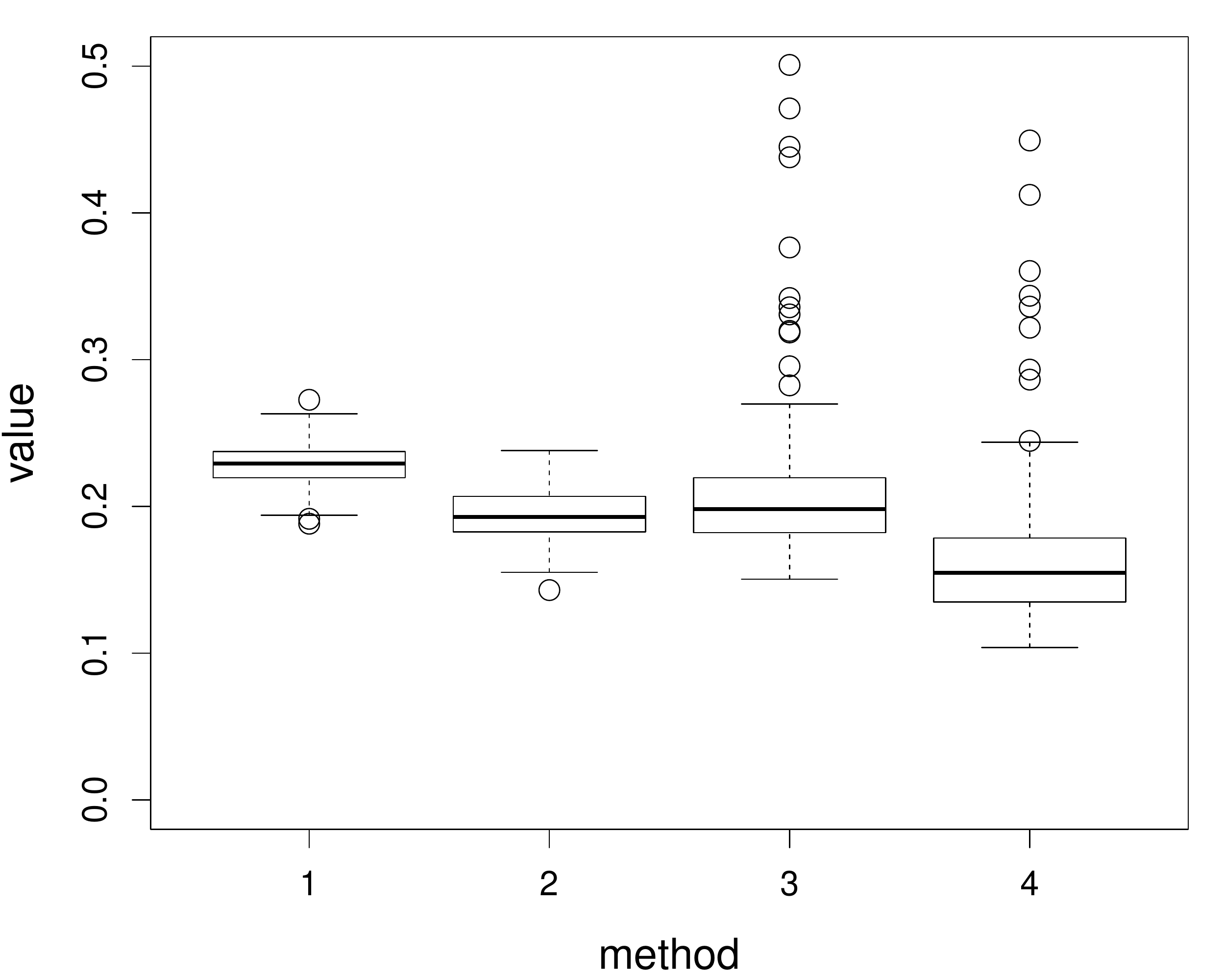} }
	\subfigure[]{ \includegraphics[width=\linewidth]{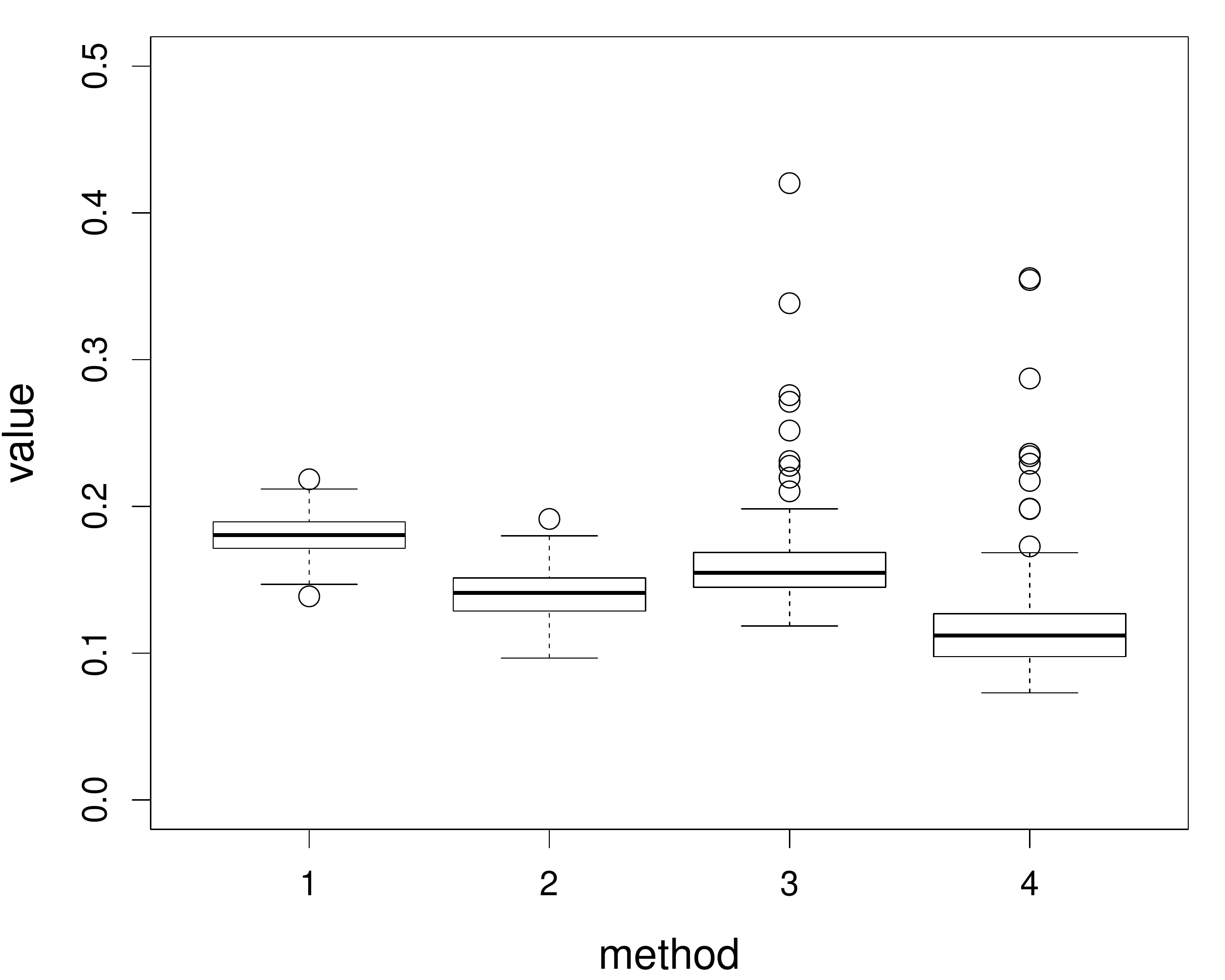} }\\
	\subfigure[]{ \includegraphics[width=\linewidth]{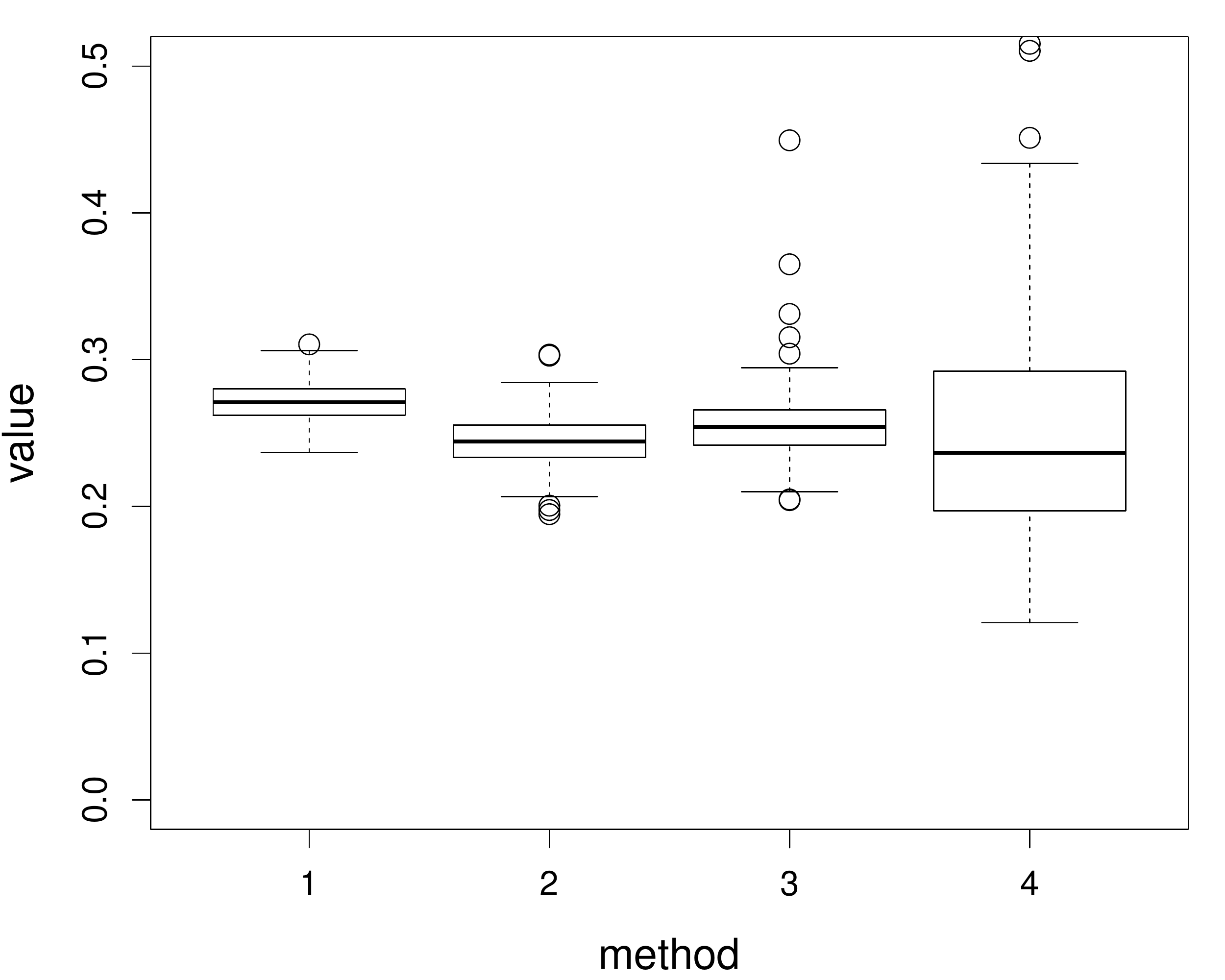} }
	\subfigure[]{ \includegraphics[width=\linewidth]{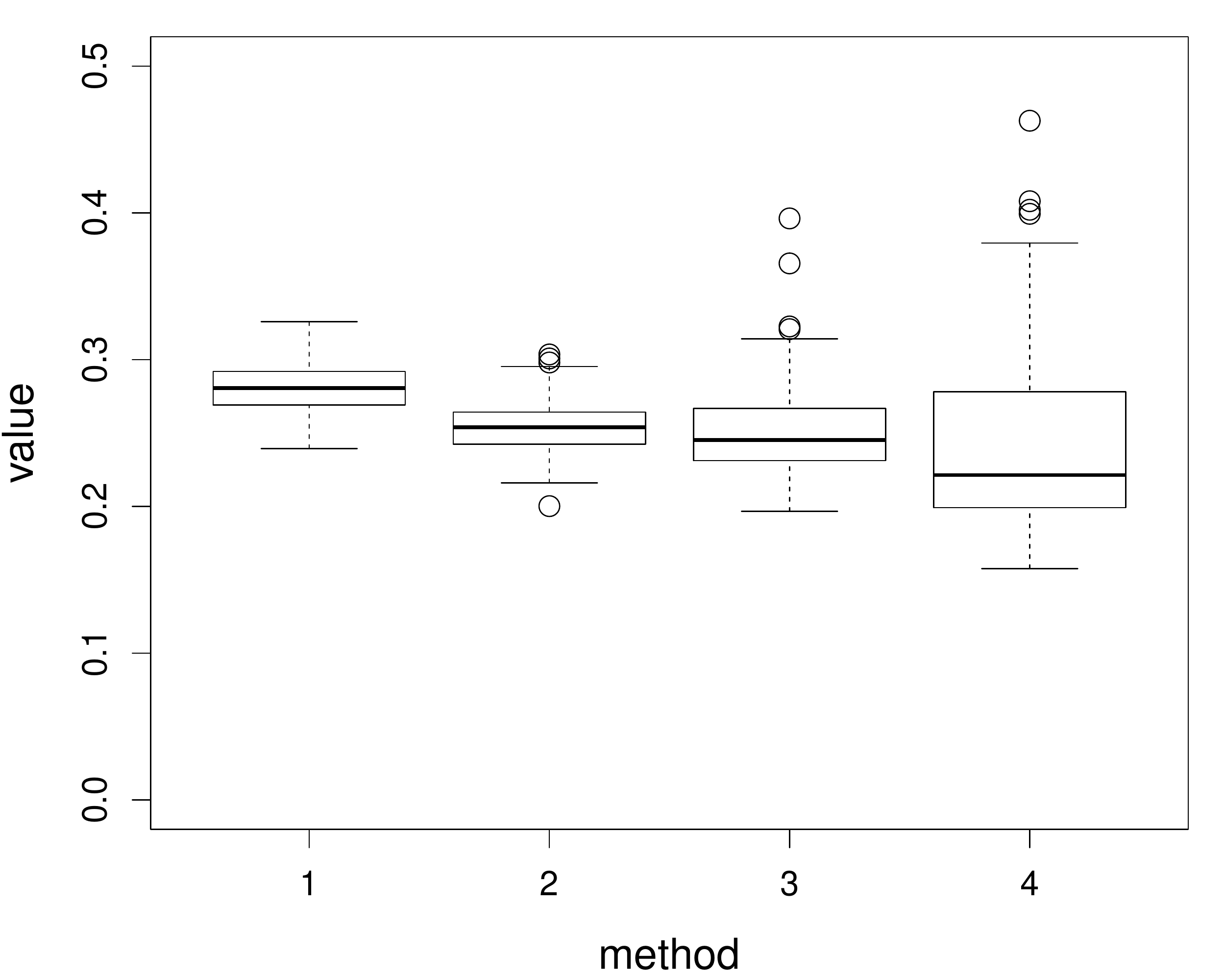} }
	\caption{Boxplots of EISE using the fully data-driven bandwidths with the cubic spline mean estimate when the primary model is (C2) and the secondary models are (a) $X\sim N(0,1)$, $U\sim \textrm{Laplace}(0, \,\sigma_u/\sqrt{2})$, $\lambda=0.8$; (b) $X\sim N(0,1)$, $U\sim \textrm{Laplace}(0, \, \sigma_u/\sqrt{2})$, $\lambda=0.9$; (c) $X\sim N(0,1)$, $U\sim N(0, \, \sigma_u^2)$, $\lambda=0.8$; (d) $X\sim \textrm{Uniform}(-2, 2)$, $U\sim \textrm{Laplace}(0, \,\sigma_u/\sqrt{2})$, $\lambda=0.8$. Method 1, 2, 3, 4 correspond to $\tilde p_1(y|x)$, $\tilde p_2(y|x)$, $\hat p_3(y|x)$, and $\hat p_4(y|x)$, respectively.} 
	\label{Sim2:box:data:spline}
\end{figure}

\begin{figure}
	\centering
	\setlength{\linewidth}{0.45\linewidth}
	\subfigure[]{ \includegraphics[width=\linewidth]{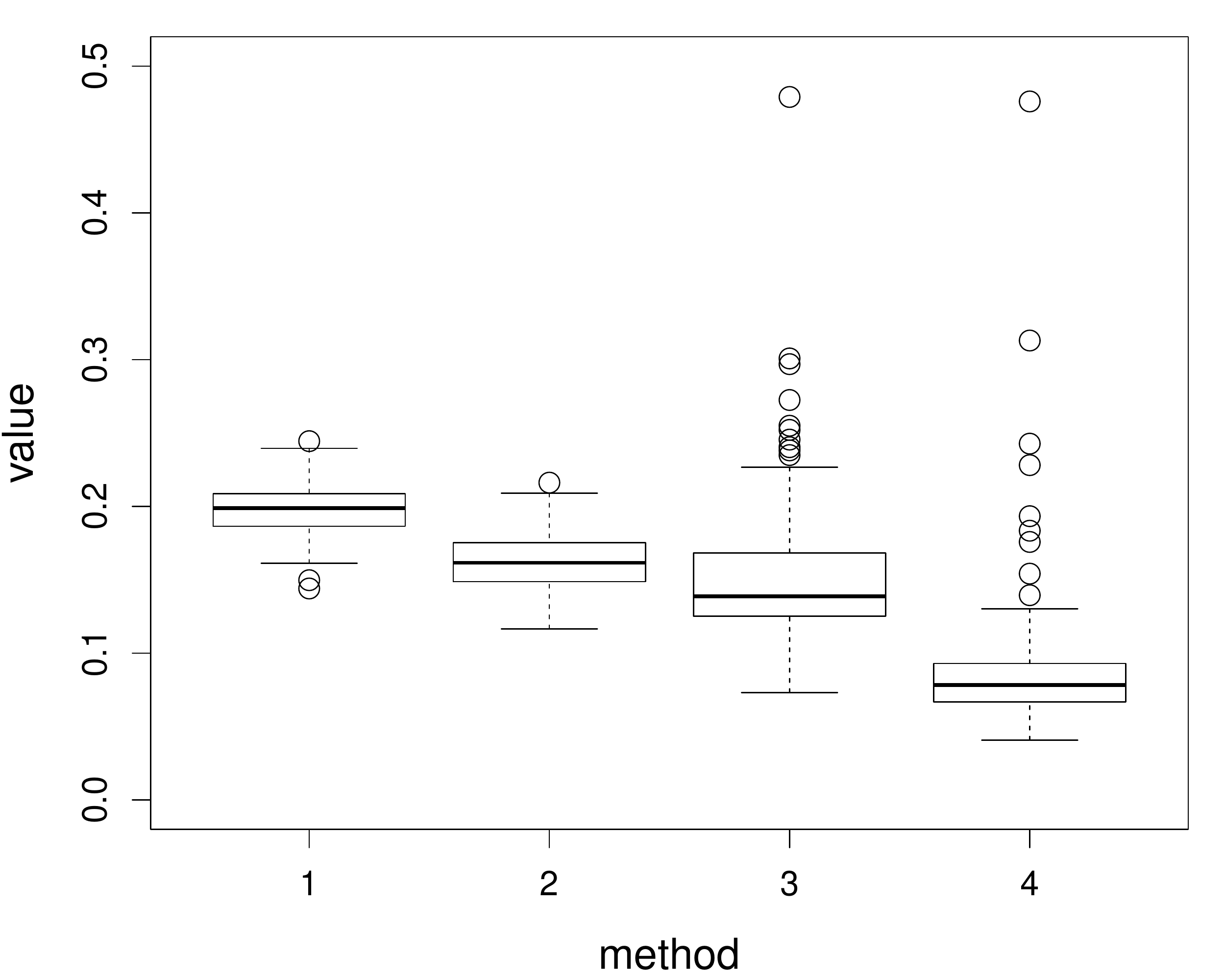} }
	\subfigure[]{ \includegraphics[width=\linewidth]{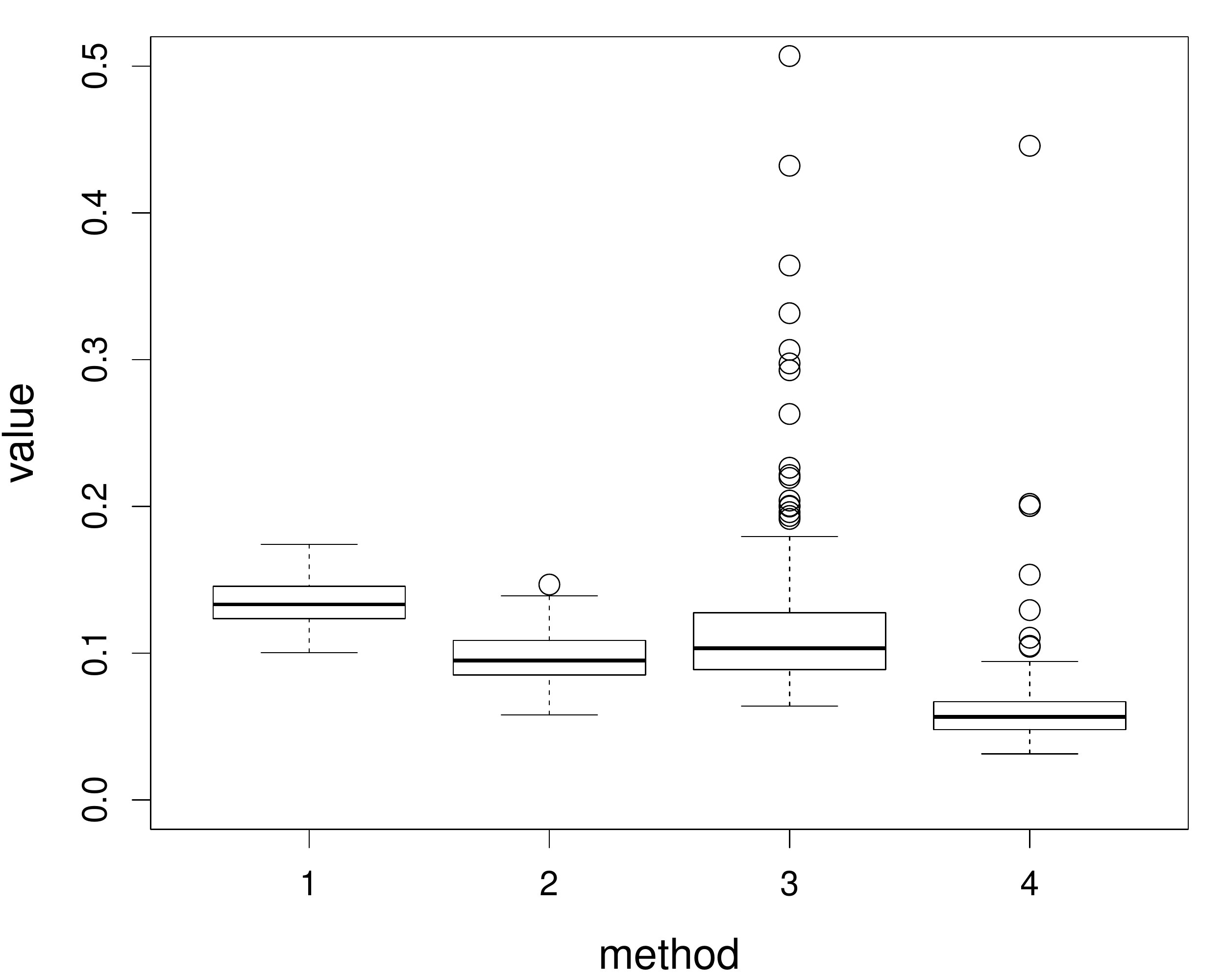} }\\
	\subfigure[]{ \includegraphics[width=\linewidth]{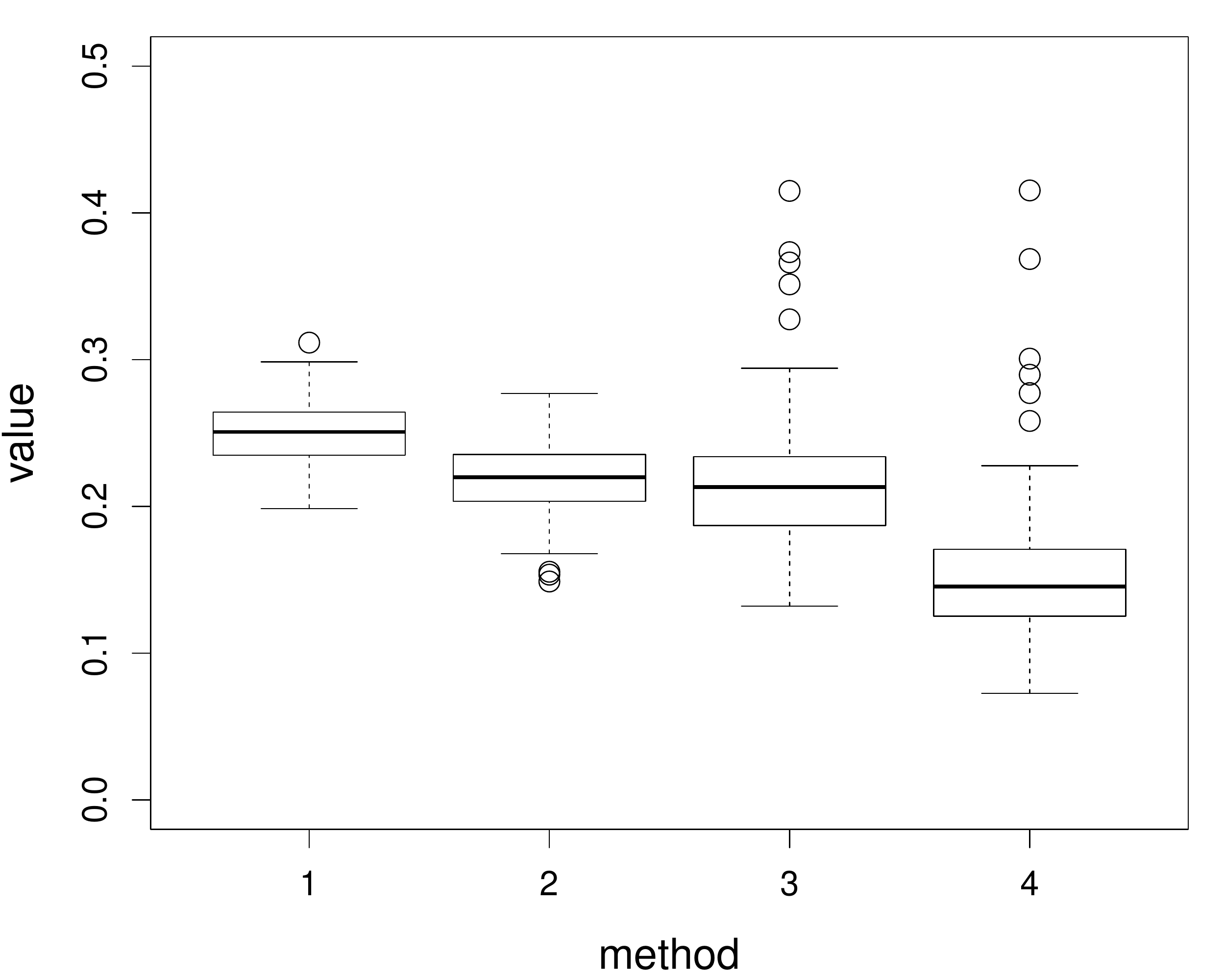} }
	\subfigure[]{ \includegraphics[width=\linewidth]{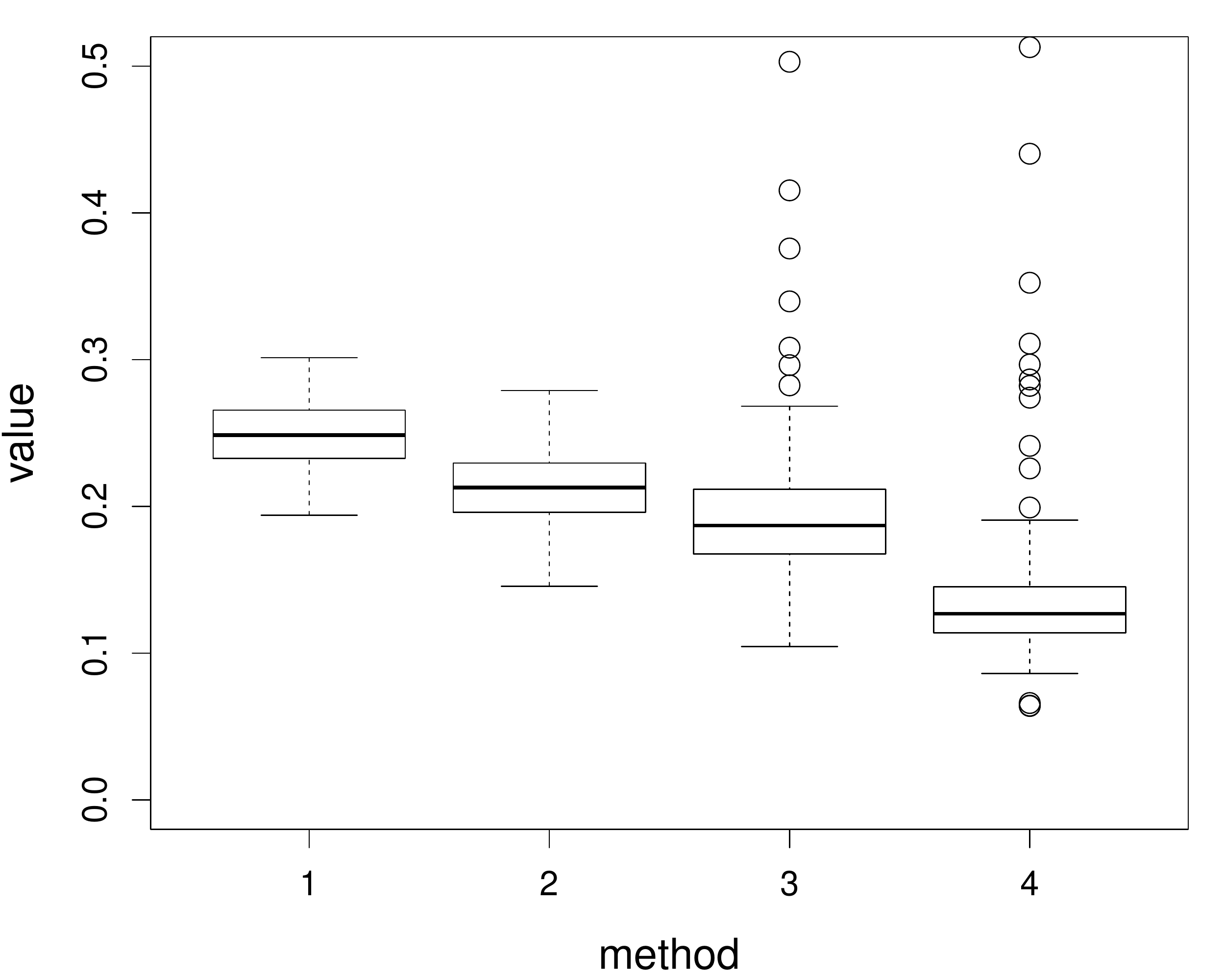} }
	\caption{Boxplots of EISE using the fully data-driven bandwidths with the cubic spline mean estimate when the primary model is (C3) and the secondary models are (a) $X\sim N(0,1)$, $U\sim \textrm{Laplace}(0, \,\sigma_u/\sqrt{2})$, $\lambda=0.8$; (b) $X\sim N(0,1)$, $U\sim \textrm{Laplace}(0, \,\sigma_u/\sqrt{2})$, $\lambda=0.9$; (c) $X\sim N(0,1)$, $U\sim N(0, \, \sigma_u^2)$, $\lambda=0.8$; (d) $X\sim \textrm{Uniform}(-2, 2)$, $U\sim \textrm{Laplace}(0, \,\sigma_u/\sqrt{2})$, $\lambda=0.8$. Method 1, 2, 3, 4 correspond to $\tilde p_1(y|x)$, $\tilde p_2(y|x)$, $\hat p_3(y|x)$, and $\hat p_4(y|x)$, respectively.} 
	\label{Sim3:box:data:spline}
\end{figure}

Figures~\ref{Sim1:box:data:spline}--\ref{Sim3:box:data:spline} are counterpart plots of Figures~\ref{Sim1:box:data}--\ref{Sim3:box:data} in the main article, where the cubic spline is used in $\tilde p_2(y|x)$ and $\hat p_4(y|x)$ to estimate $m^*(\cdot)$.  

\setcounter{equation}{0}
\setcounter{figure}{0}
\renewcommand{\theequation}{F.\arabic{equation}}
\renewcommand{\thefigure}{F.\arabic{figure}}
\renewcommand{\thesection}{F.\arabic{section}}

\section*{Appendix F: Estimated density curves using dietary data when cubic spline estimates of the mean function are used}

\begin{figure}[h!]
	\centering
	\setlength{\linewidth}{0.45\linewidth}
	\subfigure[]{ \includegraphics[width=\linewidth]{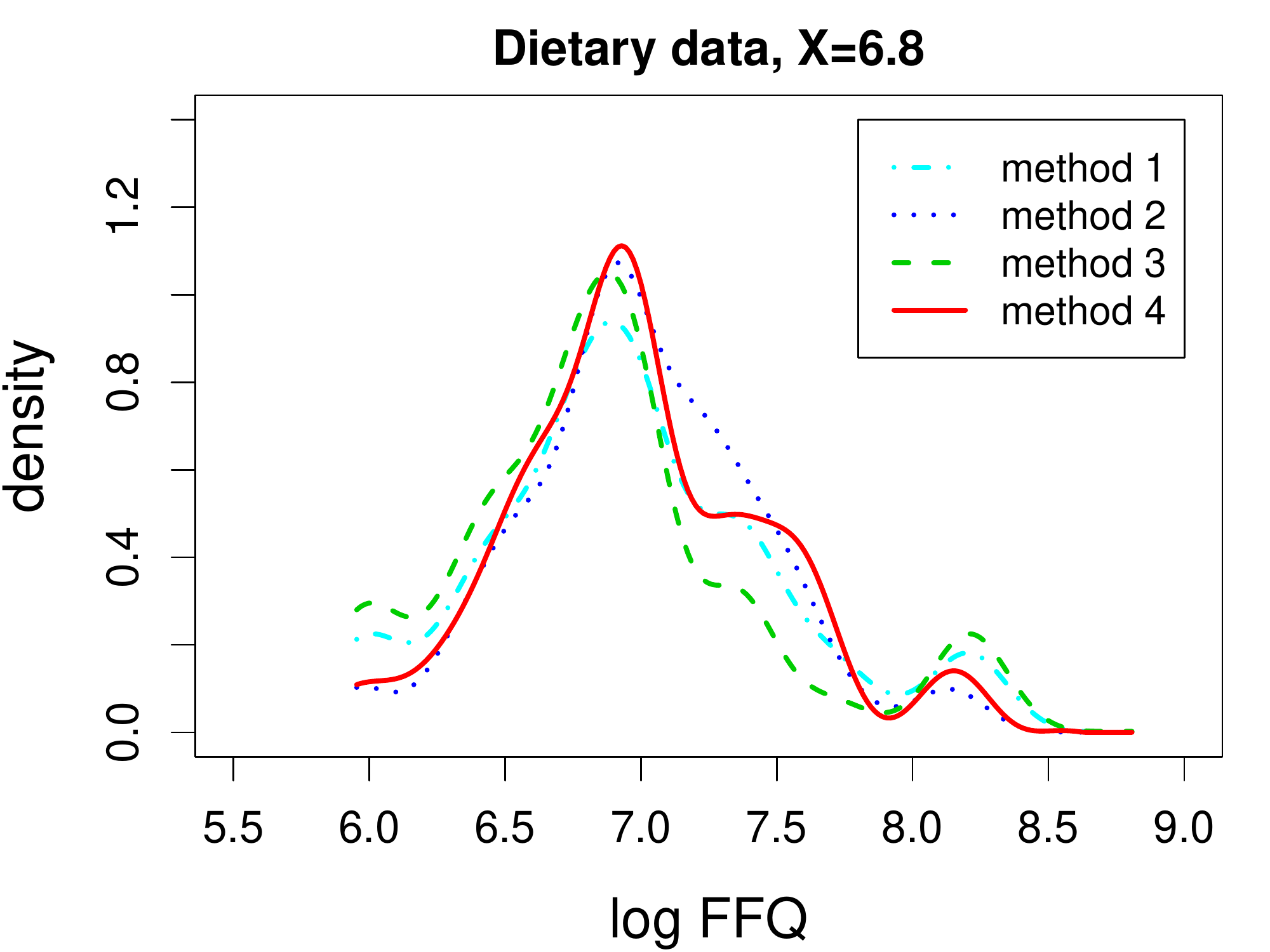} }
	\subfigure[]{ \includegraphics[width=\linewidth]{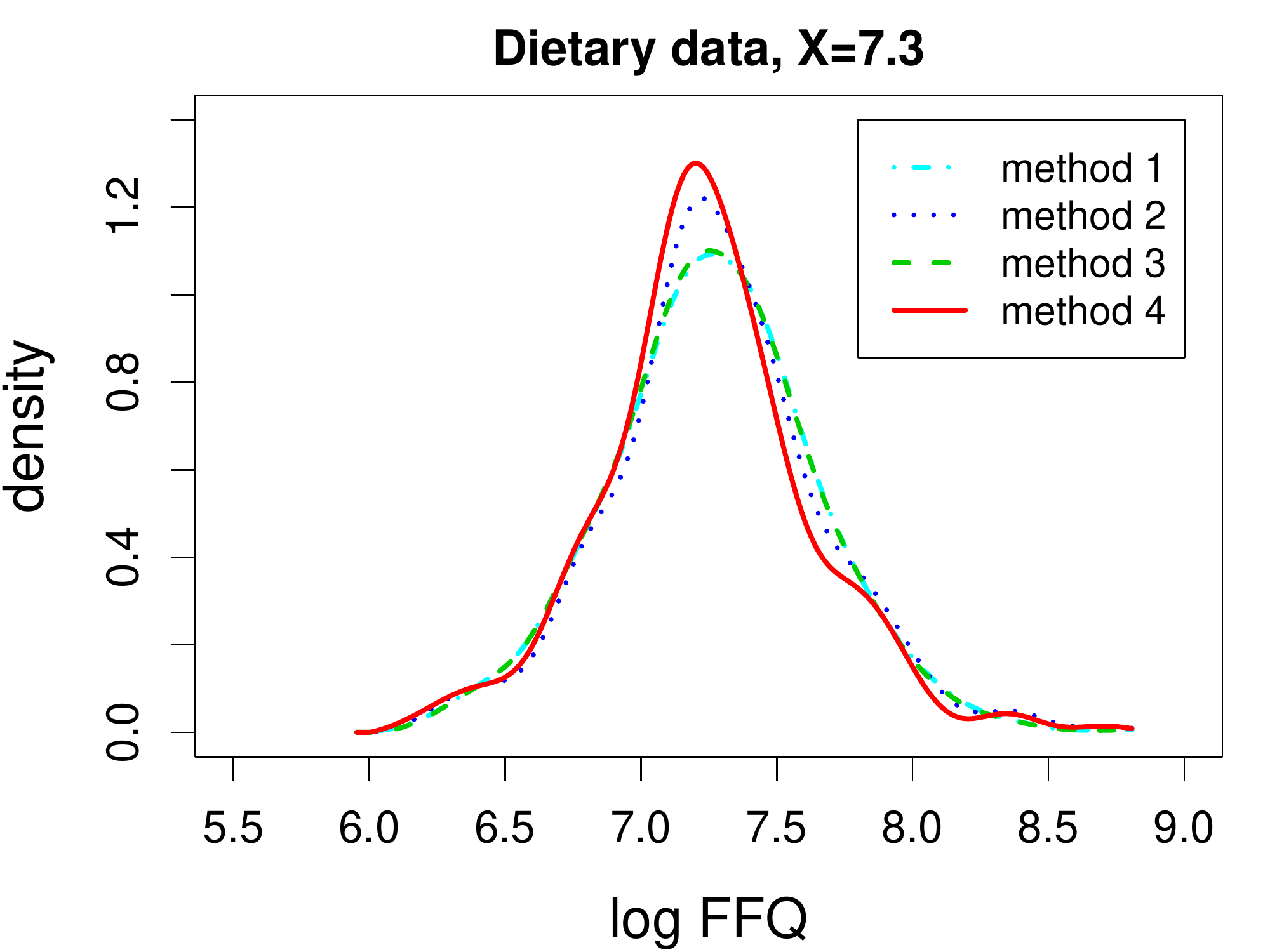} }\\
	\subfigure[]{ \includegraphics[width=\linewidth]{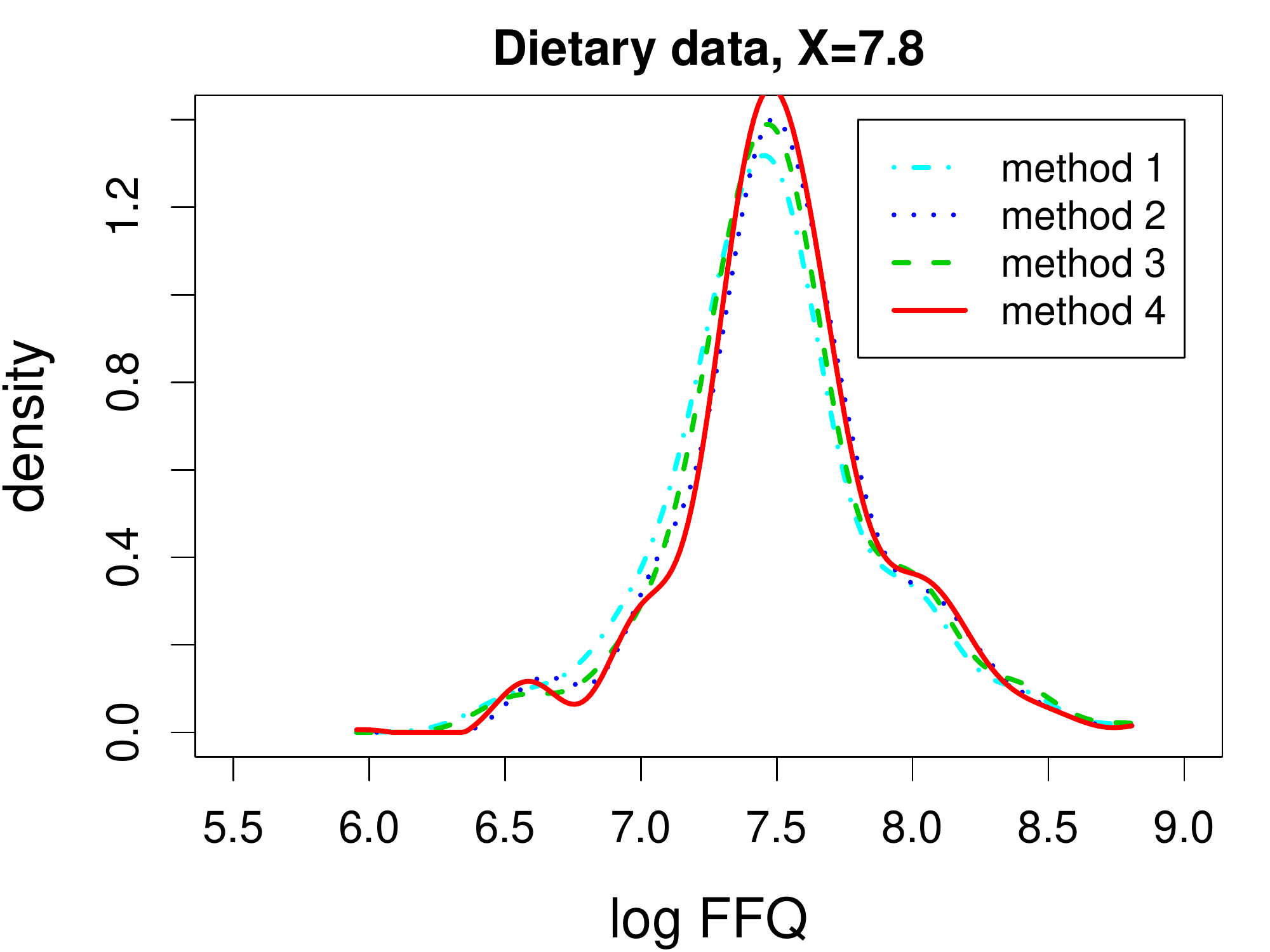} }
	\subfigure[]{ \includegraphics[width=\linewidth]{dietary-scatter.pdf} }
	\caption{Naive estimates of conditional density of the logarithm of FFQ intake corresponding to $\tilde p_1(y|x)$ (cyan dash-dotted lines) and $\tilde p_2(y|x)$ (blue dotted lines), and two non-naive density estimates, $\hat p_3(y|x)$ (green dashed lines) and $\hat p_4(y|x)$ (red solid lines) when $x=6.8$ (in panel (a)), 7.3 (in panel (b)), and 7.8 (in panel (c)), respectively. In each panel of (a)--(c), method 1, 2, 3, 4 correspond to $\tilde p_1(y|x)$, $\tilde p_2(y|x)$, $\hat p_3(y|x)$, and $\hat p_4(y|x)$, respectively. The cubic spline estimate of the mean function is used in $\tilde p_2(y|x)$ and $\hat p_4(y|x)$. The scatter plot of the observed response versus the observed covariate values is shown in panel (d), where the three values of $x$ at which $p(y|x)$ is estimated are highlighted in red dots on the horizontal axis.} 
	\label{f:dietary2}
\end{figure}

Figure~\ref{f:dietary2} is a counterpart plot of Figure~\ref{f:dietary} in the main article, where dietary data are used to estimate $p(y|x)$, but with cubic spline estimate for the mean function in $\tilde p_2(y|x)$ and $\hat p_4(y|x)$. 

\setcounter{equation}{0}
\setcounter{figure}{0}
\setcounter{table}{0}
\renewcommand{\theequation}{G.\arabic{equation}}
\renewcommand{\thefigure}{G.\arabic{figure}}
\renewcommand{\thetable}{G.\arabic{table}}
\renewcommand{\thesection}{G.\arabic{section}}

\section*{Appendix G: Boxplots of EISE associated with four density estimators when $\sigma^2_u$ is correctly specified and when it is  misspecified}
In this experiment, we generate data following the primary model specified in (C1) and the secondary model configuration (a) described in Section 5.1, where the true measurement error variance is $\sigma^2_u=0.25$, corresponding to $\lambda=0.8$. Based on each of 200 simulated data sets, each of size $n=500$, besides computing $\tilde p_1(y|x)$ and $\tilde p_2(y|x)$, we compute $\hat p_3(y|x)$ and $\hat p_4(y|x)$ while assuming $\sigma_u^2$ at its truth and three other misspecified values corresponding to $\lambda=0.7, 0.9, 0.99$. All bandwidths are chosen using the data-driven methods described in Section~\ref{s:band}.

\begin{figure}
	\centering
	\setlength{\linewidth}{0.45\linewidth}
	\subfigure[]{ \includegraphics[width=\linewidth]{Sim1-Xgau-Ulap-Lambda80-box-data.pdf} }
	\subfigure[]{ \includegraphics[width=\linewidth]{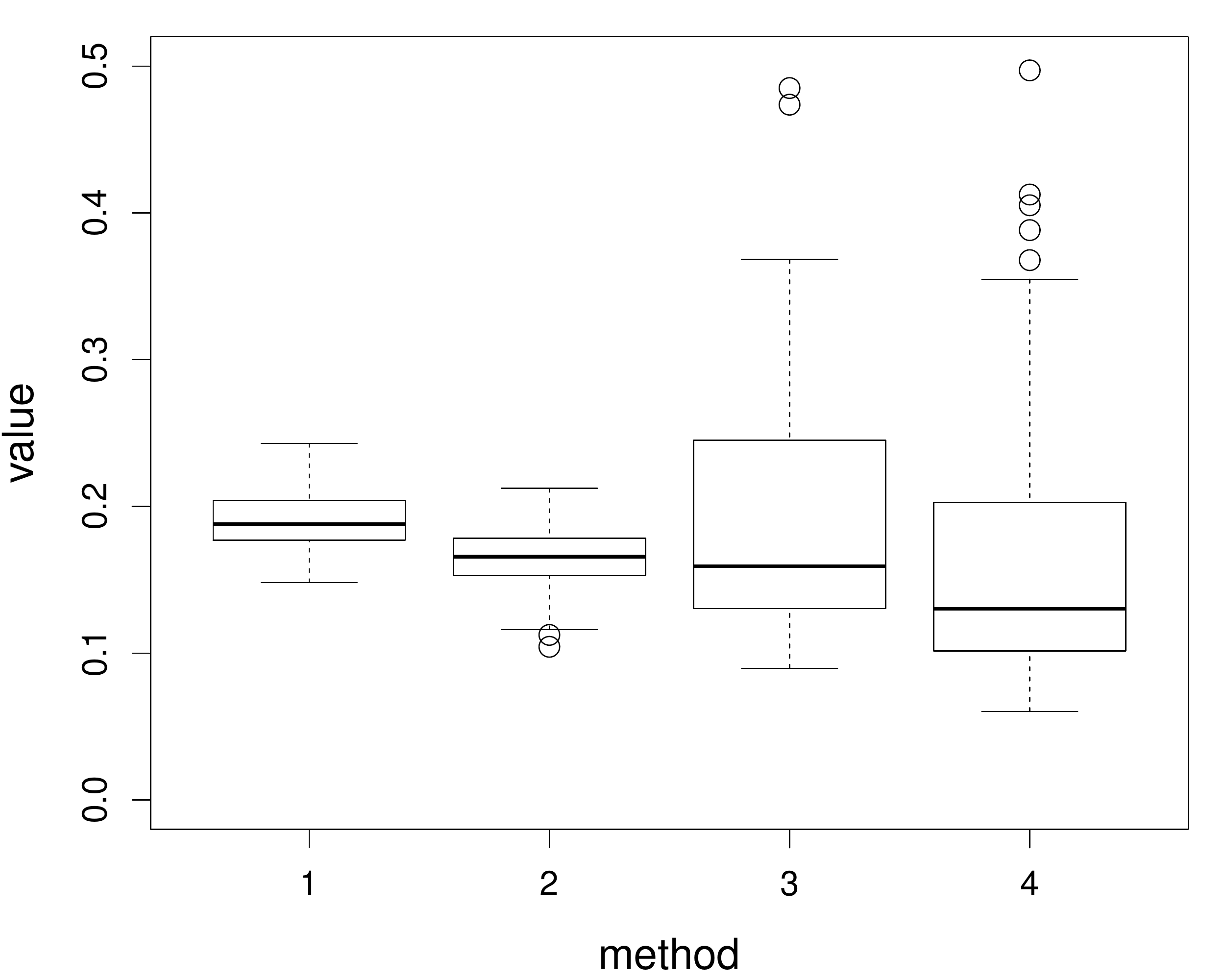} }\\
	\subfigure[]{ \includegraphics[width=\linewidth]{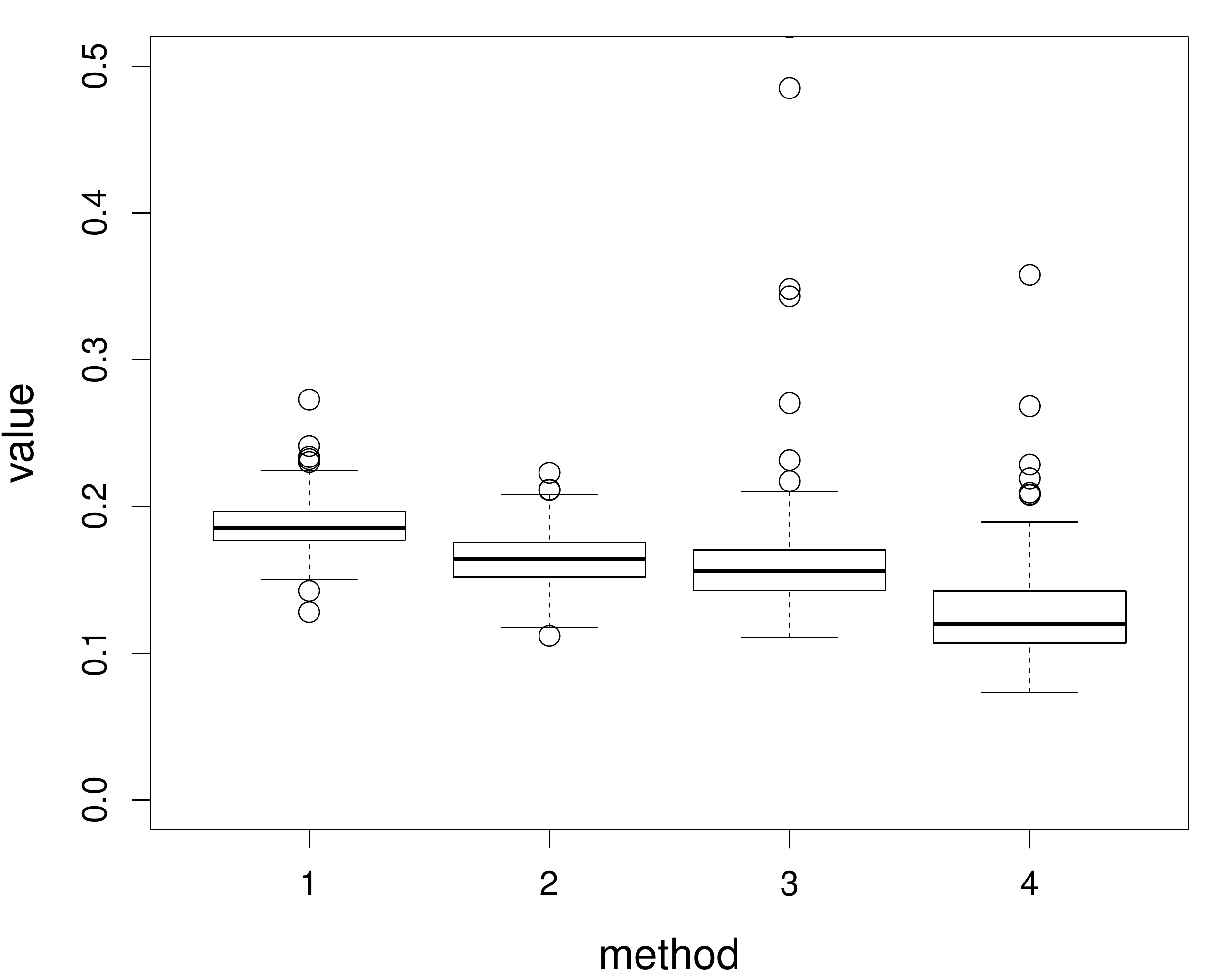} }
	\subfigure[]{ \includegraphics[width=\linewidth]{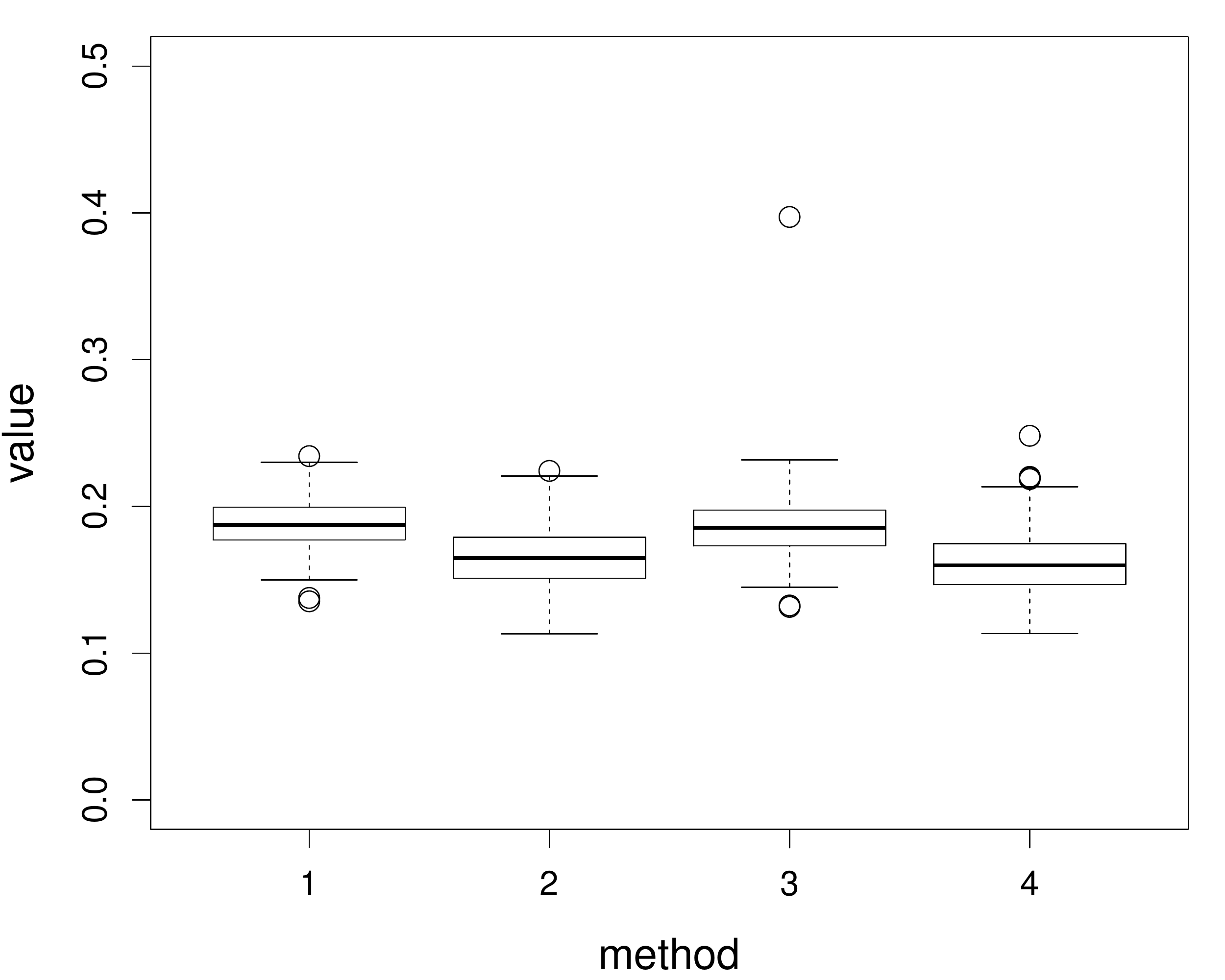} }
	\caption{Boxplots of EISE using the fully data-driven bandwidths when the primary model is (C1), the secondary model is $X\sim N(0,1)$ and $U\sim \textrm{Laplace}(0, \,\sigma_u/\sqrt{2})$, $\lambda=0.8$. Panel (a) presents the results when true $\sigma_u^2$ is used. Panels (b), (c) and (d) present the results when one misspecifies $\sigma_u^2$ such that the reliability $\lambda$ is assumed to be 0.7, 0.9 and 0.99, respectively. Method 1, 2, 3, 4 correspond to $\tilde p_1(y|x)$, $\tilde p_2(y|x)$, $\hat p_3(y|x)$, and $\hat p_4(y|x)$, respectively. The sample size is $n=500$.} 
	\label{Sim1:box:data:Lambda}
\end{figure}

Figure~\ref{Sim1:box:data:Lambda} contains boxplots of EISE across 200 Monte Carlo replicates at each assumed level of $\sigma^2_u$, i.e., at each assumed level of $\lambda(=0.8, 0.7, 0.9, 0.99)$.  When comparing with the case without misspecifying the value of $\sigma_u^2$ (in panel (a)), one can see  that even when one sets $\sigma_u^2$ at a higher level than the truth (in panel (b)) or at a lower level (in panel (c)), each proposed non-naive estimator, $\hat p_3(y|x)$ or $\hat p_4(y|x)$, still outperforms its naive counterpart, that is, $\tilde p_1(y|x)$ or $\tilde p_2(y|x)$, in the sense that the median EISE associated with $\hat{p}_3(y|x)$ or $\hat{p}_4(y|x)$ is still smaller than that of $\tilde{p}_1(y|x)$ or $\tilde{p}_2(y|x)$. The variabilities of the two proposed estimators with an assumed $\sigma^2_u$ value much larger than the truth, as the case in panel (b), are higher compared to when one uses the correct $\sigma_u^2$ or sets it at a smaller value than the truth. This can be caused by, besides involving a wrong $\phi_{\hbox {\tiny $U$}}(t)$ in the proposed estimators, one uses a larger bandwidth $h_1$ by setting $\hat \lambda$ in (\ref{eq:h1}) and (\ref{eq:h1p4}) at some smaller value than what one would use had one used the true value of $\sigma_u^2$. 

Certainly, if one assumes a low enough value for $\sigma_u^2$ such that it is close to assuming no measurement error, as in panel (d) of Figure~\ref{Sim1:box:data:Lambda}, then all four estimates behave similarly. 
Table~\ref{Sim1:table:data:Lambda} presents medians and IQRs corresponding to the EISEs depicted in Figure~\ref{Sim1:box:data:Lambda}.

\begin{table}
	\caption{Median and IQR (in parenthesis) of the EISE associated with each of the four considered estimators using the fully data-driven bandwidths under (C1) with $\sigma_u^2$ correctly specified (corresponding to panel (a) in Figure~\ref{Sim1:box:data:Lambda}) and misspecified (corresponding to panels (b)--(d) in Figure~\ref{Sim1:box:data:Lambda})}
	\label{Sim1:table:data:Lambda}
	\centering
	{
		\begin{tabular}{ccccc}
			\hline\noalign{\smallskip}
			 Method & (a) & (b)		& (c)       & (d)  \\
			\noalign{\smallskip}\hline\noalign{\smallskip}
			 $1$& 0.186 (0.028)	&   0.188 (0.027)    & 0.185 (0.020)       & 0.187 (0.023) \\
			2      & 0.163 (0.030)	& 0.167 (0.026)      & 0.164 (0.023)       & 0.165 (0.028)	\\
			3 	 & 0.151 (0.049)	& 0.159 (0.114) 	 & 0.156 (0.028)	   & 0.185 (0.023)	\\
			4 	 & 0.114 (0.059)	& 0.130 (0.100) 	 & 0.120 (0.035)	   & 0.160 (0.028)	\\		
			\hline
		\end{tabular}
	}
\end{table}

\setcounter{equation}{0}
\setcounter{figure}{0}
\renewcommand{\theequation}{H.\arabic{equation}}
\renewcommand{\thefigure}{H.\arabic{figure}}
\renewcommand{\thesection}{H.\arabic{section}}

\section*{Appendix H: An example R code for estimating conditional densities using the R package \texttt{lpme}}

For illustration purposes, we generate a data set of size $n=1000$ under the primary model configuration (C3) specified in the main article, with $X\sim N(0,1)$, $U\sim\textrm{Laplace}(0, \sigma_u/\sqrt{2})$, and $\lambda=0.8$.  The following code is used to generate data.

{\footnotesize
\begin{verbatim}
## X - True covariates
## W - Observed covariates
## Y - Individual response
rm(list=ls())
library(lpme)
## Generate Laplace random numbers
rlap = function (use.n, location = 0, scale = 1) 
{
location <- rep(location, length.out = use.n)
scale <- rep(scale, length.out = use.n)
rrrr <- runif(use.n)
location - sign(rrrr - 0.5) * scale * 
(log(2) + ifelse(rrrr < 0.5, log(rrrr), log1p(-rrrr)))
}
## Function f(y|x) to be estimated
mofx = function(x){ x }
sofx = function(x){ exp(1-x/3)/8 } 
wide	= 0.04; ymin=-4; ymax=3;
x	= seq(-2, 2, wide);
y = seq(-4, 3, wide);
nx	= length(x)
ny	= length(y);
## True density function
fy_x=function(y,x) dnorm(y, mofx(x), sofx(x));

###################  Generate data ################
set.seed(2017)
n = 1000 ## sample size:
sigma_x = 1; X = rnorm(n, 0, sigma_x); 
Y = rep(0, n);
for(i in 1:n){
Y[i] = mofx(X[i]) + rnorm(1, 0, sofx(X[i]));
}
## reliability ratio
lambda = 0.8;
sigma_u  = sqrt(1/lambda-1)*sigma_x;
W = X + rlap(n, 0, sigma_u/sqrt(2));
\end{verbatim}}
\normalsize

\begin{figure}
	\centering
	\setlength{\linewidth}{0.45\linewidth}
	\subfigure[]{ \includegraphics[width=\linewidth]{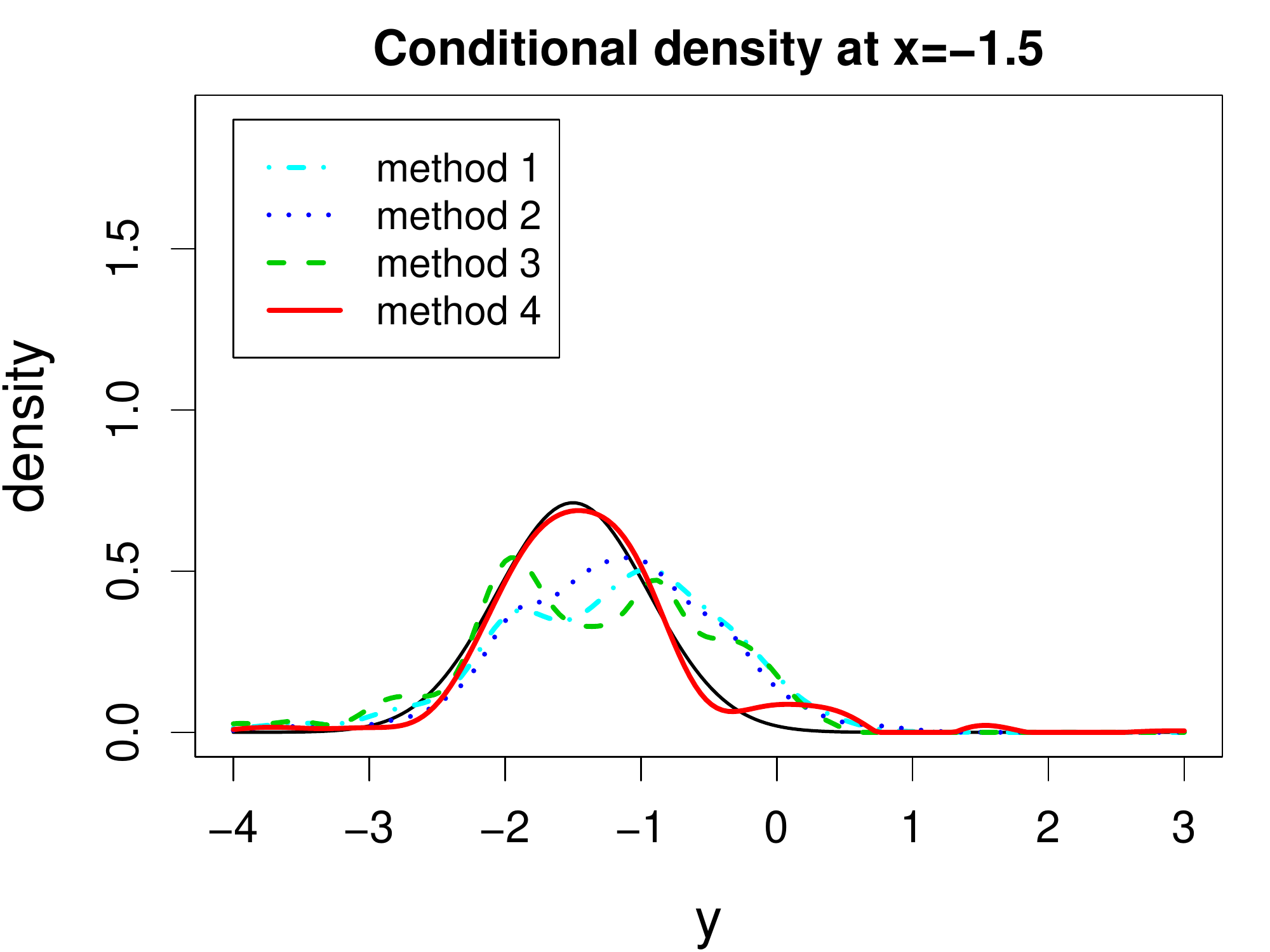} }
	\subfigure[]{ \includegraphics[width=\linewidth]{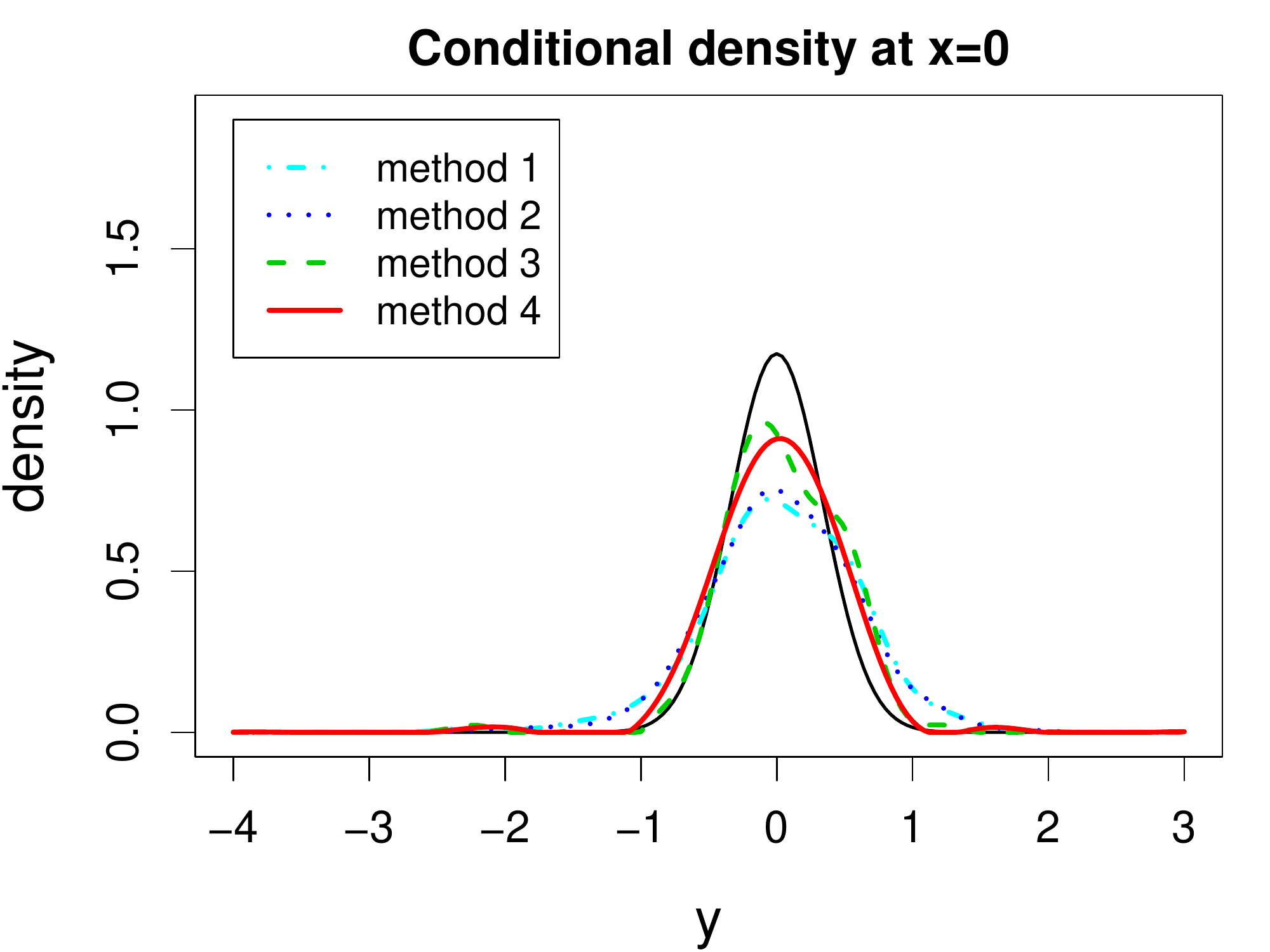} }\\
	\subfigure[]{ \includegraphics[width=\linewidth]{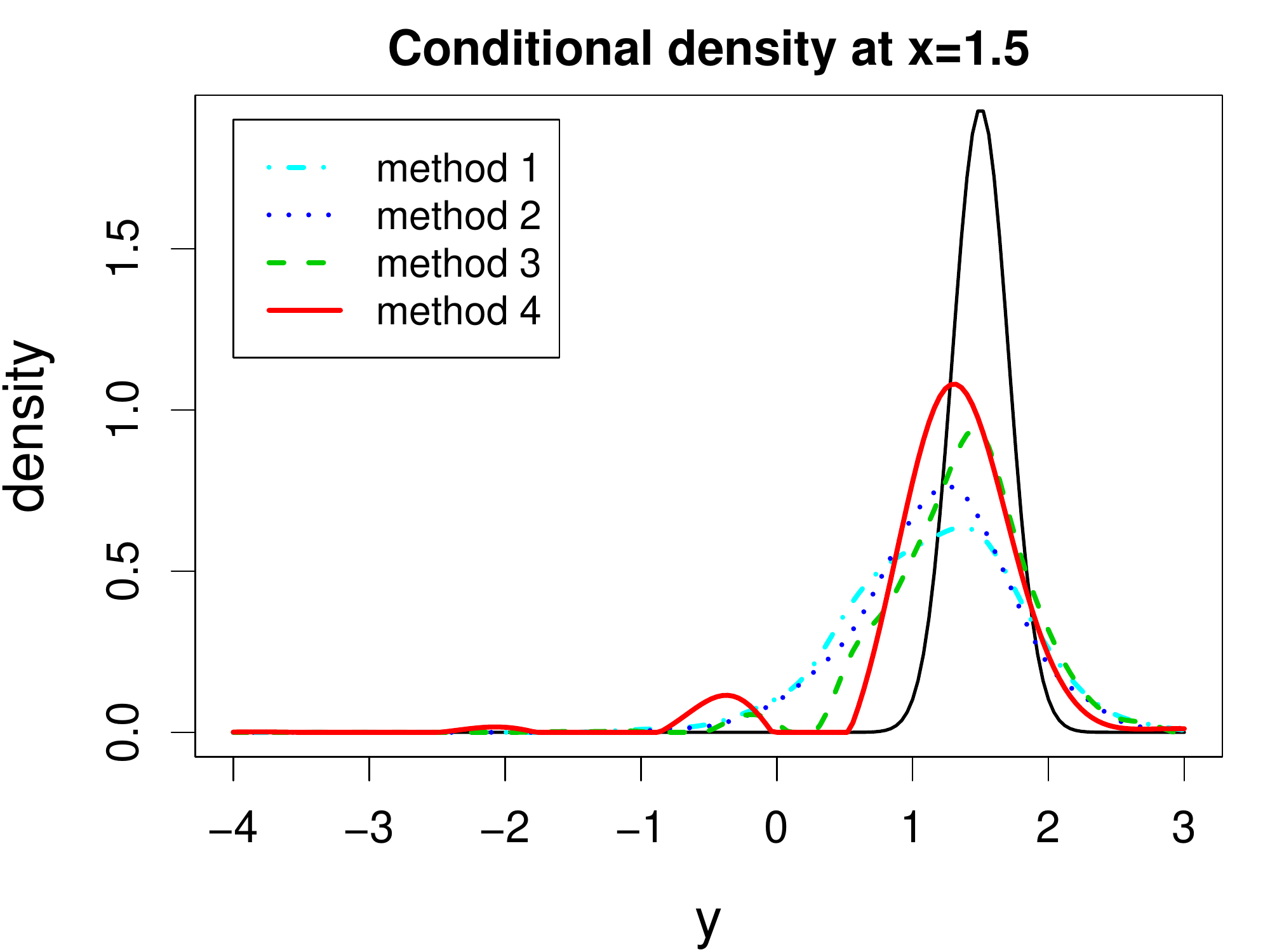} }
	\subfigure[]{ \includegraphics[width=\linewidth]{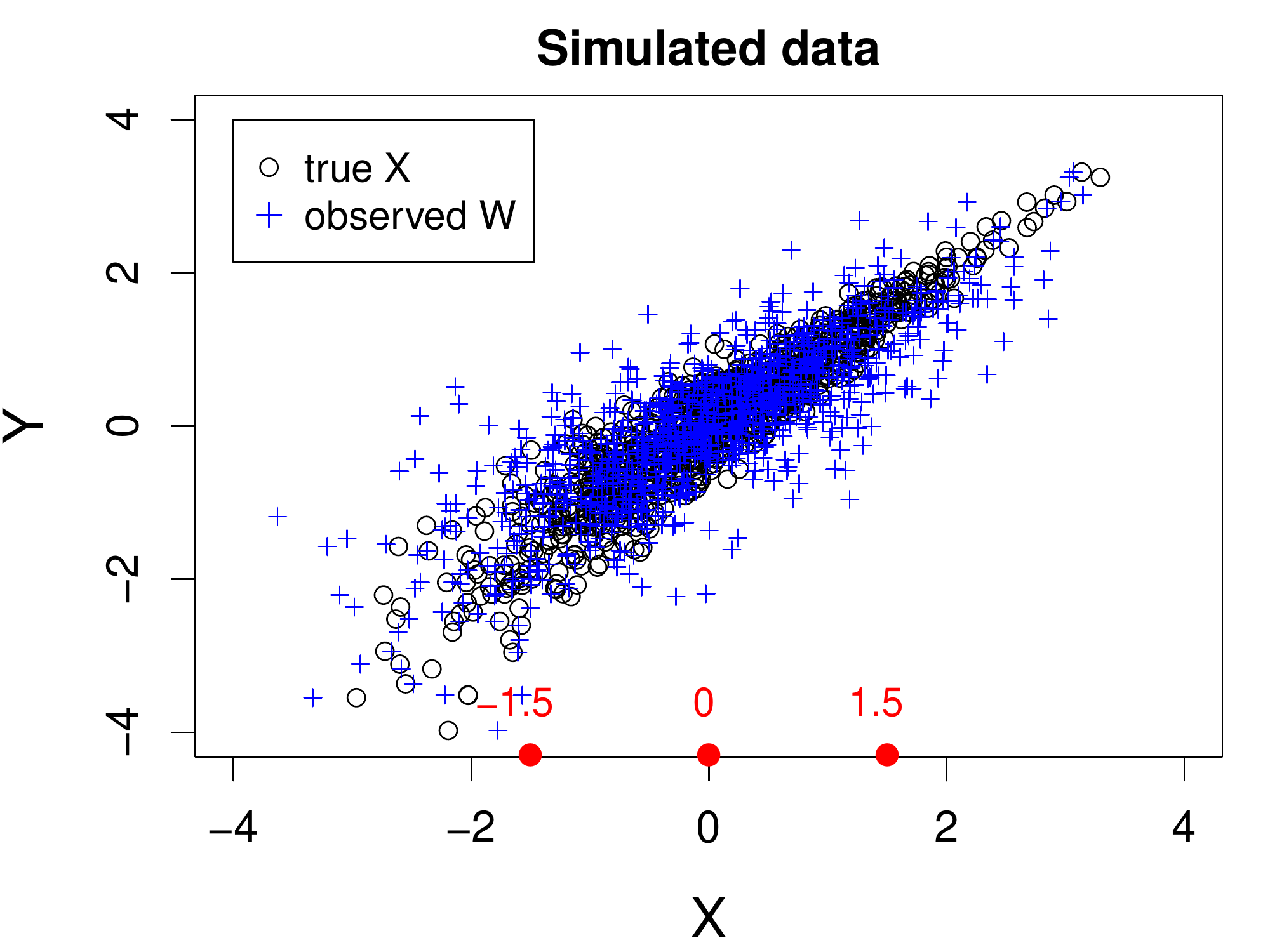} }
	\caption{Estimated conditional density curves in panels (a)--(c) obtained from the example R code, with simulated data shown in panel (d). In each panel of (a)--(c), method 1, 2, 3, 4 correspond to $\tilde p_1(y|x)$ (cyan dash-dotted lines), $\tilde p_2(y|x)$ (blue dotted lines), $\hat p_3(y|x)$ (green dashed lines), and $\hat p_4(y|x)$ (red solid lines), respectively.} 
	\label{SampleCode:curves}
\end{figure}

Panel (d) in Figure~\ref{SampleCode:curves} shows the scatter plot of the response $Y$ versus the covariate $X$, with the corresponding realizations of $W$ imposed. The following code is used to obtain the conditional density estimates, $\tilde{p}_1(y|x)$, $\tilde{p}_2(y|x)$, $\hat{p}_3(y|x)$ and $\hat{p}_4(y|x)$, at grid points \texttt{x} and \texttt{y} defined in above code. Panels (a)--(c) in Figure~\ref{SampleCode:curves} depict these four density estimates when $x=-1.5, 0, 1.5$, respectively. 

{\footnotesize
\begin{verbatim}
##----- Method 1: naive estimate without mean adjustment  ---- 
## kernel functions
K1 = "Gauss"; const1 = 1.06;
K2 = "Gauss"; const2 = 1.06;
## initial reference rule
hxyhat = c(sd(W)*const1, sd(Y)*const2)*n^(-1/5); 
## grid points for searching bandwidths
h1 = hxyhat[1]*seq(0.2, 1.5, length.out = 10 )
h2 = hxyhat[2]*seq(0.2, 1.5, length.out = 10 )
ptm<-proc.time()
fitbw1 = densityregbw(Y, W, xinterval = c(min(x), max(x)), h1 = h1, h2 = h2, 
                      K1 = K1, K2 = K2)
systime1=proc.time()-ptm; systime1;
ptm<-proc.time()
fhat1 = densityreg(Y, W, bw = fitbw1$bw, xgrid = x, ygrid = y, 
                   K1 = K1, K2 = K2);
systime11=proc.time()-ptm; systime11;
##----- Method 2: naive estimate with mean adjustment  ---- 
## kernel functions
K1 = "Gauss"; const1 = 1.06;
K2 = "Gauss"; const2 = 1.06;
## initial reference rule
hxyhat = c(sd(W)*const1, sd(Y)*const2)*n^(-1/5); 
## grid points for searching bandwidths
h1 = hxyhat[1]*seq(0.5, 3, length.out = 10 )
h2 = hxyhat[2]*seq(0.2, 1.5, length.out = 10 )
ptm<-proc.time()
fitbw2 = densityregbw(Y, W, xinterval = c(min(x), max(x)), h1 = h1, h2 = h2, 
                      K1 = K1, K2 = K2, mean.estimate = "kernel")
systime2=proc.time()-ptm; systime2;
ptm<-proc.time()
fhat2 = densityreg(Y, W, bw = fitbw2$bw, xgrid = x, ygrid = y, 
                   K1 = K1, K2 = K2, mean.estimate = "kernel");
systime22=proc.time()-ptm; systime22;

##----- Method 3: proposed method without mean adjustment  ---- 
## kernel functions
K1 = "SecOrder"; const1 = 0.427398;
K2 = "Gauss"; const2 = 1.06;
## initial reference rule
hxyhat = c(sd(W)*const1, sd(Y)*const2)*n^(-1/5); 
## grid points for searching bandwidths
h1 = hxyhat[1]*seq(0.2, 1.5, length.out = 10 )
h2 = hxyhat[2]*seq(0.2, 1.5, length.out = 10 )
ptm<-proc.time()
fitbw3 = densityregbw(Y, W, xinterval = c(min(x), max(x)), sig = sigma_u, 
                      h1 = h1, h2 = h2, K1 = K1, K2 = K2)
systime3=proc.time()-ptm; systime3;
ptm<-proc.time()
fhat3 = densityreg(Y, W, bw = fitbw3$bw, xgrid = x, ygrid = y, sig = sigma_u, 
                   K1 = K1, K2 = K2);
systime33=proc.time()-ptm; systime33;

##----- Method 4: proposed method wit mean adjustment  ---- 
## kernel functions
K1 = "SecOrder"; const1 = 0.427398;
K2 = "SecOrder"; const2 = 0.427398;
## initial reference rule
hxyhat = c(sd(W)*const1, sd(Y)*const2)*n^(-1/5); 
## grid points for searching bandwidths
h1 = hxyhat[1]*seq(0.5, 3, length.out = 10 )
h2 = hxyhat[2]*seq(0.2, 1.5, length.out = 10 )
ptm<-proc.time()
fitbw4 = densityregbw(Y, W, xinterval = c(min(x), max(x)), sig = sigma_u, 
                      h1 = h1, h2 = h2, K1 = K1, K2 = K2, mean.estimate = "kernel")
systime4=proc.time()-ptm; systime4;
ptm<-proc.time()
fhat4 = densityreg(Y, W, bw = fitbw4$bw, xgrid = x, ygrid = y, sig = sigma_u, 
                   K1 = K1, K2 = K2, mean.estimate = "kernel");
systime44=proc.time()-ptm; systime44;
\end{verbatim}
}
\normalsize

The function \texttt{densityregbw} in the R package \texttt{lpme} \citep{zhouR2017} is used for bandwidths selection. We explain five  arguments in this function next. 
\begin{enumerate}
\item[(i)] The argument \texttt{sig} allows one to specify the standard deviation of the measurement error. Its default value is \texttt{NULL}, suggesting that one assumes no measurement error. In the above code, letting \texttt{sig = NULL} or leaving it unspecified leads to the naive estimates, $\tilde p_1(y|x)$ and $\tilde p_2(y|x)$; and we set \texttt{sig = sigma\_u} with a pre-defined valeue for \texttt{sigma\_u} to obtain the non-naive estimates, $\hat p_3(y|x)$ and $\hat p_4(y|x)$.
\item[(ii)] The argument \texttt{mean.estimate} is where one specifies the type of estimates for the mean function $m^*(\cdot)$. If left unspecified, it takes the default value of \texttt{NULL}, corresponding to the density estimation methods that do not require estimating the mean function. This is value for this argument when computing $\tilde p_1(y|x)$ and $\hat p_3(y|x)$ in the above code. To compute $\tilde p_2(y|x)$ and $\hat p_4(y|x)$ in the example code, we set \texttt{mean.estimate = "kernel"} to use the local linear estimate for the mean function. For these two density estimates, one may set \texttt{mean.estimate = "spline"} to estimate the mean function using spline-based estimates, and use the argument \texttt{spline.df} to specify the order of the spline. The default value of \texttt{spline.df} is 5.
\item[(iii)] The arguments \texttt{K1} and \texttt{K2} correspond to the kernel functions $K_1(t)$ and $K_2(t)$ used in the main article. Choices for each one include the Gaussian kernel, \texttt{"Gauss"}, and the second order kernel, \texttt{"SecOrder"}, defined in (\ref{eq:sec-order}) in the main article. In the current version, not all the combinations are supported, and one will receive an error message if one chooses a combination of \texttt{K1} and \texttt{K2} that is not supported.
\item[(iv)] The arguments \texttt{h1} and \texttt{h2} are used to specify the searching grid points for bandwidths $h_1$ and $h_2$. When unspecified, bandwidths selected based on reference rules are used.
\item[(v)] The argument \texttt{xinterval} is used to specify the values $x_{\hbox {\tiny $L$}}$ and $x_{\hbox {\tiny $U$}}$ in the main article. 
\end{enumerate}
The function \texttt{densityregbw} returns an object with three variables, \texttt{bw} (selected bandwidths), \texttt{h1} (searched grid points for $h_1$), and \texttt{h2} (searched grid points for $h_2$). 

The function \texttt{densityreg} is used for density estimation. Some arguments in this function are the same as those used in \texttt{densityregbw}. Two additional arguments in this function are \texttt{xgrid} and \texttt{ygrid}, which are used to specify the grid points for $x$ and $y$ in estimating $p(y|x)$, respectively. The function \texttt{densityreg} returns an object with three variables, \texttt{fitxy} (a matrix of fitted values with rows corresponding to $x$ values), \texttt{xgrid} (grid points for $x$), and \texttt{ygrid} (grid points for $y$). 

\vskip 14pt
\noindent {\large\bf Acknowledgments}
We are grateful to the Associate Editor and referee for their constructive comments and suggestions on an earlier version of the manuscript. The first author would also like to thank Professor David W. Scott at Rice University, for insightful discussions with her during the early stage of this research project. 

\bibliographystyle{apalike}
\bibliography{references}

\end{document}